\documentclass[preprint,12pt]{elsarticle}
\usepackage[T1]{fontenc} 
\usepackage{algorithm}
\usepackage{algpseudocode}
\usepackage{mathtools}
\usepackage{cuted}
\usepackage{subfigure}
\usepackage{chemformula}
\usepackage{amssymb} 
\usepackage{amsthm}  
\usepackage{amsmath}
\usepackage{amsfonts}
\usepackage{verbatim}
\usepackage{color,soul}
\usepackage{graphicx}
\graphicspath{ {images/} } 
\usepackage{subfigure}
\usepackage[section]{placeins} 
\usepackage{lineno}  
\usepackage{hyperref}
\hypersetup{colorlinks,linkcolor={blue},citecolor={blue},urlcolor={red}} 

\journal{CMAME}
\begin{document}

\begin{highlights}
	\item An operator-splitting and matrix-based scheme is proposed.
	\item This fully decoupled scheme enhances implementability.
	\item The mix of vectorization and forward matrix assembly benefits 3D applications.
	\item A term-wise strategy is suggested to balance the computational costs and the complexity of implementation.
	\item Plenty of benchmark cases validate the accuracy and efficiency of this method.  
\end{highlights}

\begin{frontmatter}
\title{Benchmark modeling and 3D applications of solidification and macro-segregation based on an operator-splitting and fully decoupled scheme with term-wise matrix assembly} 
\author[inst1]{Xiaoyu Feng} 
\affiliation[inst1]{organization={Computational Transport Phenomena Laboratory (CTPL), Division of Physical Sciences and Engineering (PSE), King Abdullah University of Science and Technology (KAUST)},
            addressline={Thuwal}, 
            postcode={23955-6900}, 
            country={Kingdom of Saudi Arabia}}
\ead{xiaoyu.feng@kaust.edu.sa}          
\author[inst2]{Huangxin Chen}
\affiliation[inst2]{organization={School of Mathematical Sciences and Fujian Provincial Key Laboratory on Mathematical Modeling and High Performance Scientific Computing, Xiamen University},
            city={Xiamen},
            postcode={361005}, 
            country={China}}
\ead{chx@xmu.edu.cn}            
\author[inst3]{Bo Yu}
\affiliation[inst3]{organization={School of Mechanical Engineering, Beijing Institute of Petrochemical Technology},
            city={Beijing}, 
            postcode={102617}, 
            country={China}}
\ead{yubobox@vip.163.com}
\author[inst1]{Shuyu Sun\corref{cor1}} 
\ead{shuyu.sun@kaust.edu.sa}
\ead{https://ctpl.kaust.edu.sa}
\cortext[cor1]{Corresponding author}

\begin{abstract}

The solidification and macro-segregation problem involving unsteady multi-physics and multi-phase fields is typically a complex process with mass, momentum, heat, and species transfers among solid, mushy, and liquid phase regions. The quantitative prediction of phase change, chemical heterogeneities, and multi-phase and multi-component flows plays critical roles in many natural scenarios and industrial applications that involve many disciplines, like material, energy, and even planet science. In view of this, some scholars and research institutions have called for more contributors to join the benchmark analysis of solidification and segregation problems. Our work proposes an operator-splitting and matrix-based method to avoid non-linear systems. Also, the combination of vectorization and forward equation-based matrix assembly techniques enhances the implementability of extensions of 3D applications. Lastly, the novel scheme is well validated through a bunch of 2D and 3D benchmark cases. The numerical results also illustrate that this method can ensure accurate prediction and adequately capture the physical details of phenomena caused by the solutally and thermally driven flow, which include channel segregation, the formation of freckles, edge effect, aspect ratio effect, and 3D effect. 
\end{abstract}


\end{frontmatter}

\section{Introduction}

\begin{figure}
	\centering
	\includegraphics[scale=.75]{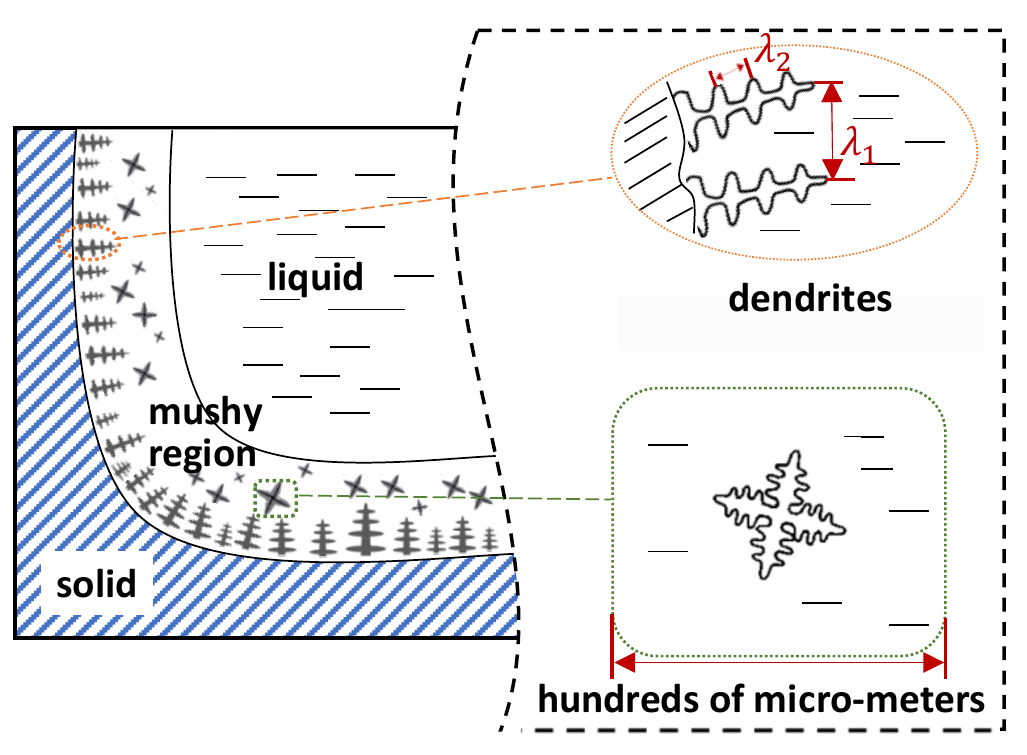}
	\caption{Schematic of solidification regions at macroscale and zooms of the dendrites at microscale.}
	\label{FIG:1}
\end{figure}

Solidification is a complicated process that involves the transfer of mass, momentum, energy, and substance species, as well as the interaction of multiple physical fields and phase change. It occurs in many natural scenarios and physical applications. On a macro-scale, some typical phenomena occur, like chemical heterogeneities, channel segregation, and the change of flow pattern among multi-phases (liquid, solid, and mushy regions, as shown in Figure \ref{FIG:1}). In the field of material research and development, segregation indeed influences the strength of materials, which is a vital factor in meeting the extremely strict requirements in aerospace and civil engineering. The industrial producers of all types of cast metal still continue to struggle with macro-segregation defects \cite{liu1998thermodynamics,pascon2007finite}. Furthermore, a better understanding of phase change problems will aid in the development of advanced phase change materials (PCMs) with increased energy storage capacity via "latent" heat \cite{hong2019solid}. Then, if we apply our vision to the energy industry, natural gas hydrate (NGH), which has enormous global reserves, has aroused a great deal of research, including the fundamental study of its formation and transport mechanism. Studies on solidification will give constructive insights about how to enhance the efficiency and economy of the subsurface recovery of hydrate \cite{song2021numerical,song2021dissociation} and how to run and manage gas pipelines properly to prevent hydrate blockage accidents \cite{duan2021simulation,zhang2019prediction}. Moreover, the wax deposition accidents in the pipeline \cite{banki2008mathematical,yu2017new} are also solidification problems. Even on an interstellar scale, the deposit of planetary components during the cooling of lava lakes or magma chambers \cite{kuritani2004magmatic,elkins2008linked} is a type of solidification and segregation in geo-dynamical processes responsible for the formation of a terrestrial planet's crust and early geological evolution \cite{shi2022nitrogen}. Basic studies on modeling solidification do benefit the progress of space exploration in ways like proving the theory about the origin of the moon and the formation of the lunar anorthosite \cite{maurice2020long,ohtake2009global}.

Lots of experimental work has been done so far with opaque metal alloys like \ch{Al-Cu} and \ch{Sn-Pb} \cite{hebditch1974observations} as well as transparent analogs like \ch{NH_{4}Cl} and \ch{Na_{2}CO_{3}} salt solutions \cite{asai1978theoretical}. Early experimental studies of solidification concluded that gravity-induced inter-dendritic flows are responsible for macro-segregation. Stewart's flow visualization experiment \cite{stewart1972fluid} for the \ch{Sn-Pb} alloy system in 1972 showed that the solid-liquid interface was mostly caused by thermal convection. Then, lots of work revealed that the combination of specific thermal convection and solutal convection configurations in the solid–liquid mush produces the channel segregation \cite{sarazin1988channel}. The buoyancy-induced flows caused by thermal and solute gradients can either complement or oppose each other, resulting in a clockwise or counterclockwise flow pattern in the liquid region. Some data and experimental results are used as a benchmark to validate numerical studies \cite{prakash1989numerical,hebditch1974observations,chen2019modeling}.

When it comes to numerical modeling, it has also significantly contributed to the understanding of melting or solidification processes, complementing experimental investigations and providing a feasible and economical way to get the details of physical processes. This method is indispensable for overcoming the inaccessibility of time scales, such as magma ocean segregation or planet cooling, or in scenarios with high temperatures and opaque flow systems in steel ingots. Solid-liquid phase change can be studied using an extensive variety of numerical methodologies. Generally, for the incompressible case:

\begin{itemize} 
\itemsep=0pt
\item The density is constant everywhere. 
\item The local thermodynamical equilibrium holds. 
\item The thermal and solutal driven forces are characterized by the buoyancy term based on the Boussinesq approximation.
\end{itemize}

The numerical methodologies to study the solidification process are commonly divided into several classes based on different principles; they include mesh-free methods and mesh-based methods. For the former, smoothed-particle hydrodynamics (SPH) \cite{monaghan2005solidification,feng2023energy} is the most classic one. For the latter, the moving grid methods, such as the Arbitrary Lagrangian-Eulerian method (ALE) \cite{bellet2004ale}, and the fixed-grid methods are possible options. Based on the fixed-grid Euler framework, some classical methods to treat the interface of multi-phase systems are well known, such as volume of fluid (VOF) \cite{hirt1981volume}, level-set \cite{osher1988fronts}, and the hybrid VOSET method \cite{sun2010coupled}. Other methods like the molecular dynamics simulation \cite{celestini2000nonequilibrium,liu2022quantitative} is also used to study solidification. The enthalpy-porosity method \cite{voller1987enthalpy} and the phase field method \cite{zhang2020fully,lin2017energetically,feng2018novel} attracted a lot of research attention as the mesh-based methods. They not only circumvent the difficulties of interface reconstruction but also maintain the reproduction of physical properties. The enthalpy-porosity method is the most popular in solid-liquid phase change simulations. It is widely utilized in commercial software packages, such as ANSYS Fluent. In this paper, we also design a numerical scheme based on this type of model.

Backtracing the development of physical models for solidification and macro-segregation, a big step forward is that the first single-domain continuum models were proposed in the 1980s by Incropera \cite{bennon1987continuum,bennon1987continuum2} and Voller \cite{voller1987enthalpy} on the basis of volume averaging techniques and classical mixture theory. There are many different mechanisms to describe the velocity transition between liquid and solid: 1.The source term method (STM), where an additional source term in the momentum equation is applied to mimic the Carman-Kozeny equations: This method was inspired by Mehrabian's idea \cite{mehrabian1970interdendritic} that the inter-dendritic flow velocities can be calculated from Darcy's law by treating the mushy region as the porous medium, namely, the typical enthalpy-porosity method; 2.the variable viscosity method (VVM), originally proposed by Gartling \cite{gartling1978finite}. This method will keep the momentum equation in its original form, whereas the viscosity of the solid is set to an extremely high value. It is less popular than the enthalpy-porosity model method due to its computational efficiency.

Based on all of the preceding fundamental works and other simplifications, such as the neglect of solute diffusion in liquids and solids at the macroscopic scale, the ``minimal" model for solidification proposed by Bellet \cite{combeau2012analysis} can be reached. A bunch of benchmark studies on solidification and segregation have been done in order to verify and compare the results of different numerical methods and schemes. In order to gain a deeper understanding and an accurate and robust simulation of this problem, it was suggested in the well-known French SMACS project and the ``Call for Contributions" paper \cite{bellet2009call} to add numerical simulations of different schemes to the benchmark data bank and to set up a way for people to work together to create thorough solidification models and efficient and accurate numerical implementation schemes.

Most numerical schemes based on the ``minimal'’ model are not implement-friendly for solving non-linear systems, or they involve iterative work without complete decoupling, causing great computational workload. In this paper, we explore the potential of a novel scheme based on operator-splitting and matrix-based ideas, which will lead to a linear solution process mathematically and make as few iterations as possible without much accuracy loss. Our scheme will be validated using benchmark analysis to demonstrate its ability to make quantitative predictions for macro-segregation and adequately capture physical properties. Also, the combination of vectorization and forward equation-based matrix assembly techniques enhances the implementability of extensions to 3D applications. To present, most studies of solidification modeling have been restricted to 2D. The computational attempts on the 3D geometry benchmark will be provided in this paper as a response to the call for contributions.

The rest of this paper is organized as follows: In Section \ref{section:Mathematical model}, we review the enthalpy-porosity model at the partial differential equation (PDE) level. In Section \ref{section:Numerical scheme}, our novel scheme based on operator-splitting and matrix-based ideas will be introduced and derived. In Section \ref{section:Algorithm implementation}, details about vectorization and forward methods, the techniques to assemble the matrix of the linear system, and the algorithm list and flowchart about the scheme will be given. In Section  \ref{section:Numerical Simulation}, plenty of numerical examples, including 2D and 3D, are presented to validate this scheme, and some interesting physical phenomena are studied. Finally, some concluding remarks are given in Section \ref{section:Conclusions}.

\section{Mathematical model} \label{section:Mathematical model}
\setcounter{equation}{0}
\renewcommand{\theequation}{\arabic{section}.\arabic{equation}}
Based on the single-domain continuum mixture model derived from the volume averaging technique and classical mixture theory, the mathematical model describing fluid and phase change mechanics, including natural convective heat and mass transfer and thermodynamic phase equilibrium relations during solidification and macro-segregation of a binary component system, will be presented in this section. The flow is laminar and Newtonian with constant viscosity, and the mushy zone is considered as an isotropic porous medium whose permeability can be determined by the Carman–Kozeny formula and whose buoyancy forces in the vertical direction are governed by the Boussinesq approximation. The solid phase remains stationary, namely $\boldsymbol{u}_s = 0$. The model equations are listed as follows: through the volume averaging and the mixture theory, we define the superficial velocity:
\begin{equation}
\boldsymbol{u}=g_s \boldsymbol{u}_s+g_l \boldsymbol{u}_l,
\end{equation}
where $\boldsymbol{u}_s$ and $\boldsymbol{u}_l$ denote the solid phase velocity and the liquid phase velocity, $g_s$ and $g_l$ are the phase volume fractions; they must add up to unity locally:
\begin{equation}
g_s+g_l=1,
\end{equation}
and density:
$$ 
\rho=g_s \rho_s+g_l \rho_l,$$
where $\rho_s$ and $\rho_l$ are phase densities.

\textbf{Mass conservation:}
\begin{equation}
\frac{\partial \rho}{\partial t} + \nabla \cdot \left( \rho \boldsymbol{u} \right)=0,
\end{equation}
with the incompressibility assumption (density $\rho=\rho_s=\rho_l$ and it is a constant), the divergence-free equation is obtained:
\begin{equation}
\nabla \cdot \boldsymbol{u}=0.
\end{equation}

\textbf{Momentum conservation:} 
\begin{equation}
\frac{\partial(\rho \boldsymbol{u})}{\partial t}+\nabla \cdot(\rho \boldsymbol{u} \otimes \boldsymbol{u})=-\nabla p+\nabla \cdot\left(\mu_l \frac{\rho}{\rho_l} \nabla \boldsymbol{u}\right)+\rho \boldsymbol{g}-\frac{\mu_l}{K} \frac{\rho}{\rho_l}\left(\boldsymbol{u}-\boldsymbol{u}_s\right),
\end{equation}	
with some additional assumptions, we can obtain the most classical two types of momentum equations, which are 
\begin{equation}
 \frac{\partial \boldsymbol{u}}{\partial t}+\frac{\boldsymbol{u}}{g_l} \cdot \nabla \boldsymbol{u} = \nabla \cdot\left(\frac{\mu_l}{\rho} \nabla \boldsymbol{u}\right) - \frac{g_l}{\rho} \nabla p -\frac{\mu_l g_l}{\rho} K^{-1} \boldsymbol{u} + g_l\boldsymbol{f},  \label{mom1}
\end{equation}
and
\begin{equation} 
\frac{\partial \boldsymbol{u} }{\partial t}+(\boldsymbol{u} \cdot \nabla) \boldsymbol{u}=\nabla \cdot\left(\frac{\mu_l}{\rho}  \nabla \boldsymbol{u}\right) - \frac{1}{\rho} \nabla p -  \frac{\mu_l}{\rho} K^{-1}\boldsymbol{u}+  \boldsymbol{f},   \label{mom2}
\end{equation}
where $\mu_l$ is the dynamic viscosity of liquid and $p$ is the pressure; $\boldsymbol{f}$ denotes the resultant driven force. These two kinds of momentum conservation expressions \cite{ahmad1998numerical,bennon1987continuum} were proposed successively, and both of them are commonly used to study solidification. The first equation belongs to the `minimal’ model system and the second expression has more assumptions for the mushy region. However, the change in $g_l$ will have the greatest impact on the flow field via the Darcy damping term. Thus, there is no considerable difference between the numerical results. Also, it is obvious that these two kinds of models are equivalent when it comes to the pure fluid or pure solid region. Their differences and the transformation between them can be referred to \cite{ganesan1990conservation}.

Moreover, a Carman–Kozeny formula is introduced to describe the flow in a mushy region, which is treated as porous media:
\begin{equation} 
K=\frac{ \lambda_2^2 g_l^3 }{  180 \left(1-g_l\right)^{2}},
\end{equation}
Here, $K$ represents the permeability of the mushy region, and $\lambda_2$ denotes the secondary dendrite arm spacing, which is one of the most important micro-structural features affecting the flow. 

Given that $\rho=\rho_s=\rho_l$, the Boussinesq approximation is necessary to account for the thermal and solute buoyancy in the vertical direction:
\begin{equation} 
\boldsymbol{f}=\boldsymbol{g}\left(1-\beta_T\left(T-T_{r e f}\right)-\beta_c\left(c_l-c_{r e f}\right)\right).
\end{equation}
Usually, we set the reference temperature and concentration as the initial uniform temperature and concentration of the system when implementing. The gravitational acceleration is denoted by $\boldsymbol{g}$. The $\beta_T$ and $\beta_c$ are the thermal and solutal expansion coefficients, respectively.

\textbf{Energy conservation:}
\begin{equation} 
\frac{\partial(\rho h)}{\partial t}+\nabla \cdot(\rho \boldsymbol{u} h)=\nabla \cdot(k \nabla T)-\nabla \cdot\left(\rho\left(h_l-h\right)\left(\boldsymbol{u}-\boldsymbol{u}_s\right)\right),
\end{equation}
where $h$ denotes the enthalpy, $k$ is thermal conductivity, and based on the definition of solid and liquid phase enthalpy:
\begin{equation} 
\left\{\begin{array}{l}
h_s=\bar{c}_{p s} T \\ 
h_l=\bar{c}_{p l} T+L,
\end{array}\right.
\end{equation}
where $\bar{c}_{p s}$ and $\bar{c}_{p l}$ are the average specific heats of solid and liquid, and $L$ is the latent heat, we have $h=g_s h_s+g_l h_l$. When Combined with relation $c_p=g_s \bar{c}_{p s}+g_l \bar{c}_{p l}$, it yields
\begin{equation} 
h=c_p T+g_l L.
\end{equation}
With the constant specific heat assumption, $\bar{c}_{p s} = \bar{c}_{p l} = c_p$, we can obtain
\begin{equation} 
\frac{\partial  T }{\partial t} + \nabla \cdot\left( \boldsymbol{u} T\right)=\nabla \cdot( \frac{k}{\rho c_p} \nabla T) - \frac{L}{c_p}\frac{\partial g_l}{\partial t} \label{T-based}.
\end{equation}
It should be noted that the energy conservation relation for the solidification process can be solved in two ways (temperature-based or enthalpy-based \cite{saad2015temperature}). As it is presented in \eqref{T-based}, the expression in temperature-based form is employed for our scheme, which is more straightforward than the enthalpy-based one.

\textbf{Solute conservation:}

Assuming the species diffusion in the solid phase to be negligible, the governing equation can be written as 
\begin{equation} 
\frac{\partial(\rho C)}{\partial t}+\nabla \cdot(\rho \boldsymbol{u} C_l )=\nabla \cdot\left(\rho g_l D_l \nabla C_l\right).
\end{equation}
Generally, at the macroscopic scale, the ``minimal'' ignores the diffusion effect in the liquid phase as well. It then reaches
\begin{equation} 
\frac{\partial  C }{\partial t}+ \boldsymbol{u} \cdot \nabla C_l =0,
\end{equation}
where the volume average concentration (or mass fraction) is denoted by $C$ and $C_l$ is the concentration of one species in the liquid phase. 

\textbf{Phase equilibrium relation:}

Considering the thermodynamic consistency of this multi-phase problem, it inevitably involves a phase equilibrium calculation based on the equation of state (EoS). The Van der Waal (VdW) EoS or Peng-Robinson (PR) EoS is widely recognized for the gas-liquid problem in academic research or industrial applications \cite{liu2020computational,feng2022fully}. While there is no widely accepted EoS model for the solid-liquid problem, the empirical phase diagram from experiments is still used as the basis for simulation. For the binary system, a linear approximation can be applied to get the mathematically formulated phase equilibrium relation. The lever rule can be derived from mass conservation as follows:
\begin{equation} 
C_A + C_B = C_l + C_s = 1,
\end{equation}
where $C$ is the mass fraction, such that $C_{A}=\frac{m_{A}}{m}$ and $C_l=\frac{m_l}{m}$. Usually, the species with less content in the binary system should be treated as the primary component when calculating. Assuming we have one kind of alloy $\text{B-10wt\%A}$, then $A$ should be selected as the primary one, where $C_s^A$ represents the mass fraction of the species A in the solid phase and $C_l^A$ stands for the mass fraction of the species A in the liquid phase, we have
\begin{equation} 
m_A = C_A m = C^A_l m_l + C^A_s \left(m-m_l\right).	
\end{equation}
Finally, the lever rule relation can be derived:
\begin{equation} 
\left\{\begin{array}{l}
m^l \left(C_l^A-C_s^A\right) = m \left(C_A-C_s^A\right) \\ 
m^s \left(C_l^A-C_s^A\right) = m \left(C_l^A-C_A\right).
\end{array}\right.
\end{equation}
\begin{figure}[h]
	\centering
	\includegraphics[scale=.5]{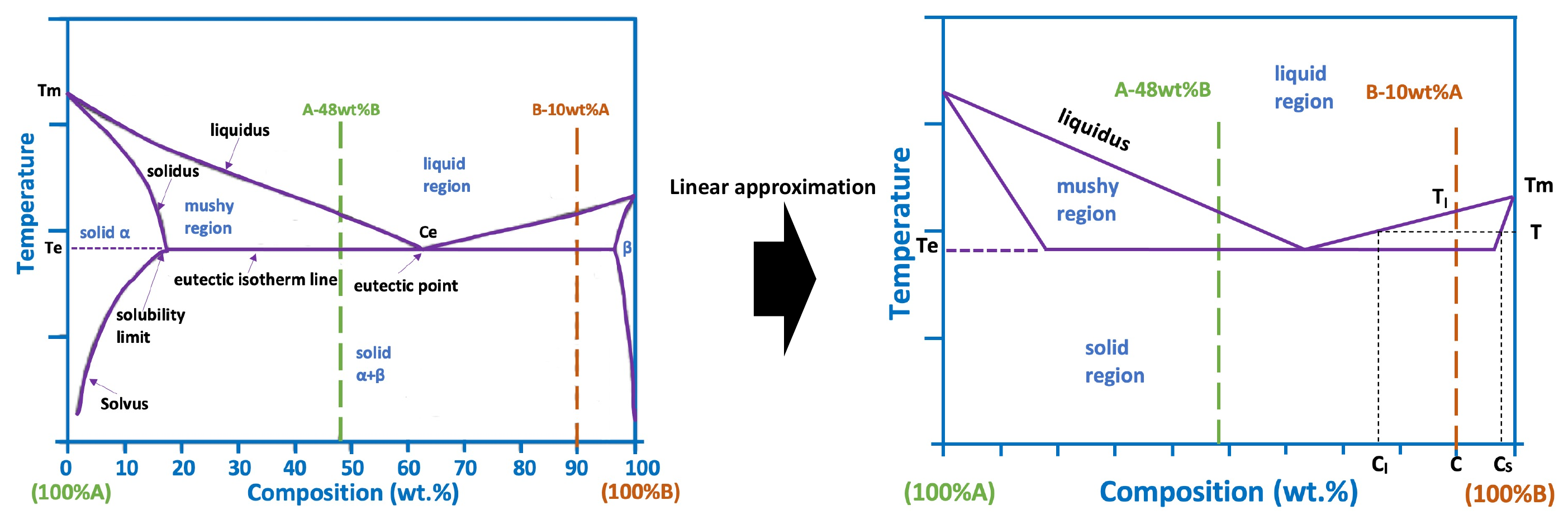}
	\caption{The linear approximation of the phase diagram}
	\label{FIG:2}
\end{figure}
The implication of `lever rule' can be confirmed through the phase diagram after linearization in Figure \ref{FIG:2}. The colored dash line represents a solidification process by decreasing temperature. $C_e$ and $T_e$ indicate the composition and temperature of the eutectic point. If the composition proportion changes, the selection of the primary component in the binary system will change as well. Given the $\text{B-10wt\%A}$ alloy, for the sake of brevity and conciseness of mathematical expressions, the notation $C$ in the following part will stand for $C_A$. Notations $C^A_l$ and $C^A_s$ are replaced by $C_l$ and $C_s$, respectively. This type of notation is always used to describe this sense in the following content. The phase equilibrium relations are finalized as follows:
\begin{equation} 
C =g_l C_l+g_s C_s = \left(g_l+k_{p}\left(1-g_l\right)\right) C_l,	
\label{EQequilibrium1}
\end{equation}
and
\begin{equation} 
T=T_{m}+ m_{sp} C_l,
\label{EQequilibrium2}
\end{equation}
where $k_{p}$ represents the partition coefficient and $k_{p} = C_s/C_l$. The slope of linear liquidus is denoted by $m_{sp}$, and the melting point of material is denoted by $T_{m}$. 

\section{Numerical scheme} \label{section:Numerical scheme}
\setcounter{equation}{0}
\renewcommand{\theequation}{\arabic{section}.\arabic{equation}}
On the basis of the aforementioned mathematical model, a fully decoupled discretization is proposed. In this section, the operator-splitting and matrix-based techniques are employed for constructing the scheme. In the first place, the temporal discretization is given:

\textbf{The semi-implicit pressure correction method for momentum equation:}
\begin{equation} 
\frac{\boldsymbol{u}^{n+1}-\boldsymbol{u}^{n}}{\Delta t}+ \boldsymbol{u}^{n} \cdot \nabla \boldsymbol{u}^{n+1}=
\nabla \cdot\left(\frac{\mu_l}{\rho} \nabla \boldsymbol{u}^{n+1}\right)-\frac{1}{\rho} \nabla p^{n+1}-\frac{\mu_l}{\rho} K^{-1} \boldsymbol{u}^{n+1}+\boldsymbol{f},
\end{equation}
\begin{equation} 
\nabla \cdot \boldsymbol{u}^{n+1} =0.
\end{equation}
The superscripts $n+1$ and $n$ represent implicit and explicit terms, respectively. Given these explicit terms, the scheme determines these implicit terms. $\Delta t$ is the discrete time step. The pressure correction scheme is used for solving velocity. Combining the divergence-free condition with the momentum equation gives
\begin{equation} 
\frac{\boldsymbol{u}^{*}-\boldsymbol{u}^{n}}{\Delta t}+ \boldsymbol{u}^{n} \cdot \nabla \boldsymbol{u}^{*}=
\nabla \cdot\left(\frac{\mu_l}{\rho} \nabla \boldsymbol{u}^{*}\right)-\frac{1}{\rho} \nabla p^{n}-\frac{\mu_l}{\rho} K^{-1} \boldsymbol{u}^{*}+\boldsymbol{f}, \label{mom}
\end{equation}
and the pressure correction Poisson equation
\begin{equation} 
\nabla^{2} \left(p^{n+1}-p^{n}\right)=\frac{\rho}{\Delta t} \nabla \cdot \boldsymbol{u}^{*}. \label{Poisson}
\end{equation}
Then the velocity can be updated through pressure correction:
\begin{equation} 
\frac{\boldsymbol{u}^{n+1}-\boldsymbol{u}^{*}}{\Delta t}=-\frac{1}{\rho} \nabla\left(p^{n+1}-p^{n}\right)	
\label{UPu}
\end{equation}

\textbf{The matrix-based technique for solute transport equation:}

The ``minimal'' model reads as:
\begin{equation} 
\frac{C^{n+1}-C^n}{\Delta t}+ \boldsymbol{u}^{n+1} \cdot  \nabla C_l^{n+1} = 0, \label{transport}
\end{equation}
If we consider the diffusion effect in the liquid phase as well, we have
\begin{equation} 
\frac{C^{n+1}-C^n}{\Delta t}+ \boldsymbol{u}^{n+1} \cdot  \nabla C_l^{n+1} = \nabla \cdot\left( g_l^{n} D_l \nabla C_l^{n+1}\right).
\end{equation}
It must be noted that convection terms in the momentum equation, transport equation, and energy equation are all treated with an upwind implicit scheme. However, the main difficulty caused by the transport equation comes from $C_l^{n+1}$ in the convection and diffusion terms, which differ from $C^{n+1}$ in the temporal term. According to the explicit relationship between $C_l$ and $C$:
\begin{equation} 
C_l^{n+1} = \frac{ C^{n+1} }{ 1-(1-k_p)(1-g_l^{n}) },
\end{equation}
in matrix form like:
\begin{equation} 
C_l^{n+1} = \boldsymbol{A}_g C^{n+1}	,
\end{equation}
where $\boldsymbol{A}_g$ is a diagonal matrix. Given the  the upwind coefficient matrix $\boldsymbol{A}_{con}$ and the Laplacian coefficient matrix $\boldsymbol{A}_{lap}$, the temporal discrete transport equation in matrix and vector form can be expressed as:
\begin{equation}
\left( 
\frac{1}{\Delta t}\boldsymbol{I}  + \boldsymbol{A}_g*\boldsymbol{A}_{con} - \boldsymbol{A}_g*\boldsymbol{A}_{lap} 
\right) C^{n+1} = \boldsymbol{rhs},
\end{equation}
where $\boldsymbol{I}$ is the identity matrix. The notation $\boldsymbol{rhs}$ denotes the right-hand side vector of the linear system. This matrix-based method avoids auxiliary terms in traditional methods:
$$
\frac{\partial C}{\partial t}+\nabla \cdot( \boldsymbol{u} C) =  \nabla \cdot\left( g_l D_l \nabla C\right)+\nabla \cdot\left[g_l D_l \nabla\left(C_l-C\right)\right] -\nabla \cdot\left[ \left(C_l-C\right)\boldsymbol{u} \right].
$$

\textbf{The operator-splitting method for energy equation:}

\begin{figure*}[!h]
	\centering
	\includegraphics[scale=0.12]{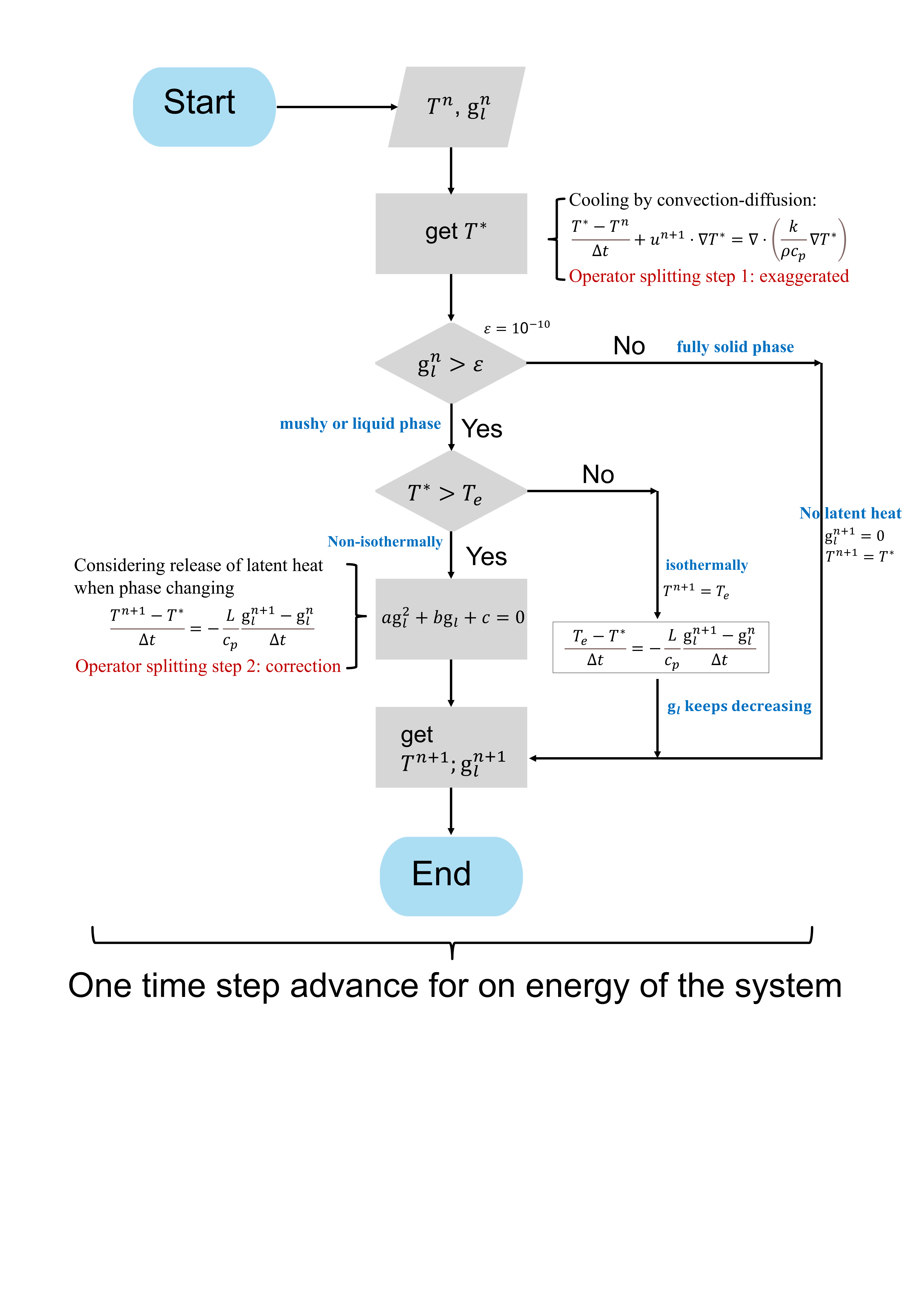}
	\caption{The flowchart for determining the energy equation in one timestep}
	\label{FIG:flowchart}
\end{figure*}

In this part, the solution procedure of the energy equation is split into two steps: the first step takes account for the heat convection–diffusion,
\begin{equation} 
\frac{T^{*} - T^n}{\Delta t} + \boldsymbol{u}^{n+1} \cdot \nabla T^{*} = \nabla \cdot\left(\frac{k}{\rho c_p} \nabla T^{*} \right). \label{T1}
\end{equation}
The second step is responsible for the correction of latent heat when a phase transition occurs,
\begin{equation} 
\frac{T^{n+1}-T^{*}}{\Delta t} = - \frac{L}{c_p} \frac{ g_l^{n+1} - g_l^{n}}{\Delta t}.  \label{T2}
\end{equation}
The linear assumption of the phase equilibrium relation \eqref{EQequilibrium2}yields:
\begin{equation} 
T_{l} = T_{m}+m_{s p} C, \label{CforTl}
\end{equation}
where $T_{l}$ corresponds to the liquidus temperature at the current mass fraction C. Because of the segregation process, the deviation of concentration $C^{n+1}$ from the uniform will result in a range of $T_{l}(C^{n+1})$ across the entire computational domain. A relationship between $g_l^{n+1}$ and $T^{n+1}$ yields:

\begin{equation} 
g_l^{n+1} = 1-\frac{1}{1-k_p} \frac{T^{n+1}- T_{l}(C^{n+1}) } {T^{n+1}-T_m}.   \label{Tlforgl}
\end{equation}
Then, by substituting it in the second step of operator-splitting expressions, a quadratic equation of the liquid fraction $g_l^{n+1}$ will be found as follows:
\begin{equation} 
a \cdot (g_l^{n+1})^2+b \cdot g_l^{n+1}+c=0, \label{quadratic}
\end{equation}
where coefficients are
\begin{equation} 
\left\{\begin{array}{l}
a= 1 - k_{p}, \\
b= k_p-g_l^{n}\left(1-k_{p}\right)+\frac{c_p}{L}\left(1-k_{p}\right)\left(T_m-T^*\right), \\
c= -\left(f_l^n+\frac{c_p}{L} T^*\right) k_{p} + \frac{c_p}{L}\left(T_{l} - \left(1-k_{p}\right) T_m\right).
\end{array}\right.
\end{equation}
By employing the limiting condition $[0,1]$ for the phase fraction and the truncation of the larger root, the unique solution for $g_l$ will be obtained. It must be emphasized that the second step is crucial for accurately quantifying the change in temperature when it involves the phase change and the latent heat releases. The second step will also determine $T^{n+1}$. It is worth remarking that when it comes to the eutectic temperature, phase change will occur isothermally; that is, the temperature at any point can only decrease once the material at that point has solidified completely. Figure \ref{FIG:flowchart} depicts the flowchart for determining the energy equation in a single time step. 

\textbf{Updating the concentration:}

With the determined $g_l^{n+1}$, we can calculate the $C_l^{n+1}$ and $C_s^{n+1}$ for each phase,
\begin{equation} 
C_l^{n+1}=\frac{C^{n+1}}{1 + \left( 1-g_l^{n+1} \right) \left( k_{p} - 1\right)},	\label{UPCl}
\end{equation}
and
\begin{equation} 
C_s^{n+1} = k_{p}C_l^{n+1}. \label{UPCs}
\end{equation}
One step of progress has now been completed. Our fully decoupled scheme based on operator splitting and matrix-based techniques can be solved without any non-linear systems and auxiliary items. It really benefits implementability and makes it friendly for programming.

\textbf{Spatial discretization:}

\begin{figure}[!h]
	\centering
	\includegraphics[scale=.65]{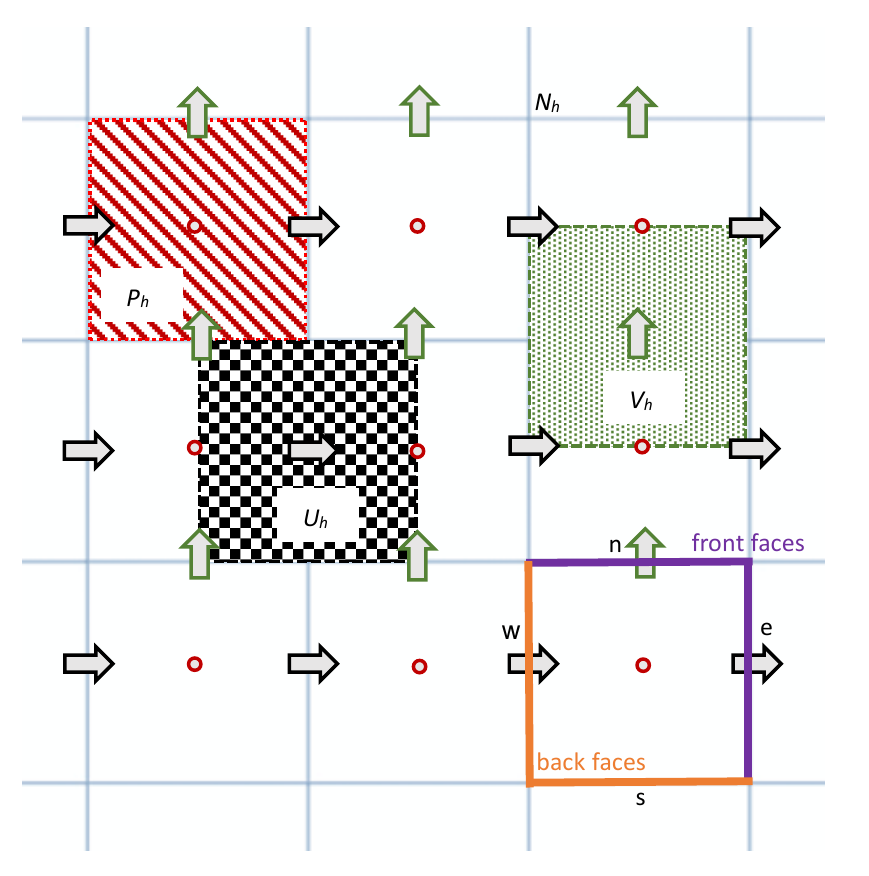}
	\caption{The spatial discretization based on the staggered grid}
	\label{FIG:Staggered}
\end{figure}

For spatial discretization, a finite difference method based on staggered grids as shown in Figure \ref{FIG:Staggered} is applied. The computational domain $\Omega=\left[0, l_x\right] \times\left[0, l_y\right]$ is used as an example of the mathematical notations of spatial discretization, where $l_x$ and $l_y$ are lengths of the physical domain. The subdivision of the computational domain into a finite number of rectangular meshes and the mesh vertex points are located at:
\begin{equation} 
x_i = i * h_x, \quad i=0, 1, \cdots, n_x,
\end{equation}
\begin{equation} 
y_j = j * h_y, \quad j=0, 1, \cdots, n_y.
\end{equation}
where $n_x$ and $n_y$ are the number of meshes in each direction and $h_x=l_x / n_x$ and $h_y=l_y / n_y$ are mesh sizes. Four sets of mesh points are defined:
\begin{equation} 
E^{we}= \left\{\left(x_i, \frac{y_{j-1} + y_j}{2} \right) \mid i=0, 1, \ldots, n_x ; j=1, 2, \ldots, n_y\right\},
\end{equation}
\begin{equation} 
E^{sn}= \left\{\left(\frac{x_{i-1} + x_i}{2}, y_j \right) \mid i=1, 2, \ldots, n_x ; j=0, 1, \ldots, n_y\right\},
\end{equation}
\begin{equation}
E^{c}= \left\{ \left(\frac{x_{i-1}+x_i}{2}, \frac{y_{j-1}+y_j}{2}\right) \mid i=1, 2, \ldots, n_x; j=1, 2, \ldots, n_y\right\},
\end{equation}
\begin{equation}
E^{v}=\left\{ \left(x_i, y_j \right) \mid i=0, 1, \cdots, n_x; j=0, 1, \cdots, n_y \right\}.
\end{equation}
where $E^{we}$ is the set of west-east edge points, $E^{sn}$ is the set of south-north edge points, $E^{c}$ is the set of cell-centered points, and $E^{v}$ is the set of vertex points. The east and north interfaces of one cell are considered the front interfaces. On the contrary, the west and south interfaces are the back interfaces. Some boundary points can follow the above rules by using ghost cells or apply high-order schemes by using half of the cell, and then we have the discretized spaces:
$$
U_h=\left\{u: E^{we} \rightarrow \mathbb{R}\right\}, \quad V_h=\left\{v: E^{sn} \rightarrow \mathbb{R}\right\},
$$
$$
P_h=\{P: E^{c} \rightarrow \mathbb{R}\}, \quad N_h=\{interpolation: E^{v} \rightarrow \mathbb{R}\}.
$$

Since the velocity vector is composed of orthogonal components as $\boldsymbol{u} = u\boldsymbol{i} + v\boldsymbol{j}$, the horizontal component $u$ is defined on $U_h$ and the vertical component $v$ is defined on $V_h$. All other physical scalar quantities (pressure $P$, temperature $T$, concentration $C$, phase fraction $g_l$ and $g_s$, etc.) are initially defined on the cell center, namely $P_h$. The space $N_h$ is for the interpolation of coefficients such as the dynamic viscosity $\mu$. With the help of other operators, like the difference operator and average operator, the spatial discretization can be accomplished.

It must be noted that these multi-physical governing equations all obey the general convection-diffusion form:
\begin{equation}
\frac{\partial \Phi}{\partial t}+\nabla \cdot(\boldsymbol{u} \Phi)=\nabla \cdot(D \nabla \Phi)+S_{\Phi}, 
\label{Generalform}
\end{equation}
where $\Phi$ denotes a general physical variable and $D$ is the general diffusion coefficient; the convection velocity is denoted by $\boldsymbol{u}$ and $S_{\Phi}$ represents the source term.

It should be mentioned that the central-difference and fully implicit scheme is applied to all diffusion terms,
\begin{equation}
\left\{\begin{array}{l}
\left.  D\frac{\partial^2 \Phi}{\partial x^2}\right|_{h_x}=\frac{D_e \frac{\Phi_{i+1,j}-\Phi_{i,j}}{h x}-D_w \frac{\Phi_{i,j}-\Phi_{i-1, j}}{h_x}}{h_x}
\\
\left.  D\frac{\partial^2 \Phi}{\partial y^2}\right|_{h_y}=\frac{D_n \frac{\Phi_{i,j+1}-\Phi_{i,j}}{h y}-D_s \frac{\Phi_{i,j}-\Phi_{i,j-1}}{h_y}}{h_y},
\end{array}\right. \label{DIFFUSION}
\end{equation}
The upwind scheme is used for all convection terms, if we take x-direction as an example, we have
\begin{equation}
\left. u \frac{\partial \Phi}{\partial x} \right|_{h_x} = \frac{ (\Phi u)_e - (\Phi u)_w }{h_x},
\end{equation}
and
\begin{equation}
(\Phi u)_e = \begin{cases} \Phi_i     u_e, & u_e \geqslant 0 \\ \Phi_{i+1} u_e, & u_e<0  \end{cases} ; \quad
(\Phi u)_w = \begin{cases} \Phi_{i-1} u_w, & u_w \geqslant 0 \\ \Phi_i     u_w, & u_w<0  \end{cases} ,
\label{CONVECTION}
\end{equation}
where the subscripts $e, w, n, s$ represent the east, west, south, and north interfaces of a single control volume as it shown in Figure \ref{FIG:Staggered}. 

\section{Algorithm implementation} \label{section:Algorithm implementation}
\setcounter{equation}{0}
\renewcommand{\theequation}{\arabic{section}.\arabic{equation}}
The fundamental target of matrix assembly is to transfer computational information from the actual physical distribution space to the mathematical space via the matrix and vector formats. The combination of two kinds of matrix assembly strategies, vectorization and the forward method, can achieve this task with better computational efficiency and less implementation complexity. This concept also applies to all of these multi-physical governing equations because they all follow the general convection-diffusion form as \eqref{Generalform}:

According to the fully decoupled scheme and the spatial discretization described above, all of the coefficient matrices $\mathbf{A}$ of each linear system $\mathbf{Ax} = \mathbf{b}$ for unknown variables $\mathbf{u}, P, T, C$ share the same nonzero element distribution pattern. The sparse coefficient matrix $\mathbf{A}$ has five diagonals for the 2D domain (7 diagonals for the 3D case) as follows:
\begin{equation}
\mathbf{A}=\left(\begin{array}{cccccccc} A_0 & A_1 & \ldots & A_{nx} & 0 & 0 & \ldots & 0 \\
A_{-1} & A_0 & A_1 & \ldots & A_{nx} & 0 & \ldots & 0 \\ 
\vdots & A_{-1} & A_0 & \ddots & \ddots & A_{nx} & \ddots & \vdots \\
 A_{-nx} & \ddots & \ddots & \ddots & A_1 & \ddots & \ddots & 0 \\ 
0 & A_{-nx} & \ddots & \ddots & A_0 & A_1 & \ddots & A_{nx} \\
0 & \ddots & \ddots & \ddots & A_{-1} & A_0 & \ddots & \vdots \\
\vdots & \ddots & \ddots & \ddots & \ddots & \ddots & \ddots & A_1 \\ 
0 & 0 & \ldots & 0 & A_{-nx} & \ldots & A_{-1} & A_0
\end{array}\right);
\end{equation}
and
\begin{equation}
\mathbf{x}=
\left(\begin{array}{c}
\Phi(1,1)\\
\Phi(2,1)\\
$\vdots$\\
\Phi(i,j)\\
$\vdots$\\
\Phi(nx-1,ny)\\
\Phi(nx,ny)
\end{array}\right);
\quad 
\mathbf{b}=
\left(\begin{array}{c}
b(1,1)\\
b(2,1)\\
$\vdots$\\
b(i,j)\\
$\vdots$\\
b(nx-1,ny)\\
b(nx,ny)
\end{array}\right),
\end{equation}
 where diagonal $A_0$ represents the coefficient of the current cell, diagonals $A_{-1}$ and $A_{1}$ correspond to the coefficients of neighboring cells at the x-direction from the west and east sides, and diagonals $A_{-nx}$ and $A_{nx}$ correspond to the coefficients of neighboring cells at the y-direction from the south and north sides. Thus, we can also conclude that the upper part and the lower part of $\boldsymbol{A}$ are contributed from the front direction and the back direction in the real physical domain, as shown in Figure \ref{FIG:Staggered}, respectively. Similarly, the vector $\boldsymbol{b}$ on the right side can be decomposed into four parts, one from each side of the boundary. These explicit terms in discrete governing equation also contribute to the vector $\boldsymbol{b}$ as well. In the following content, a brief description of these two kinds of techniques will be presented.

\textbf{Vectorization method:}

\begin{figure*}[t]
	\centering
	\includegraphics[width=1.0\textwidth,height=0.5\textheight]{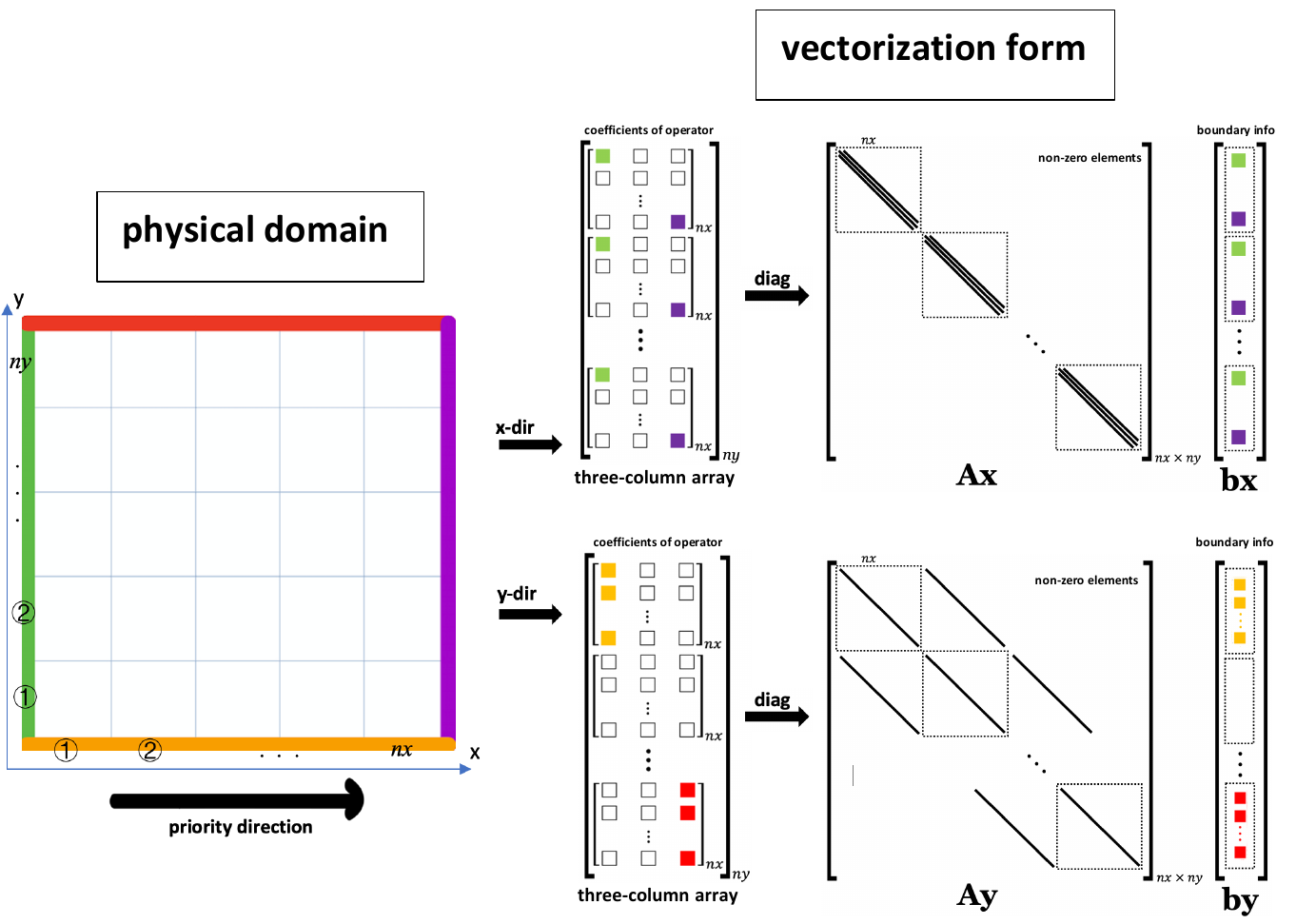}
	\caption{Schematic of matrix assembly by vectorization method}
	\label{FIG:vectorization}
\end{figure*}
The vectorization method directly assigns these non-zero values of the coefficient matrix to these diagonals. Here is the brief matrix assembly process, as shown in Figure \ref{FIG:vectorization}. If we directly create a three-column array to store the coefficient series for one direction. Then, based on the discrete expression of diffusion term \eqref{DIFFUSION}, we have the following coefficient series of this operator at one row in the matrix $\boldsymbol{A}$:
\begin{equation}
\begin{cases} 
[D_w ;-D_w-D_e ; D_e]/{h_x}^2 = [D_w ; -D_w ; 0]/{h_x}^2 + [0 ; -D_e ; D_e]/{h_x}^2     \\ 
[D_s ;-D_s -D_n ; D_n]/{h_y}^2 = [D_s ; -D_s ; 0]/{h_y}^2 + [0 ; -D_n ; D_n]/{h_y}^2.
\end{cases} 
\end{equation}
Similarly, for the convection term, we get
\begin{equation}
\begin{cases} 
\left[-\max \left(u_{\text {w}}, 0\right); \max \left(u_{\text{e}}, 0\right)-\min \left(u_{\text {w}}, 0\right); \min \left(u_{\text {e}}, 0\right)\right]/h_x     \\ 
\left[-\max \left(v_{\text {s}}, 0\right); \max \left(v_{\text{n}}, 0\right)-\min \left(v_{\text {s}}, 0\right); \min \left(v_{\text {n}}, 0\right)\right]/h_y
\end{cases} 
\end{equation}

where the subscripts $e, w, n, s$ represent the east, west, south, and north interfaces of a single control volume. Then, we can transfer these three columns into three diagonals in the matrix. As we mentioned before, the $\boldsymbol{A}$ can be decomposed into the upper part and the lower part, which can be used to multiply with the corresponding diffusion coefficients on the front edge or back edge of one control volume easily. By putting all these boundary condition information and the explicit terms in the right row of the right-hand side vector $\boldsymbol{b}$, the linear system can be obtained successfully by the vectorization method.

\textbf{Forward method:}

\begin{figure*}[t]
	\centering
	\includegraphics[width=1.0\textwidth,height=0.5\textheight]{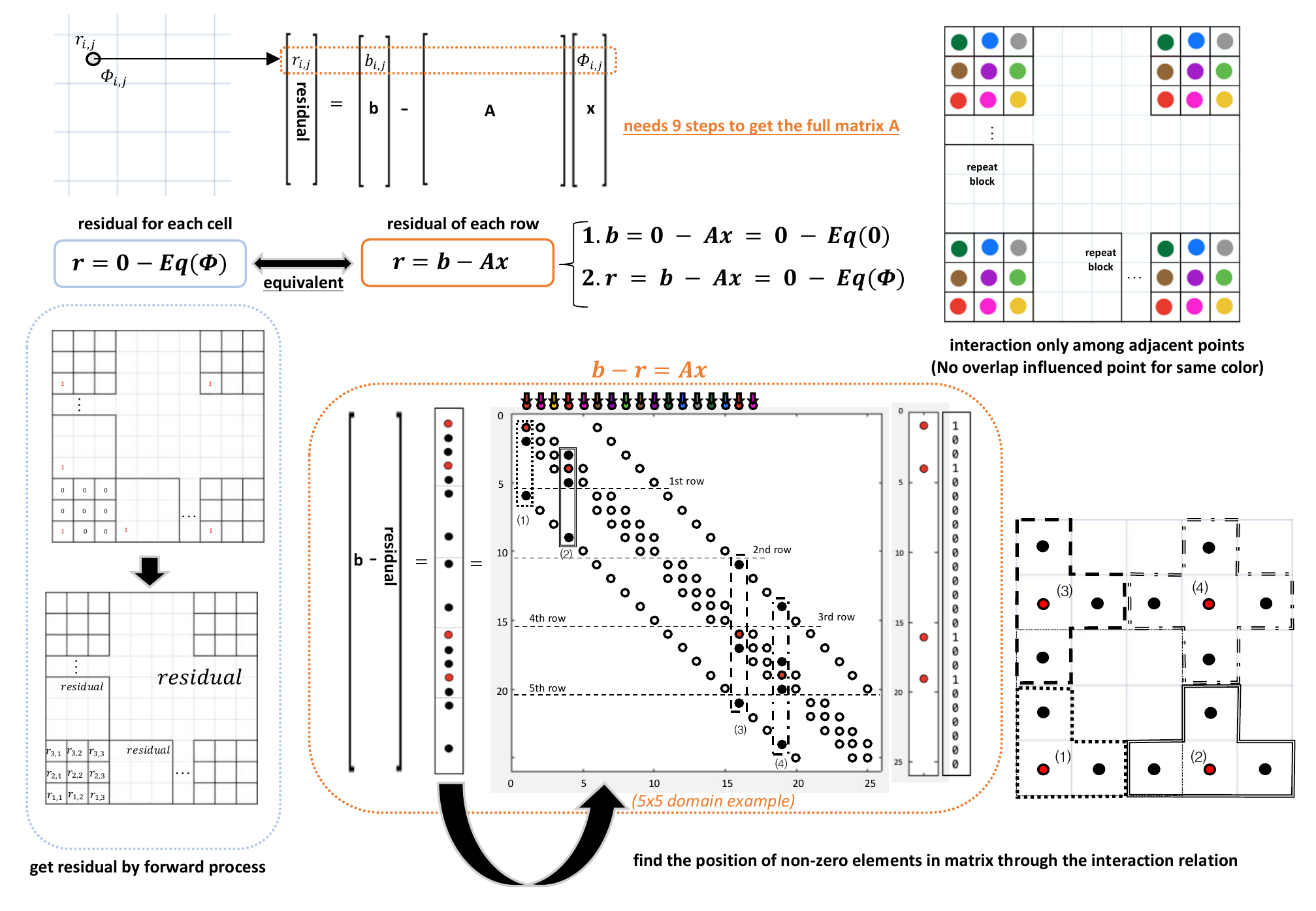}
	\caption{Schematic of matrix assembly by forward method}
	\label{FIG:forward}
\end{figure*}

When it comes to the equation based forward method for assembling matrices, it is essential to note that the resulting matrix will be identical to that obtained through direct construction by setting coefficients via vectorization, with the exception of any rounding errors introduced by the computer. The basic idea is to get these coefficient values based on the residual of a linear system and find the original position of these non-zero elements through the interaction relation. The details can be referred to Figure \ref{FIG:forward}. Through the following equivalent relation \eqref{equivalent}, the vector $\mathbf{b}$ of a linear system and the residual $\mathbf{r}$ can be obtained by a forward calculation of the residual $r_{i,j}$ of each cell on a real physical domain.
\begin{equation} 
\mathbf{r}=\mathbf{0}-\operatorname{Eq}(\boldsymbol{\Phi}) = \mathbf{r}=\mathbf{b}-\mathbf{A} \mathbf{x}, \label{equivalent}
\end{equation}
where the $\operatorname{Eq}$ denotes the forward plain discrete expression \eqref{equivalent} for one point and $\boldsymbol{\Phi}$ in $\operatorname{Eq}(\boldsymbol{\Phi})$ actually represents these implicit $\Phi^{n+1}$.  
\begin{equation}
\begin{split}
&\frac{\Phi_{i,j}^{n+1} - \Phi_{i,j}^{n}}{\Delta t} + 
\frac{ (\Phi^{n+1} u)_e - (\Phi^{n+1} u)_w }{h_x}  +
\frac{ (\Phi^{n+1} v)_n - (\Phi^{n+1} v)_s }{h_y} \\
&=
\frac{D_e \frac{\Phi_{i+1, j}^{n+1}-\Phi_{i, j}^{n+1}}{h x}-D_w \frac{\Phi_{i, j}^{n+1}-\Phi_{i-1, j}^{n+1}}{h_x}}{h_x} +
\frac{D_n \frac{\Phi_{i, j+1}^{n+1}-\Phi_{i, j}^{n+1}}{h y}-D_s \frac{\Phi_{i, j}^{n+1}-\Phi_{i, j-1}^{n+1}}{h_y}}{h_y}
+
S(\Phi_{i, j}^{n+1}). \label{Eqstraight}
\end{split}
\end{equation}

At the first step, the vector $\mathbf{b}$ can be obtained from the residual by assigning zero values to all unknown variables.
\begin{equation}
\mathbf{b}=\mathbf{0}-\mathbf{A x}=\mathbf{0}-\mathrm{Eq}(\mathbf{0})	
\end{equation}

Next, by assigning some unknown variables to ones, some elements in the determined the residual vector $\mathbf{r}$ will be equal to non-zero matrix elements in coefficent matrix $\mathbf{A}$. The original position information of these non-zero matrix elements can be derived from the spatial discretization interaction relation shown in Figure \ref{FIG:forward}. To obtain the one coefficient matrix in 2D cases, the second step must be repeated at least nine times. For 3D cases, it will take 27 times. This method is easy to implement because we just need to code these discretized expressions \eqref{Eqstraight} straightforwardly as a function module and finish all commands on the array. The shortcoming is that whenever a linear system needs to be constructed, a 10-time calculation of residuals is inevitable, which does increase the computational costs. 

Therefore, a term-wise strategy to combine two kinds of matrix assembly methods is proposed here. For these mathematical terms with a constant coefficient, such as the diffusion terms, in an unsteady process, the corresponding matrix is obtained in an offline way at a one-time computational cost. The direct vectorization method is used for convection terms whose coefficients change with the evolving flow field. This term-wise combination results in an optimally efficient matrix assembly strategy that benefits both the multi-physical problem and its 3D application.

\textbf{Detailed procedure:} we go over steps of the complete algorithm \ref{algorithm}.
\begin{algorithm}
  \caption{Steps for the decoupled scheme} \label{algorithm}
    Given initial fields and boundary conditions for $\mathbf{u}^{n},P^{n},T^{n},C^{n},gl^{n}$
	\begin{algorithmic}
	\For{$timestep \gets$ $1$ to $N_{total}$}                    
        \State {\text {1: }solve the intermediate $\mathbf{u}^{*}$ velocity by \eqref{mom}};
        \State {\text {2: }get the pressure correction by Poisson relation \eqref{Poisson}};
        \State {\text {3: }update the velocity form $\mathbf{u}^{n}$ to $\mathbf{u}^{n+1}$ by \eqref{UPu}};
        \State {\text {4: }solve the transport \eqref{transport} for $C^{n+1}$ using matrix-based technique};   
        \State {\text {5: }get the corresponding $T_{l}$ from $C^{n+1}$ through equilibrium relation as \eqref{CforTl}};
        \State {\text {6: }solve intermediate temperature $T^{*}$ by \eqref{T1}, considering the heat convection and diffusion};
        \State {\text {7: }get the temperature correction accounting for the release of latent heat $L$ caused by the phase transition as \eqref{T2} and get the $gl^{n+1}$ from the quadratic relation as \eqref{quadratic}, then update temperature $T^{n+1}$};
        \State {\text {8: }determine the $C_l^{n+1}$ and $C_s^{n+1}$}.   
    \EndFor
	\end{algorithmic}
\end{algorithm}

\section{Numerical Simulation} \label{section:Numerical Simulation}
In this section, some 2D and 3D numerical benchmark cases are simulated and analyzed based on the proposed method. The solidification and macro-segregation processes of some common binary alloy systems \ch{Sn}-10{\%Pb} and \ch{Pb}-48\ch{\%Sn} are investigated in the following examples. Data on the physical properties of these two kinds of alloys are given in Table \ref{alloy}. Findings about physical phenomena and the evolution of the velocity, temperature, and chemical species distributions are reported. All these numerical cases are directly simulated on the MacOS Mojave system with a 2.5GHz quad-core Intel Core i7 processor. Codes are written in Matlab from scratch.

%
\begin{table*}[!t]
\label{alloy}
\centering
\small
\begin{tabular}{lllll}
\hline Property & Symbol & Units & \ch{Pb}-48\ch{\%Sn} & \ch{Sn}-10{\%Pb} \\
\hline Specific heat & $c_{p}$ & $\mathrm{J}(\mathrm{kg} \cdot \mathrm{K})^{-1}$ & $200$ & $260$ \\
Thermal conductivity & $k$ & $\mathrm{W} (\mathrm{m} \cdot \mathrm{K})^{-1}$ & $50.0$ & $55.0$ \\
Reference density & $\rho_{\mathrm{o}}$ & $\mathrm{Kg} \cdot \mathrm{m}^{-3}$ & $9000$ & $7000$ \\
Latent heat of fusion & $\mathrm{J}$ & $\mathrm{J} \cdot \mathrm{kg}^{-1}$ & $5.35 \times 10^4$ & $6.1 \times 10^4$ \\
Liquid dynamic viscosity & $\mu_{\mathrm{l}}$ & $\mathrm{Pa} \cdot \mathrm{s}$ & $1.0 \times 10^{-3}$ & $1.0 \times 10^{-3}$ \\
Liquid thermal expansion coefficient & $\beta_{\mathrm{T}}$ & $\mathrm{K}^{-1}$ & $1.0\times 10^{-4}$ & $6.0 \times 10^{-5}$ \\
Liquid solutal expansion coefficient & $\beta_{\mathrm{C}}$ & $(\mathrm{wt} \%)^{-1}$ & $4.5 \times 10^{-3}$ & $-5.3 \times 10^{-3}$ \\
Secondary dendrite arm spacing & $\lambda_2$ & $\mathrm{~m}$ & $40.0 \times 10^{-6}$ & $65.0 \times 10^{-6}$ \\
Melting point at $C=0$ & $T_{m}$ & ${ }^{\circ} \mathrm{C}$ & $327.5$ & $232.0$ \\
Eutectic concentration & $C_{\mathrm{e}}$ & $\mathrm{wt} \%$ & $61.9$ & $38.1$ \\
Equilibrium partition coefficient & $k_{p}$ & & $0.307$ & $0.0656$ \\
Liquidus slope & $m_{sp}$ & ${ }^{\circ} \mathrm{C} \cdot (\mathrm{wt} \%)^{-1}$ & $-2.334$ & $-1.286$ \\
Nominal concentration & $C$ & $\mathrm{wt}\%$ & $48.0$ & $10.0$ \\
Initial temperature & $T_0=T_{\mathrm{l}}$ & ${ }^{\circ} \mathrm{C}$ & $215.44$ & $219.14$ \\
Heat transfer coefficient & $h$ & $\mathrm{W} \cdot \mathrm{m}^{-2} \hspace{0.3em} \mathrm{K}^{-1}$ & $400$ & $400$ \\
External temperature & $T_{\mathrm{ext}}$ & ${ }^{\circ}\mathrm{C}$ & $25$ & $25$ \\
\hline
\end{tabular}
\caption{Physical properties of alloy \ch{Pb}-48\ch{\%Sn} and alloy \ch{Sn}-10{\%Pb} }
\end{table*} 

\textbf{Example 1: solidification of \ch{Sn}-10{\%Pb} alloy in a 2D domain}

\begin{figure*}[!t]
	\centering
	\includegraphics[width=0.4\textwidth]{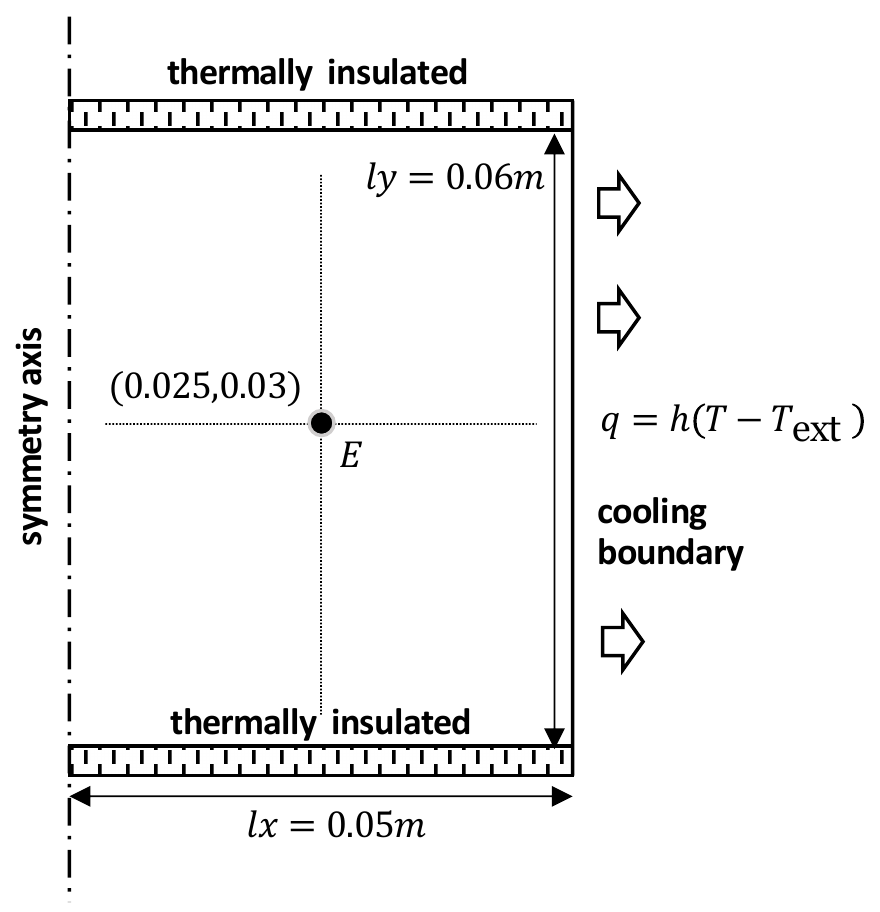}
	\caption{Schematic of physical settings in Example 1}
	\label{FIG:case1geo}
\end{figure*}

This benchmark case considers the solidification of a binary alloy \ch{Sn}-10{\%Pb} in a rectangular cavity, which is proposed by Bellet \cite{combeau2012analysis}. The computational domain, initial conditions and boundary conditions are shown in Figure \ref{FIG:case1geo}. The cavity is filled with a still liquid alloy at a uniform temperature $T_0 = T_{l}$ and with a uniform chemical species concentration $C_0$. At time $t=0$, solidification process begins by cooling the left and right walls of the enclosure through a forced convection mechanism with an overall heat transfer coefficient $h$, which can be described by the following relation:
$$
q=h\left(T-T_{\mathrm{ext}}\right).
$$
The top and bottom walls of the cavity are thermally insulated, and it is assumed that the cavity walls are rigid and nonslip. The phase transition occurs, and the inner flow is induced by the thermal and solutal buoyancy. A central observation point $E$ is set to check the evolution of flow and concentration and to justify the accuracy of our scheme. 
 \begin{figure*}[!h]
    \centering \subfigure[$t=5$]{
    \begin{minipage}[b]{0.24\textwidth}
    \centering
    \includegraphics[width=1.0\textwidth]{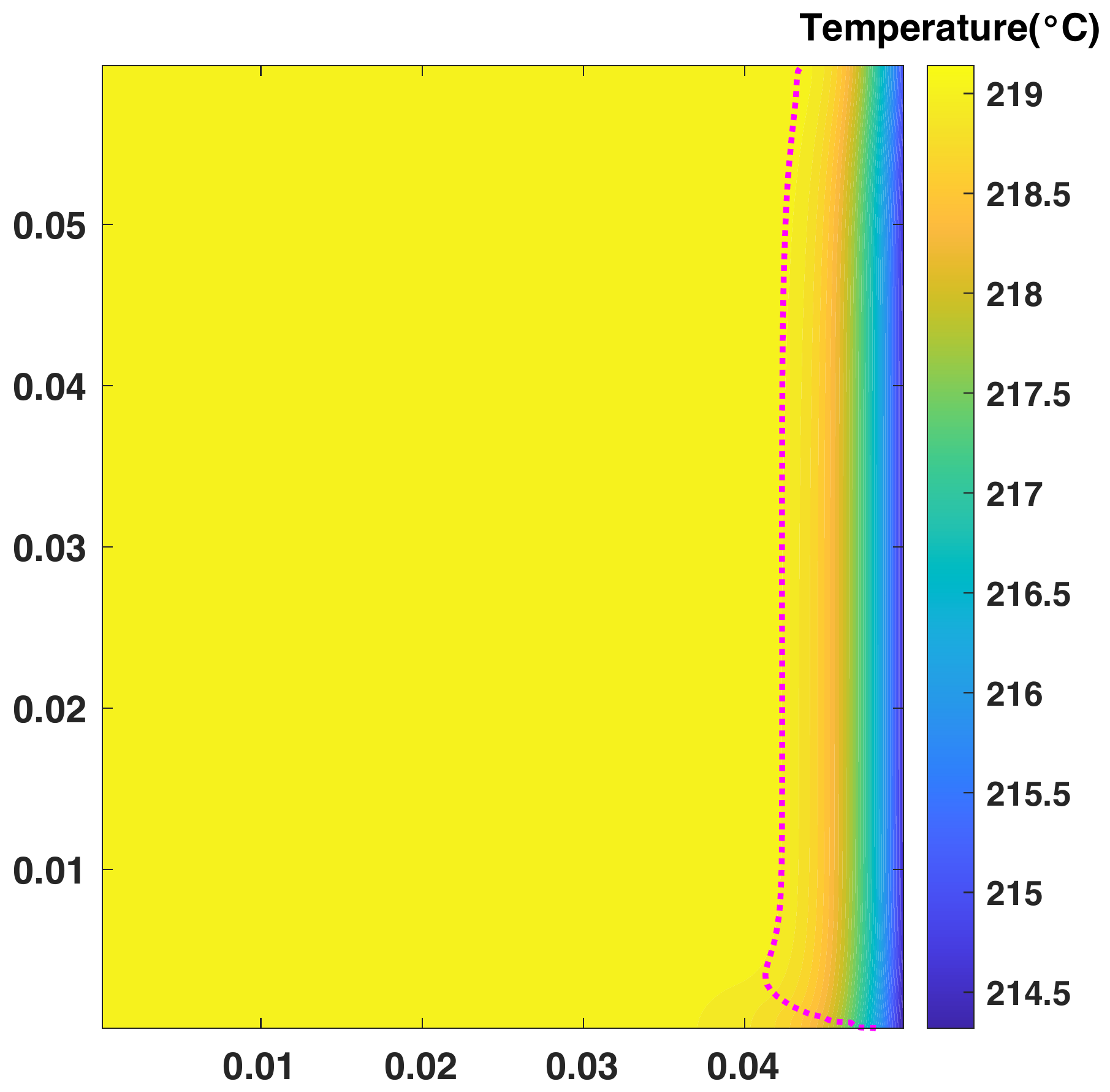}
    \end{minipage}   \label{ex_1_T5}
    }
    \centering \subfigure[$t=38$]{
    \begin{minipage}[b]{0.24\textwidth}
    \centering
    \includegraphics[width=1.0\textwidth]{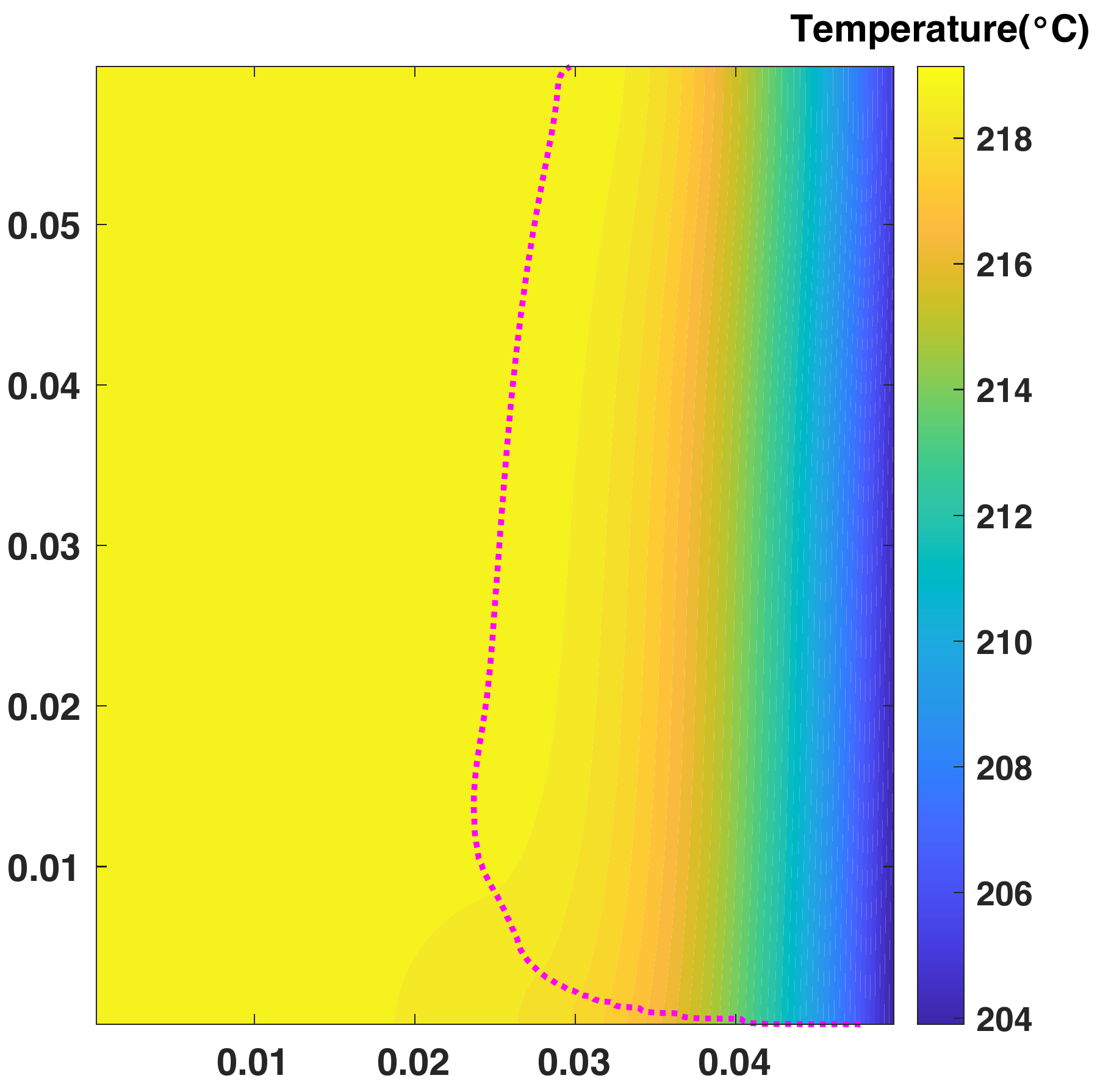}
    \end{minipage}   \label{ex_1_T38}
    }
    \centering \subfigure[$t=168$]{
    \begin{minipage}[b]{0.24\textwidth}
    \centering
    \includegraphics[width=1.0\textwidth]{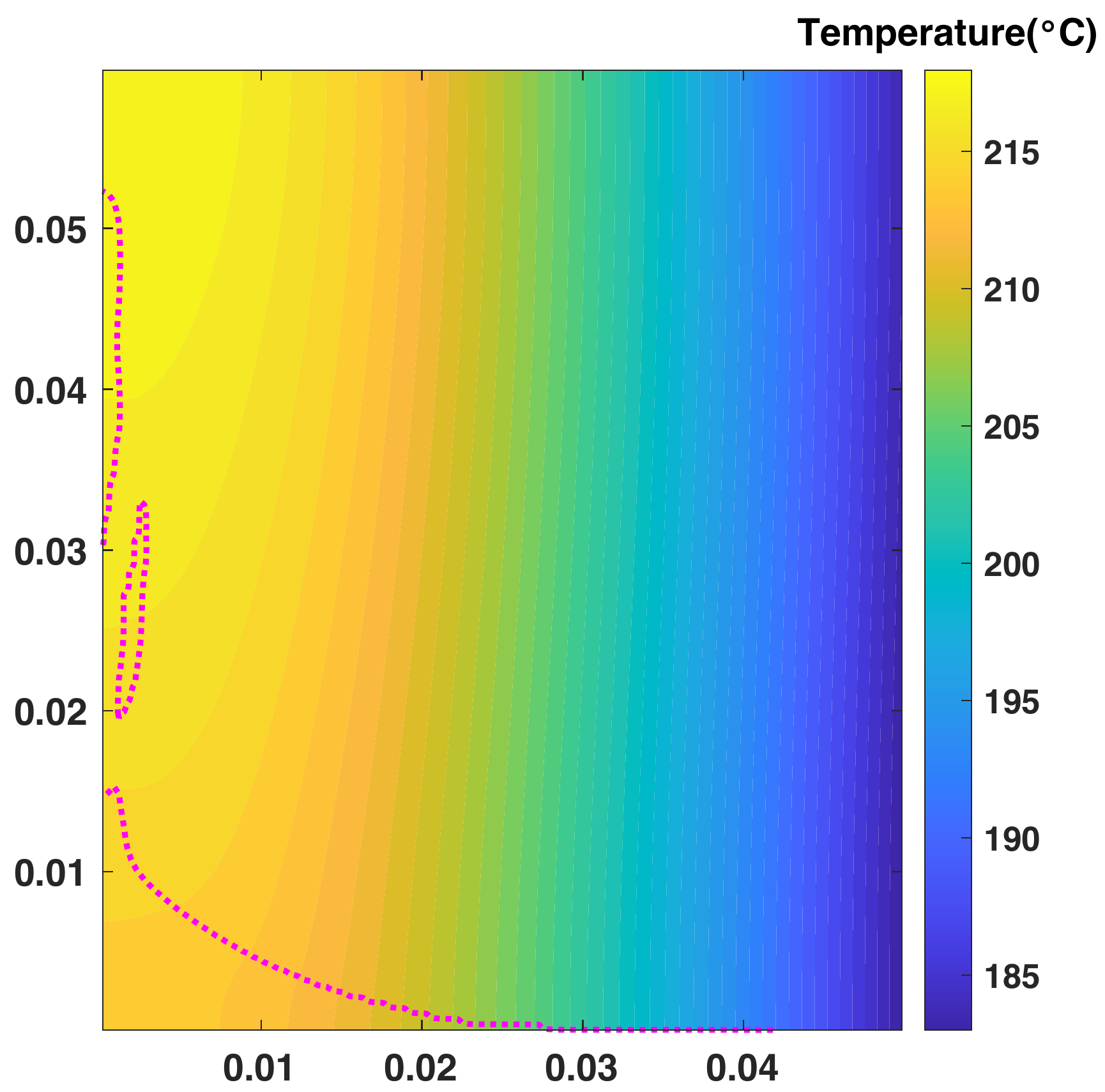}
    \end{minipage}   \label{ex_1_T168}
    }
    \centering \subfigure[$t=250$]{
    \begin{minipage}[b]{0.24\textwidth}
    \centering
    \includegraphics[width=1.0\textwidth]{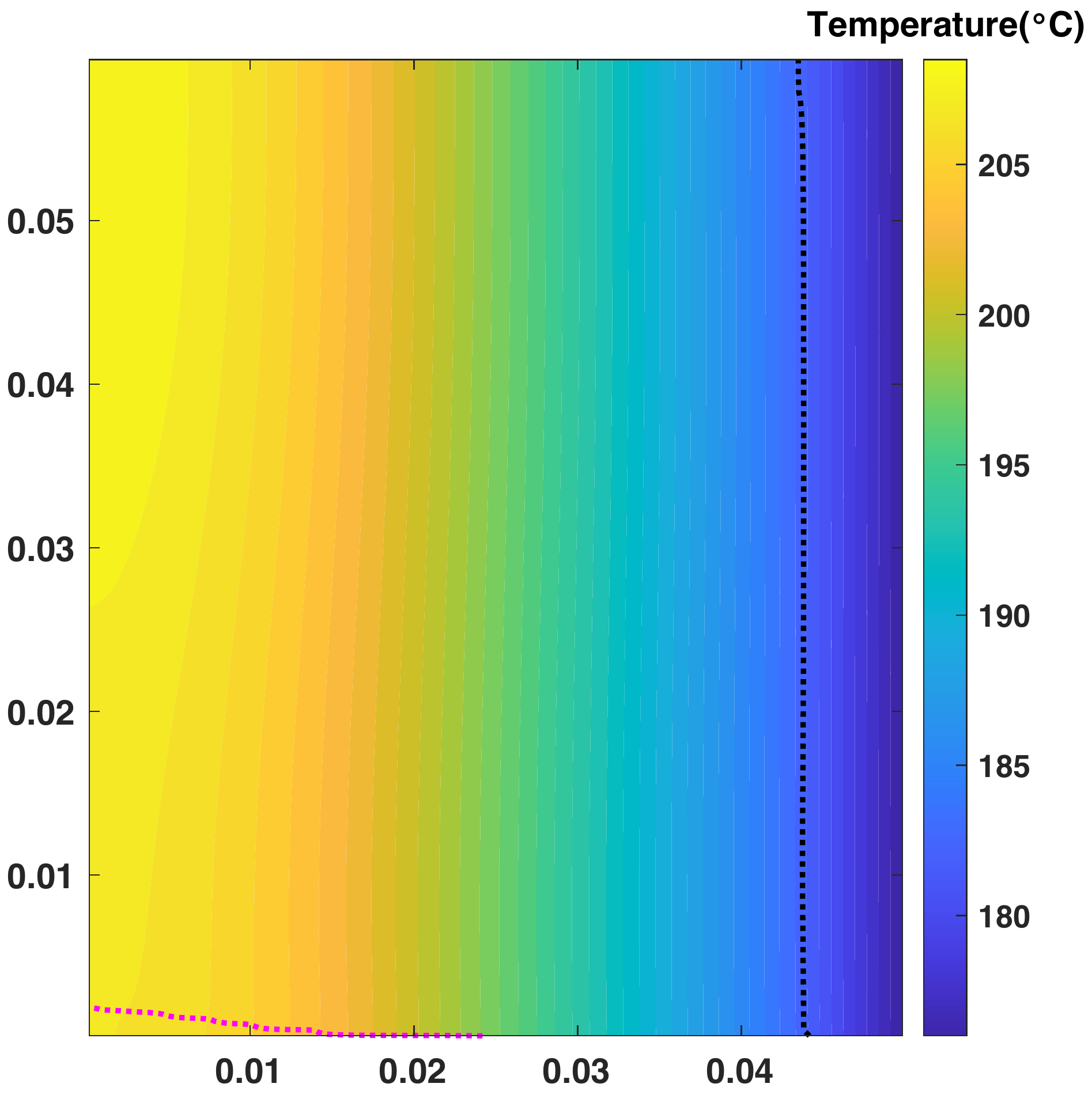}
    \end{minipage}  \label{ex_1_T250}
    }
    \centering \subfigure[$t=350$]{
    \begin{minipage}[b]{0.24\textwidth}
    \centering
    \includegraphics[width=1.0\textwidth]{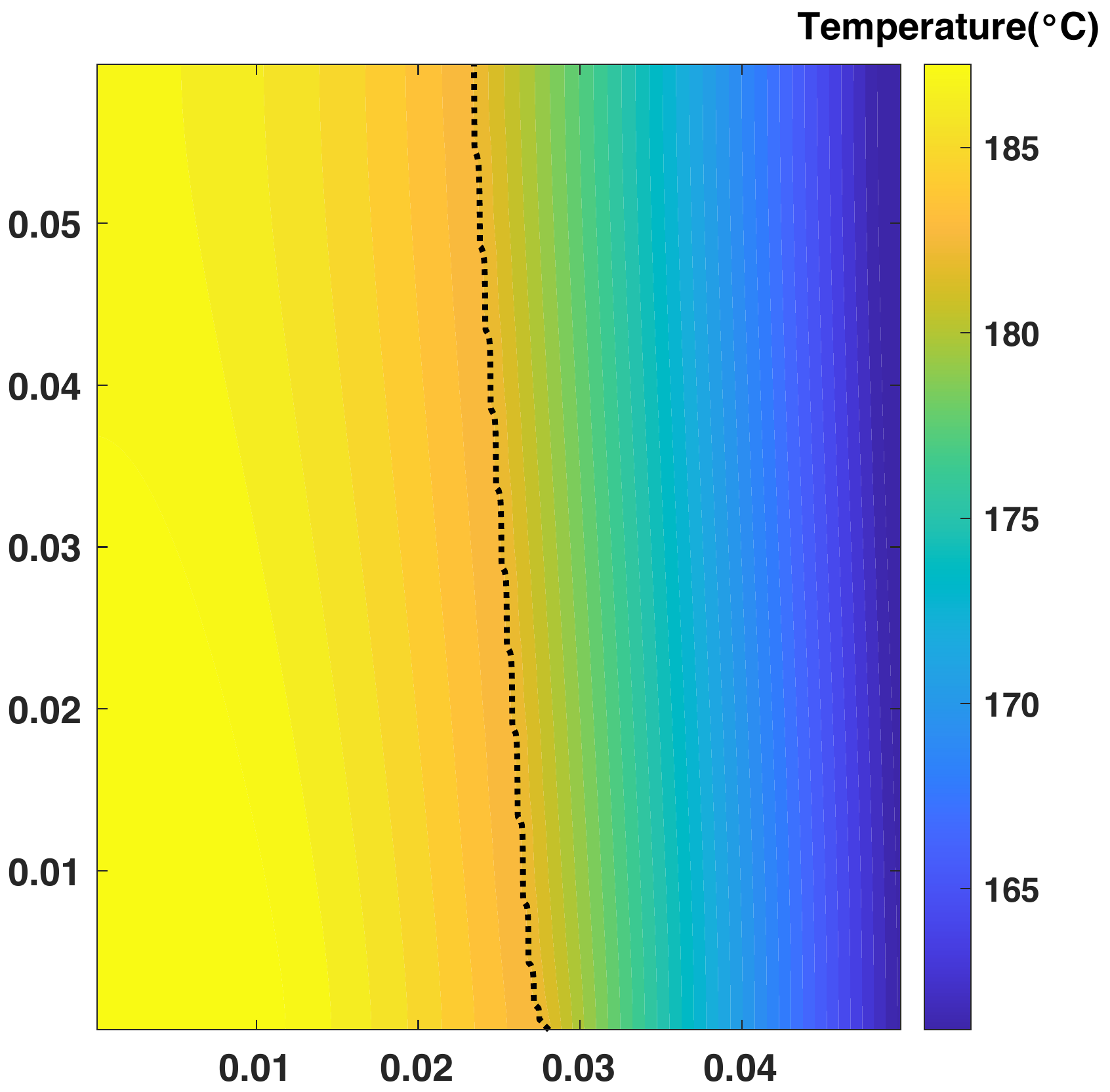}
    \end{minipage}  \label{ex_1_T350}
    }
    \caption{Evolution of the temperature field with time in Example 1; the liquid/mush interface (magenta dotted line); the mush/solid interface (black dotted line); temperature (colorbar)}
    \label{ex_1_T}
 \end{figure*}
  \begin{figure*}[!h]
    \centering \subfigure[$t=5$]{
    \begin{minipage}[b]{0.24\textwidth}
    \centering
    \includegraphics[width=1.0\textwidth]{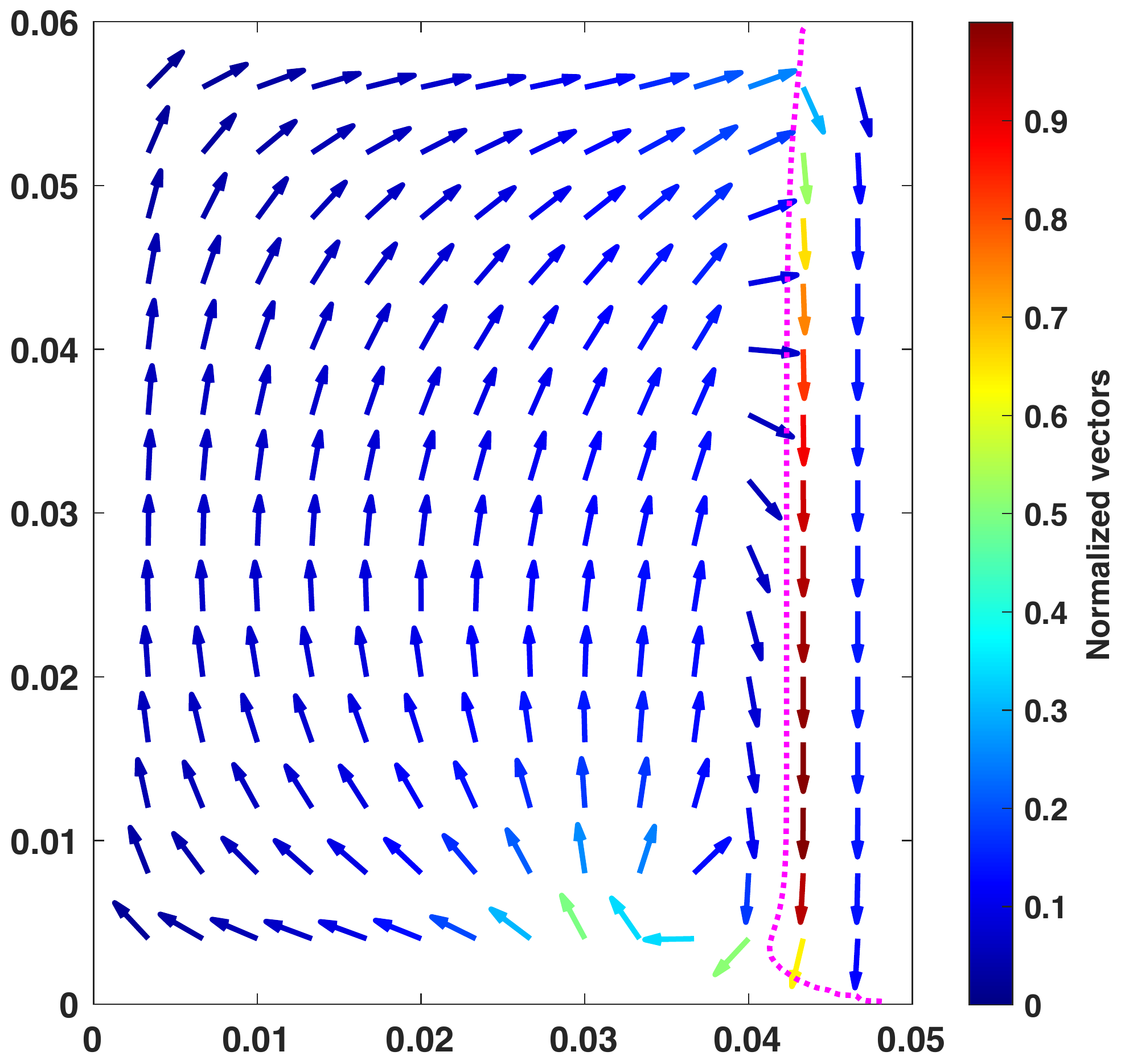}
    \end{minipage}  \label{ex_1_color5}
    }
    \centering \subfigure[$t=38$]{
    \begin{minipage}[b]{0.24\textwidth}
    \centering
    \includegraphics[width=1.0\textwidth]{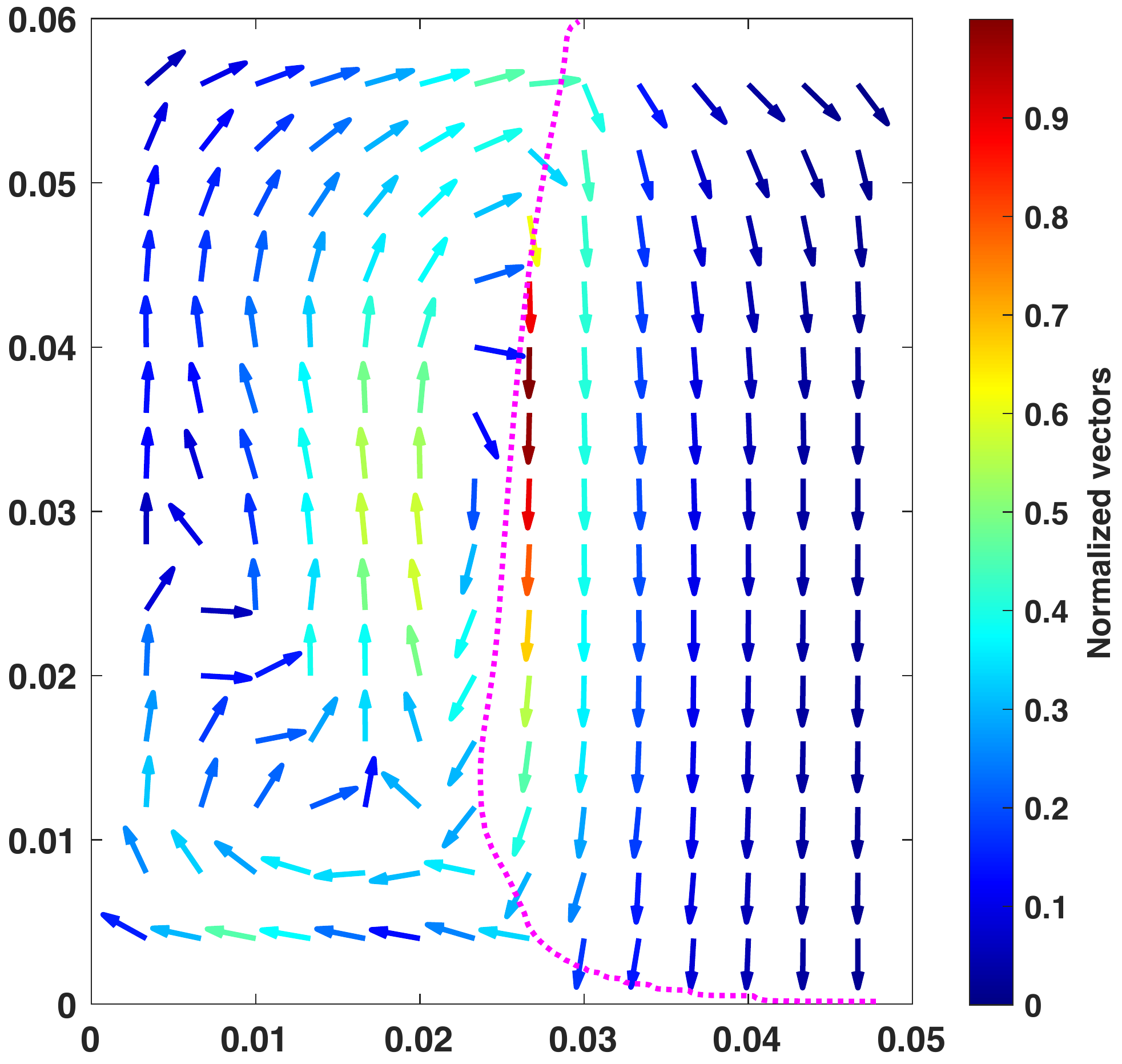}
    \end{minipage} \label{ex_1_color38}
    }
    \centering \subfigure[$t=168$]{
    \begin{minipage}[b]{0.24\textwidth}
    \centering
    \includegraphics[width=1.0\textwidth]{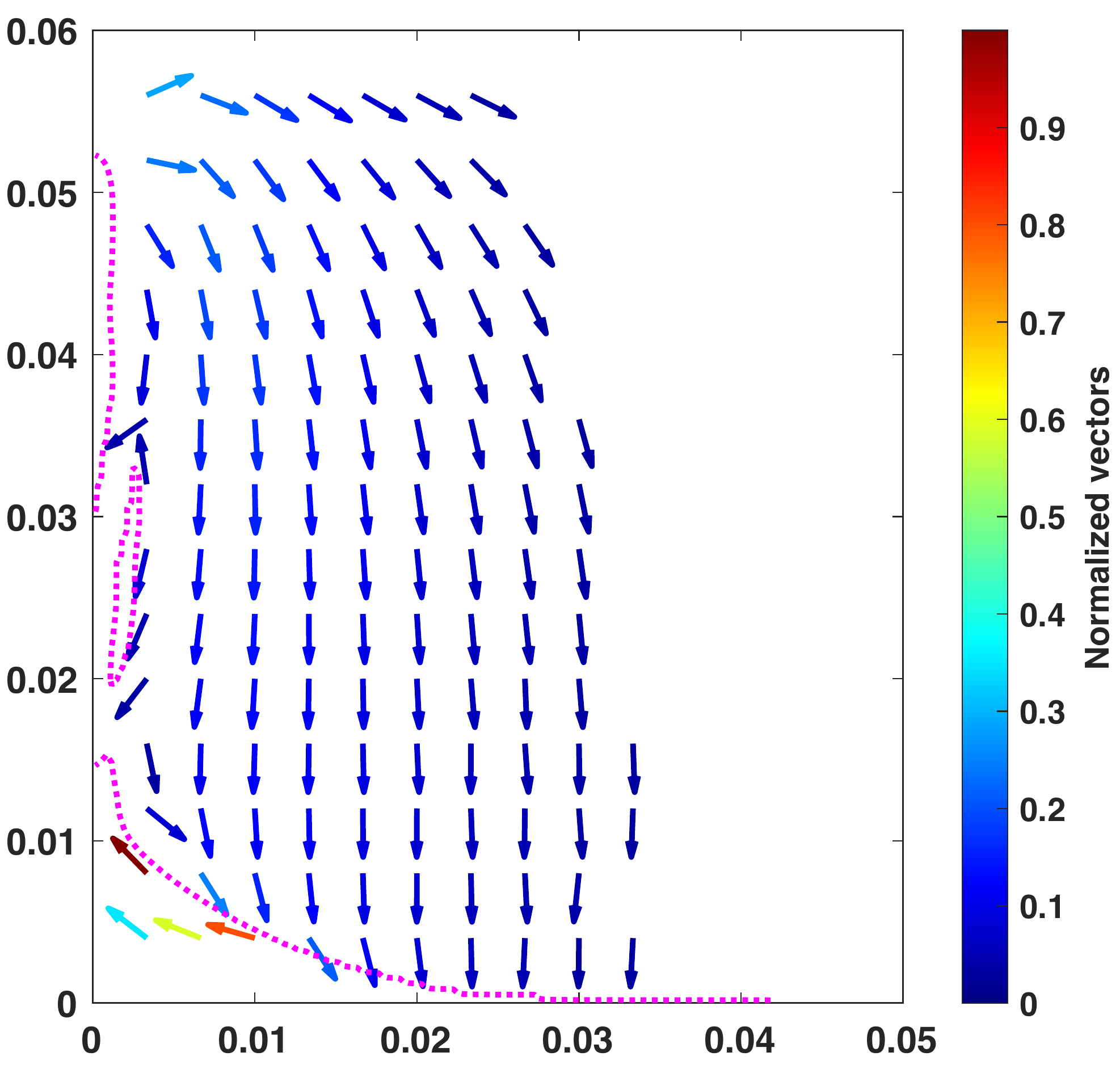}
    \end{minipage} \label{ex_1_color168}
    }
    \centering \subfigure[$t=250$]{
    \begin{minipage}[b]{0.24\textwidth}
    \centering
    \includegraphics[width=1.0\textwidth]{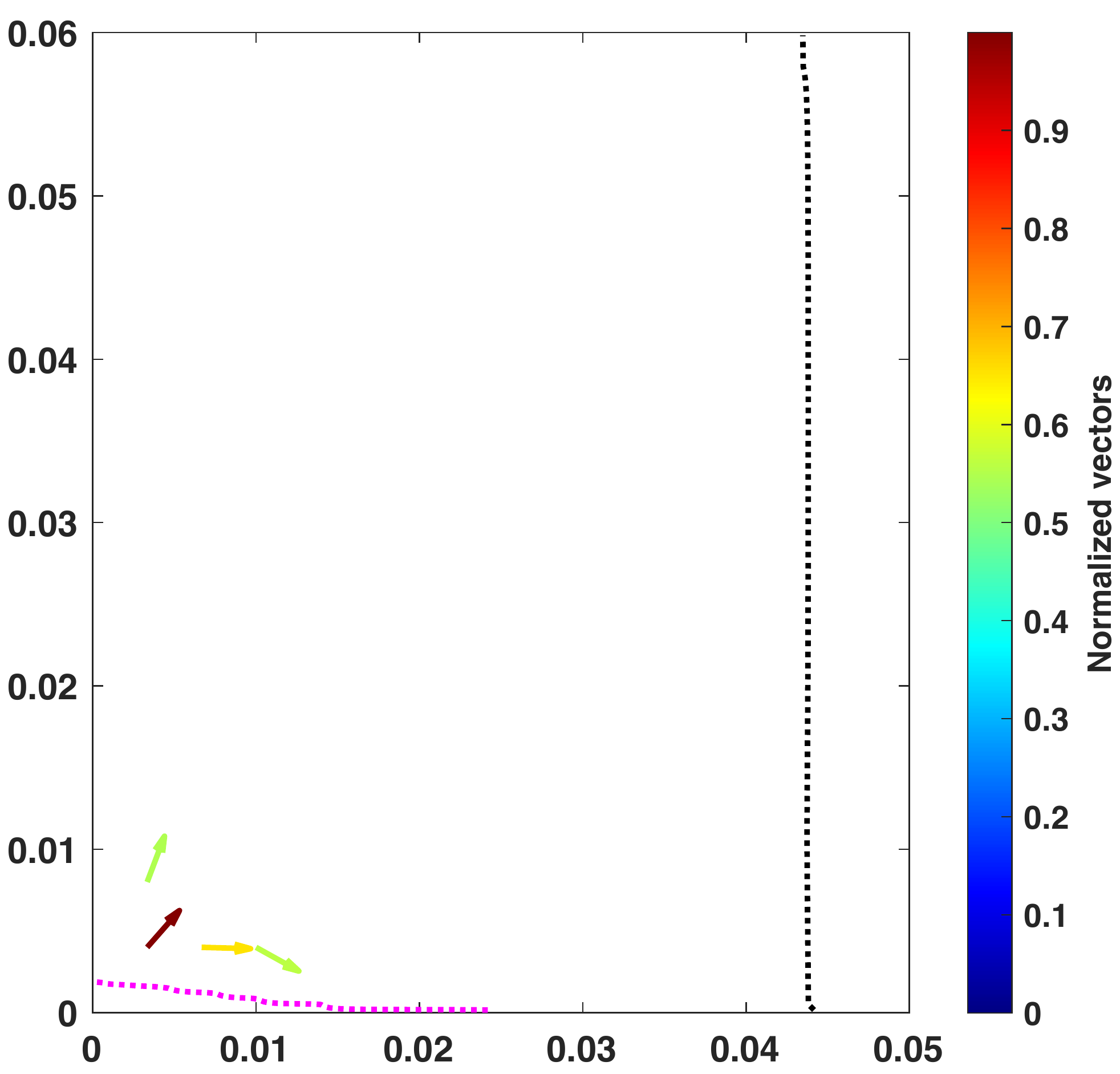}
    \end{minipage} \label{ex_1_color250}
    }
    \centering \subfigure[$t=350$]{
    \begin{minipage}[b]{0.24\textwidth}
    \centering
    \includegraphics[width=1.0\textwidth]{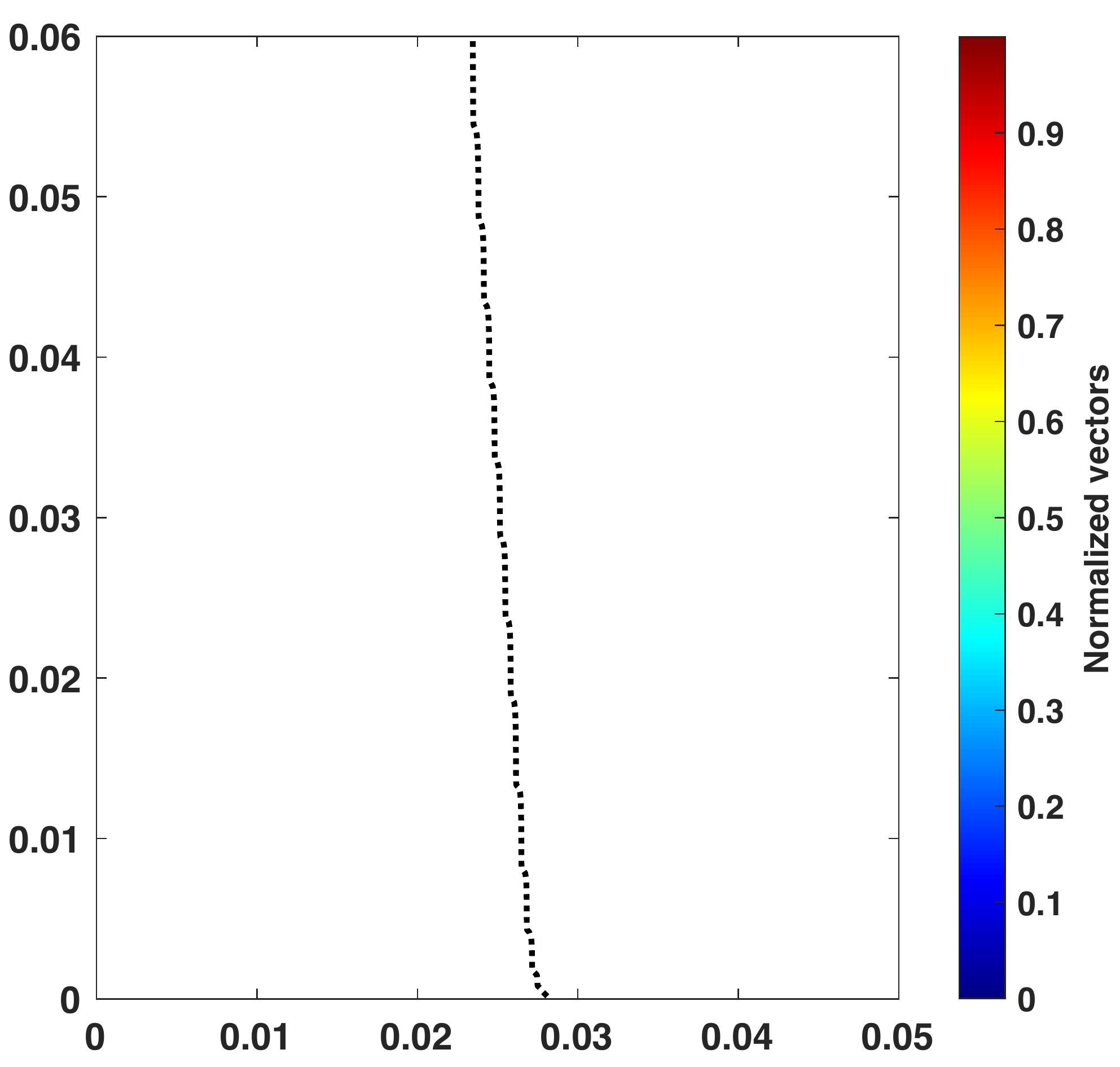}
    \end{minipage} \label{ex_1_color350}
    }
    \caption{Evolution of the velocity field with time in Example 1; the liquid/mush interface (magenta dotted line); the mush/solid interface (black dotted line); magnitude of velocity norm (colorbar)}
    \label{ex_1_color}
 \end{figure*}
 \begin{figure*}[!h]
    \centering \subfigure[$t=5$]{
    \begin{minipage}[b]{0.24\textwidth}
    \centering
    \includegraphics[width=1.0\textwidth]{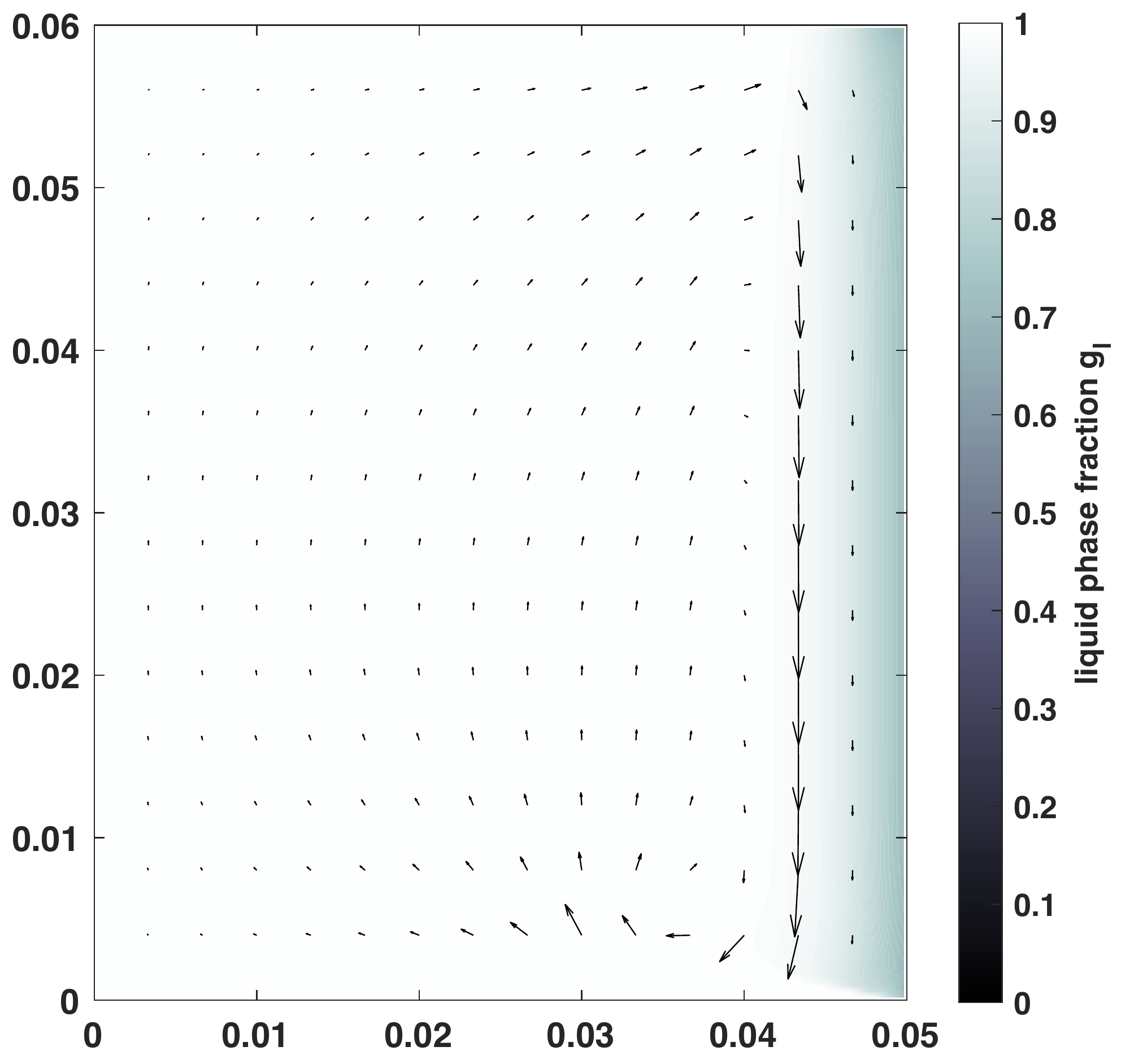}
    \end{minipage}  \label{ex_1_gl5}
    }
    \centering \subfigure[$t=38$]{
    \begin{minipage}[b]{0.24\textwidth}
    \centering
    \includegraphics[width=1.0\textwidth]{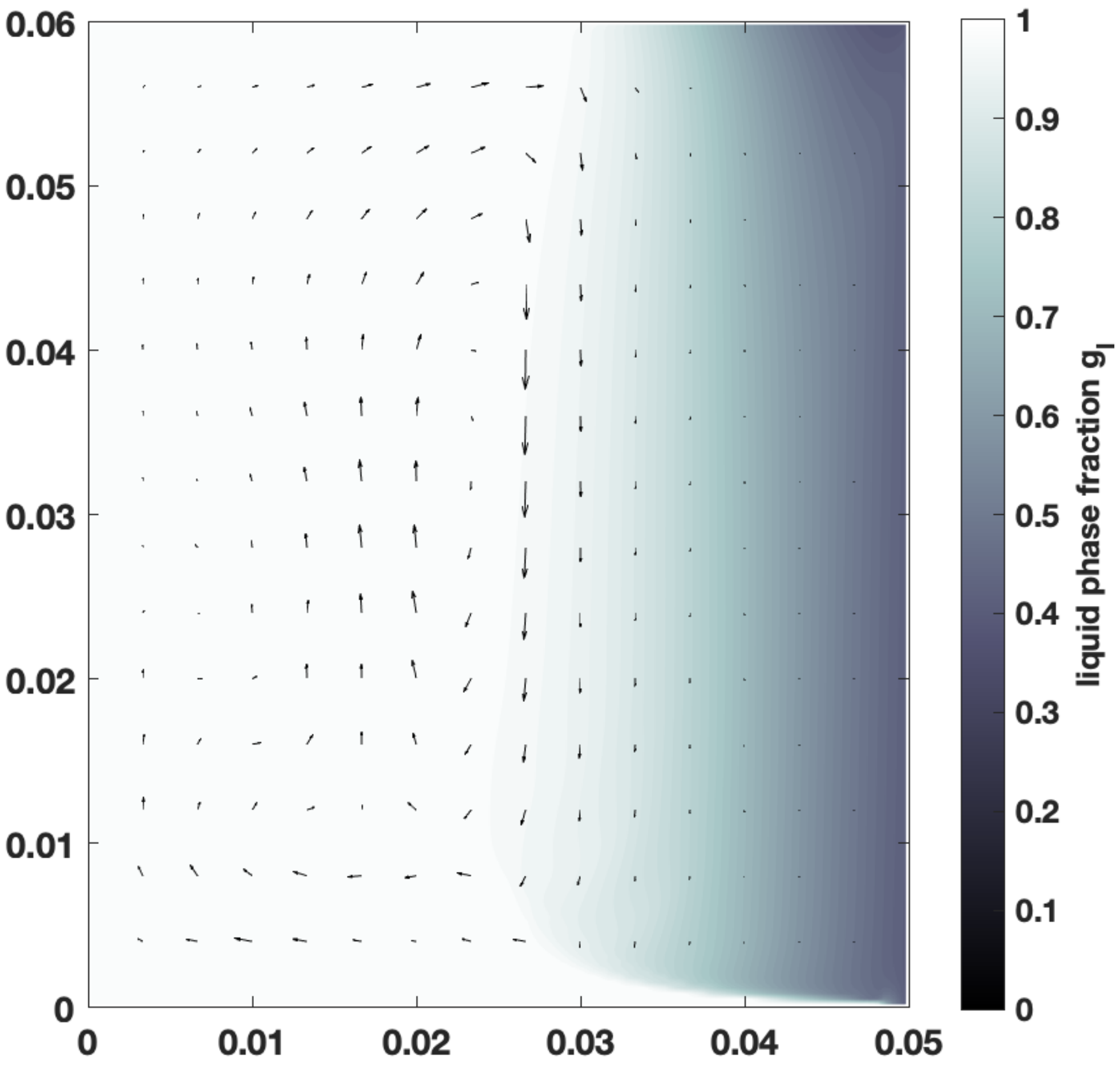}
    \end{minipage}  \label{ex_1_gl38}
    }
    \centering \subfigure[$t=168$]{
    \begin{minipage}[b]{0.24\textwidth}
    \centering
    \includegraphics[width=1.0\textwidth]{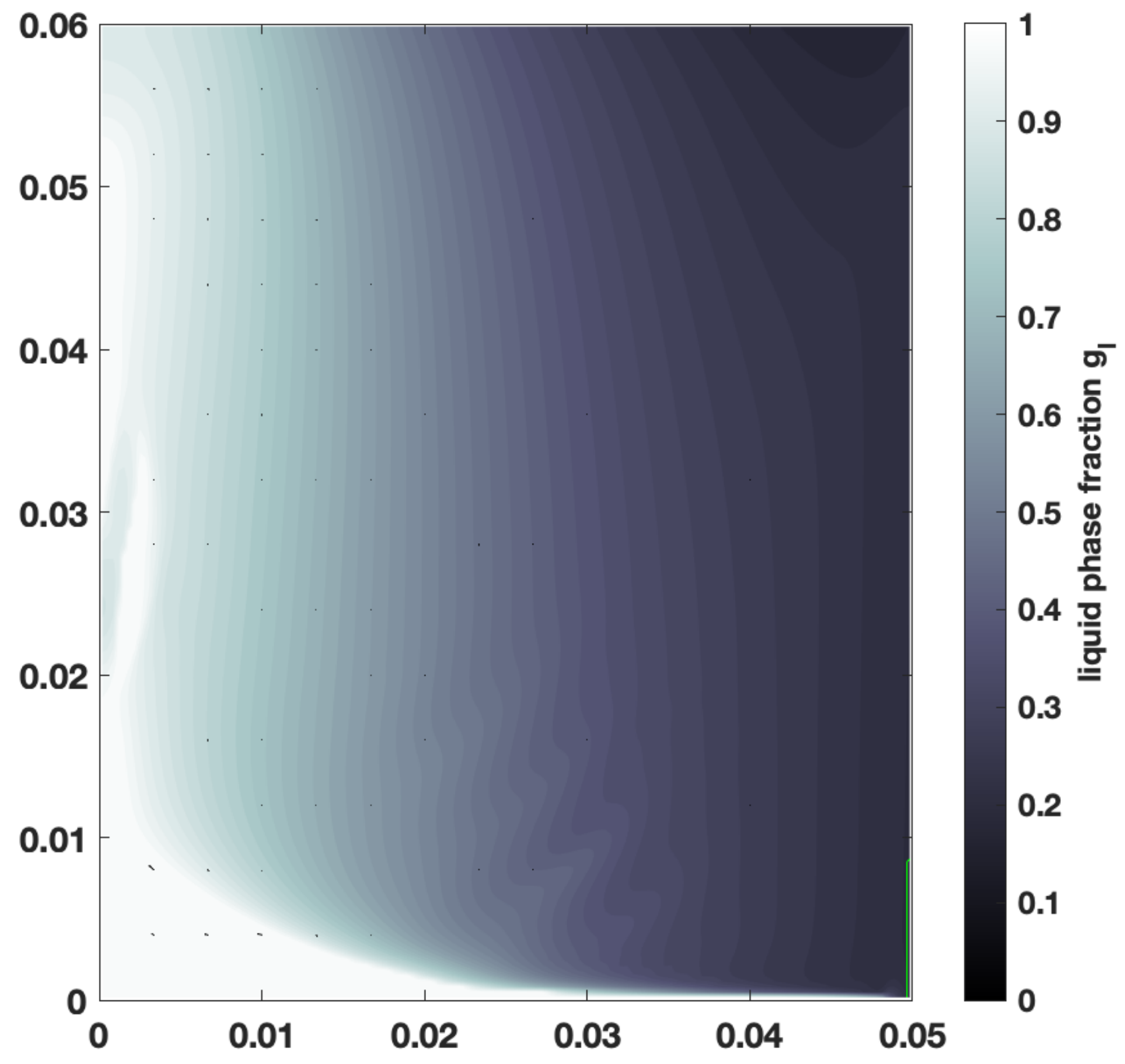}
    \end{minipage}  \label{ex_1_gl168}
    }
    \centering \subfigure[$t=250$]{
    \begin{minipage}[b]{0.24\textwidth}
    \centering
    \includegraphics[width=1.0\textwidth]{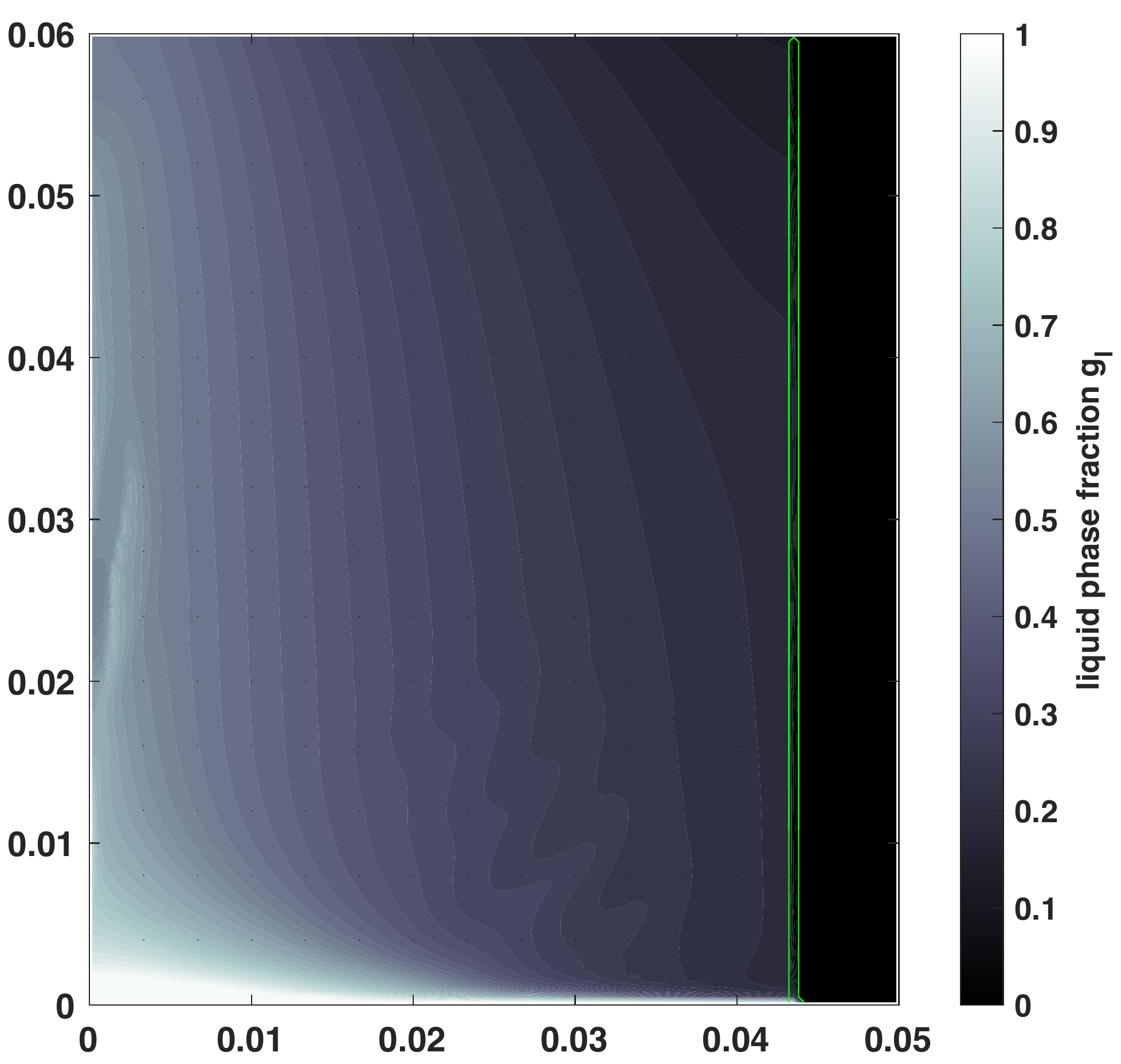}
    \end{minipage}  \label{ex_1_gl250}
    }
    \centering \subfigure[$t=350$]{
    \begin{minipage}[b]{0.24\textwidth}
    \centering
    \includegraphics[width=1.0\textwidth]{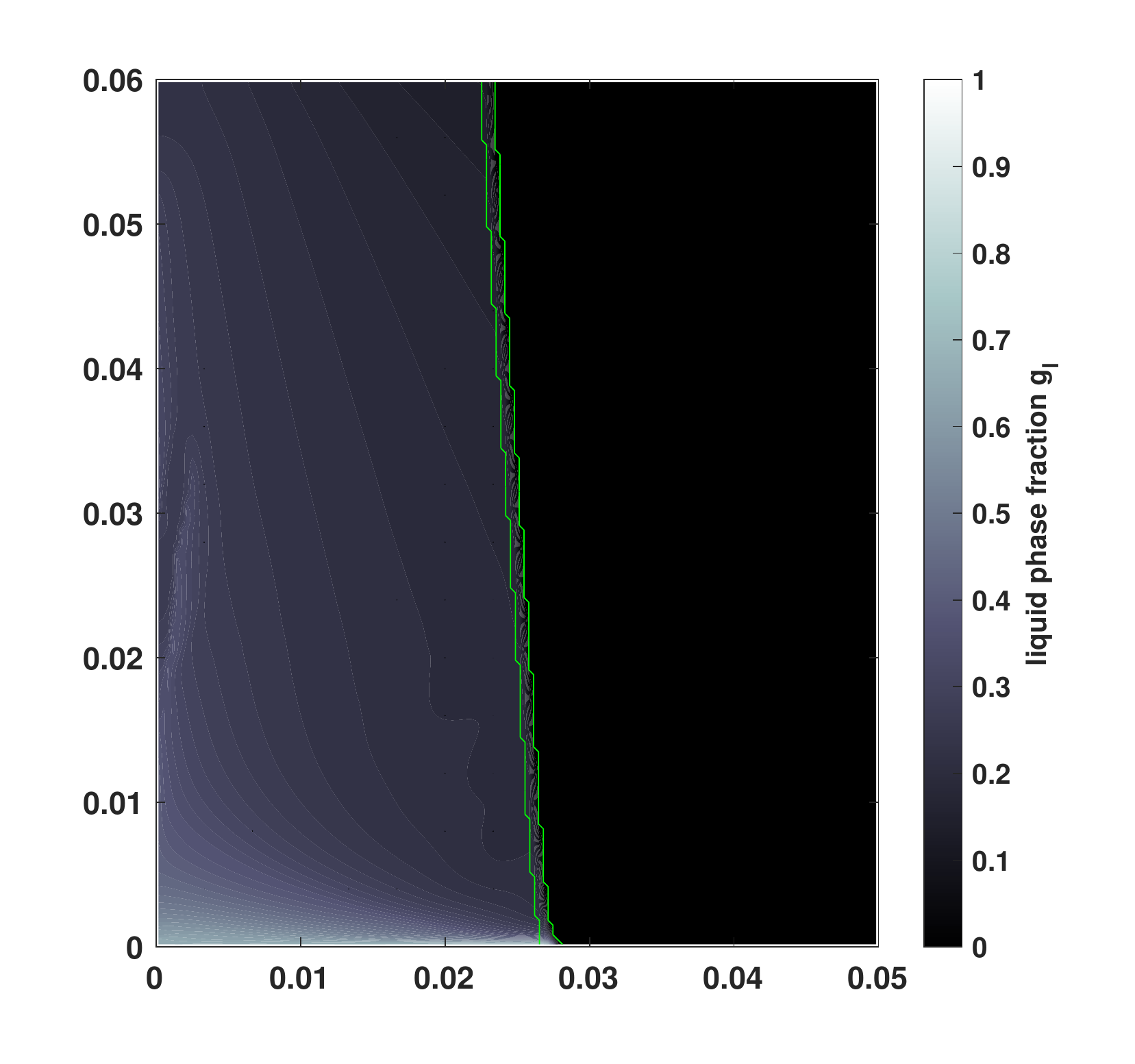}
    \end{minipage}  \label{ex_1_gl350}
    }
    \caption{Evolution of liquid fraction with time in Example 1; the velocity field characterized by quivers with lengths; the iso-thermal phase transition layer (enclosed by green solid line); liquid fraction (colorbar).}
    \label{ex_1_gl}
 \end{figure*}
 \begin{figure*}[!h]
    \centering \subfigure[$t=5$]{
    \begin{minipage}[b]{0.24\textwidth}
    \centering
    \includegraphics[width=1.0\textwidth]{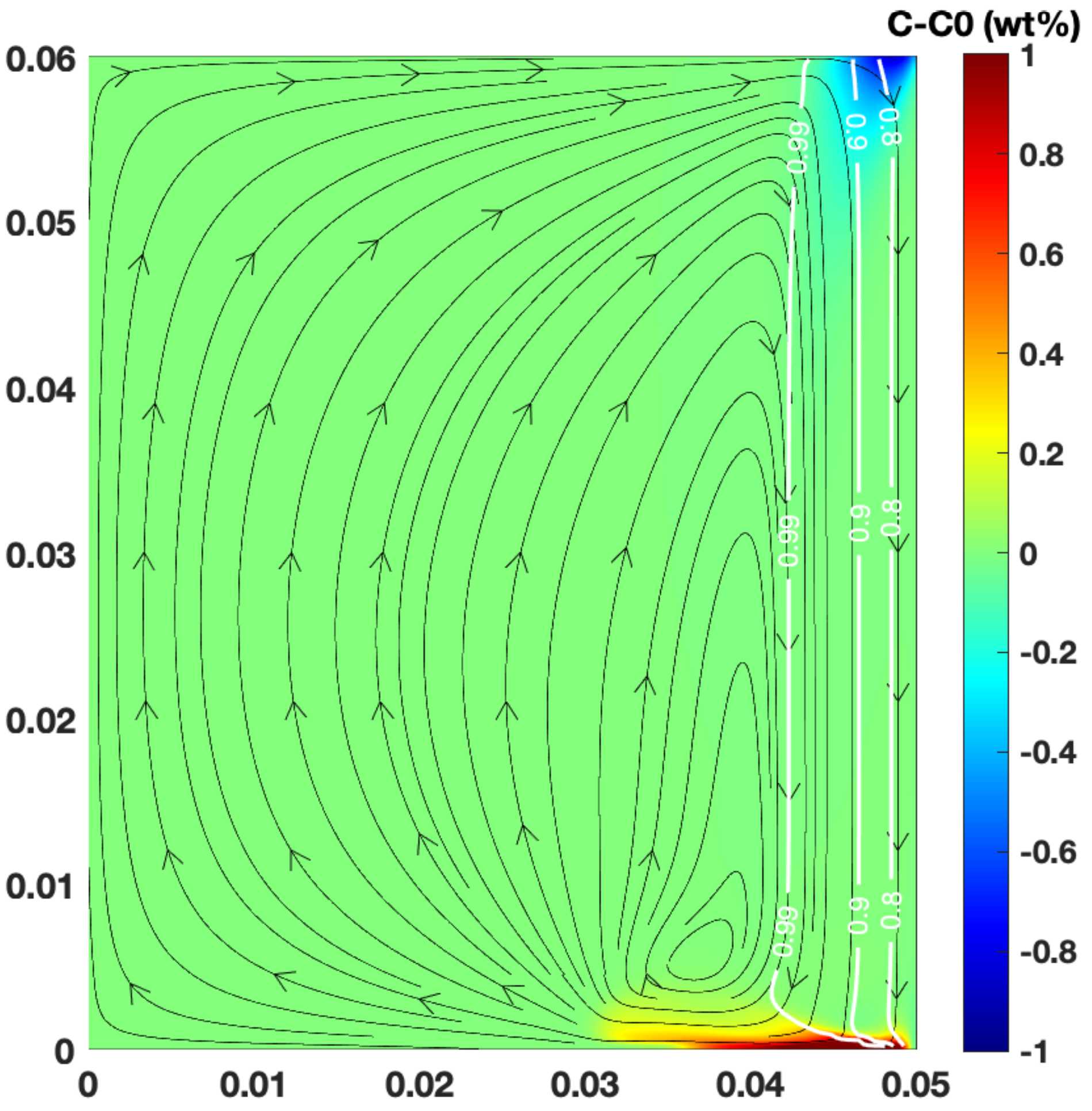}
    \end{minipage}  \label{ex_1_C5}
    }
    \centering \subfigure[$t=38$]{
    \begin{minipage}[b]{0.24\textwidth}
    \centering
    \includegraphics[width=1.0\textwidth]{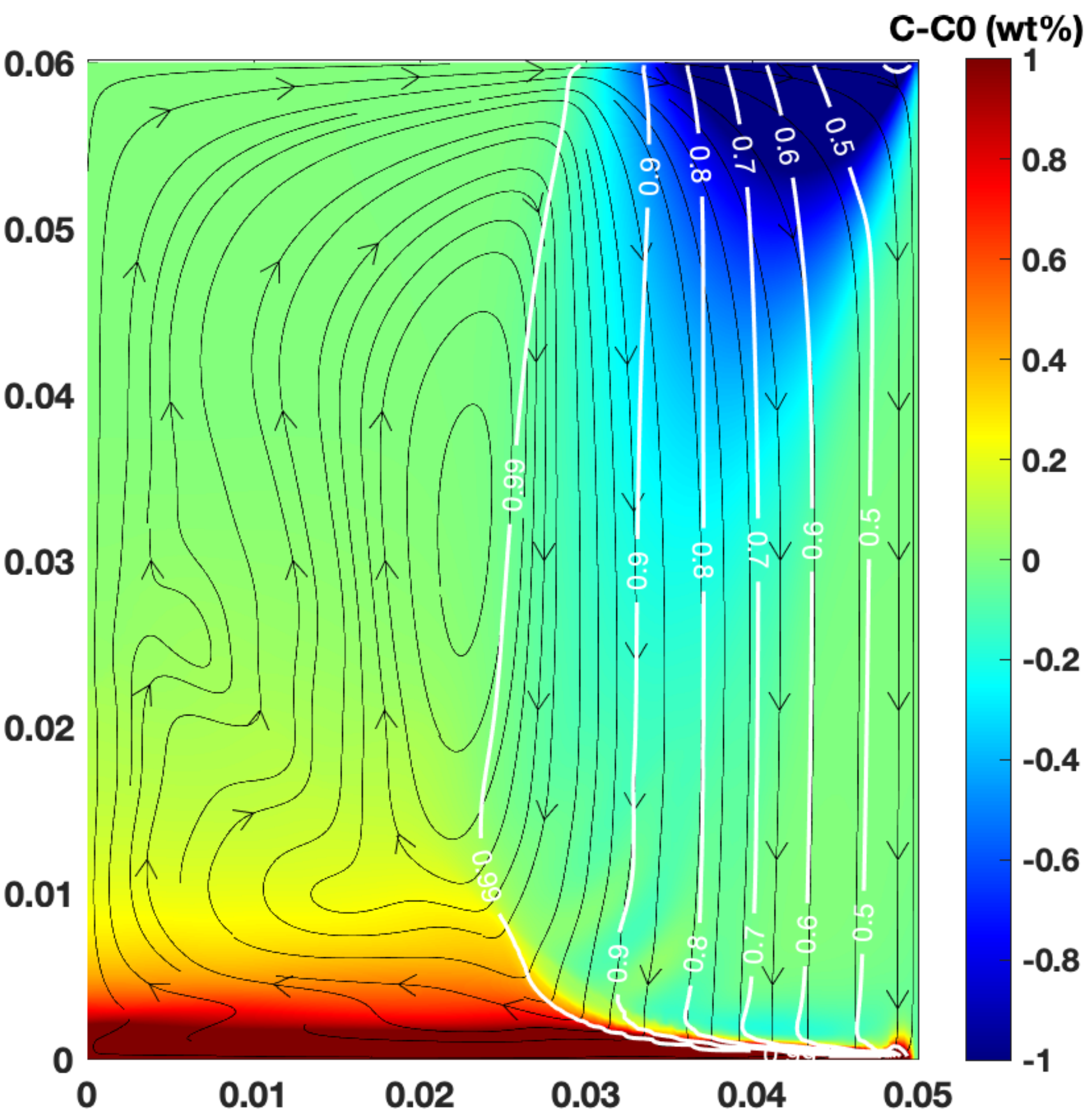}
    \end{minipage}  \label{ex_1_C38}
    }
    \centering \subfigure[$t=168$]{
    \begin{minipage}[b]{0.24\textwidth}
    \centering
    \includegraphics[width=1.0\textwidth]{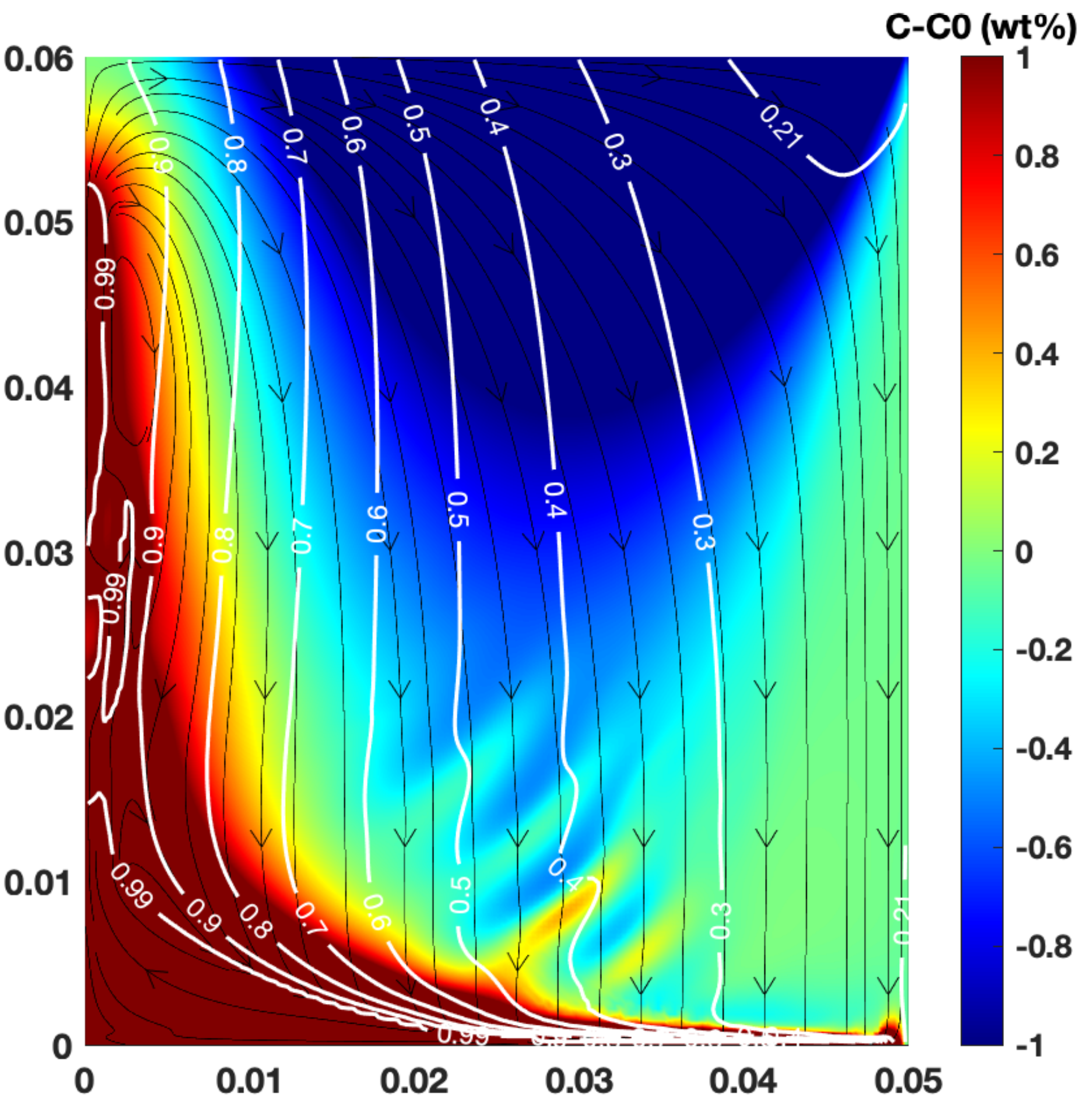}
    \end{minipage} \label{ex_1_C168}
    }
    \centering \subfigure[$t=250$]{
    \begin{minipage}[b]{0.24\textwidth}
    \centering
    \includegraphics[width=1.0\textwidth]{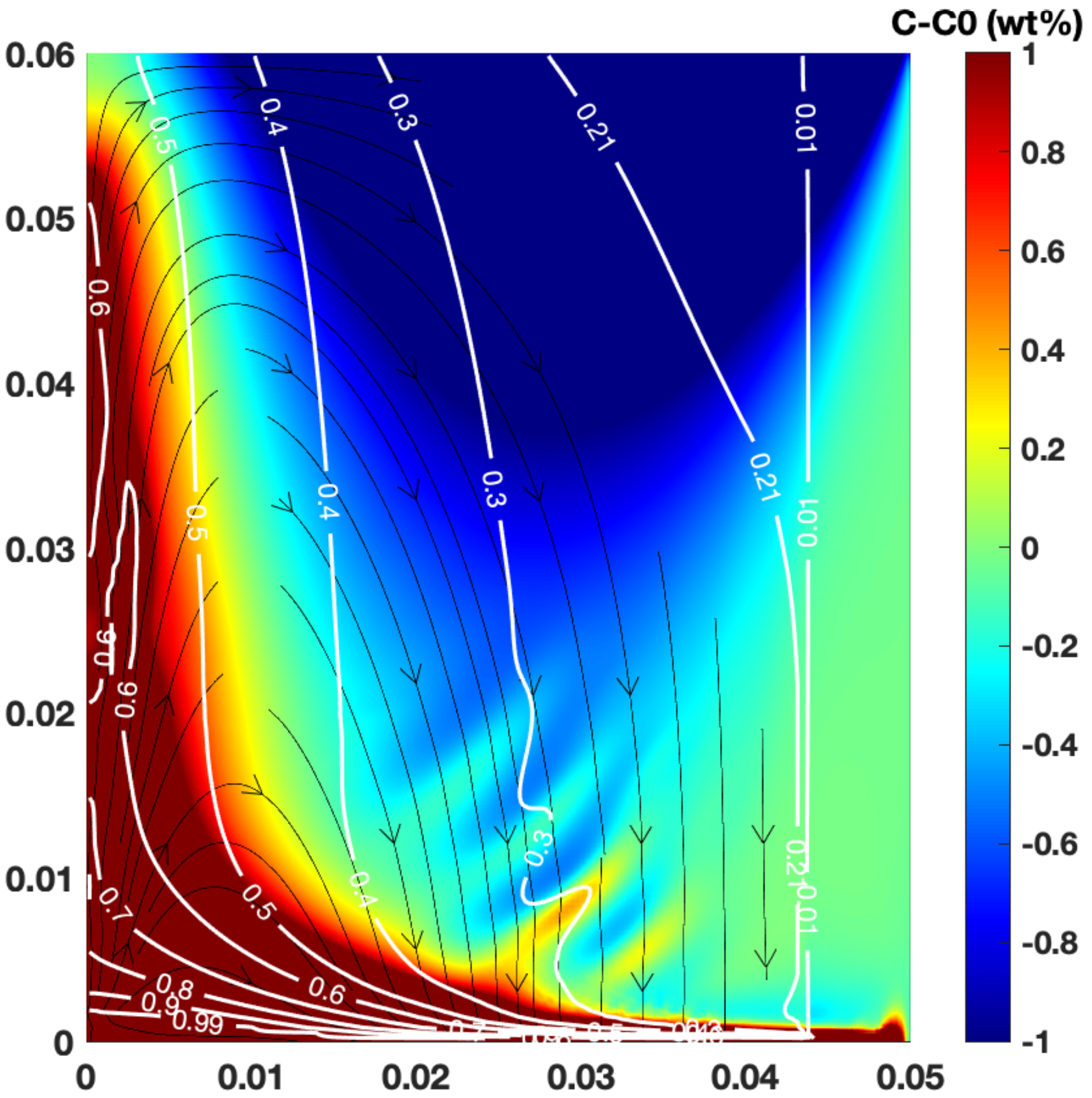}
    \end{minipage}  \label{ex_1_C250}
    }
    \centering \subfigure[$t=350$]{
    \begin{minipage}[b]{0.24\textwidth}
    \centering
    \includegraphics[width=1.0\textwidth]{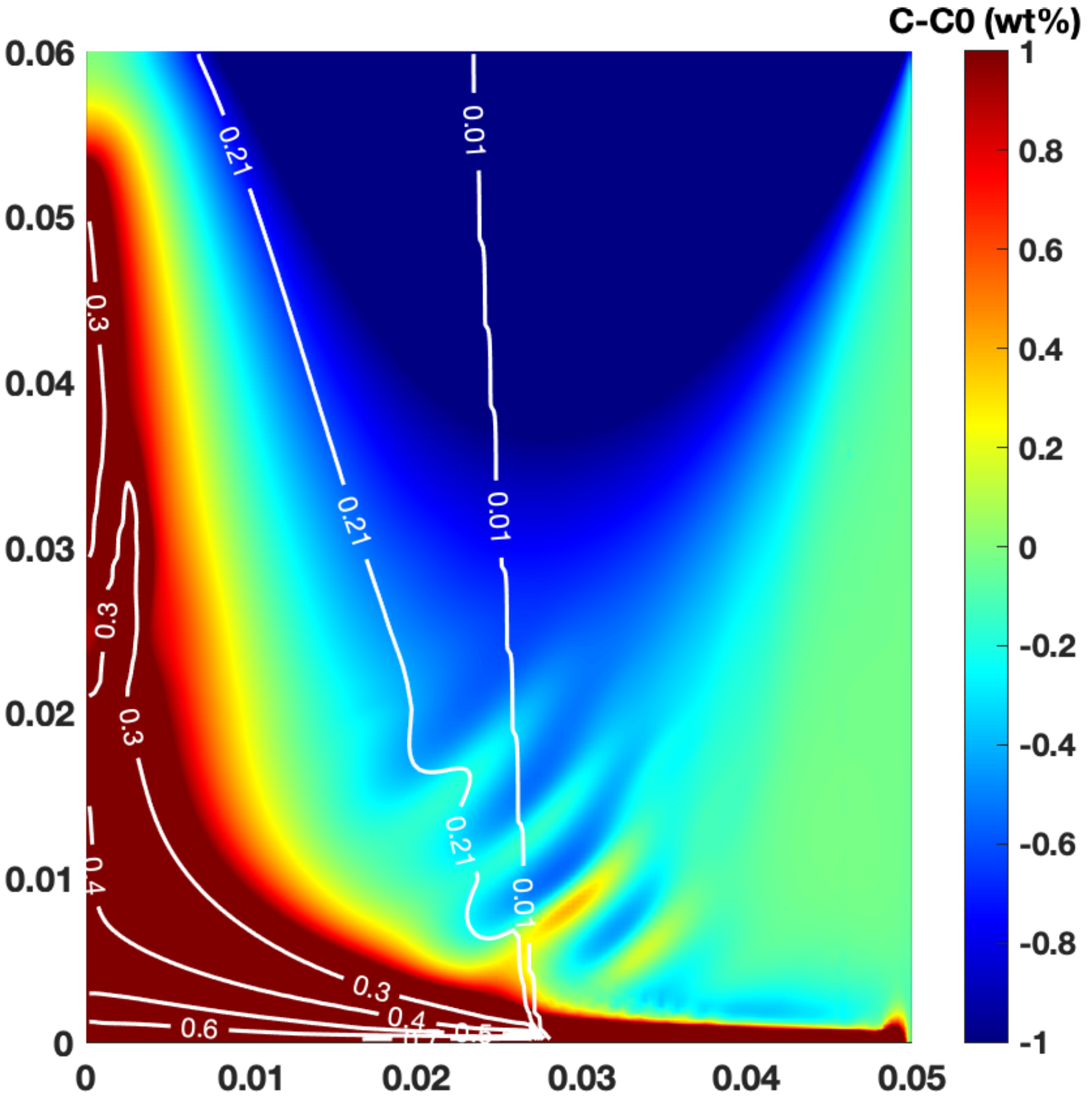}
    \end{minipage}  \label{ex_1_C350} 
    }
    \centering \subfigure[$Shen$ \cite{wensheng2012response}]{
    \begin{minipage}[b]{0.24\textwidth}
    \centering
    \includegraphics[width=1.3\textwidth,height=0.17\textheight]{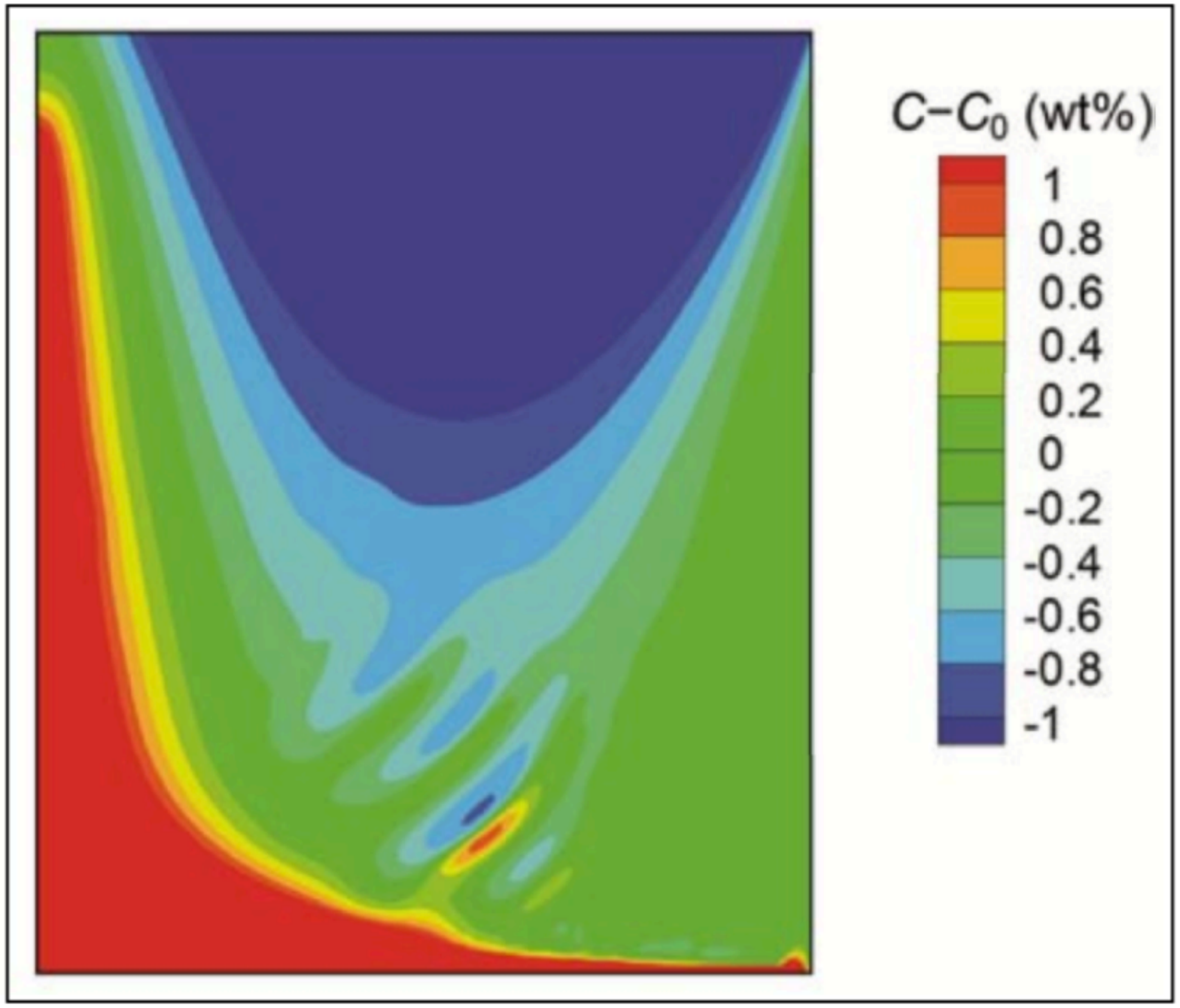}
    \end{minipage}
    }
    \caption{Evolution of concentration variation $C-C_0$ with time in Example 1; the liquid phase fraction contour line (white solid line); the stream line (black solid line with direction arrow); $C-C_0$ (colorbar); final segregation map from \cite{wensheng2012response}}
    \label{ex_1_C}
 \end{figure*}
 
To study the formation of macro-segregation with channel segregation, numerical results are obtained based on the $150 \times 180$ structured mesh and time step $\Delta t = 5\times10^{-3}s$. Figures depict solidification process predictions at time $t = 5s, 38s, 168s, 250s, 350s$ and physical configurations including (1) temperature; (2) concentration; (3) liquid fraction; (4) velocity quivers (normalized or with lengths); (5) streamlines pattern; (6) interfaces between phases (liquid/mush/solid). In general, the contour lines of $g_l=0.99$ and $g_l=0.01$ are thought to be the interfaces between liquid and mush and between mush and solid, respectively.

At the initial stage of the solidification ($t = 5s$ and $t = 38s$), Figures \ref{ex_1_T5} and \ref{ex_1_color5} indicate that thermal buoyancy force dominates, and the higher density of the cooling melt induces the most intense downward flow near the liquid/mush interface. Moreover, it results in a clockwise circulation flow inside the liquid zone, referring to Figures \ref{ex_1_color5} and \ref{ex_1_color38}. The \ch{Pb}-enriched melt is heavier, so the solutal buoyancy enhances the downward flow as well.

At $t = 5s$, the liquid/mush interface almost follow the one temperature contour as it shown in Figure \ref{ex_1_T5}. The composition of \ch{Pb} enriches at the bottom area over time, referring to Figure \ref{ex_1_C38}. Since a higher concentration causes a lower liquidus temperature, it gradually leads to the deviation of the liquid/mush interface from the temperature contour line at $t = 38s$ as shown in Figure \ref{ex_1_T38}.

From time $t = 38s$ to $t = 168s$, by observing Figure \ref{ex_1_C38} and Figure \ref{ex_1_C168}, it can be found that the high compositional concentration of Pb causes the lower corresponding liquidus temperature. It naturally makes the bottom section be the last section to reach the mushy state; thus, severe flow transports the composition Pb fast. This kind of flow and transport generates a platform-shaped Pb-enriched area in the liquid region. The circle flow gradually raises it along the symmetric axis (namely, the vertical centerline of the cavity). 

In addition, the combined effect of species transport and temperature gradient variation causes the bending of the liquid fraction contour lines (Figure \ref{ex_1_gl168}), which results in the occurrence of the channel segregation phenomenon. The entire procedure is analogous to drop-shaped melts embedding in a solid and losing their mobility. The \ch{Pb} enrichment and dilution layers alternate, and the channel segregates at a 45-degree inclined angle, connecting the positive and negative segregation regions. At about $t = 168s$, the channels are well developed. Most of the domain now is in a mushy state except for the left-bottom corner, as shown in Figure \ref{ex_1_color168}. A very thin isothermal phase change layer appears at the right-bottom corner (see Figure \ref{ex_1_C168}), and the local temperature $T=T_e$, indicating that a fully solid region is coming soon.

\begin{figure*}[!h]
\centering
\subfigure[liquid fraction $g_l$ at $E$]{
\begin{minipage}[b]{.45\textwidth}
\includegraphics[width=1\textwidth]{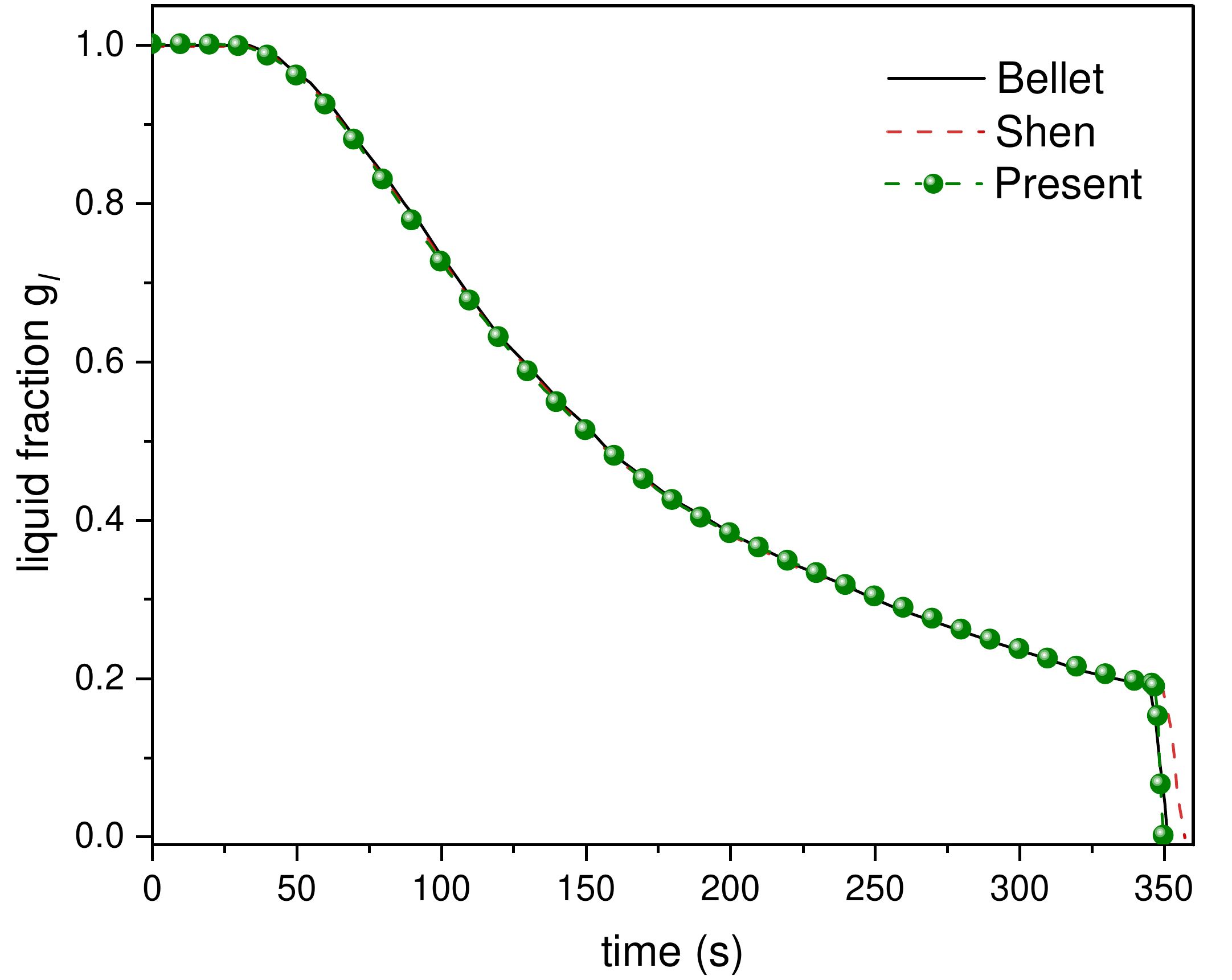}     \label{FIG:glE}
\end{minipage}
}
\subfigure[velocity magnitude at $E$]{
\begin{minipage}[b]{.45\textwidth}
\includegraphics[width=1.\textwidth]{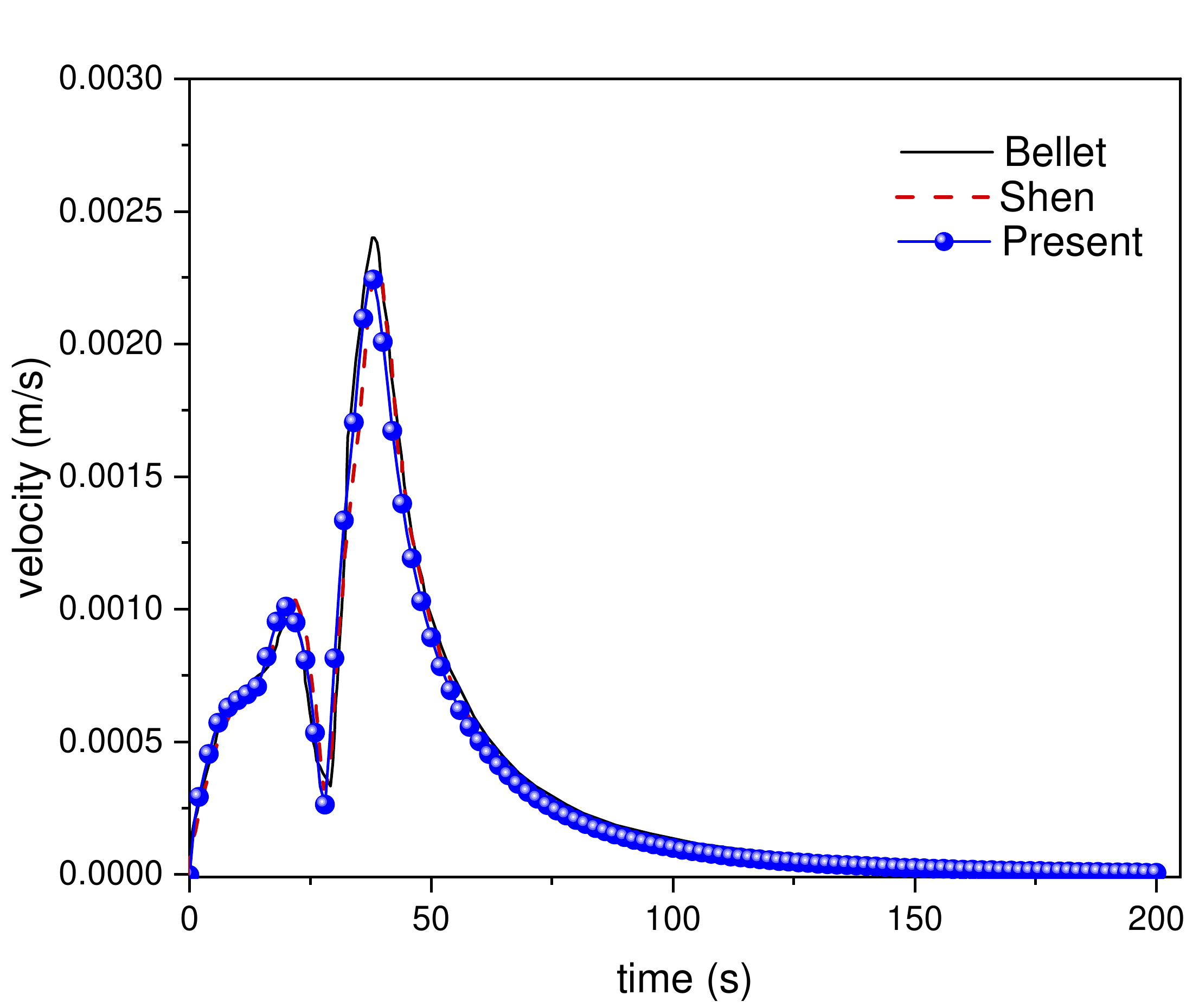}   \label{FIG:velE}
\end{minipage}
}
\caption{(a) The time-dependent profile of liquid fraction $g_l$ at central sample point $E$; (b) the time-dependent profile of velocity magnitude at central sample point $E$.}
\end{figure*}

With the mush/solid interface pushing towards the regional center, we can find the evolution of the isothermal phase change layer in Figures \ref{ex_1_gl250} and Figure \ref{ex_1_gl350}. The fully solid front reaches the central point $E$ at about $t = 350s$. The correctness of this observation can be validated as well via the ending point of the time-dependent profile of the liquid fraction in Figure \ref{FIG:glE}. No more noticeable changes can be found in the concentration distribution after $t = 350s$ until the entire domain finishes the solidification process at about $t = 450s$. The channel segregation configuration is nearly identical to Shen's reference results \cite{wensheng2012response}. 

\begin{figure*}[!h]
\centering
\subfigure[max$(C-C_0)$]{
\begin{minipage}[b]{0.45\textwidth}
\includegraphics[width=1\textwidth]{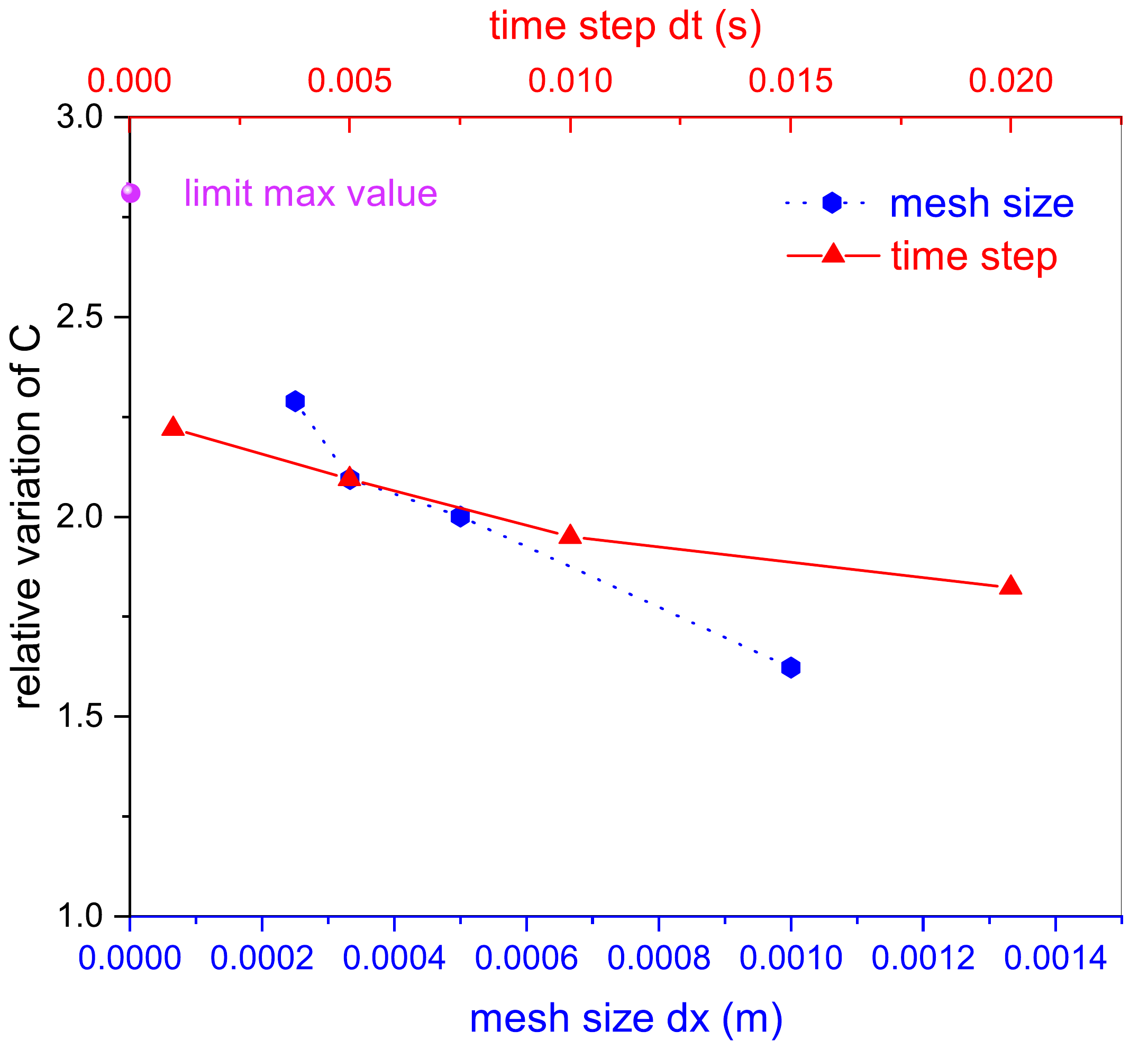}     \label{FIG:maxC}
\end{minipage}
}
\subfigure[min$(C-C_0)$]{
\begin{minipage}[b]{0.45\textwidth}
\includegraphics[width=1.\textwidth]{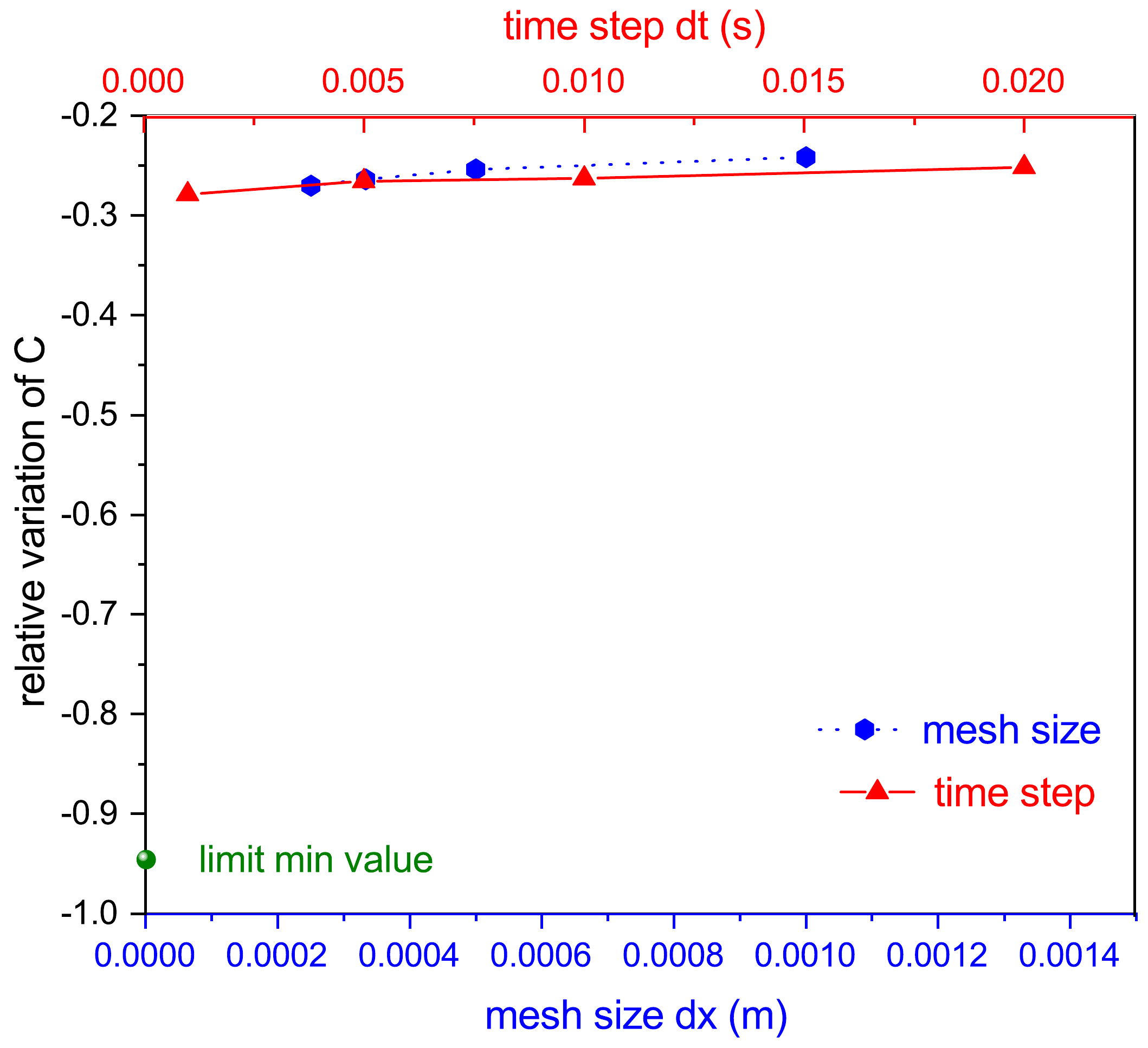}   \label{FIG:minC}
\end{minipage}
}
\caption{The sensitivity plot of the mesh size and the time step to (a) the maximum concentration and (b) the minimum concentration in the field at the end of solidification}
\end{figure*}

The curve in Figure \ref{FIG:glE} depicts the time-dependent profile of the liquid fraction at central point $E$. The results agree fairly well with the reference result calculated by Bellet \cite{combeau2011numerical} and Shen \cite{wensheng2012response}. The time when the liquid fraction profile line suddenly drops corresponds to a local phase transition at point $E$ in an isothermal process, which is when the large amount of latent heat is released. It confirms that our scheme is capable of capturing this important property. The time is about $ t = 350s$ when the point $E$ turns into a fully solid point, which is also almost the same as other benchmark predictions.

Figure \ref{FIG:velE} exhibits the change of velocity magnitude of point $E$, noteworthy is that our prediction is in good agreement as well. They have the same tendency: two peaks and one local minimum. Compared with the time-dependent profile of $g_l$, it can be found that the second peak (the higher one) corresponds to the turning point from liquid to mush. The velocity reaches its maximum due to the work of the buoyancy-driven forces, and the point $E$ is now near the mushy front with the most intense flow. Surprisingly, the magnitude of velocity rises and falls for the first time when the vortex crosses the point $E$; the minimum occurs when the point $E$ is in the center of the vortex.

The mesh and time step sensitivity are also investigated through our scheme, referring to Figures \ref{FIG:maxC} and \ref{FIG:minC}. With the refinement of time step and mesh size, the maximum concentration and minimum concentration in the numerical simulation are approaching theoretical limits. The sensitivity plot is similar to the tendency from \cite{combeau2011numerical}. It can be concluded that appropriate mesh size and time step ensure the concentration varies within a physically reasonable bound range. 

\newpage

\textbf{Example 2: Formation of the freckle in alloy}

\begin{figure}[!h]\centering
   \begin{minipage}{0.49\textwidth}
     \centering
     \frame{\includegraphics[width=0.8\linewidth]{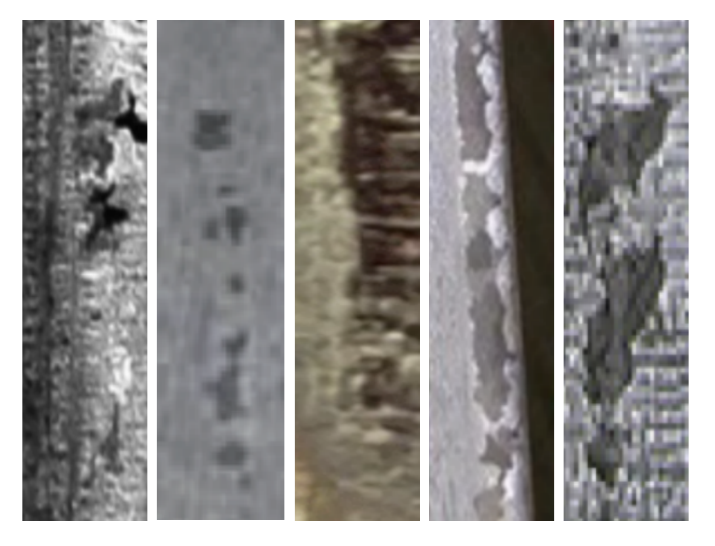}}
     \caption{Freckle defects in alloy} 
     \label{FIG:freckles}
   \end{minipage}
   \begin {minipage}{0.49\textwidth}
     \centering
     \includegraphics[width=1.2\linewidth]{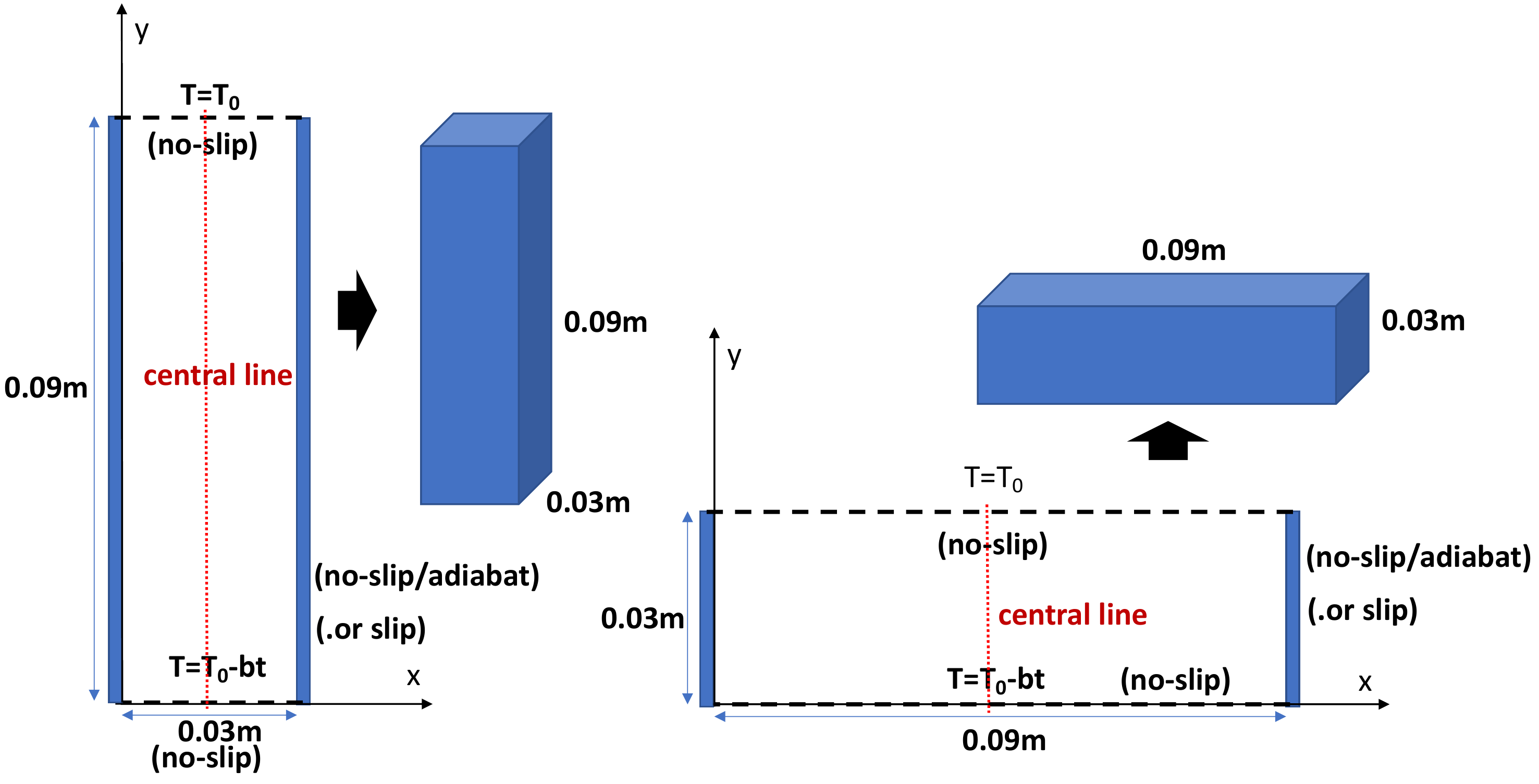} 
     \caption{Physical settings in Example 2}
     \label{FIG:settings2} 
   \end{minipage}
\end{figure}

In this part, we model another phenomenon of the solidification of a binary alloy: the formation of freckles. Also, the aspect ratio effect of the casting space will be investigated. It is acknowledged that freckles are developed by thermo-solutal convection, which is driven by density inversion in the mushy region. The interdendritic liquid in the formed channel tends to ``jet" upward. This type of undesirable macro-segregation, if not properly controlled, can pose a risk to engineering projects and inferior components, as shown in Figure \ref{FIG:freckles}. The direct simulation provides a possible way to predict the formation and explore the physical details of this non-transparent alloy system. The ``aspect ratio effect'' and the ``edge effect'' are studied as well through different 2D and 3D simulation cases.

Different from the boundary condition of Example 1, the physical settings here are shown in Figure \ref{FIG:settings2}, where bottom-up cooling is provided by a wall temperature that changes linearly with time, obeying $T(t)=T_0-b t$ ($T_0 = 216^{\circ} \mathrm{C}$ and $b=0.05^{\circ} \mathrm{C} / \mathrm{s}$). The material is binary alloy \ch{Pb}-48\ch{\%Sn} and the numerical experiments are conducted in rectangle cavities with different aspect ratios and different boundary conditions, including free slip and non-slip.

The material is the binary alloy \ch{Pb}-48\ch{\%Sn} and the numerical experiments are conducted in rectangle cavities with different aspect ratios and different boundary conditions, including free slip and non-slip. We define the aspect ratio $R$ as length/height in 2D and length/width/height in 3D. Initially, a perturbation of concentration $C = 1.0001C_0$ is posed at the central line to generate the density inversion, where $C_0$ is the initial uniform concentration. The constant temperature $T_0$ is maintained by the top boundary. For 2D examples with $R = 1:3$ and $R = 3:1$, the computations are performed on structured meshes of $30 \times 90$ and $90 \times 30$, respectively. The meshes of $30 \times 30 \times 90$ and $90 \times 30 \times 30$ is set for 3D examples with $R = 1:1:3$ and $R = 3:1:1$. A time step $\Delta t=0.01$ is used for all. 

 \begin{figure*}[!t]
    \centering \subfigure[t=5s]{
    \begin{minipage}[b]{0.17\textwidth}
    \centering
    \includegraphics[width=1.1\textwidth]{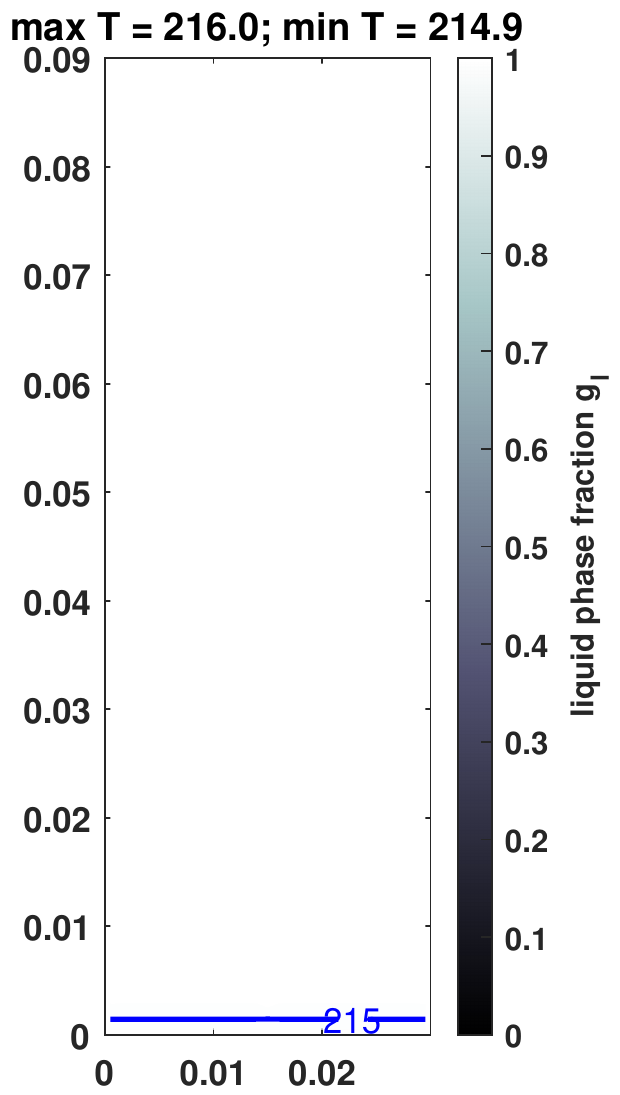}
    \end{minipage}
    }
    \centering \subfigure[t=50s]{
    \begin{minipage}[b]{0.17\textwidth}
    \centering
    \includegraphics[width=1.1\textwidth]{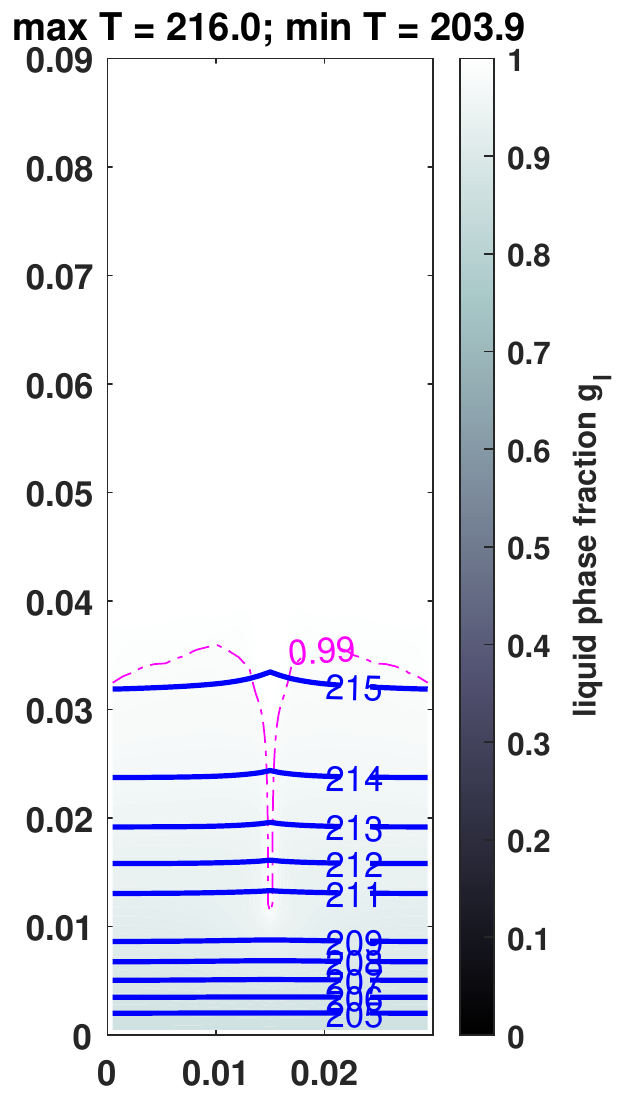}
    \end{minipage}
    }
    \centering \subfigure[t=100s]{
    \begin{minipage}[b]{0.17\textwidth}
    \centering
    \includegraphics[width=1.1\textwidth]{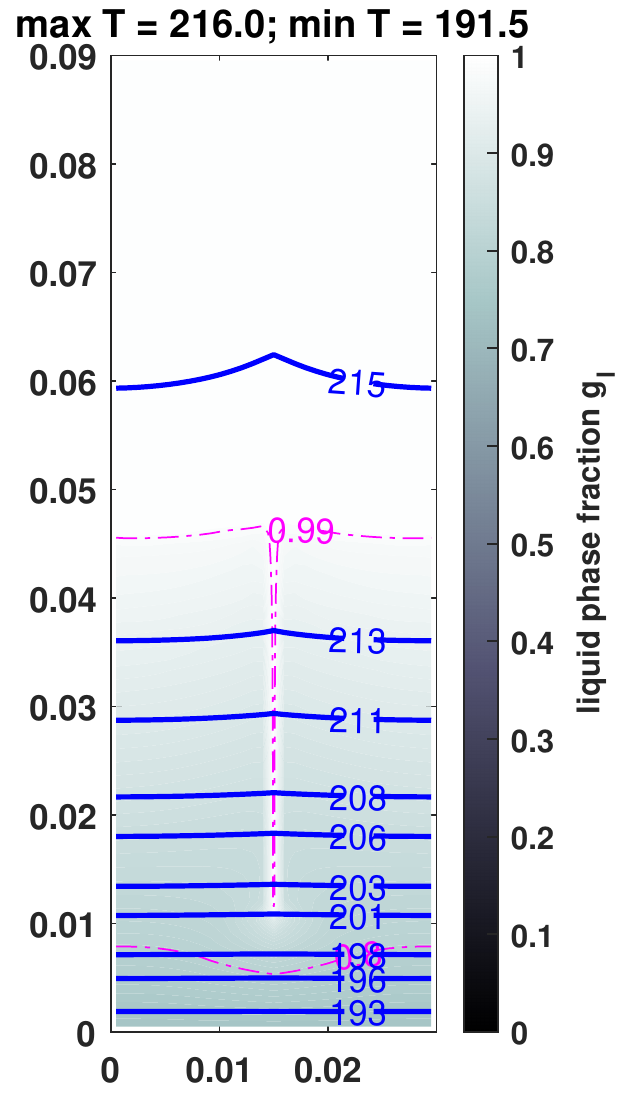}
    \end{minipage}
    }
    \centering \subfigure[t=150s]{
    \begin{minipage}[b]{0.17\textwidth}
    \centering
    \includegraphics[width=1.1\textwidth]{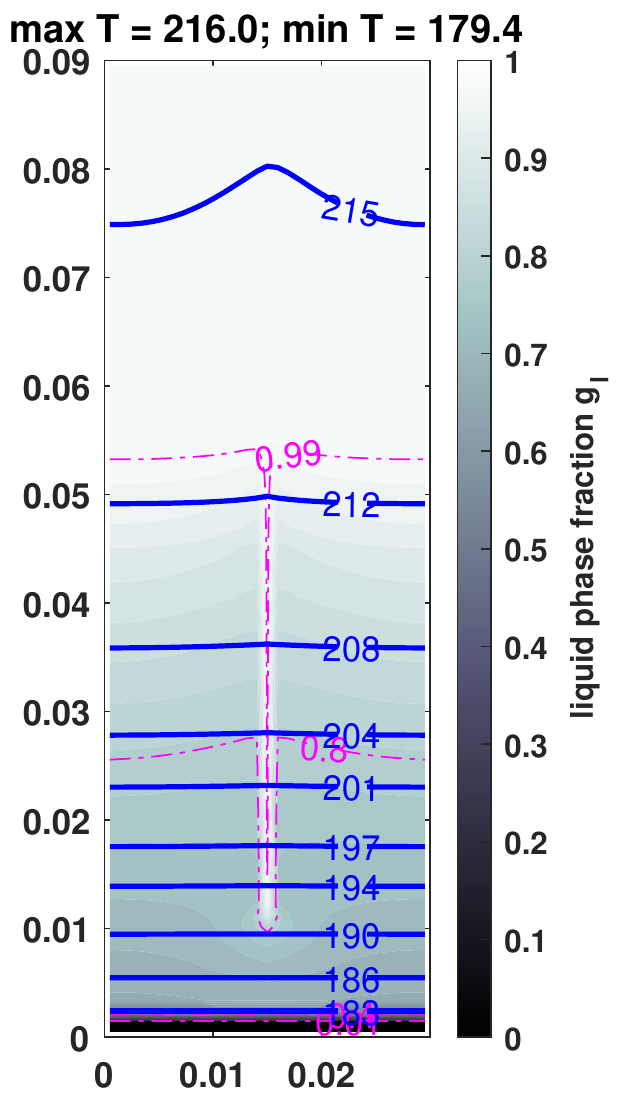}
    \end{minipage}
    }
    \centering \subfigure[t=200s]{
    \begin{minipage}[b]{0.17\textwidth}
    \centering
    \includegraphics[width=1.1\textwidth]{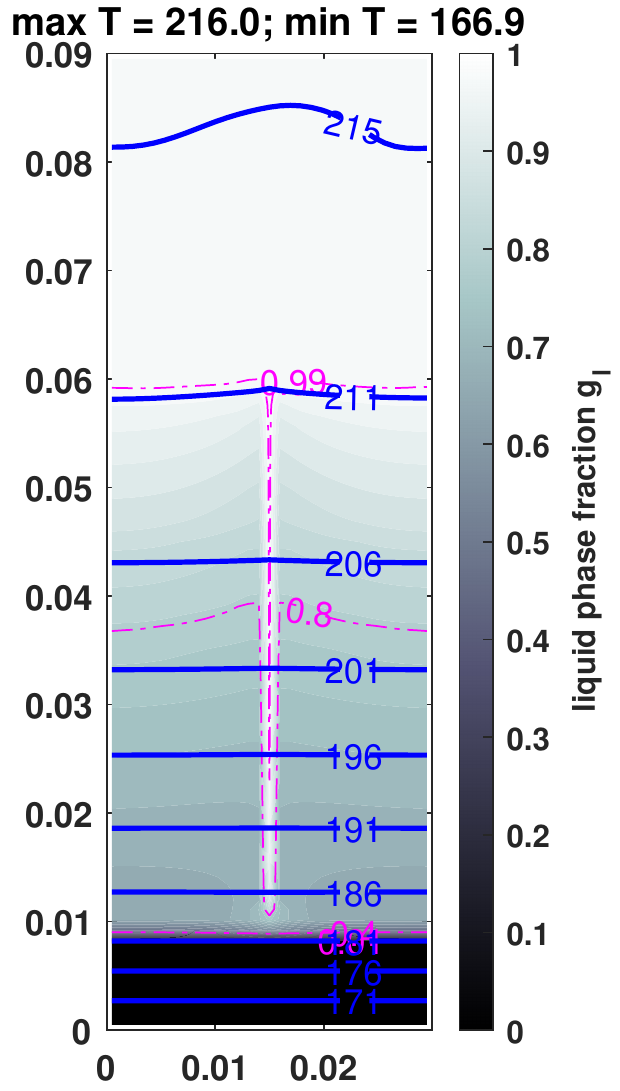}
    \end{minipage}
    }
    \caption{Evolution of liquid fraction in Example 2 with $R=1:3$; the temperature contour line (blue solid lines); the liquid fraction contour line (magenta dash-dotted line); liquid fraction (colorbar)}
    \label{example2_T}
 \end{figure*}
  \begin{figure*}[!b]
    \centering \subfigure[t=5s]{
    \begin{minipage}[b]{0.17\textwidth}
    \centering
    \includegraphics[width=0.8\textwidth]{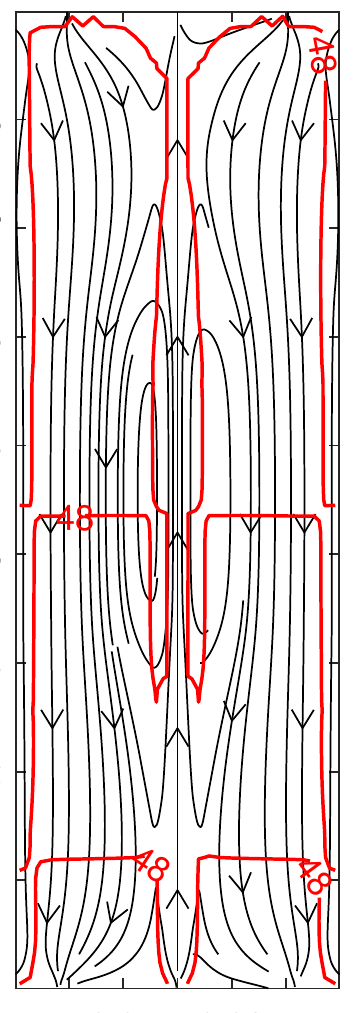}
    \end{minipage}
    }
    \centering \subfigure[t=50s]{
    \begin{minipage}[b]{0.17\textwidth}
    \centering
    \includegraphics[width=0.8\textwidth]{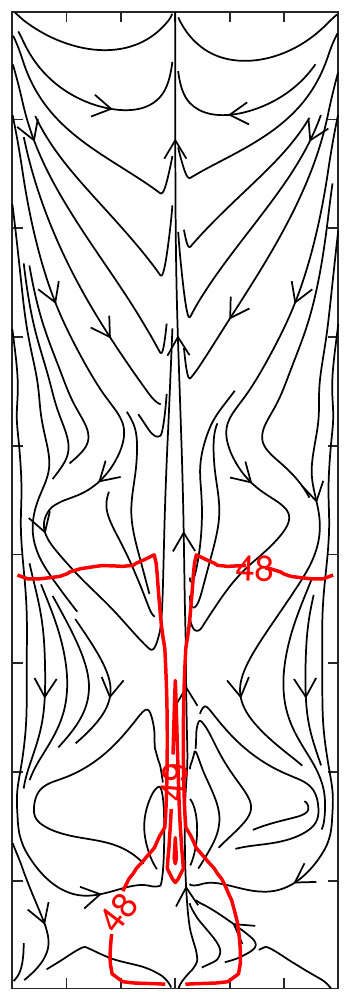}
    \end{minipage}
    }
    \centering \subfigure[t=100s]{
    \begin{minipage}[b]{0.17\textwidth}
    \centering
    \includegraphics[width=0.8\textwidth]{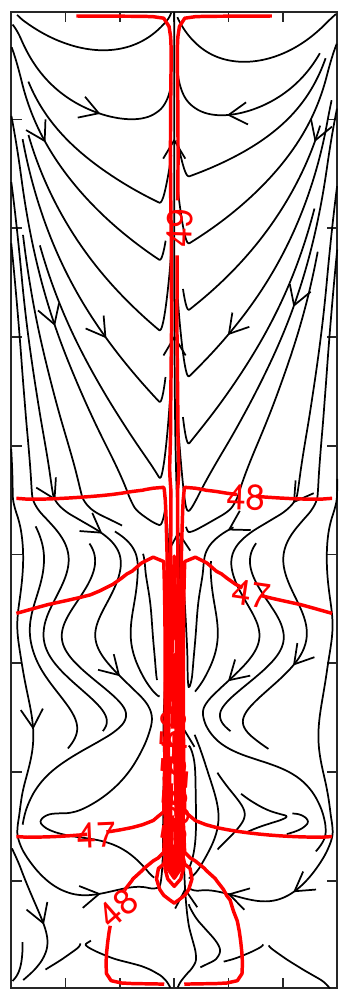}
    \end{minipage}
    }
    \centering \subfigure[t=150s]{
    \begin{minipage}[b]{0.17\textwidth}
    \centering
    \includegraphics[width=0.8\textwidth]{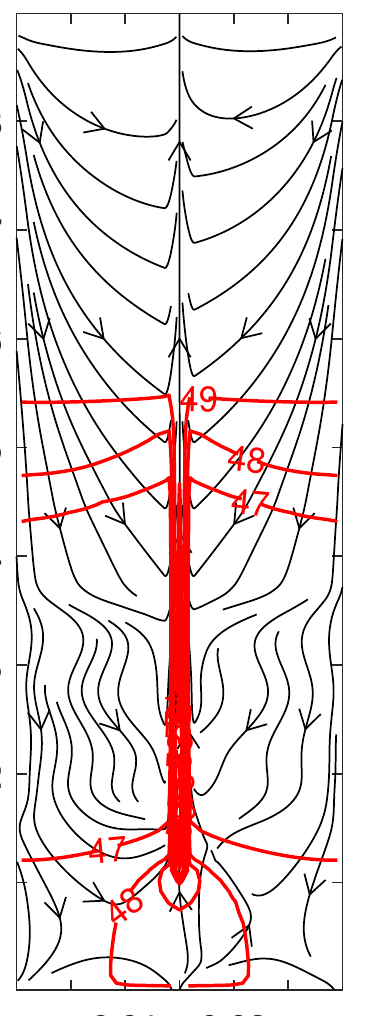}
    \end{minipage}
    }
    \centering \subfigure[t=200s]{
    \begin{minipage}[b]{0.17\textwidth}
    \centering
    \includegraphics[width=0.8\textwidth]{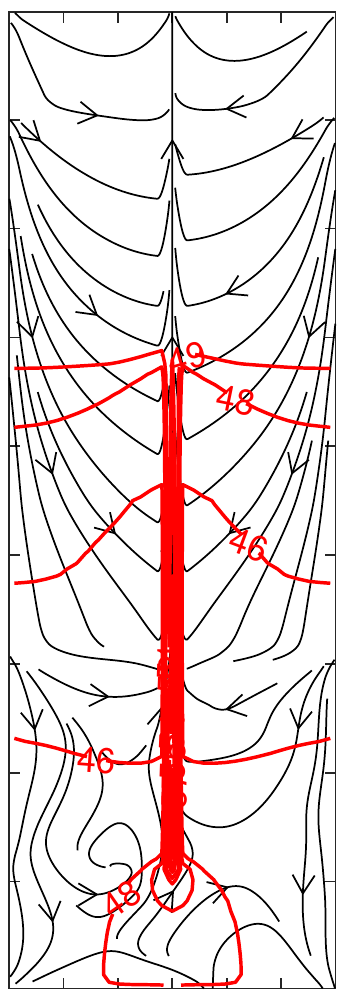}
    \end{minipage}
    }
    \caption{Evolution of stream lines in Example 2 with $R=1:3$; the stream line (black solid line with direction arrow); the concentration contour line (red solid line)}
    \label{example2_S}
 \end{figure*}
\begin{figure*}[!h]
    \centering \subfigure[$t=5s$]{
    \begin{minipage}[b]{0.17\textwidth}
    \centering
    \includegraphics[width=1.1\textwidth]{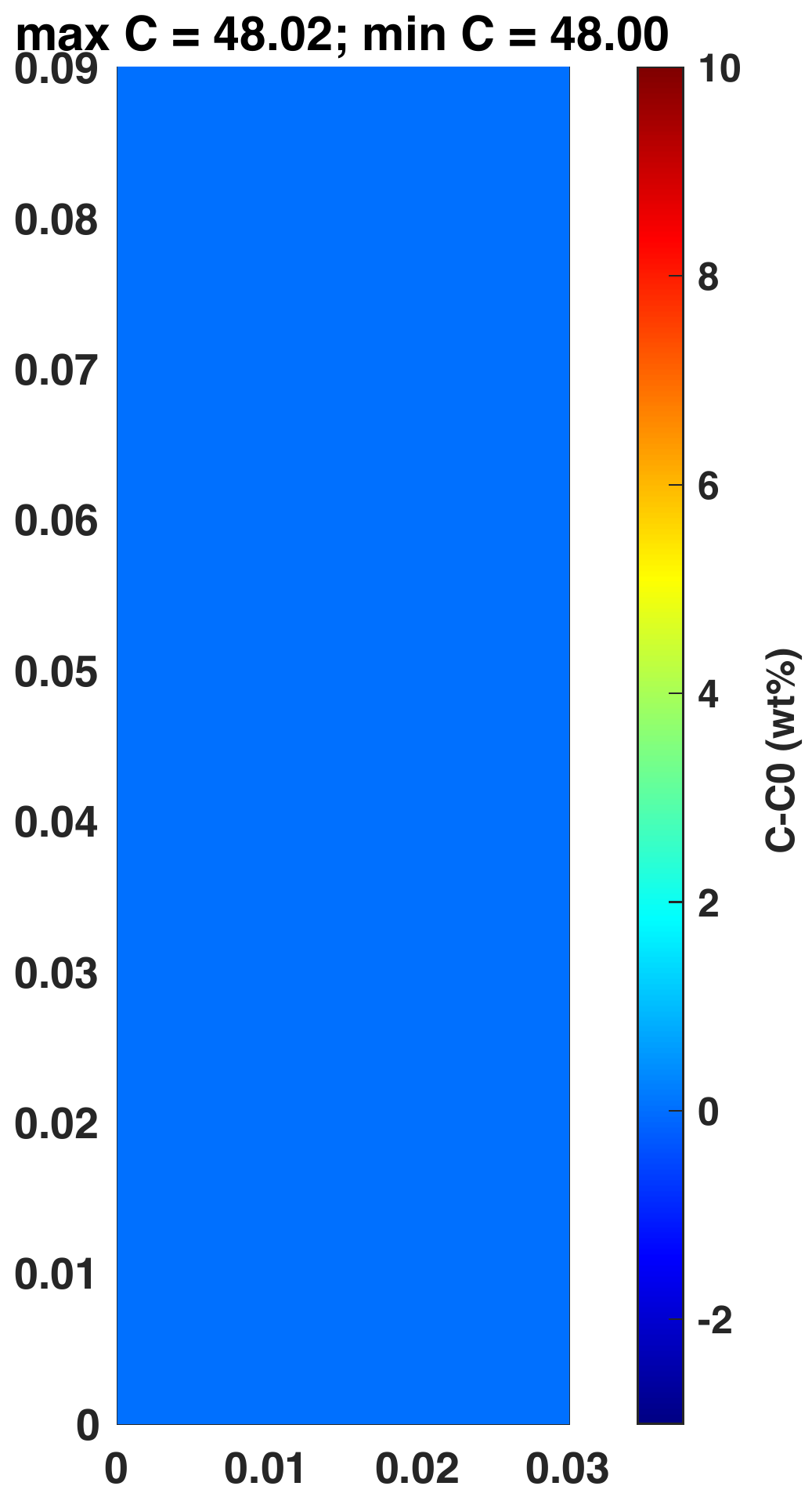}
    \end{minipage}
    }
    \centering \subfigure[$t=50s$]{
    \begin{minipage}[b]{0.17\textwidth}
    \centering
    \includegraphics[width=1.1\textwidth]{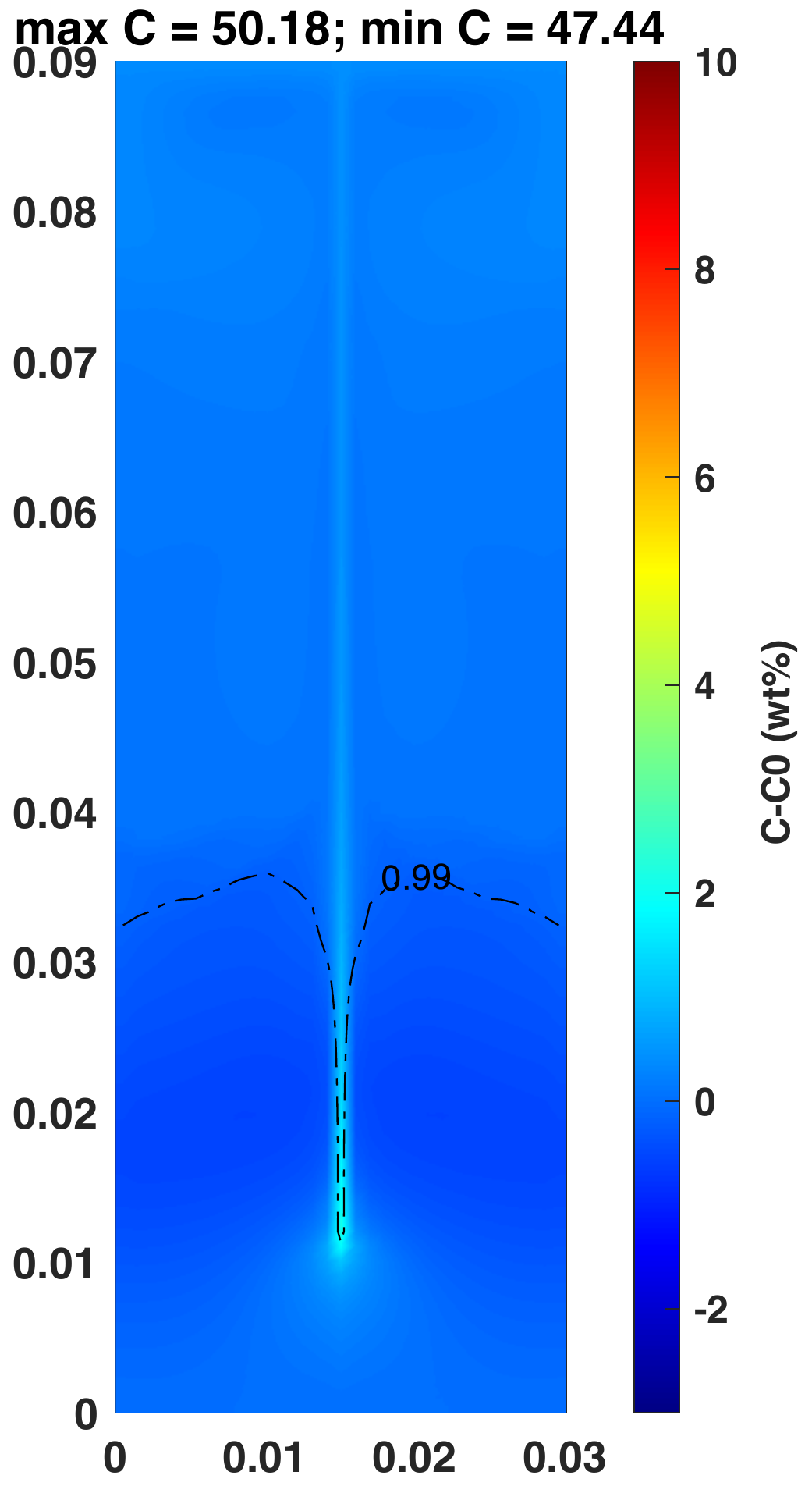}
    \end{minipage}
    }
    \centering \subfigure[$t=100s$]{
    \begin{minipage}[b]{0.17\textwidth}
    \centering
    \includegraphics[width=1.1\textwidth]{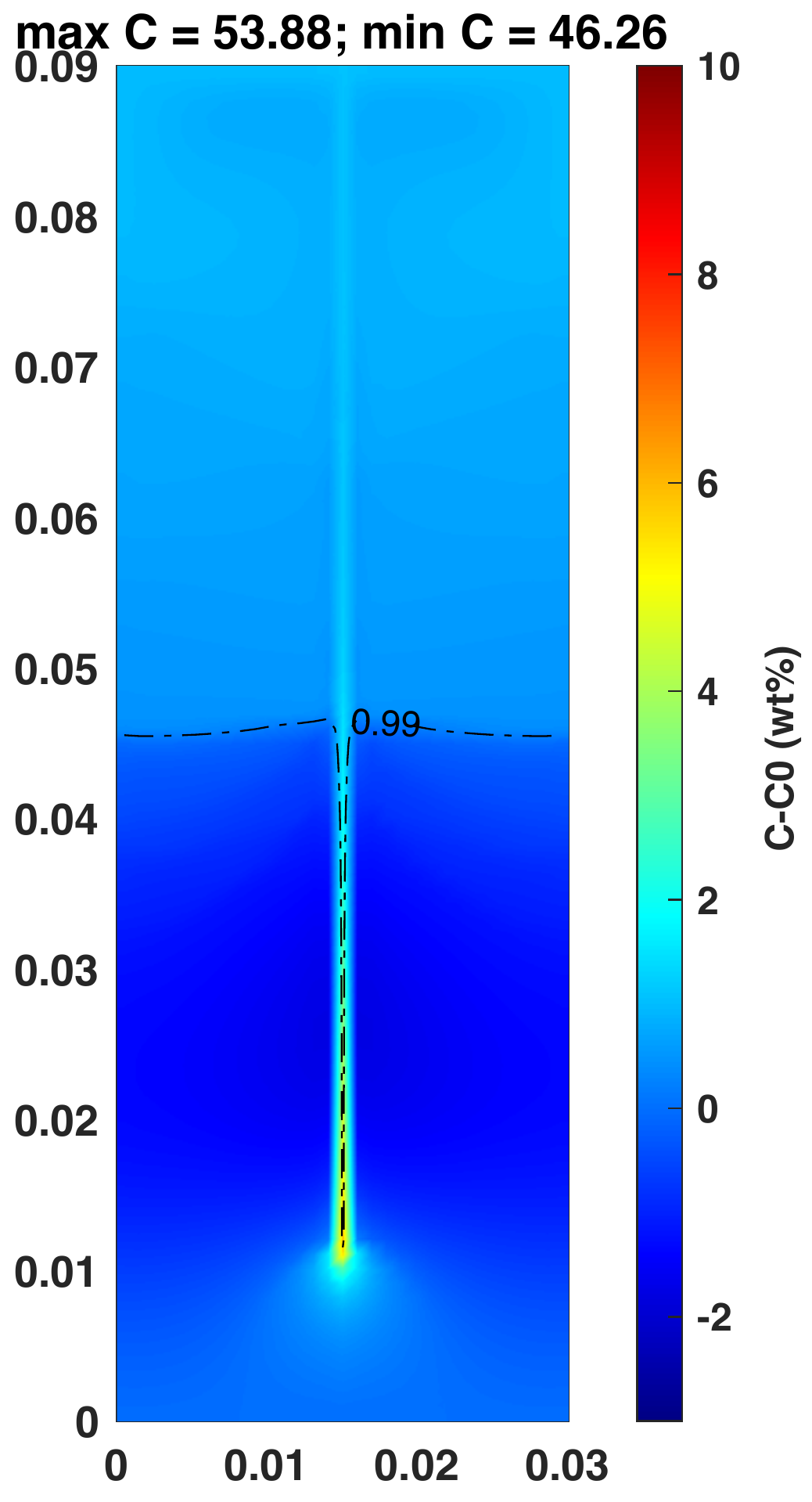}
    \end{minipage}
    }
    \centering \subfigure[$t=150s$]{
    \begin{minipage}[b]{0.17\textwidth}
    \centering
    \includegraphics[width=1.1\textwidth]{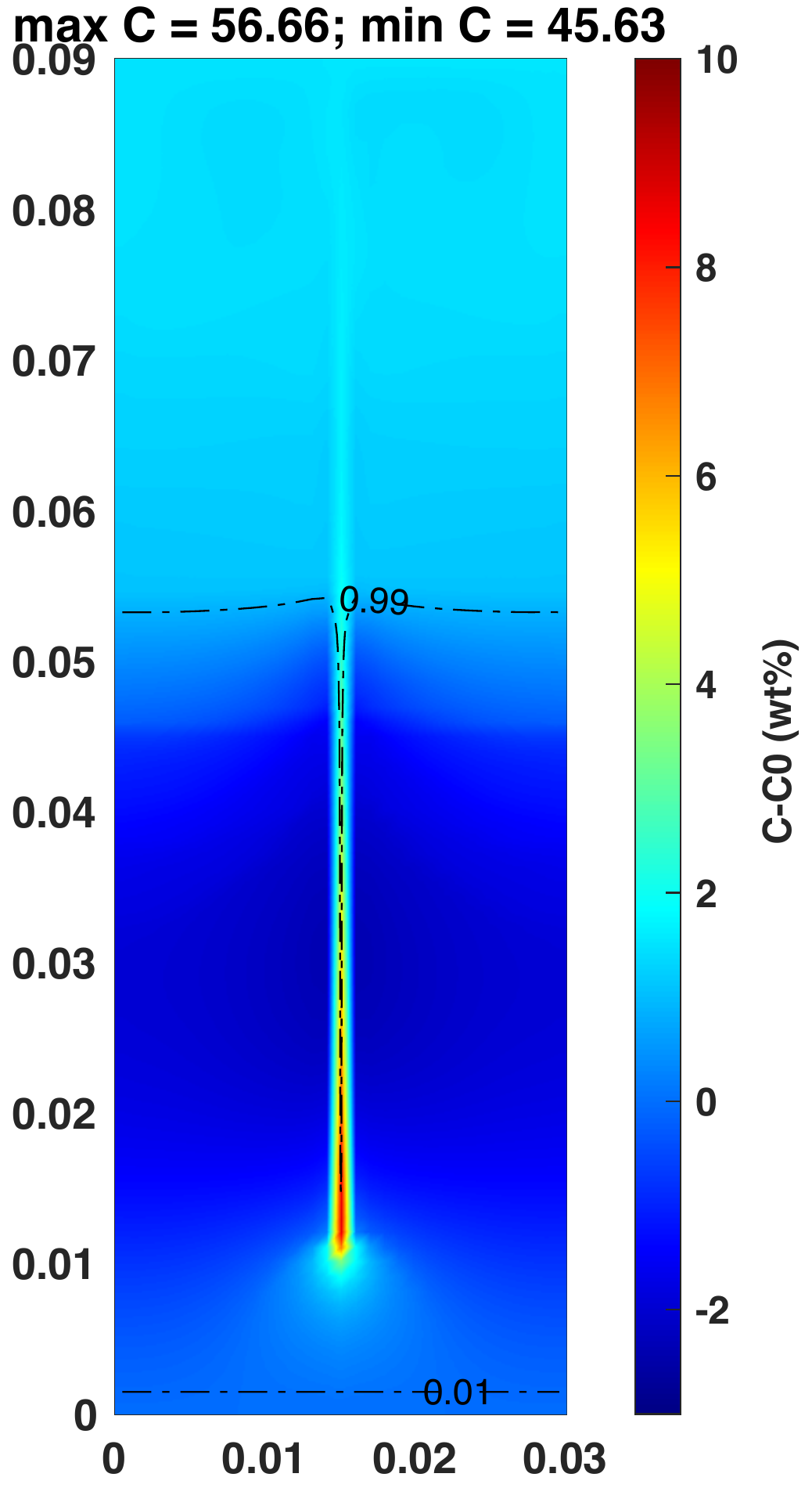}
    \end{minipage}
    }
    \centering \subfigure[$t=200s$]{
    \begin{minipage}[b]{0.17\textwidth}
    \centering
    \includegraphics[width=1.1\textwidth]{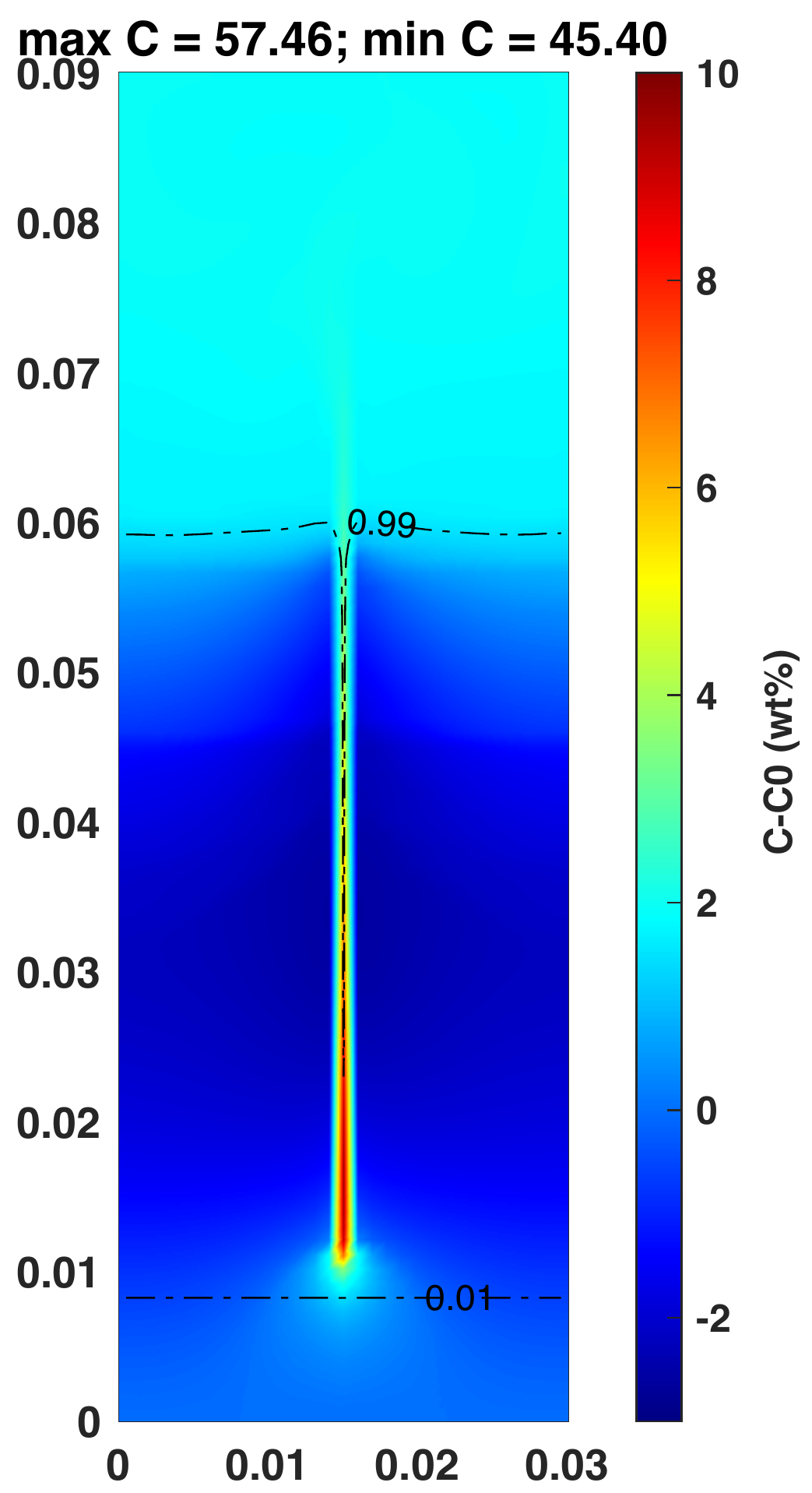}
    \end{minipage}
    }
    \caption{Evolution of $C-C_0$ contour map in Example 2 with $R=1:3$; the liquid/mush interface and the mush/solid interface (black dash-dotted line); $C-C_0$ (colorbar)}
    \label{example2_C}
 \end{figure*}
 \begin{figure*}[!h]
    \centering \subfigure[$t=5s$]{
    \begin{minipage}[b]{0.17\textwidth}
    \centering
    \includegraphics[width=1.1\textwidth]{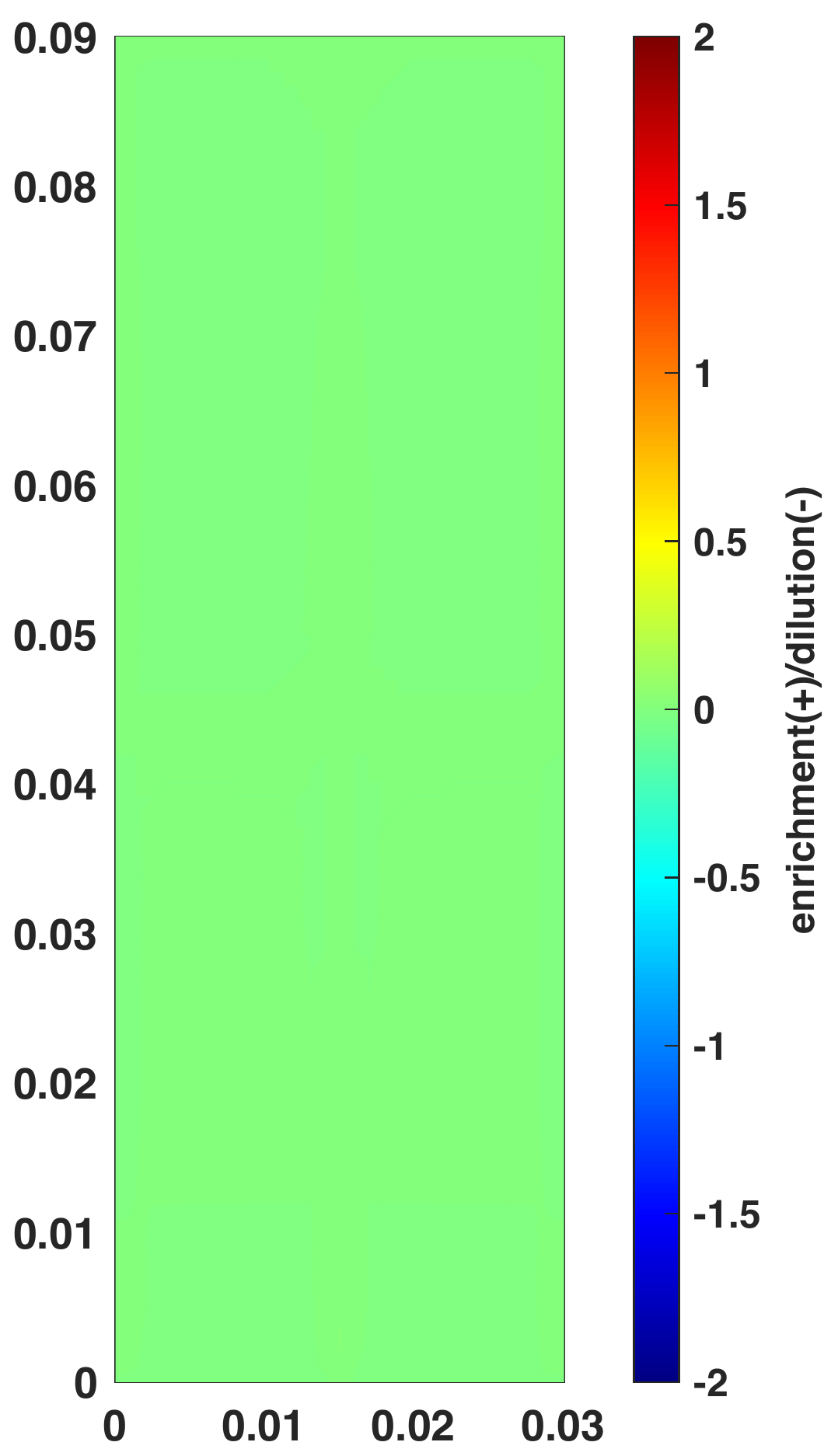}
    \end{minipage}
    }
    \centering \subfigure[$t=50s$]{
    \begin{minipage}[b]{0.17\textwidth}
    \centering
    \includegraphics[width=1.1\textwidth]{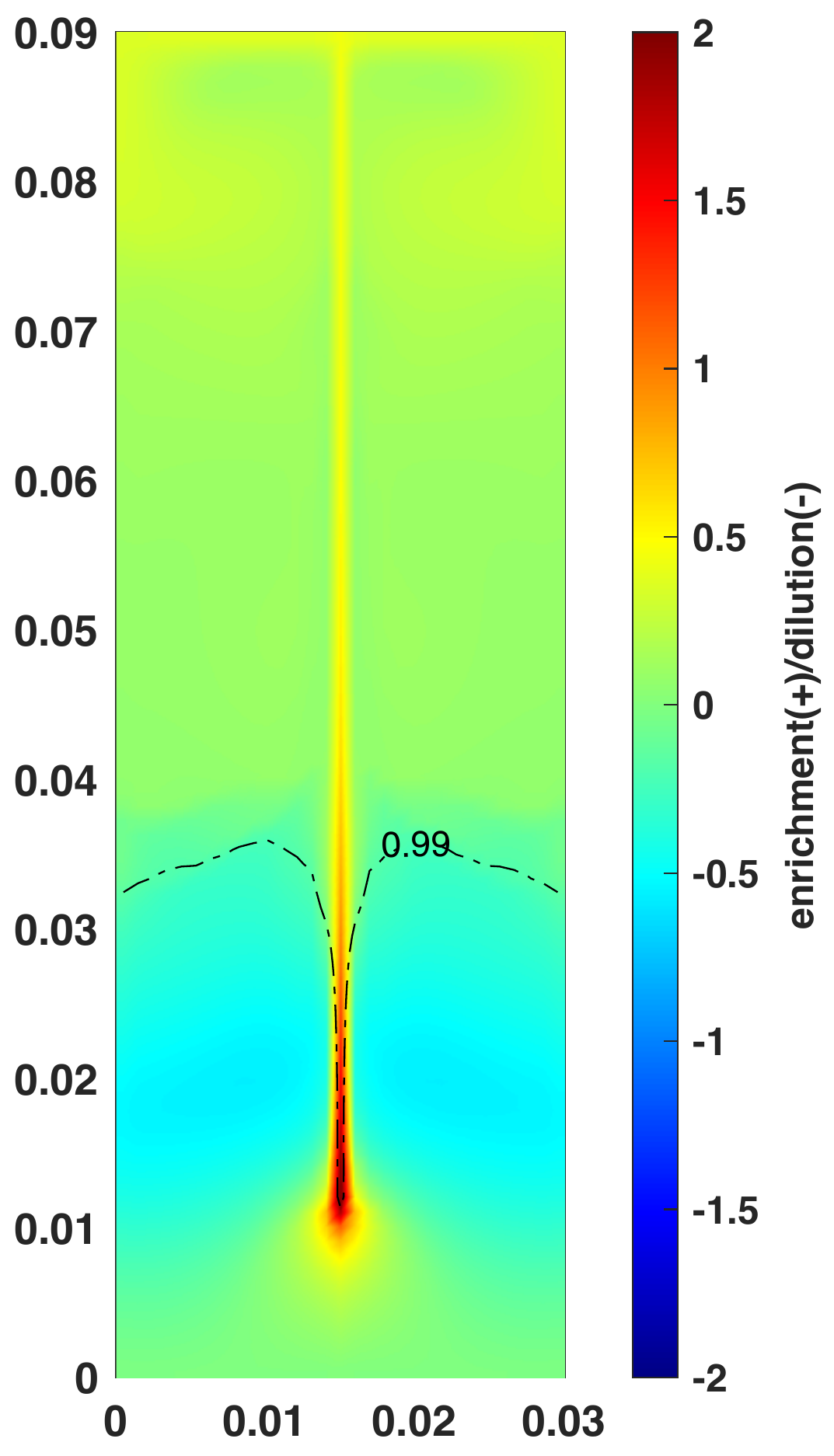}
    \end{minipage}
    }
    \centering \subfigure[$t=100s$]{
    \begin{minipage}[b]{0.17\textwidth}
    \centering
    \includegraphics[width=1.1\textwidth]{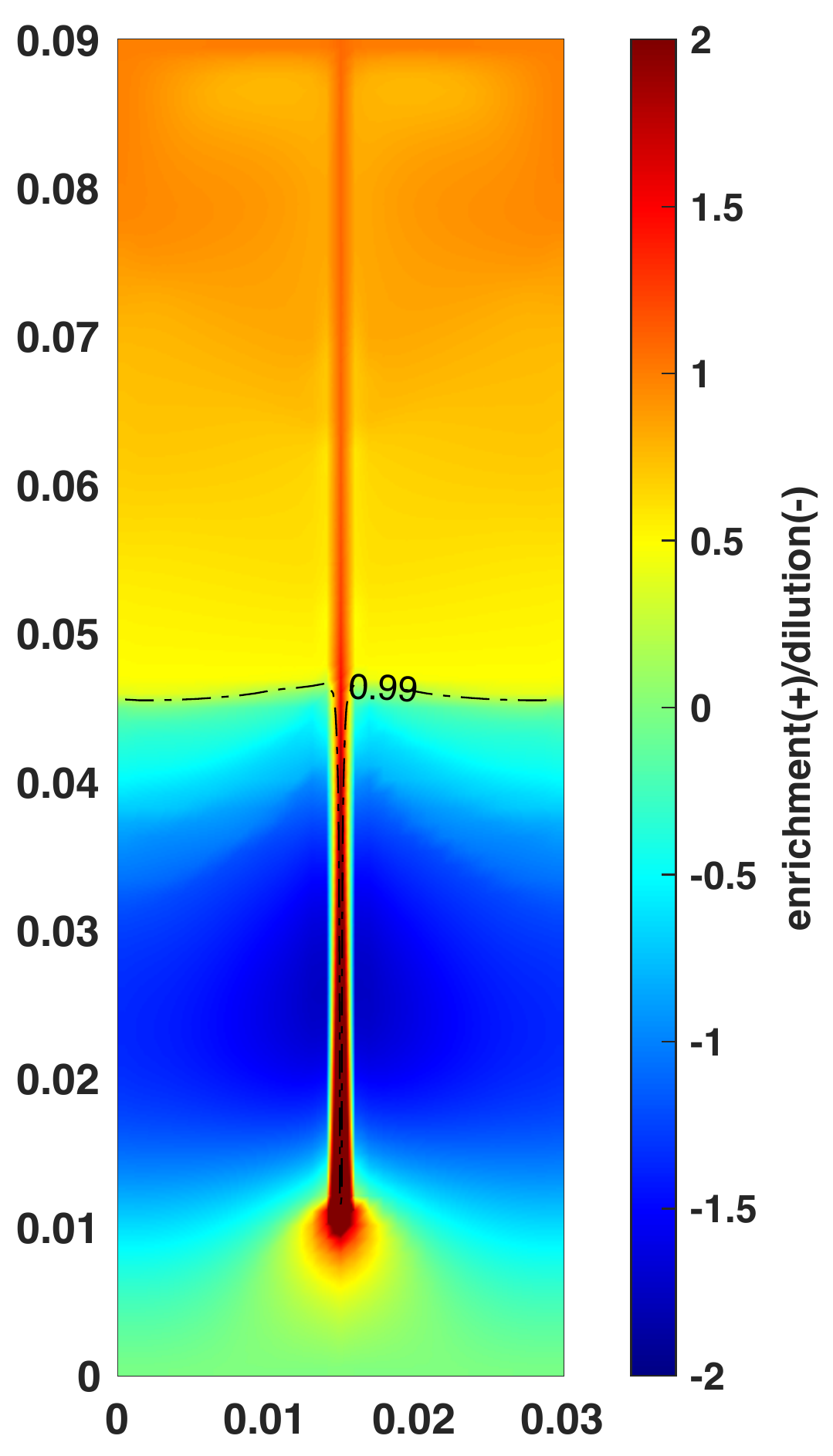}
    \end{minipage}
    }
    \centering \subfigure[$t=150s$]{
    \begin{minipage}[b]{0.17\textwidth}
    \centering
    \includegraphics[width=1.1\textwidth]{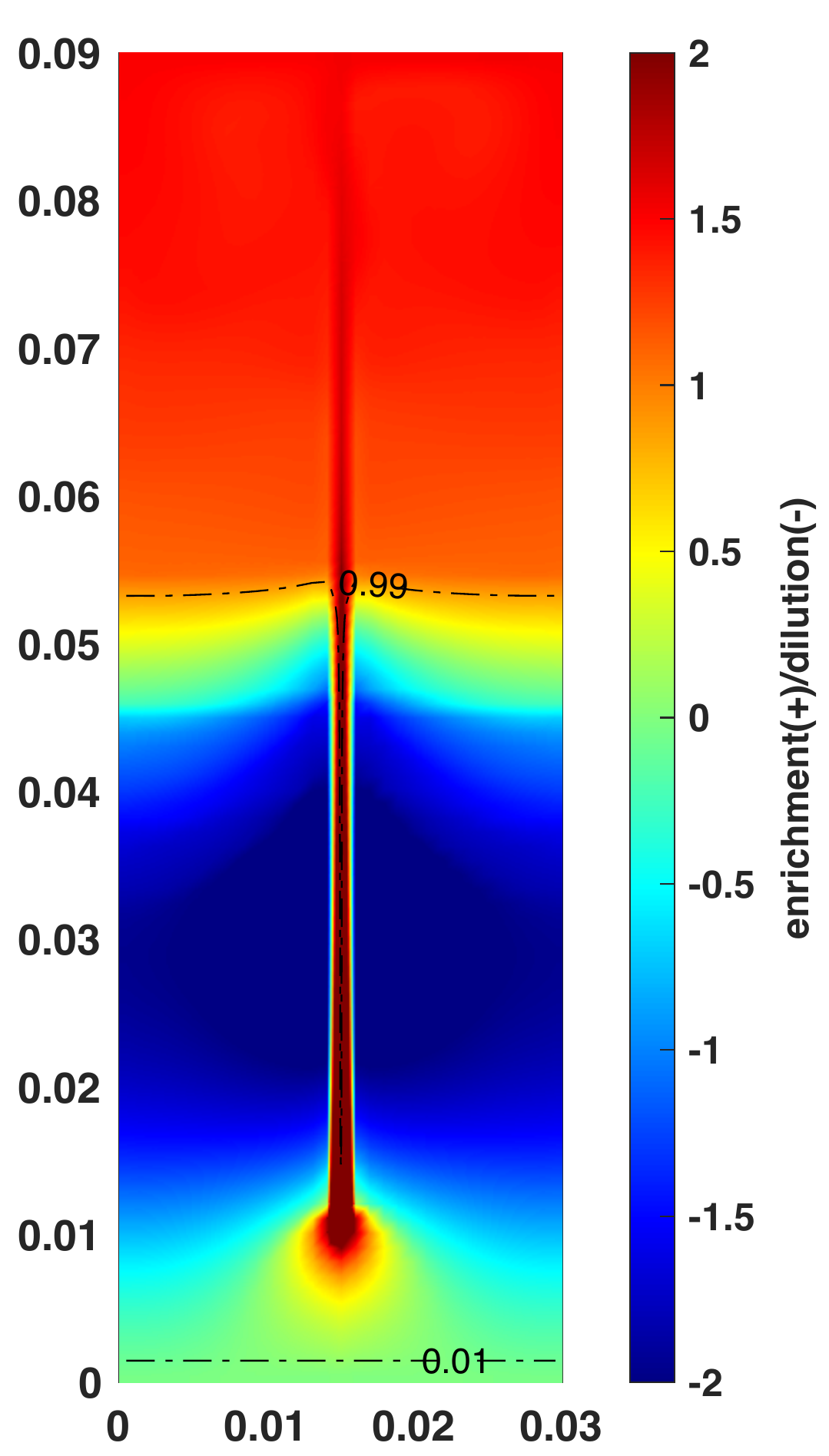}
    \end{minipage}
    }
    \centering \subfigure[$t=200s$]{
    \begin{minipage}[b]{0.17\textwidth}
    \centering
    \includegraphics[width=1.1\textwidth]{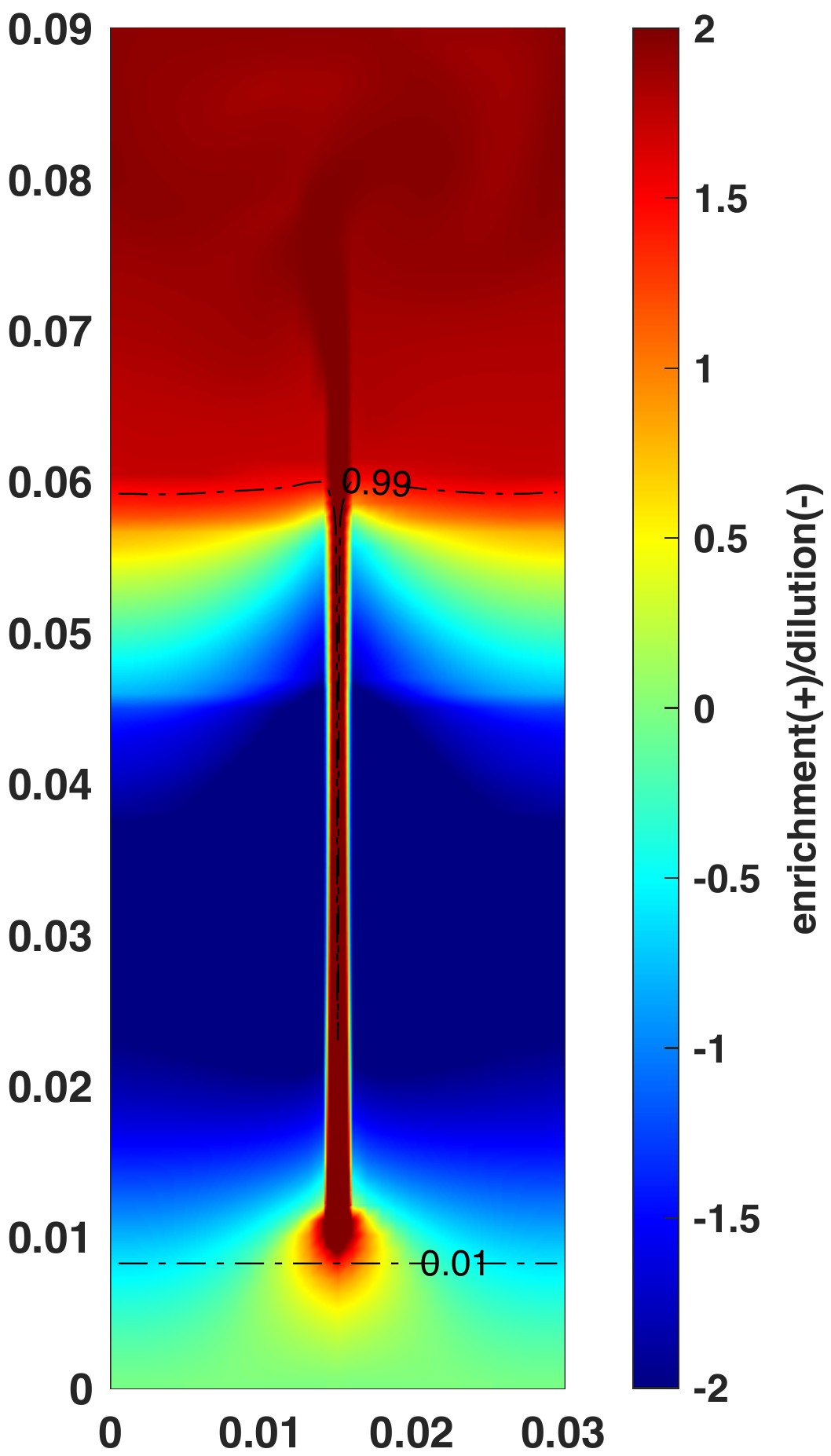}
    \end{minipage}
    }
    \caption{Evolution of enrichment and dilution region of composition \ch{Sn} in example 2 with $R=3:1$}
    \label{example2_ED}
 \end{figure*}
%

In 2D cases, the non-slip boundary condition is employed, and the evolution of physical fields is all presented at time $t=5, 50, 100, 150, 200s$. 

For the 2D case with an aspect ratio $R=1:3$, initially the perturbation of concentration is strong enough to overcome the ``edge effect'' of the non-slip vertical side walls, then it triggers the circular flow pattern. With cooling from the bottom, the combination of solute and thermal convection results in the formation of the freckle and chimney-like narrow channel across the mushy region. The stratification of temperature with a bump of the isotherms and a deep valley-shaped sink of the liquid/mush interface at the central part of the domain are observed in Figure \ref{example2_T}. The fully solid block appears at the bottom when $t = 150s$; In Figure \ref{example2_S}, streamlines of the circular flow and the iso-concentration lines show good symmetry as expected. And they exhibit the same tendencies and patterns predicted by \cite{cabrales2017mathematical}. The lighter composition \ch{Sn} in the liquid phase flows from the cool bottom to the upper, warmer liquid region through a narrow channel in the mush, referring to Figure \ref{example2_C}. Figure \ref{example2_ED} shows that the channel connects the freckle and the upper liquid region with the enrichment of the composition, and the channel penetrates the \ch{Sn} diluted region. It works like a ``volcano'' in the internal structure of the alloy. At the outlet of the channel, the obvious plume flow and instability can be observed over the mushy platform and inside the liquid region at time $t = 200s$.
 \begin{figure*}[!b]
    \centering \subfigure[t=5s]{
    \begin{minipage}[b]{0.45\textwidth}
    \centering
    \includegraphics[width=1.0\textwidth]{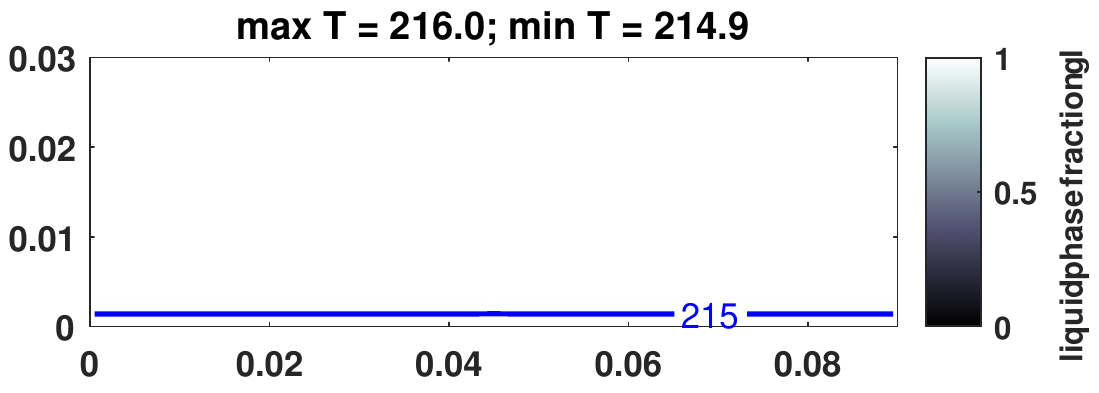}
    \end{minipage}
    }
    \centering \subfigure[t=50s]{
    \begin{minipage}[b]{0.45\textwidth}
    \centering
    \includegraphics[width=1.0\textwidth]{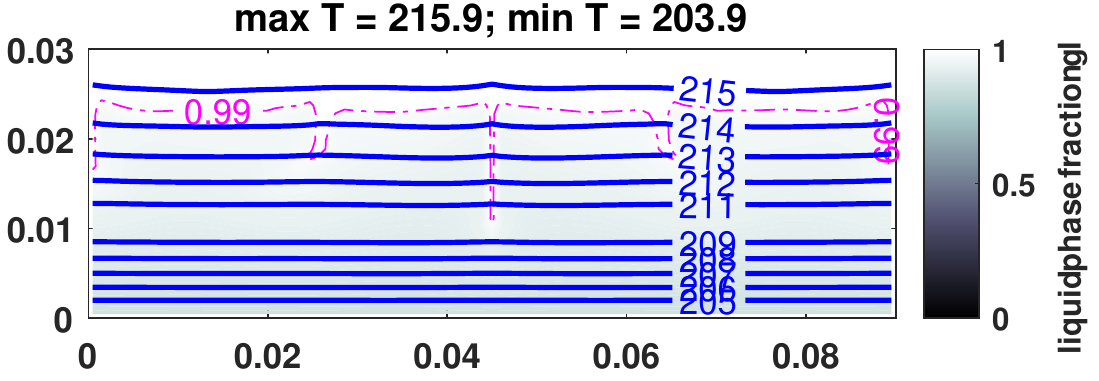}
    \end{minipage}
    }
    \centering \subfigure[t=100s]{
    \begin{minipage}[b]{0.45\textwidth}
    \centering
    \includegraphics[width=1.0\textwidth]{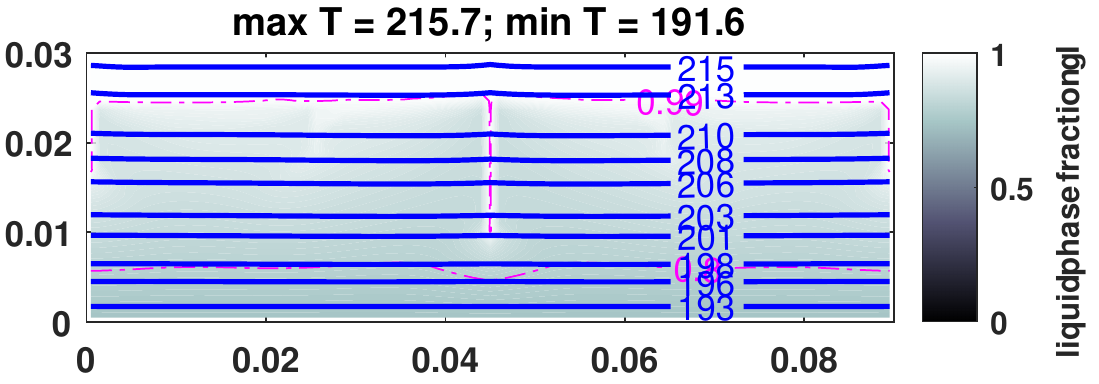}
    \end{minipage}
    }
    \centering \subfigure[t=200s]{
    \begin{minipage}[b]{0.45\textwidth}
    \centering
    \includegraphics[width=1.0\textwidth]{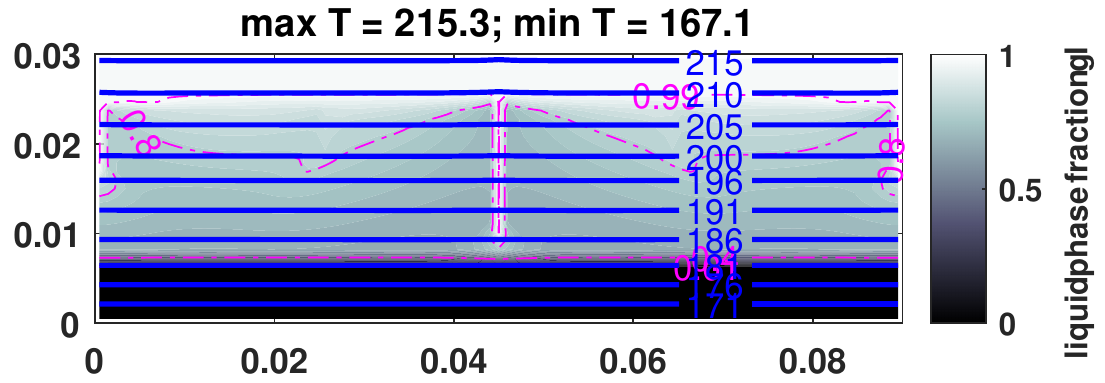}
    \end{minipage}
    }
    \caption{Evolution of liquid fraction in Example 2 with $R=3:1$; the temperature contour line (blue solid lines); the liquid fraction contour line (magenta dash-dotted line); liquid fraction (colorbar)}
    \label{example2_TT}
 \end{figure*}
 \begin{figure*}[!h]
    \centering \subfigure[t=5s]{
    \begin{minipage}[b]{0.35\textwidth}
    \centering
    \includegraphics[width=1.0\textwidth]{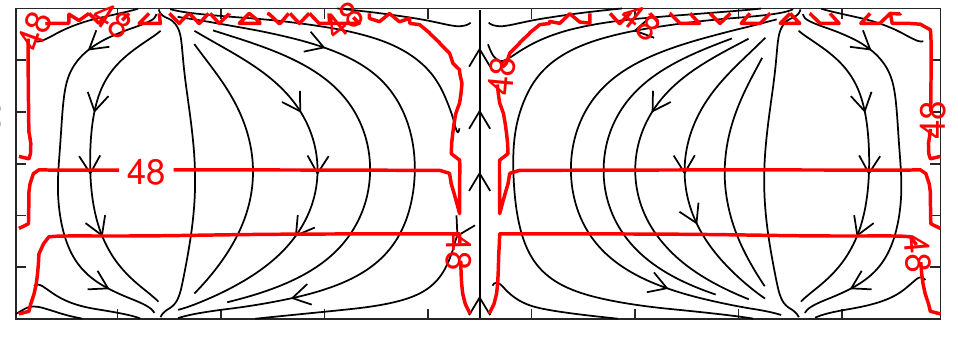}
    \end{minipage}
    } \hspace{1.5cm}
    \centering \subfigure[t=50s]{
    \begin{minipage}[b]{0.35\textwidth}
    \centering
    \includegraphics[width=1.0\textwidth]{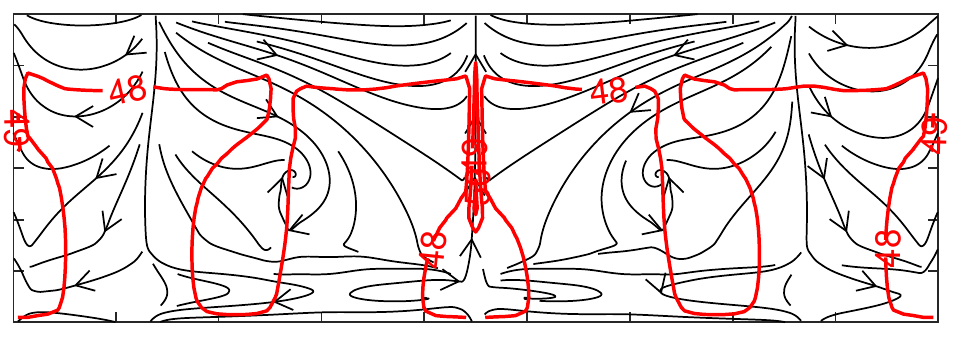}
    \end{minipage}
    } \hspace{1.5cm}
    \centering \subfigure[t=100s]{
    \begin{minipage}[b]{0.35\textwidth}
    \centering
    \includegraphics[width=1.0\textwidth]{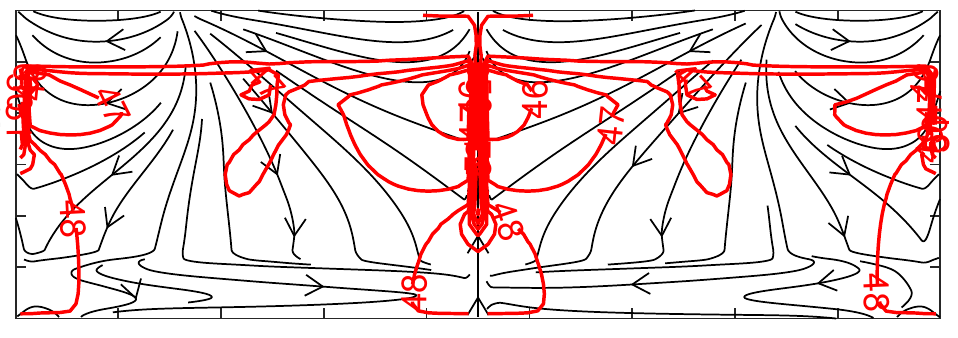}
    \end{minipage}
    } \hspace{1.5cm}
    \centering \subfigure[t=200s]{
    \begin{minipage}[b]{0.35\textwidth}
    \centering
    \includegraphics[width=1.0\textwidth]{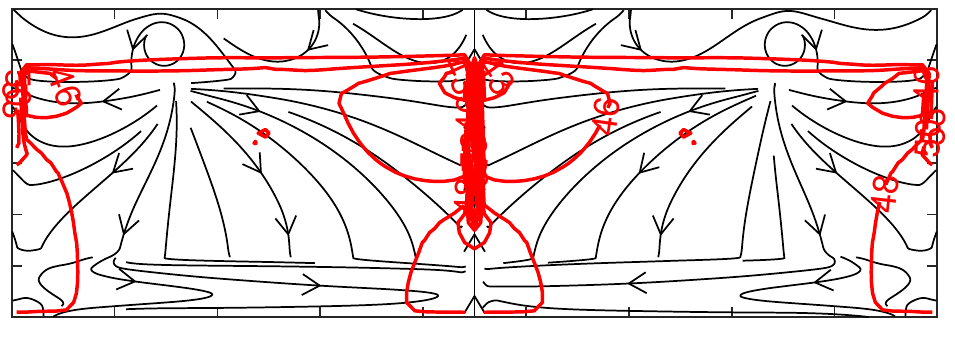}
    \end{minipage}
    }
    \caption{Evolution of stream lines in Example 2 with $R=3:1$; the stream line (black solid line with direction arrow); the concentration contour line (red solid line)}
    \label{example2_SS}
 \end{figure*}
  \begin{figure*}[!h]
    \centering \subfigure[t=5s]{
    \begin{minipage}[b]{0.45\textwidth}
    \centering
    \includegraphics[width=1.0\textwidth]{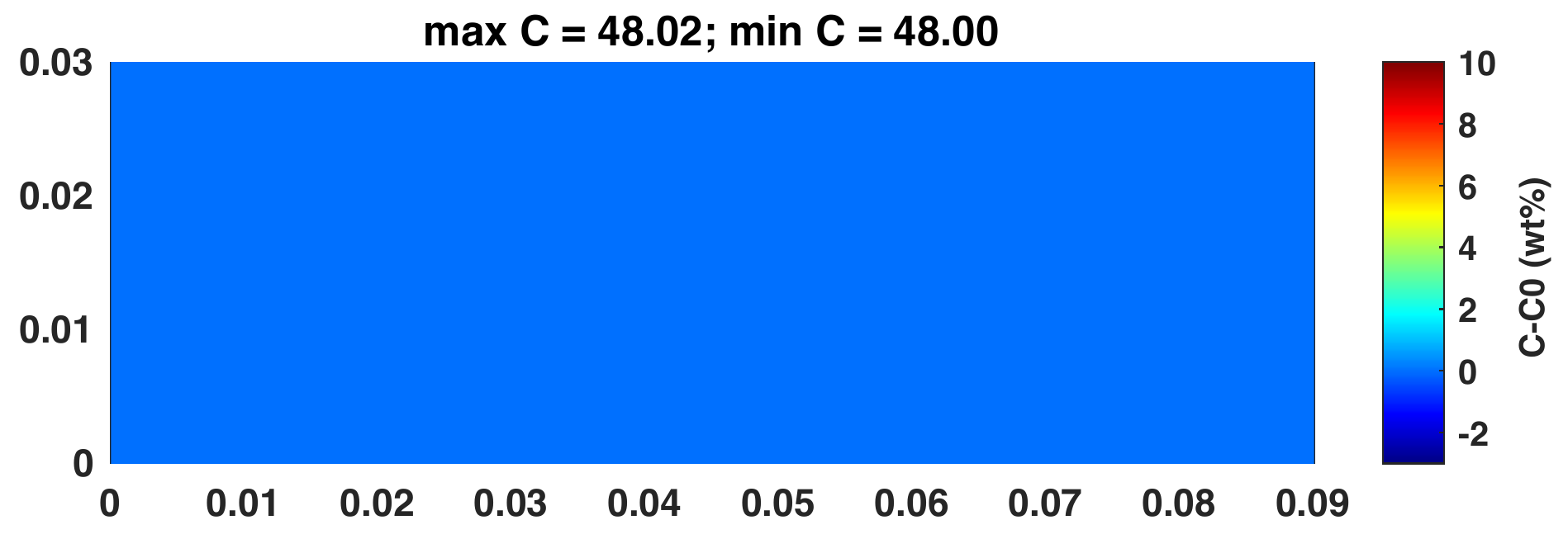}
    \end{minipage}
    }
    \centering \subfigure[t=50s]{
    \begin{minipage}[b]{0.45\textwidth}
    \centering
    \includegraphics[width=1.0\textwidth]{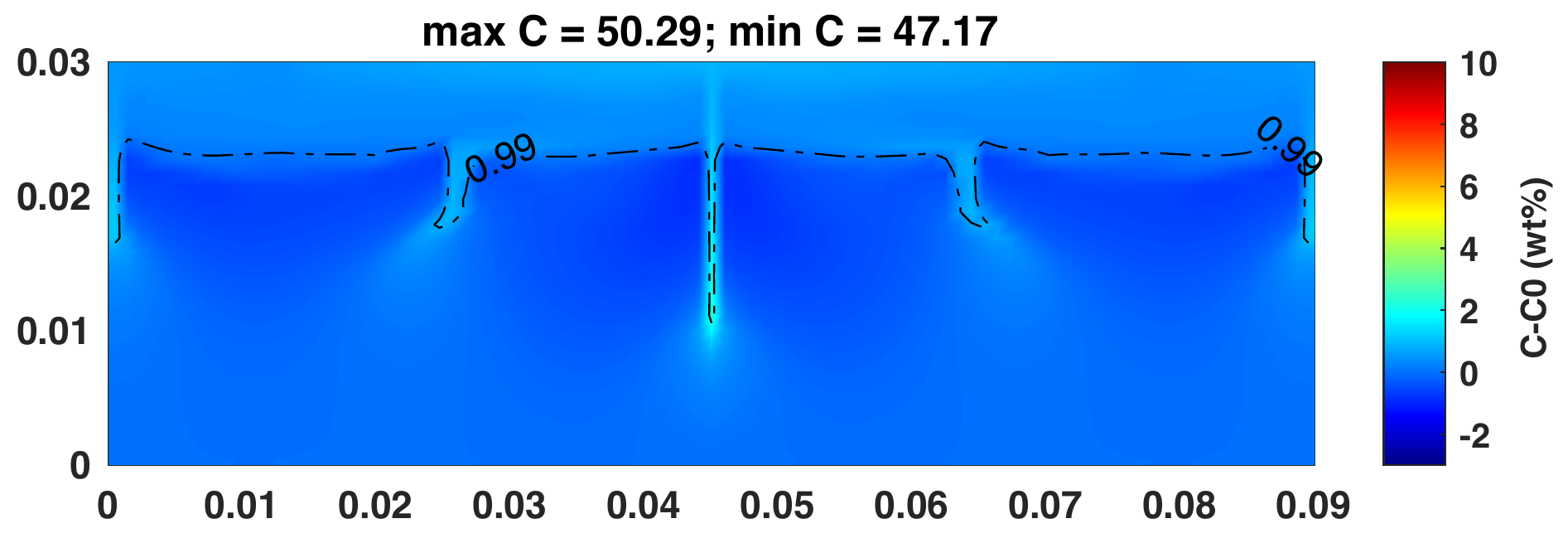}
    \end{minipage}
    }
    \centering \subfigure[t=100s]{
    \begin{minipage}[b]{0.45\textwidth}
    \centering
    \includegraphics[width=1.0\textwidth]{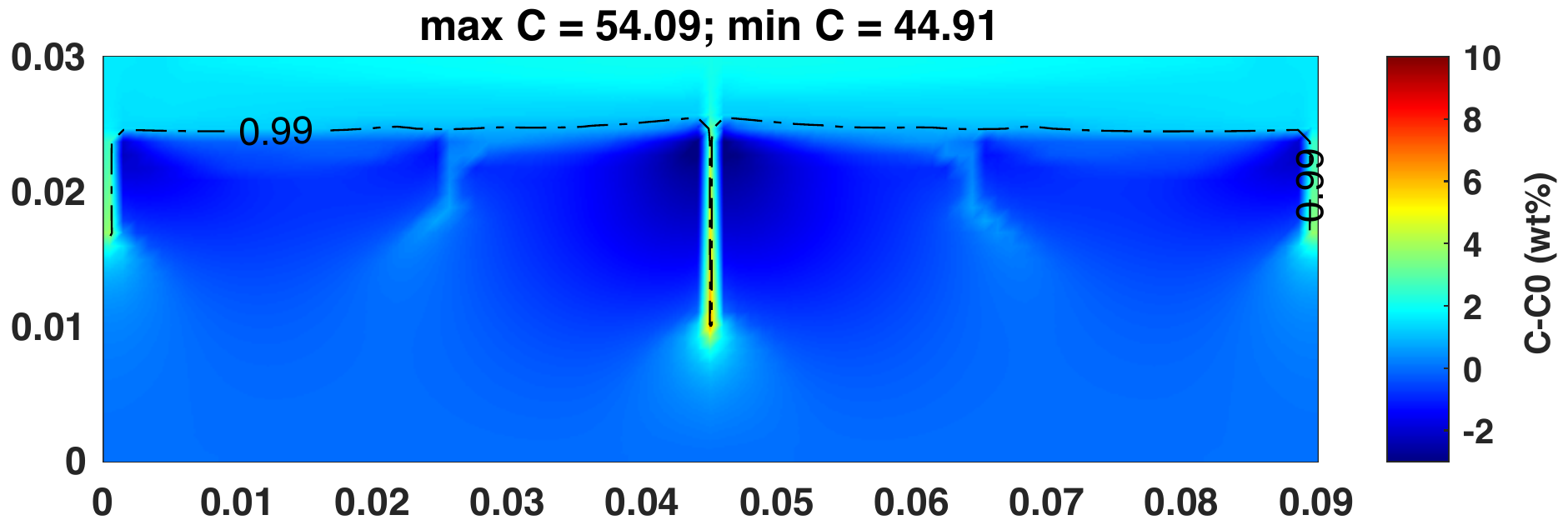}
    \end{minipage}
    }
    \centering \subfigure[t=200s]{
    \begin{minipage}[b]{0.45\textwidth}
    \centering
    \includegraphics[width=1.0\textwidth]{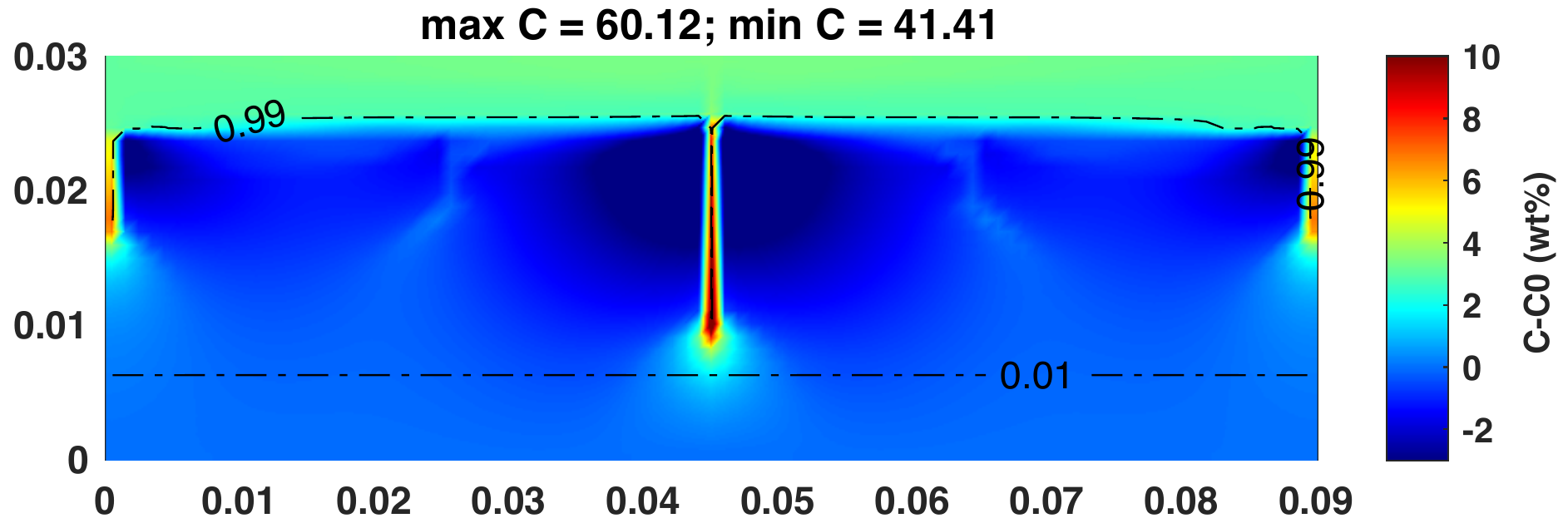}
    \end{minipage}
    }
    \caption{Evolution of $C-C_0$ contour map in Example 2 with $R=3:1$; the liquid/mush interface and the mush/solid interface (black dash-dotted line); $C-C_0$ (colorbar)}
    \label{example2_CC}
 \end{figure*}
 
The same concentration perturbation is imposed for the cavity with an aspect ratio $R = 3:1$, the much shorter central line will make the perturbation effect dissipate rapidly. Naturally, the ``edge effect'' will play an important role in this case as well. As the results show in Figures \ref{example2_TT}, \ref{example2_SS} and \ref{example2_CC}. Freckles not only form at the central line but also stick to the vertical boundary. This phenomenon often occurs both in real experiments and in other simulation results. The main reasons for this are: (1) the offset effect of the no-slip boundary on the initial downward flow induced by the concentration perturbation (edge effect); and (2) the central solutal-driven force has not a sufficiently strong influence on this space aspect ratio (aspect ratio effect) to maintain a steady circle flow. All these factors induce the upward flow along the vertical edge and eventually lead to the formation of freckles attached to the wall.

 \begin{figure*}[!b]
    \centering \subfigure[t=30s]{
    \begin{minipage}[b]{0.17\textwidth }
    \centering
    \includegraphics[width=1.1\textwidth,height=0.23\textheight]{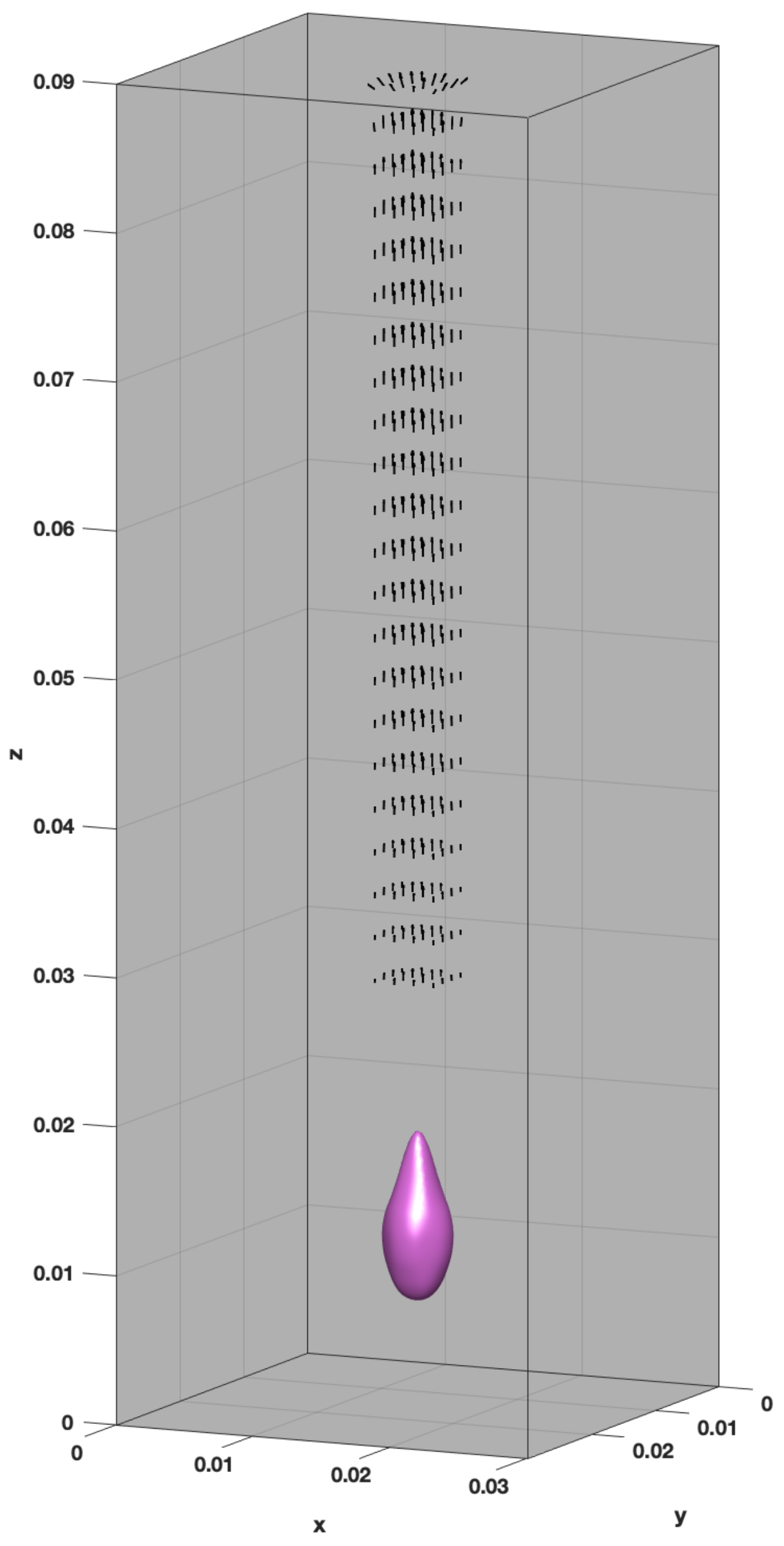}
    \end{minipage}
    }
    \centering \subfigure[t=50s]{
    \begin{minipage}[b]{0.17\textwidth}
    \centering
    \includegraphics[width=1.1\textwidth,height=0.23\textheight]{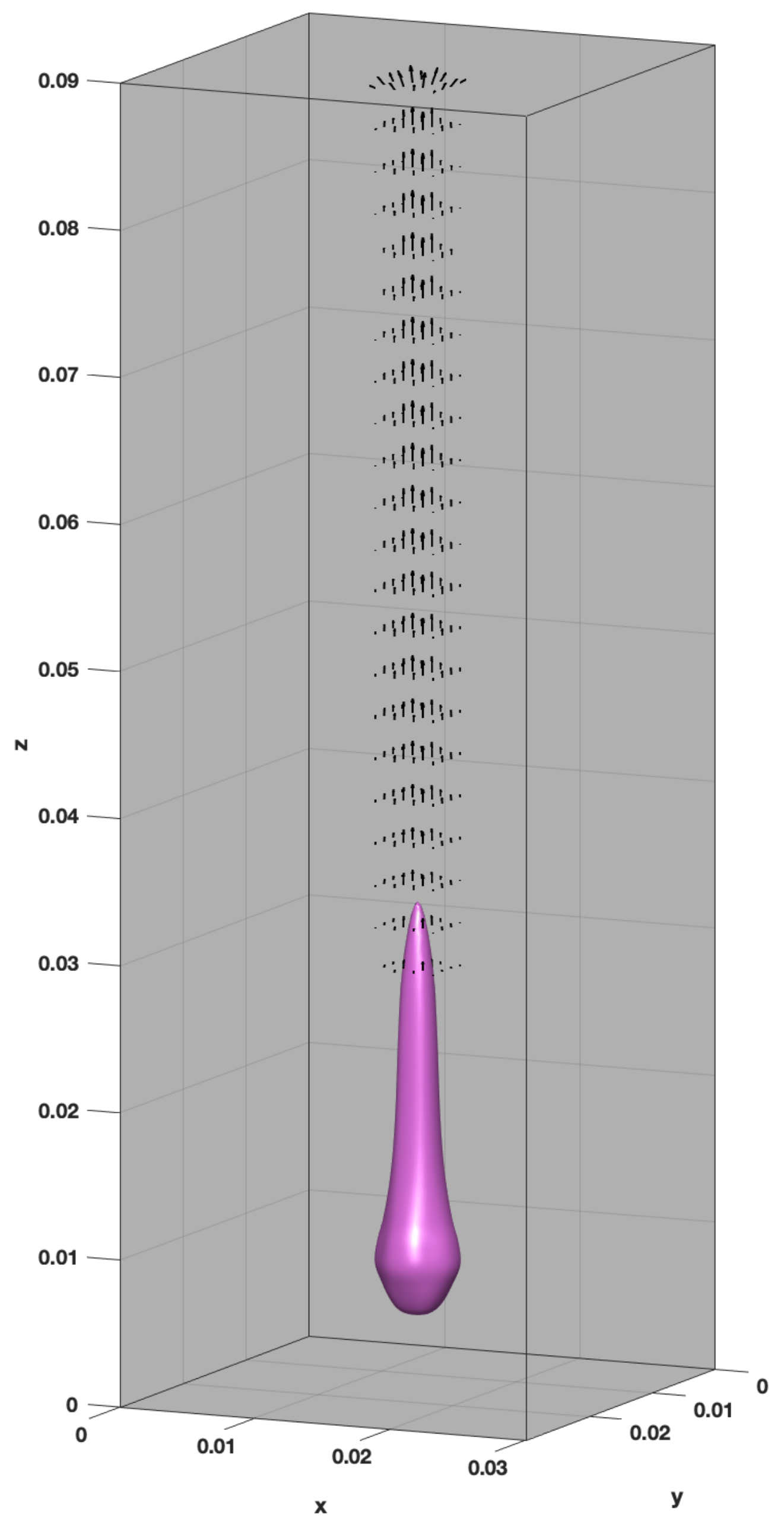}
    \end{minipage}
    }
    \centering \subfigure[t=100s]{
    \begin{minipage}[b]{0.17\textwidth}
    \centering
    \includegraphics[width=1.1\textwidth,height=0.23\textheight]{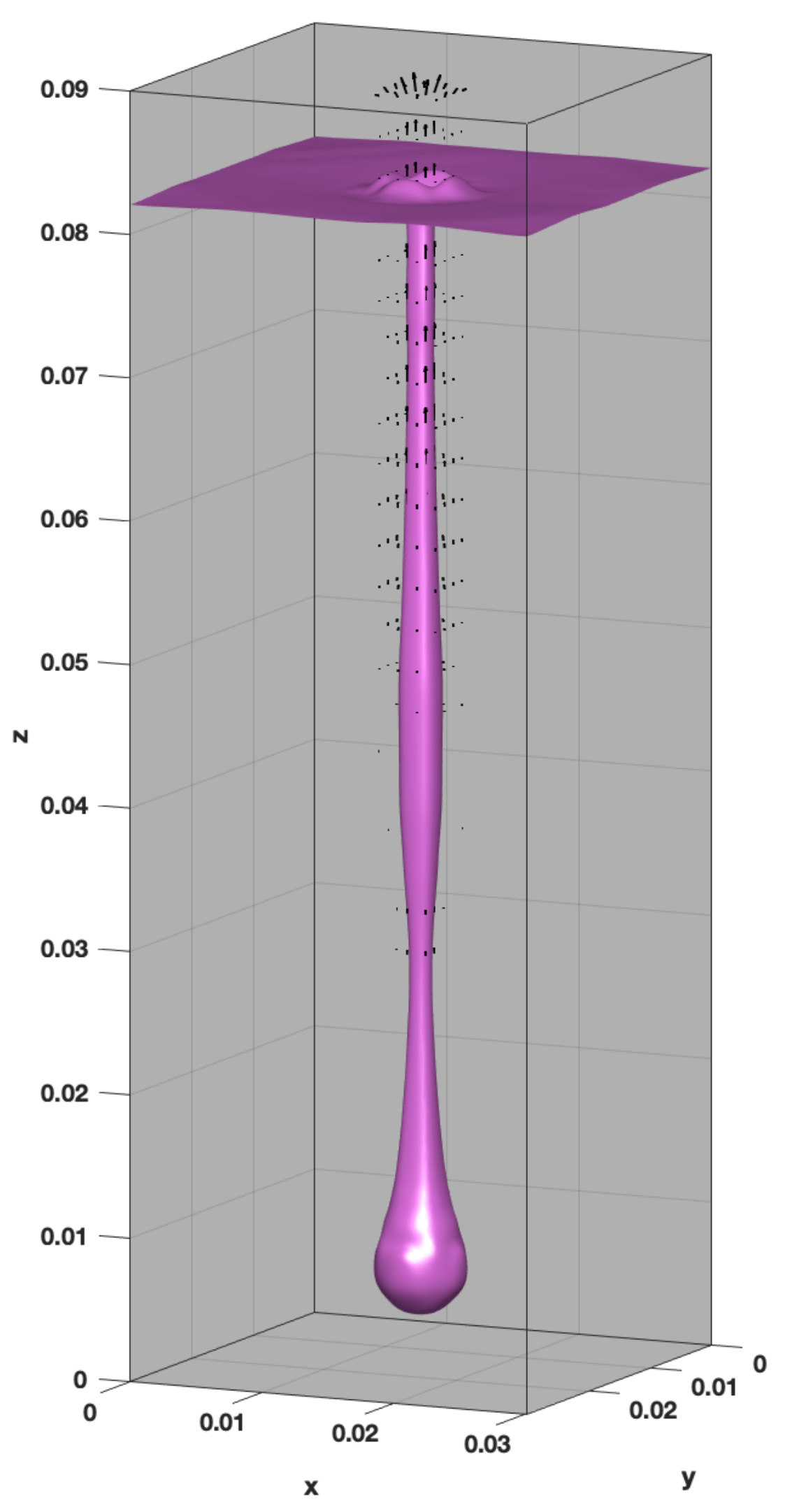}
    \end{minipage}
    }
    \centering \subfigure[t=150s]{
    \begin{minipage}[b]{0.17\textwidth}
    \centering
    \includegraphics[width=1.1\textwidth,height=0.23\textheight]{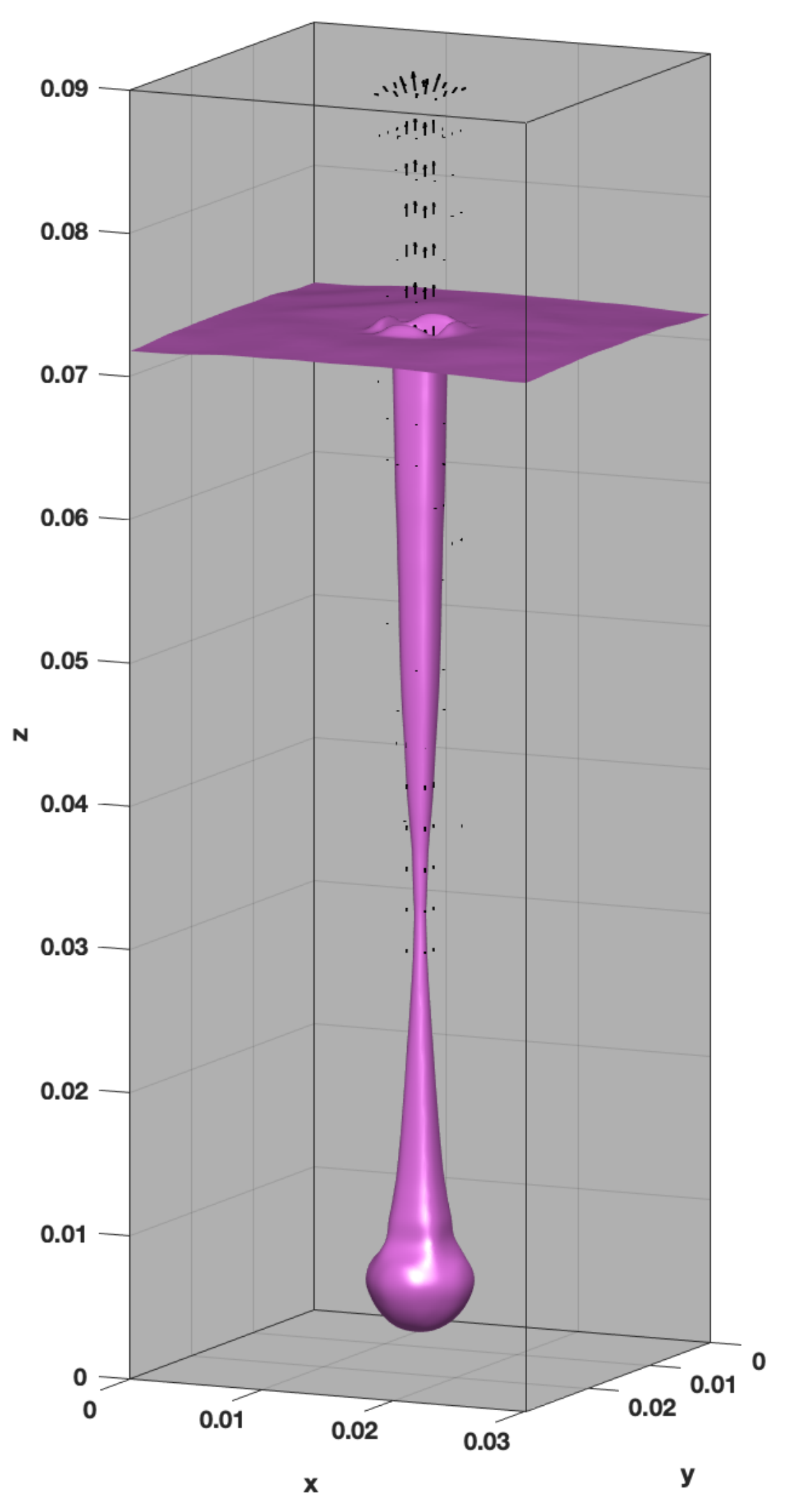}
    \end{minipage}
    }
    \centering \subfigure[t=200s]{
    \begin{minipage}[b]{0.17\textwidth}
    \centering
    \includegraphics[width=1.1\textwidth,height=0.23\textheight]{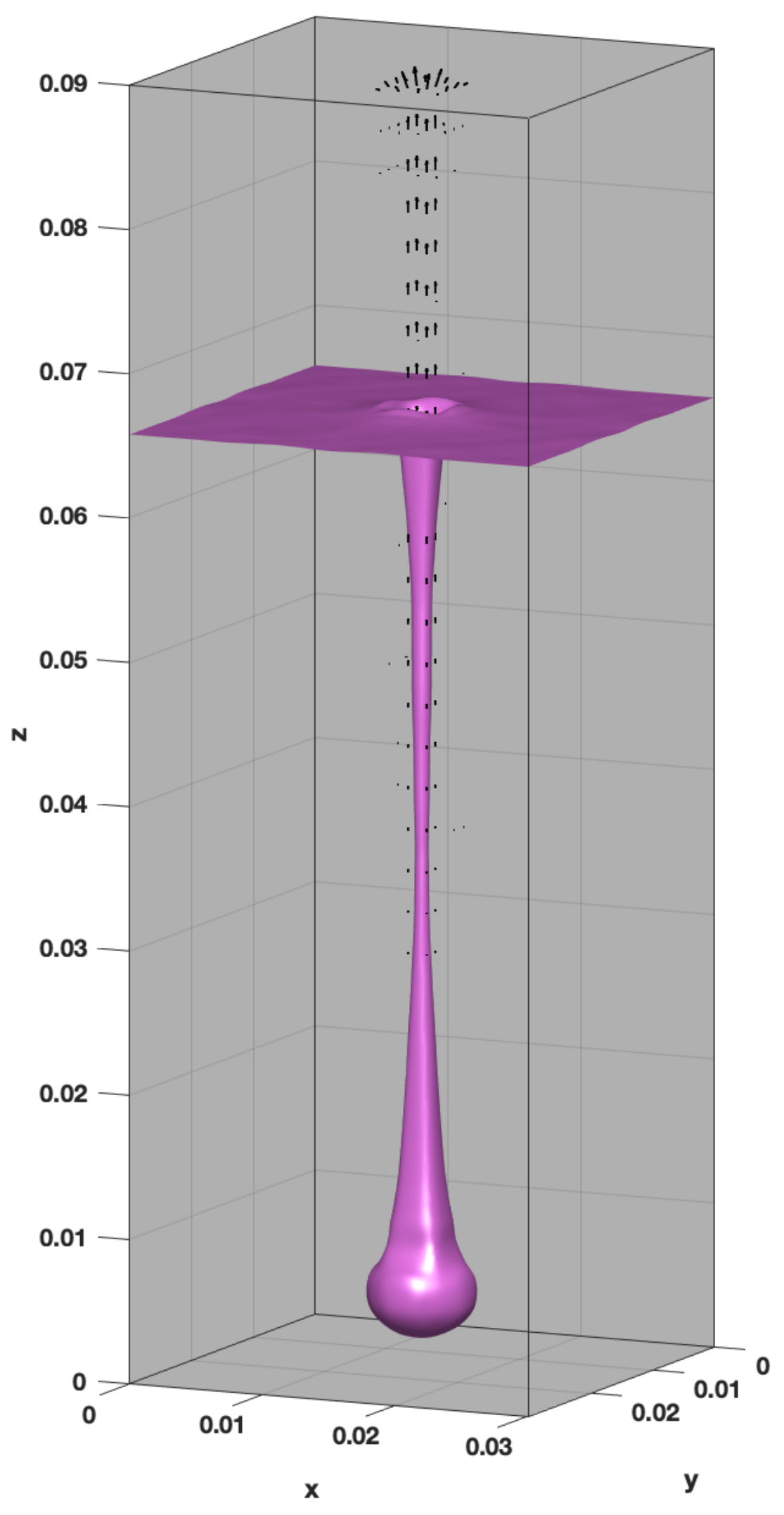}
    \end{minipage}
    }
    \caption{Evolution of the isosurface of $C-C_0=1$ over time in Example 2 with $R=1:1:3$ and free-slip vertical boundaries}
    \label{example2_3diso}
 \end{figure*}

Next, the simulation extends to the 3D application with an aspect ratio $R=1:1:3$ and the same initial perturbation at the vertical central line. The "edge effect" in 3D space will be more considerable. In order to make the freckle form at the central line as well, the slip boundaries are applied to these four vertical facades, and no-slip boundaries are set for the top face and bottom face.

\begin{figure*}[!t]
    \centering \subfigure[temperature]{
    \begin{minipage}[b]{0.22\textwidth }
    \centering
    \includegraphics[width=1.1\textwidth,height=0.3\textheight]{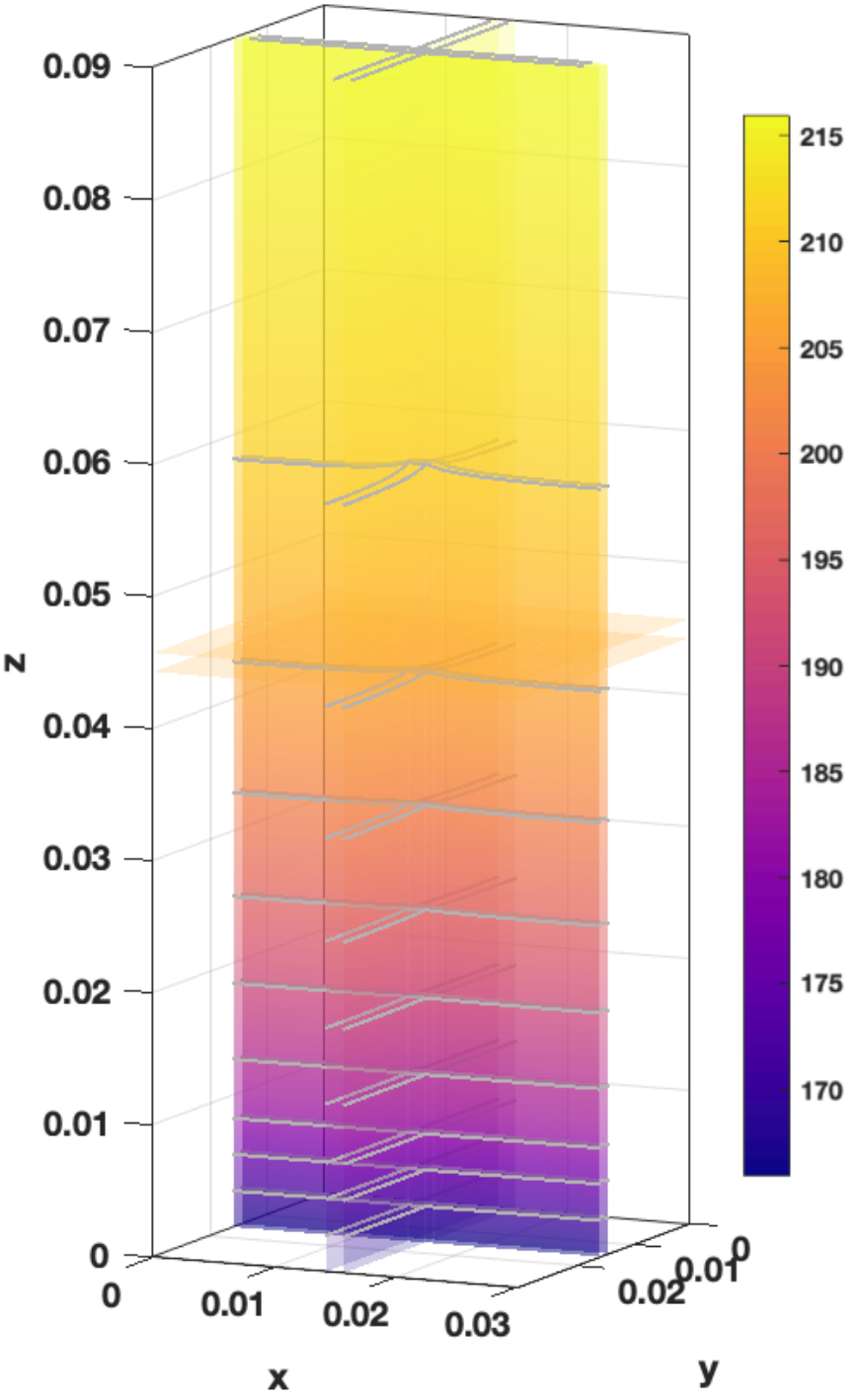} \label{ex2_200T}
    \end{minipage}
    }
    \centering \subfigure[streamlines]{
    \begin{minipage}[b]{0.22\textwidth}
    \centering
    \includegraphics[width=1.1\textwidth,height=0.3\textheight]{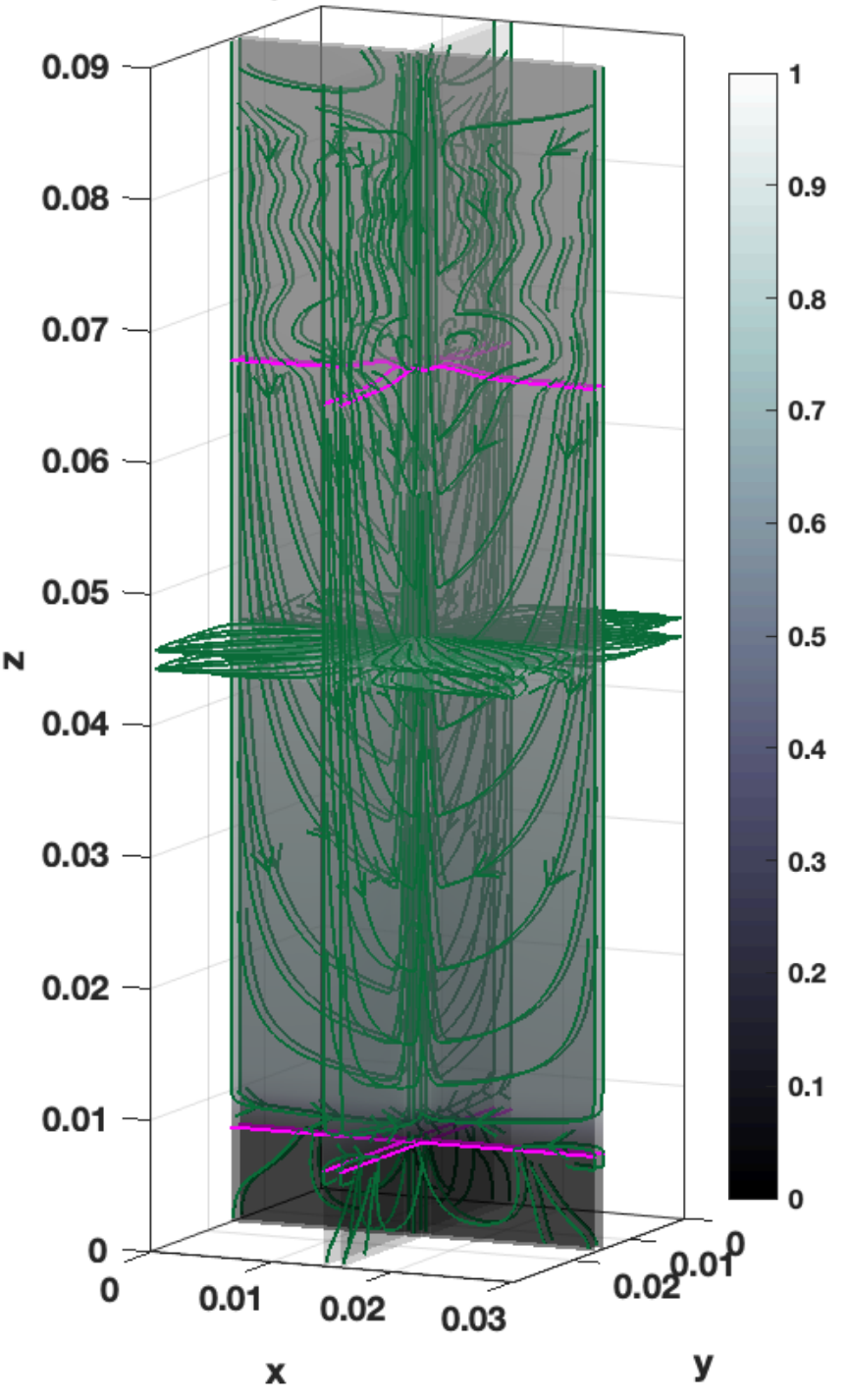} \label{ex2_200S}
    \end{minipage}
    }
    \centering \subfigure[concentration]{
    \begin{minipage}[b]{0.22\textwidth}
    \centering
    \includegraphics[width=1.1\textwidth,height=0.3\textheight]{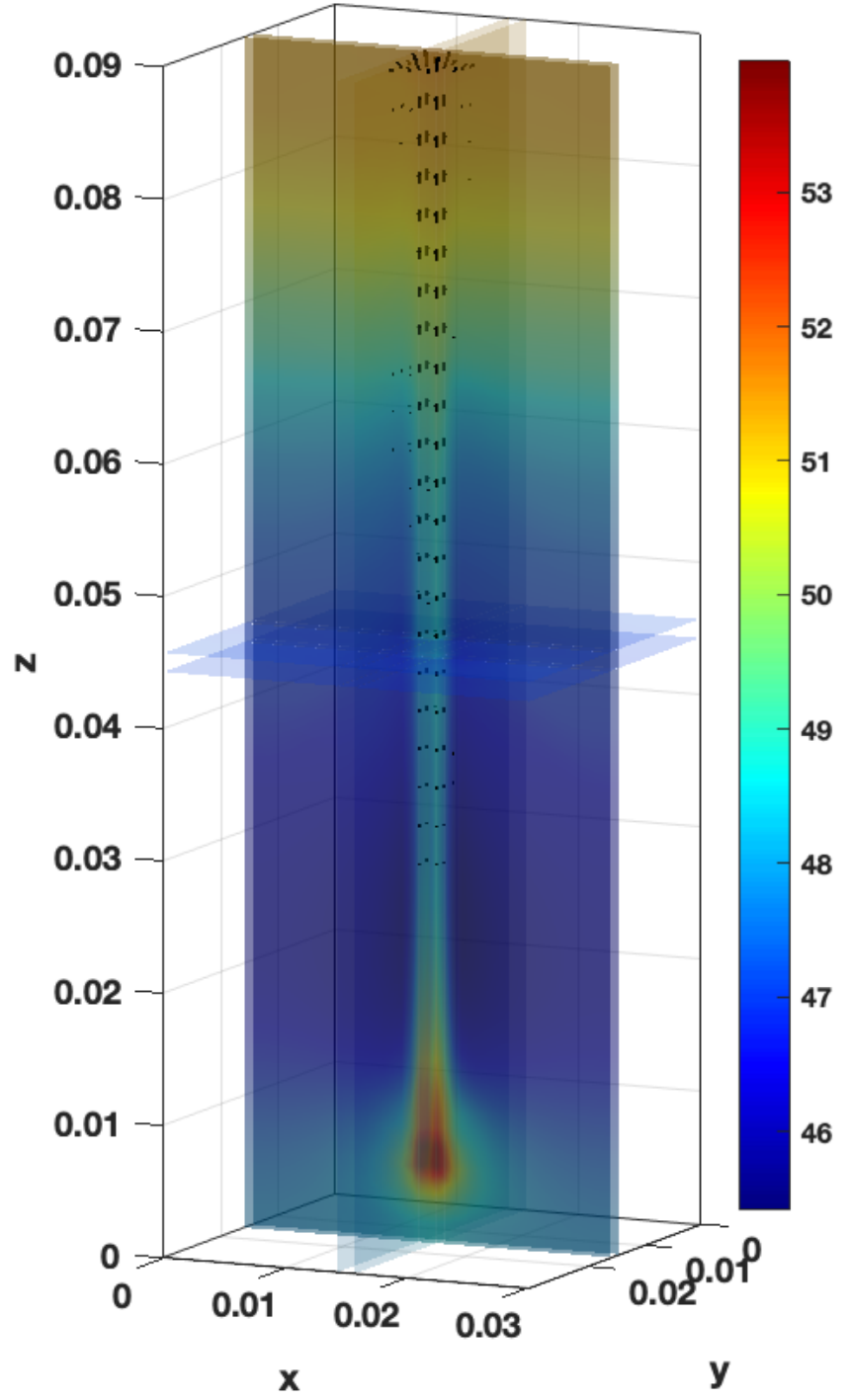} \label{ex2_200C}
    \end{minipage}
    }
    \centering \subfigure[$C-C_0$ isosurfaces]{
    \begin{minipage}[b]{0.22\textwidth}
    \centering
    \includegraphics[width=1.1\textwidth,height=0.3\textheight]{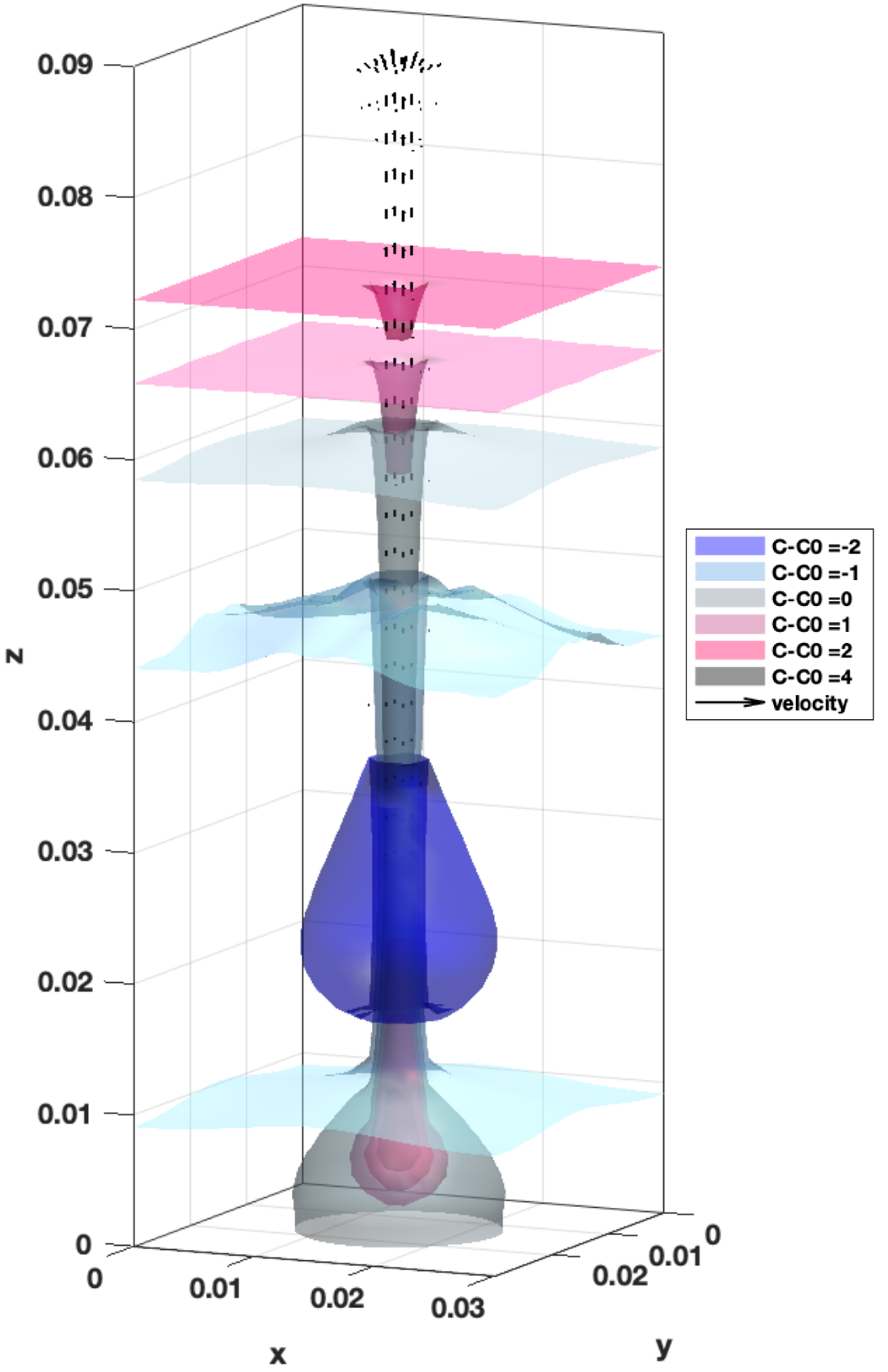} \label{ex2_200I}
    \end{minipage}
    }
    \caption{The physical fields in Example 2 with aspect ratio $R=1:1:3$ and free-slip vertical boundaries at time $t=200s$}
    \label{example2_3dphy}
 \end{figure*}
\begin{figure*}[!t]
    \centering \subfigure[0.5z-plane]{
    \begin{minipage}[b]{0.4\textwidth}
    \centering
    \includegraphics[width=1.1\textwidth,height=0.3\textheight]{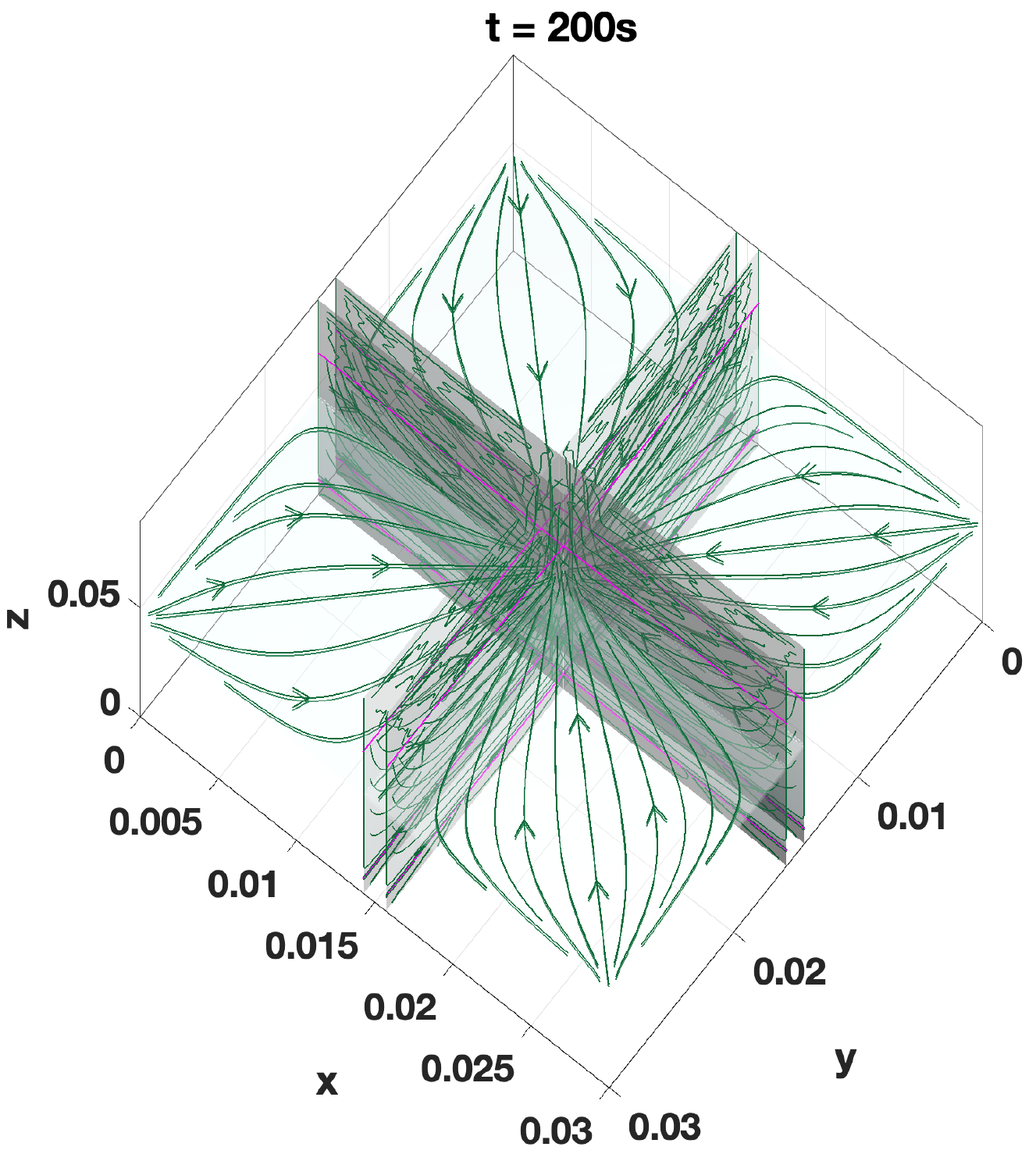}
    \label{3dzoom0.5Z}
    \end{minipage}
    }
    \centering \subfigure[Outlet of the mushy region]{
    \begin{minipage}[b]{0.4\textwidth}
    \centering
    \includegraphics[width=0.8\textwidth,height=0.3\textheight]{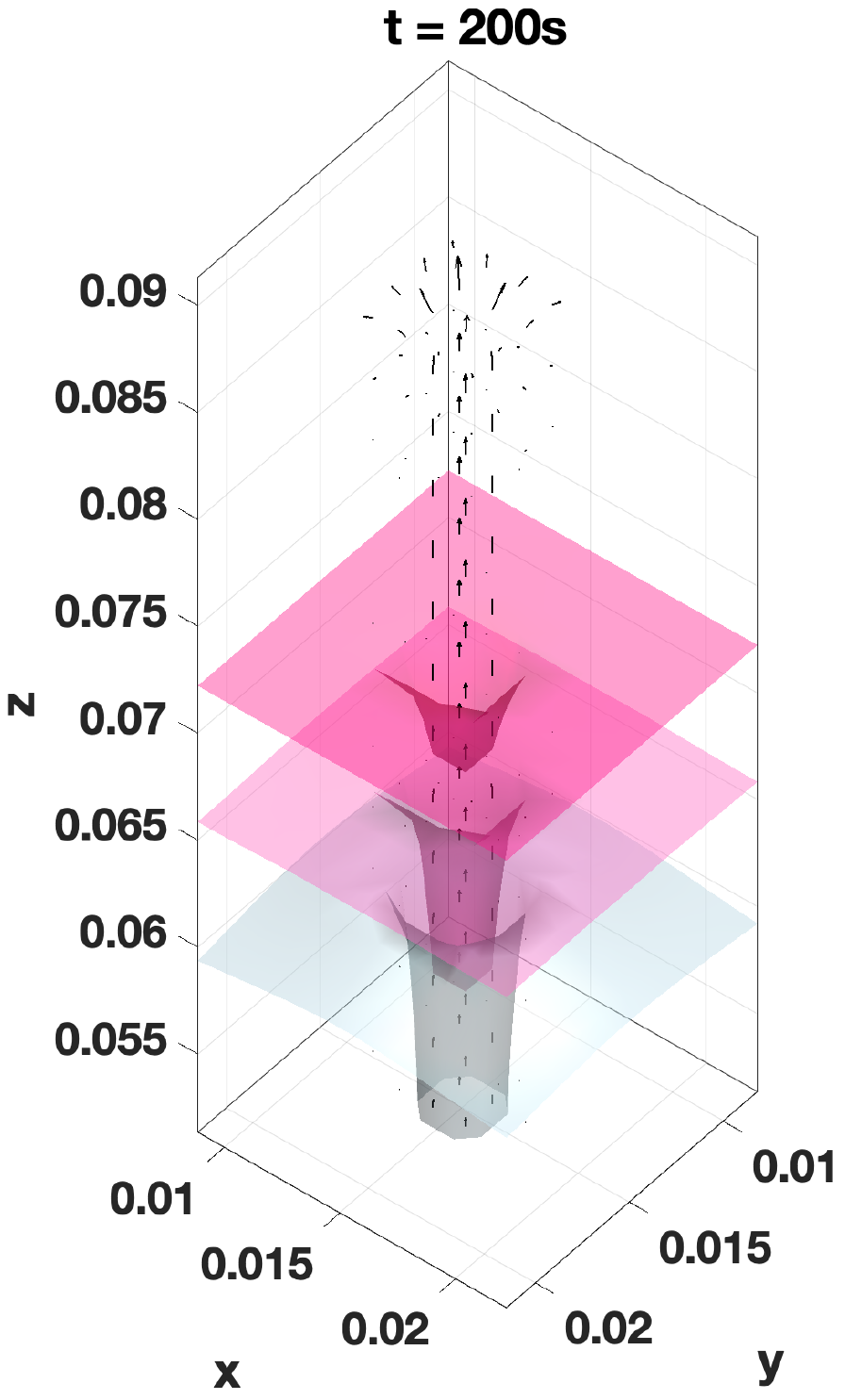}
    \label{3dzoomOutlet}
    \end{minipage}
    }
    \caption{The zoom windows of (1) streamlines on the $0.5z$-plane; (2) the platform-shaped concentration isosurfaces and the outlet of the upward flow channel at $t = 200s$}
 \end{figure*}

As shown in Figure \ref{example2_3diso}, the simulation results are similar with the 2D results. A round droplet-shape freckle forms at the initial stage, then the upward thin channel gradually grow. With the enrichment of composition \ch{Sn} in the upper liquid region, the platform-shaped isosurface layer $C-C_0=1$ is pushed and moved down. Finally, this layer is embedded in the solidifying mushy region and becomes still. Moreover, it can be observed that the shape of the central channel varies with time, the possible reason can be inferred from the velocity field around the central line in Figure \ref{example2_3diso}, also streamlines and liquid fraction in \ref{ex2_200S}. In the late of solidification, the bottom of the cavity is fully solidified and the composition \ch{Sn} is supplemented into the channel not from the bottom any more, but only from the dilution region through the stem of the channel. The process is illustrated directly by the streamlines on the plane at $0.5-Z$ in Figure \ref{3dzoom0.5Z}. In Figure \ref{3dzoomOutlet}, the zoom window of the velocity field at the outlet of the chimney flow over the liquid/mush interface platform indicates where the intense flow occurs, which naturally causes the plumes to flow in the liquid region. The temperature isosurface, streamline pattern, concentration distribution, and $C-C_0$ isosurface at $t = 200s$ in Figure \ref{example2_3dphy} all look very similar to the results from the 2D case with $R = 1:3$.
\begin{figure*}[!t]
    \centering \subfigure[$R=1:1:3$]{
    \begin{minipage}[b]{0.25\textwidth}
    \centering
    \includegraphics[width=1.\textwidth,height=0.3\textheight]{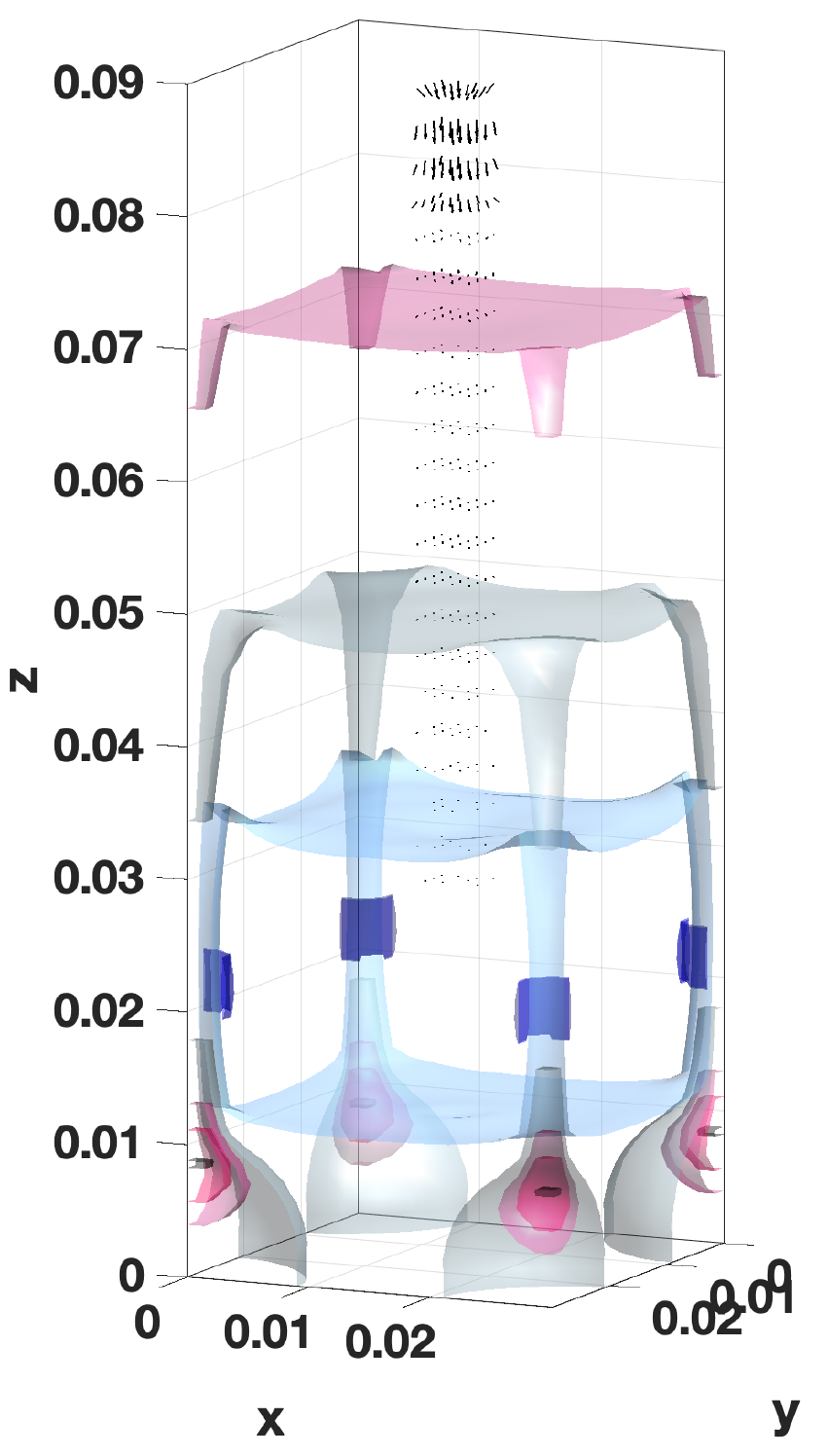}
    \end{minipage}
    }
    \centering {
    \begin{minipage}[b]{0.1\textwidth}
    \centering
    \includegraphics[width=1.\textwidth,height=0.3\textheight]{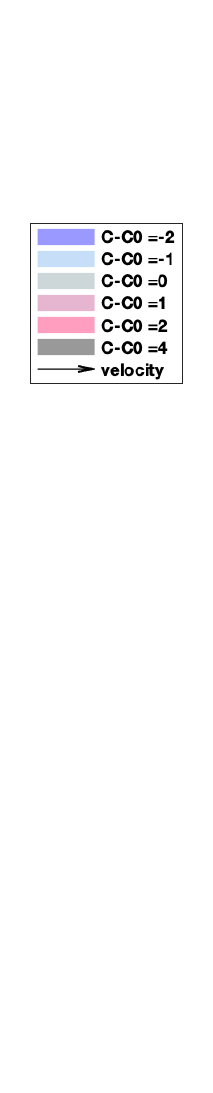}
    \end{minipage}
    }
    \centering \subfigure[$R=3:1:1$]{
    \begin{minipage}[b]{0.4\textwidth}
    \centering
    \includegraphics[width=1.1\textwidth,height=0.3\textheight]{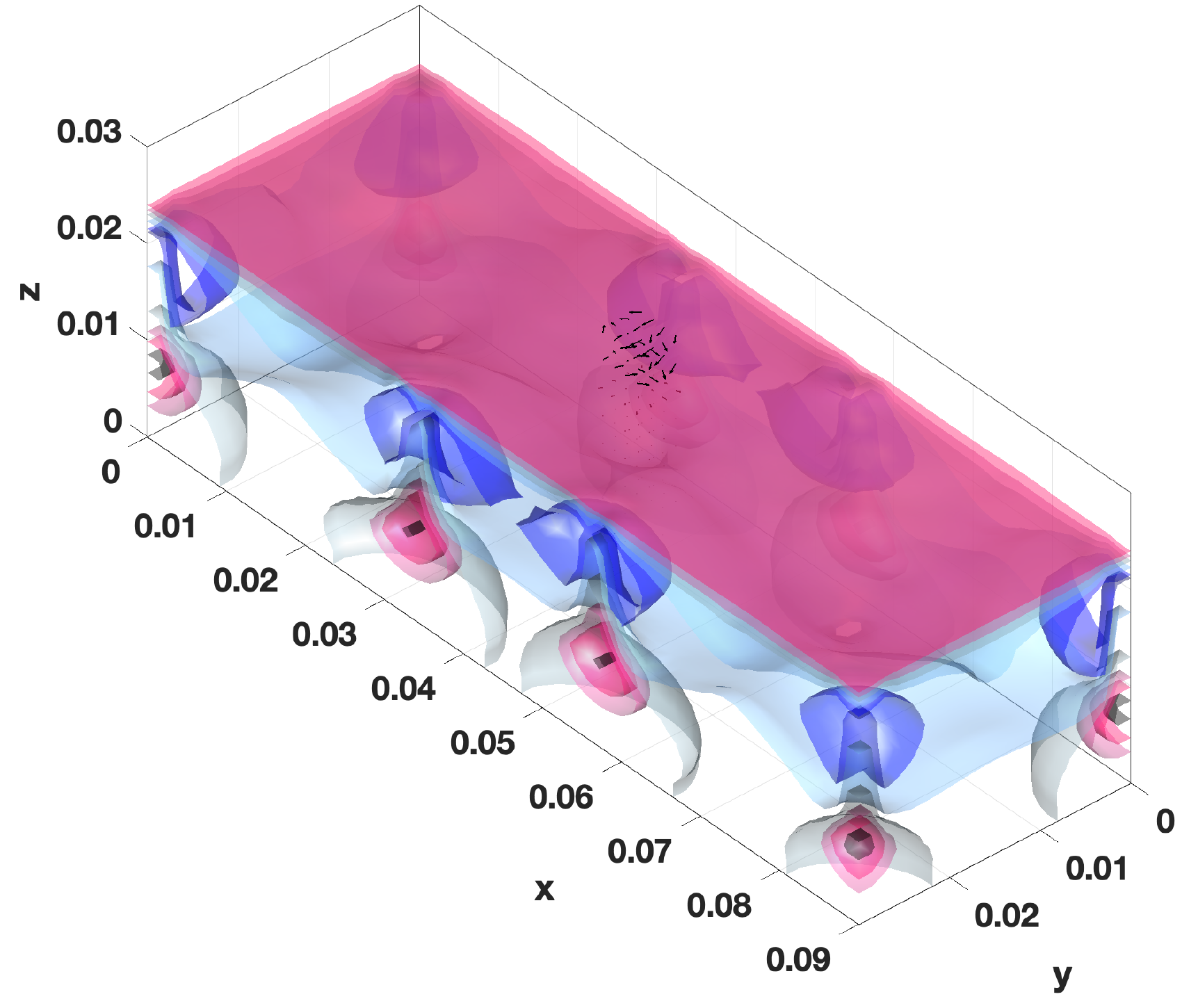}
    \end{minipage}
    }
    \caption{The isosurfaces of $C-C_0$ at $t = 100s$ in Example 2 with no-slip vertical boundaries and aspect ratio (a) $R=1:1:3$ and (b) $R=3:1:1$}
    \label{example_2_no-slip}
 \end{figure*}

In 3D cases, the no-slip boundary conditions are also employed for the aspect ratios $R = 1:1:3$ and $R = 3:1:1$ for capturing the considerable ``edge effect''. The same concentration perturbation is imposed initially, while the interesting thing is that no freckles or vertical channels are found in the vertical central line, and the direction of flow at the central part will change from upward to downward, as shown in Figure \ref{example_2_no-slip}. The possible reason is that the ``edge effect'' plays a dominant role. The flow induced by walls and edges cancels out the influence of the initial perturbation completely. In the $R = 1:1:3$ case, channels prefer to form to cling to the four upright edges, connecting these corners vertically, due to the strengthening effect of the joint action of two walls. In the $R = 3:1:1$ case, due to the aspect ratio effect, the flow induced by these four vertical edges is not enough to overcome the ``aspect ratio effect'' from two long facades to maintain one complete circle flow. Another four freckles generate and stick to the wall faces with long lengths, which is very different from the non-slip 2D case. This distinction is also known as the ``3D effect''. This kind of ``3D effect'' will be further investigated in the following Example 3.

\newpage

\textbf{Case 3: Numerical studies compared with H-H experiment and 3D effect}
\begin{figure}[b]
    \centering
    \includegraphics[width=1.1\textwidth]{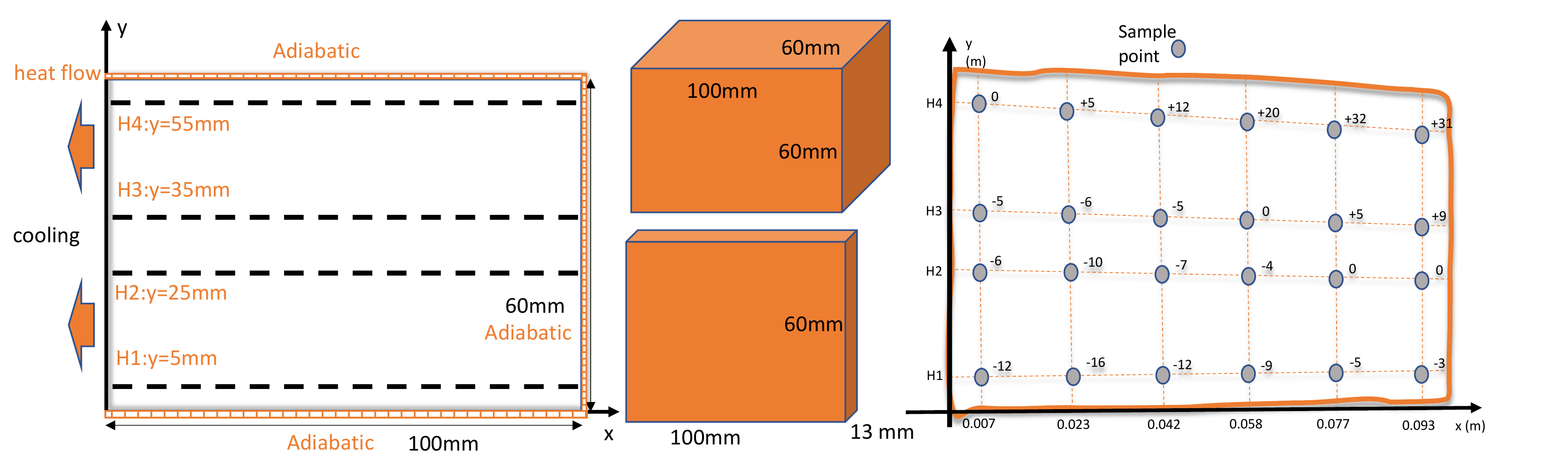}
    \caption{Schematic of the solidification block with sampling points and its geometries in Example 3}
    \label{example_3_schematic}
\end{figure}
 
In the classic Hebditch and Hunt (H-H) solidification experiment \cite{hebditch1974observations}, the material \ch{Pb}-48\ch{\%Sn} was solidified in a cavity $60mm$ high, $100mm$ long, and $13mm$ thick. This cavity is thermally insulated on all surfaces except for the westmost thin surface, and the melt in the cavity was cooled by natural convection. In the original H-H experimental work, some sample points were set to measure the concentration distribution after the fully solidification as Figure \ref{example_3_schematic}. Four horizontal lines (H1=5mm, H2=25mm, H3=35mm, H4=55mm) and serval points at theses horizontal lines were selected as the sample points. The physical properties and setting can be referred to Table \ref{alloy} as well. The numerical studies based on our method are not only conducted in 2D but also in 3D with different thicknesses. One is completely consistent with the experimental geometry, while the other has a larger thickness to validate the ``3D effect''. Simulations are implemented with $100 \times 60$ 2D mesh, $100 \times 60 \times 60$ 3D mesh, $100 \times 13 \times 60$ 3D mesh and time step $\Delta t = 0.005s$. 

\begin{figure*}[!h]
    \centering \subfigure[reference results for the 2D case of Example 3 provided in \cite{ahmad1998numerical}]{
    \begin{minipage}[b]{1.0\textwidth }
    \centering
    \includegraphics[width=1.0\textwidth]{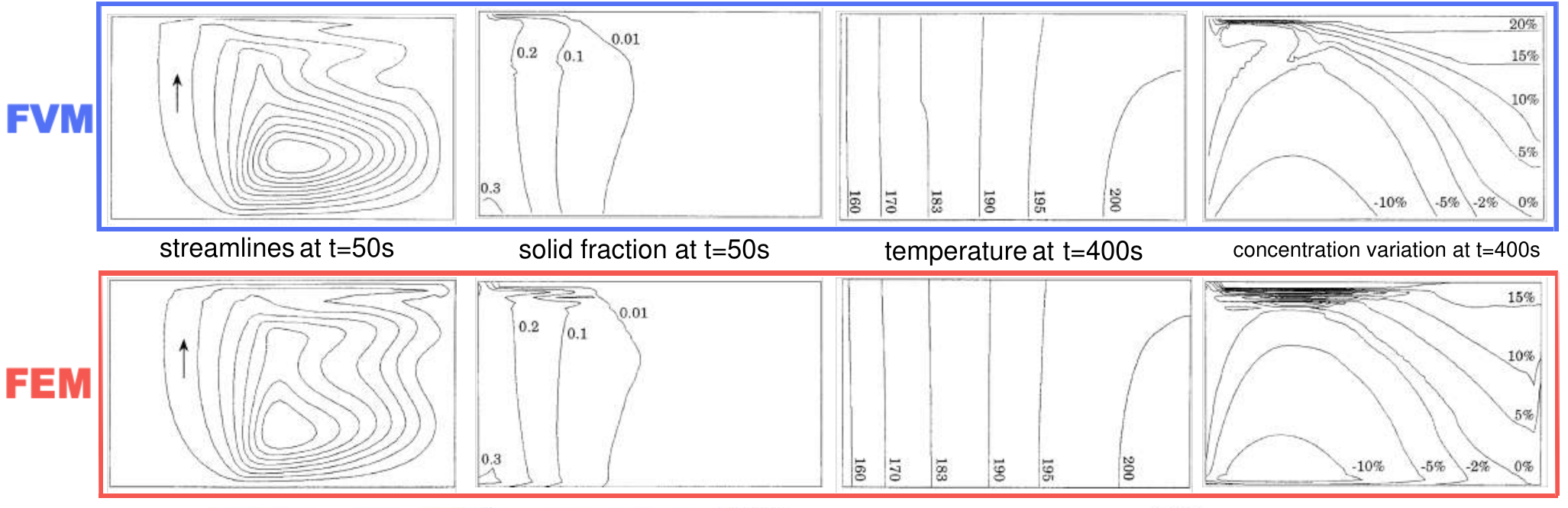} \label{example_3_2dCompareRef}
    \end{minipage}
    }
    \centering \subfigure[velocity field (black quivers); liquid fraction (colorbar)]{
    \begin{minipage}[b]{0.45\textwidth}
    \centering
    \includegraphics[width=1.0\textwidth]{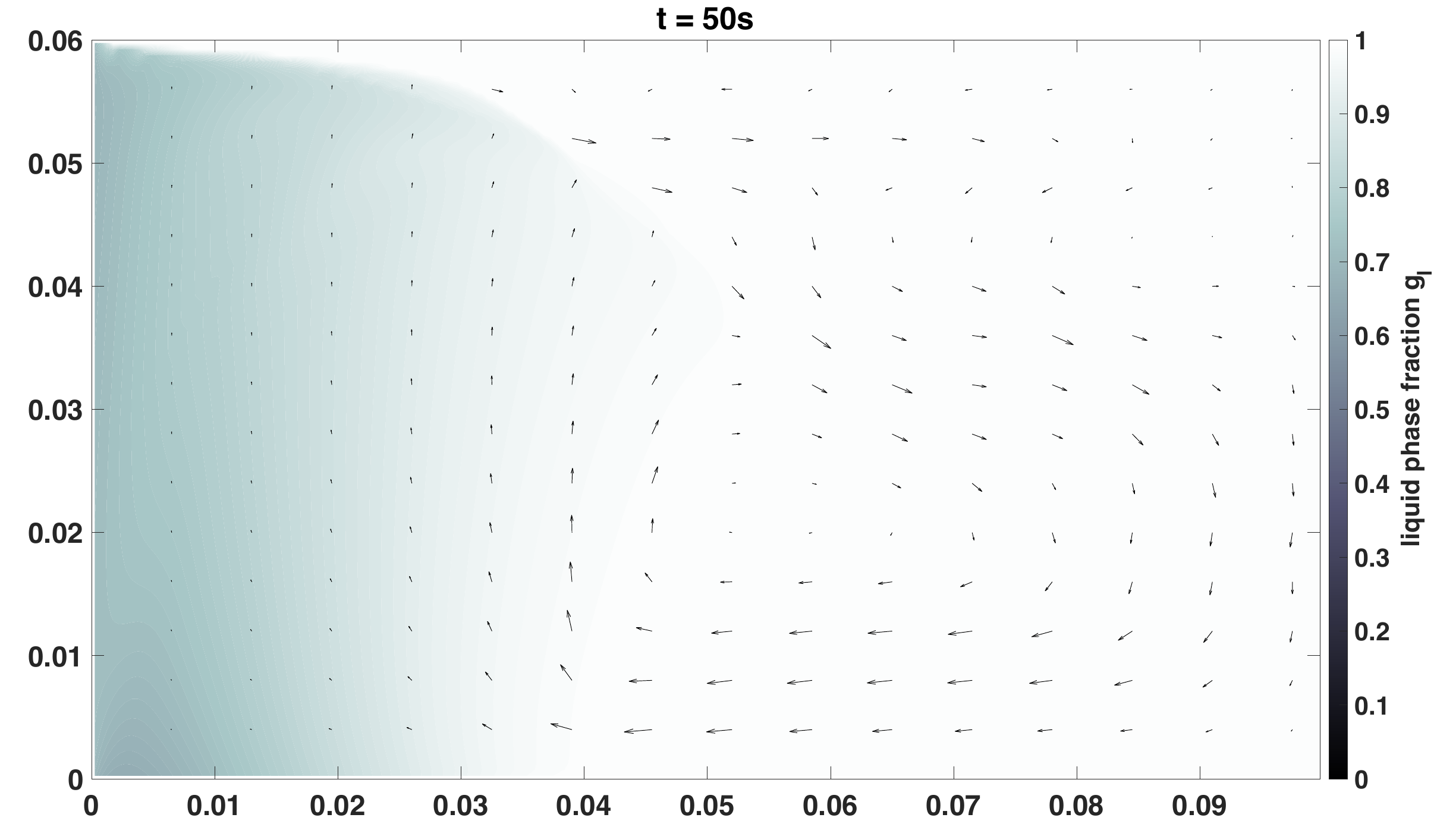} \label{example_3_2dgl}
    \end{minipage}
    }  
    \hspace{0.5cm} 
    \centering \subfigure[solid fraction (white solid line); streamlines (black solid line with arrow); concentration relative variation (colorbar)]{
    \begin{minipage}[b]{0.45\textwidth}
    \centering
    \includegraphics[width=1.0\textwidth]{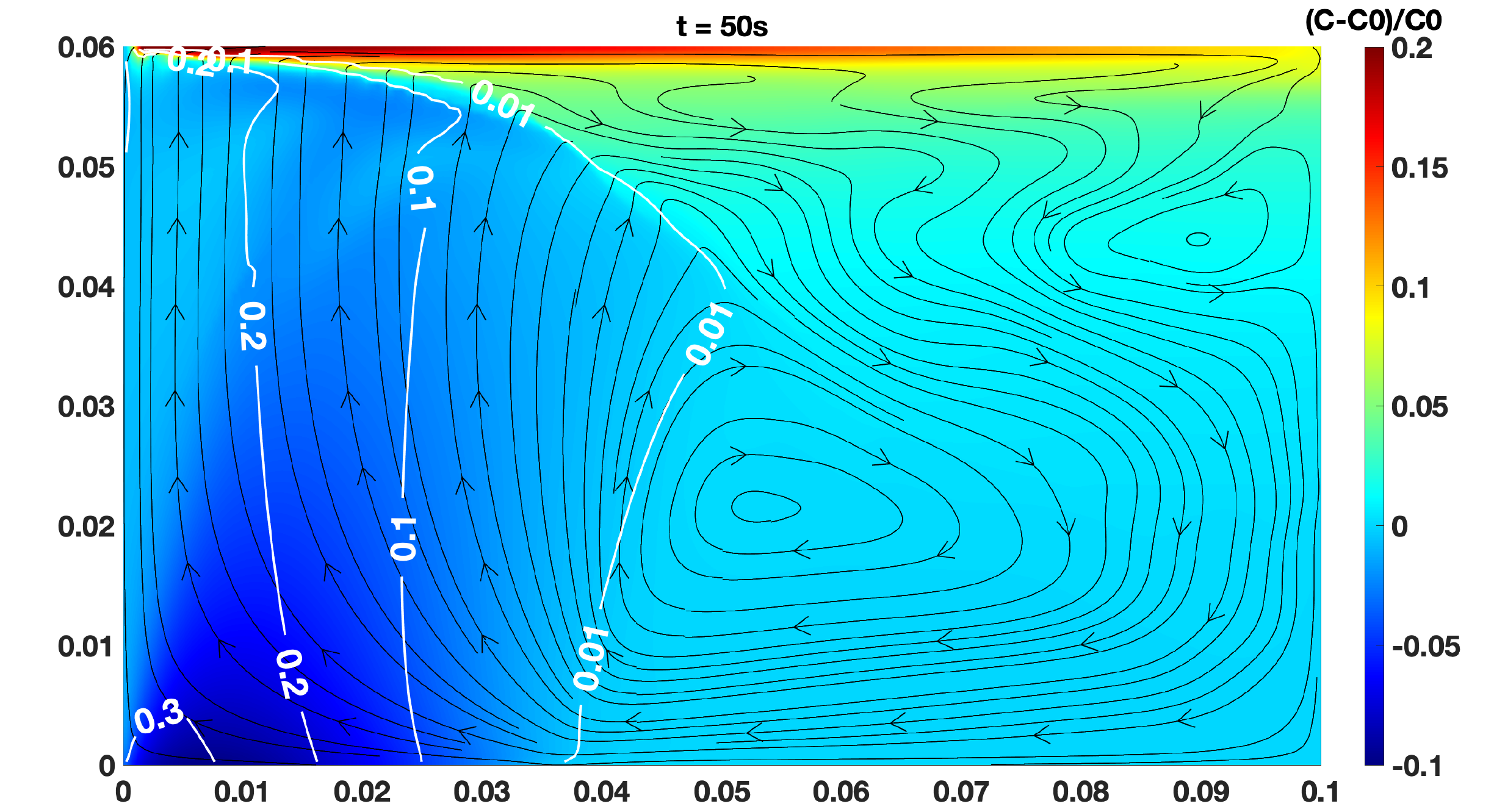} \label{example_3_2dstreamC}
    \end{minipage}
    }
    \centering \subfigure[liquid/mush interface (magenta dot line); mush/solid interface (black dot line); normalized velocity (colored vectors)]{
    \begin{minipage}[b]{0.45\textwidth}
    \centering
    \includegraphics[width=1.0\textwidth]{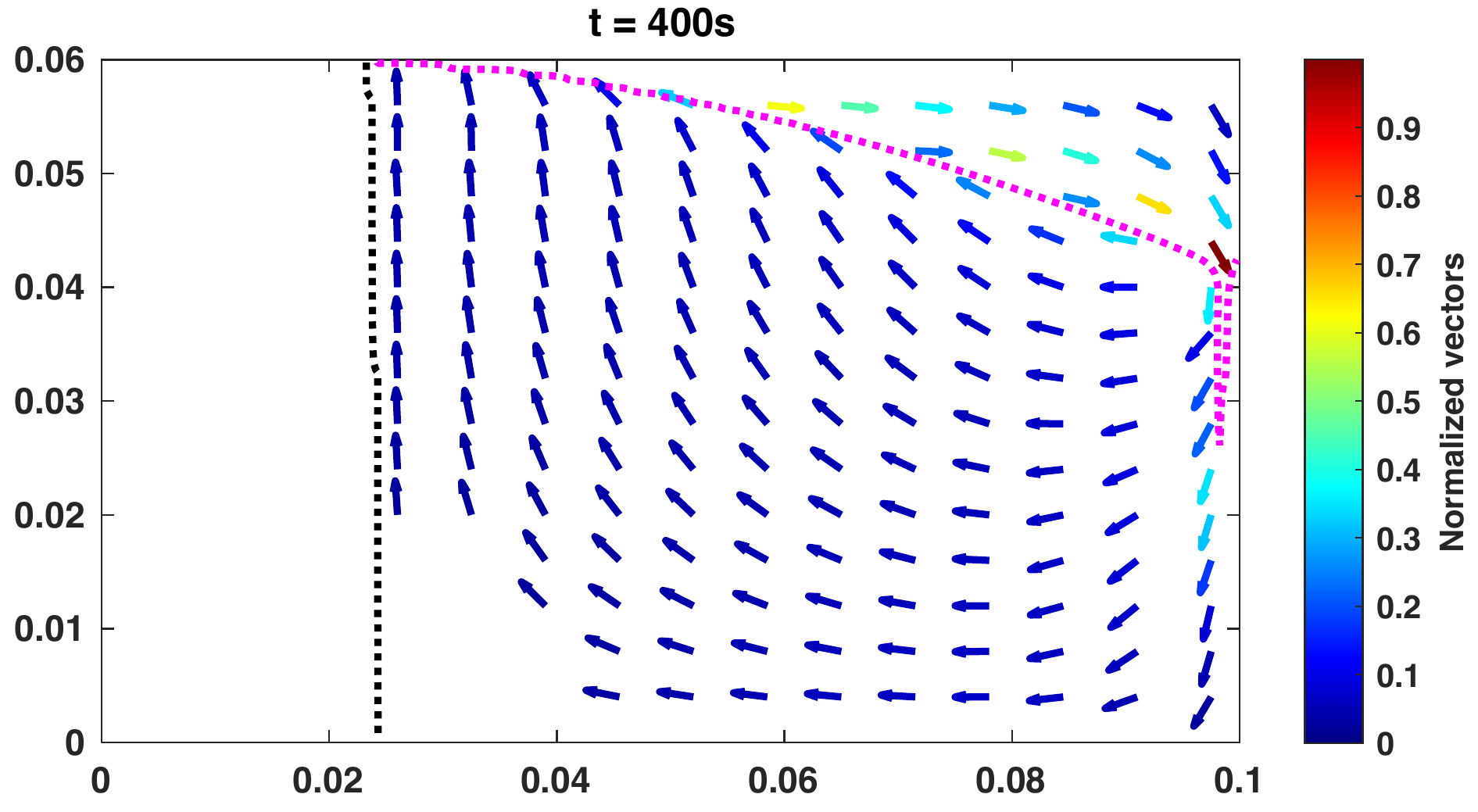} \label{example_3_2dcolor}
    \end{minipage}
    }
    \hspace{0.5cm} 
    \centering \subfigure[$(C-C0)/C0$ (magenta dash-dotted line); temperature contour line (white solid line); temperature (colorbar)]{
    \begin{minipage}[b]{0.45\textwidth}
    \centering
    \includegraphics[width=1.0\textwidth]{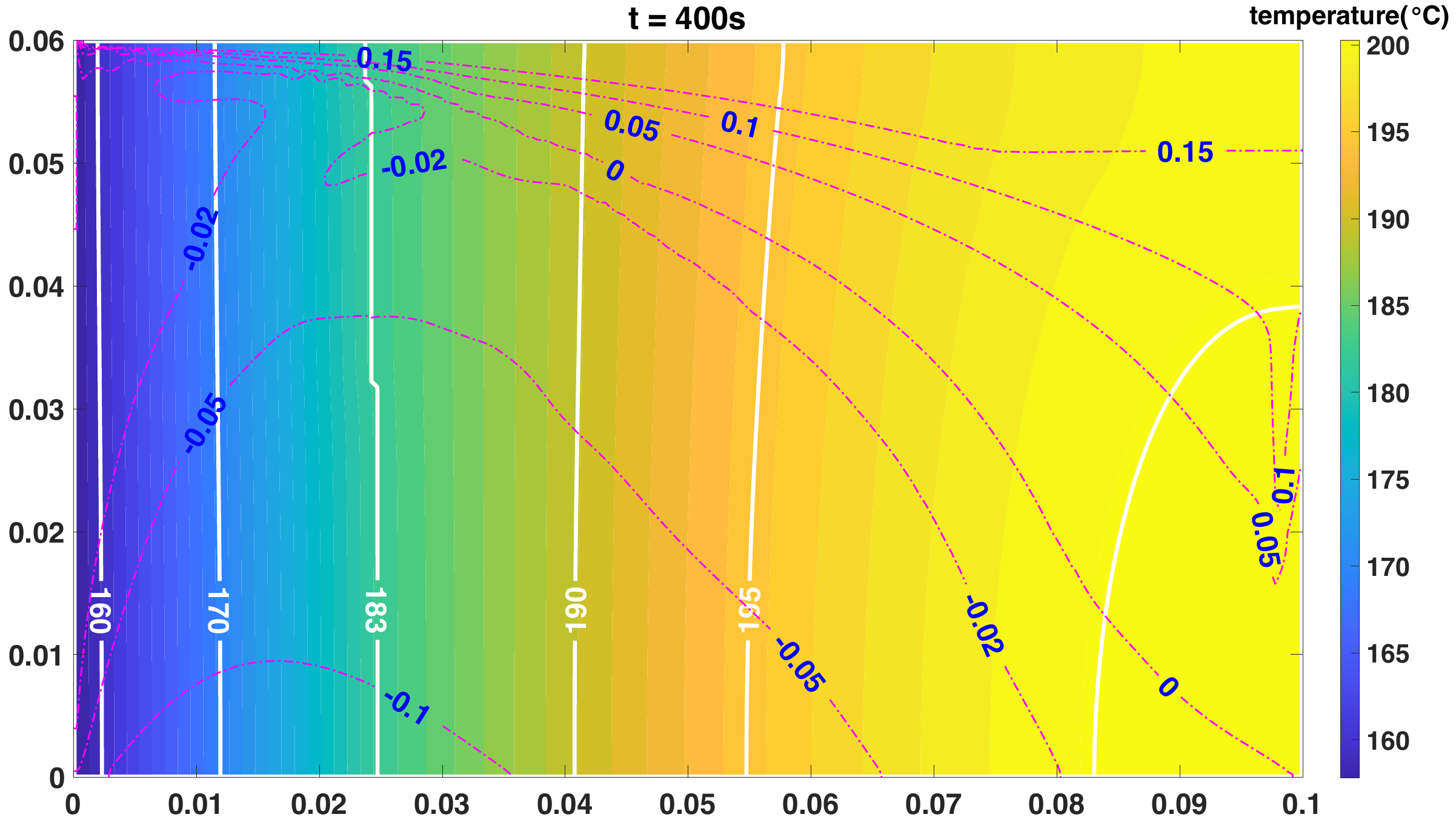} \label{example_3_2dTC}
    \end{minipage}
    }
    \caption{Reference numerical results and the results predicted by this scheme for the 2D case of Example 3}
    \label{example_3_2dCompare}
\end{figure*}

\begin{figure}[!h]
    \centering
    \includegraphics[width=0.5\textwidth]{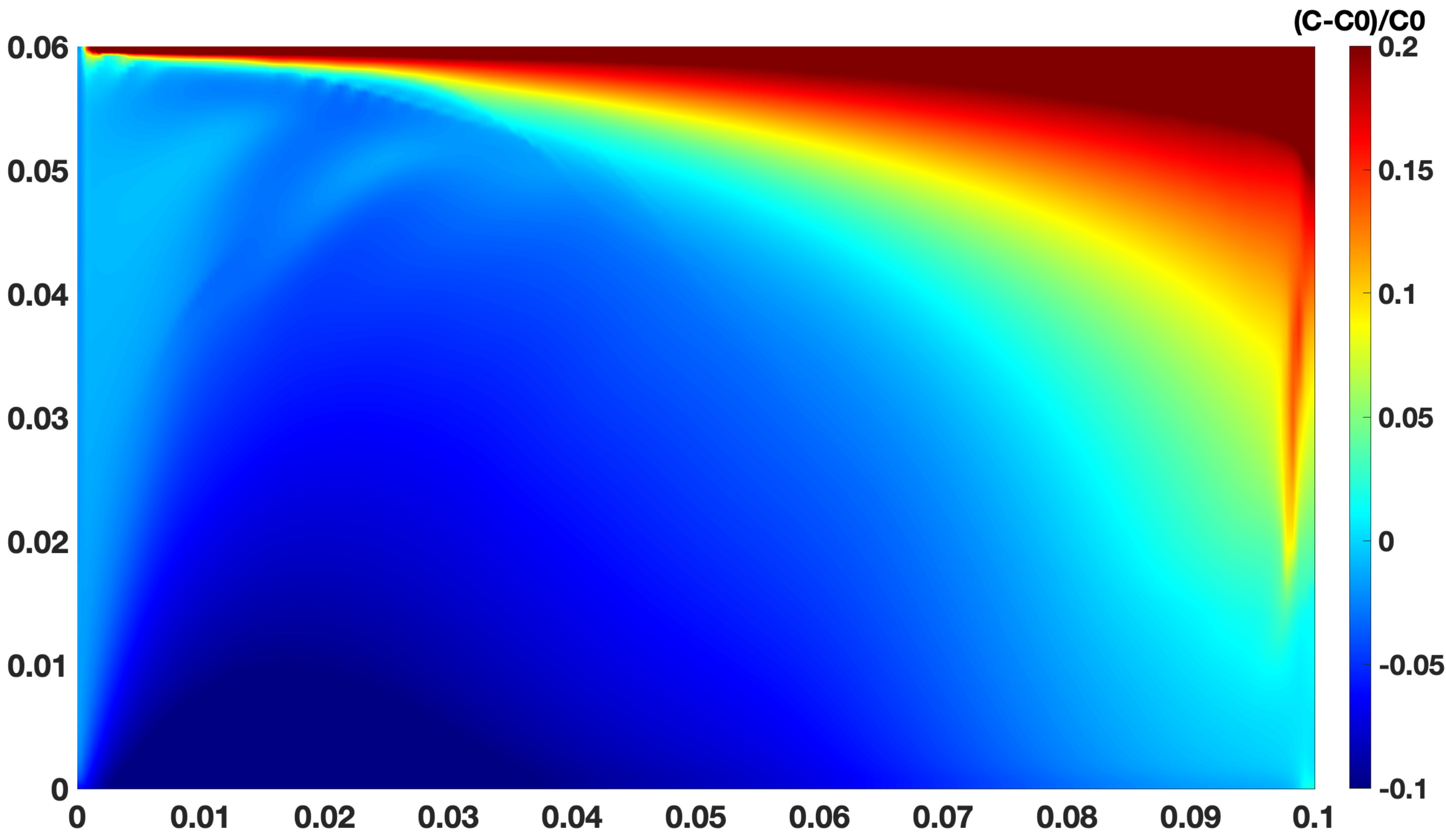}
    \caption{The segregation map after fully solidification}
    \label{example_3_fullySolid}
\end{figure}

In the 2D case, our results are compared with the reference results computed by the classic FVM and FEM methods \cite{ahmad1998numerical}, as shown in Figure \ref{example_3_2dCompare}. The results show pretty good agreement at $t = 50s$ and $ t = 400s$ based on the (1) temperature contour map, (2) concentration distribution, and (3) streamline pattern. The comparison demonstrates good accuracy in describing the complex and unsteady process. The segregation pattern after fully solidification computed by our scheme is also presented in Figure \ref{example_3_fullySolid} and it clearly indicates the distribution of the dilution and enrichment of composition regions and captures the details of the segregation phenomenon.

\begin{figure*}[!h]
    \centering \subfigure[5mm]{
    \begin{minipage}[b]{0.35\textwidth }
    \centering
    \includegraphics[width=1.0\textwidth]{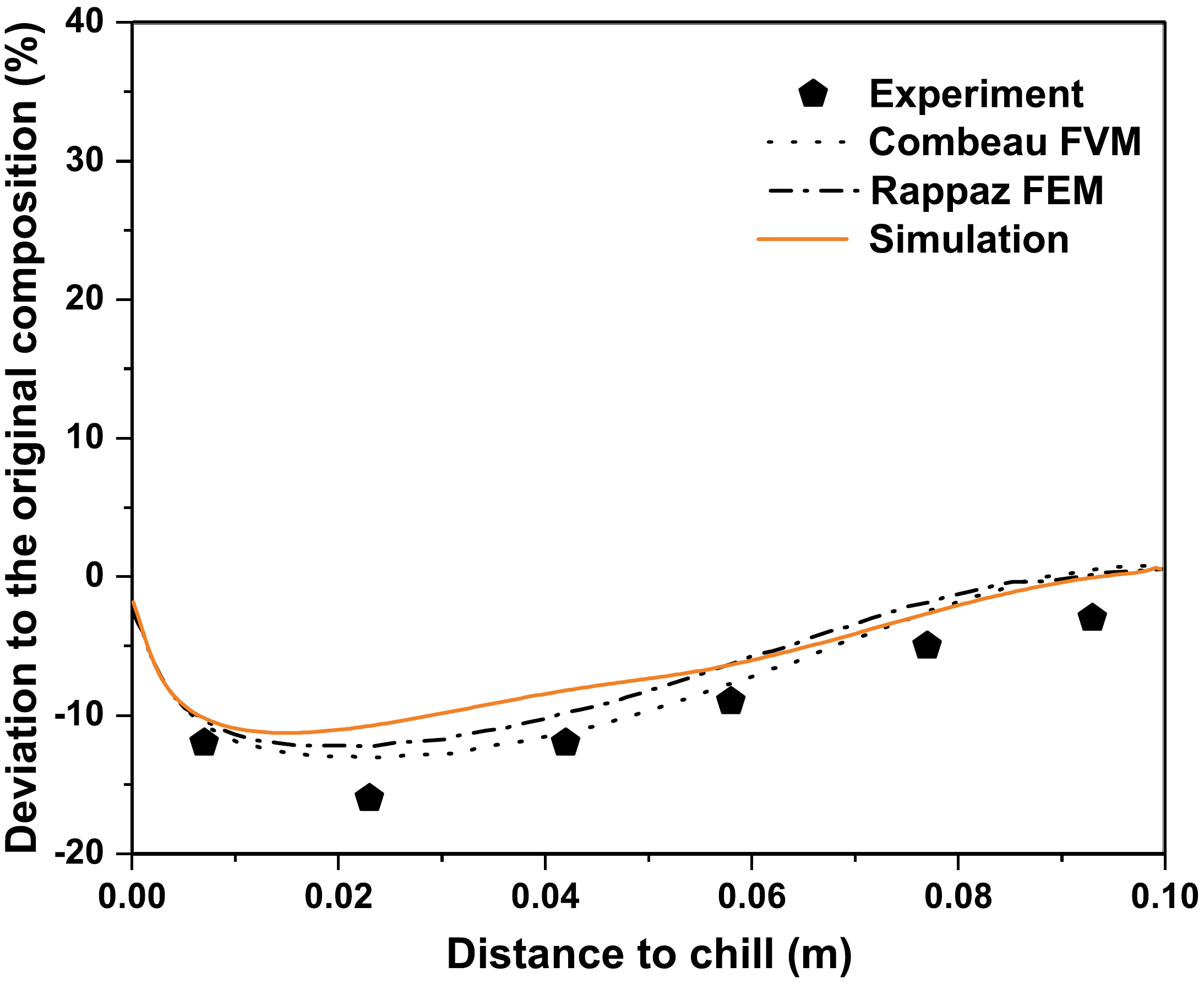}
    \end{minipage}
    }
    \centering \subfigure[25mm]{
    \begin{minipage}[b]{0.35\textwidth}
    \centering
    \includegraphics[width=1.0\textwidth]{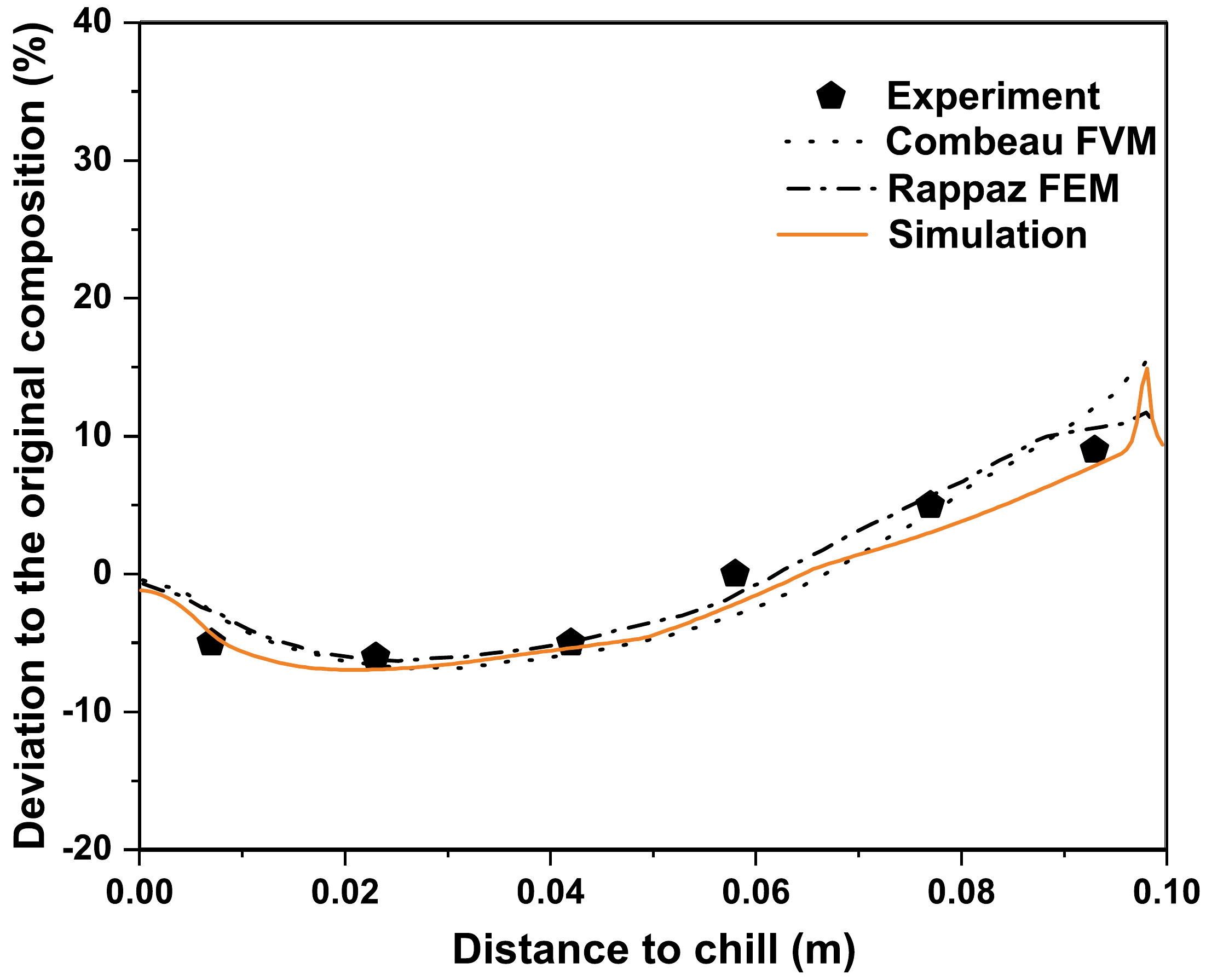}
    \end{minipage}
    }
    \centering \subfigure[35mm]{
    \begin{minipage}[b]{0.35\textwidth}
    \centering
    \includegraphics[width=1.0\textwidth]{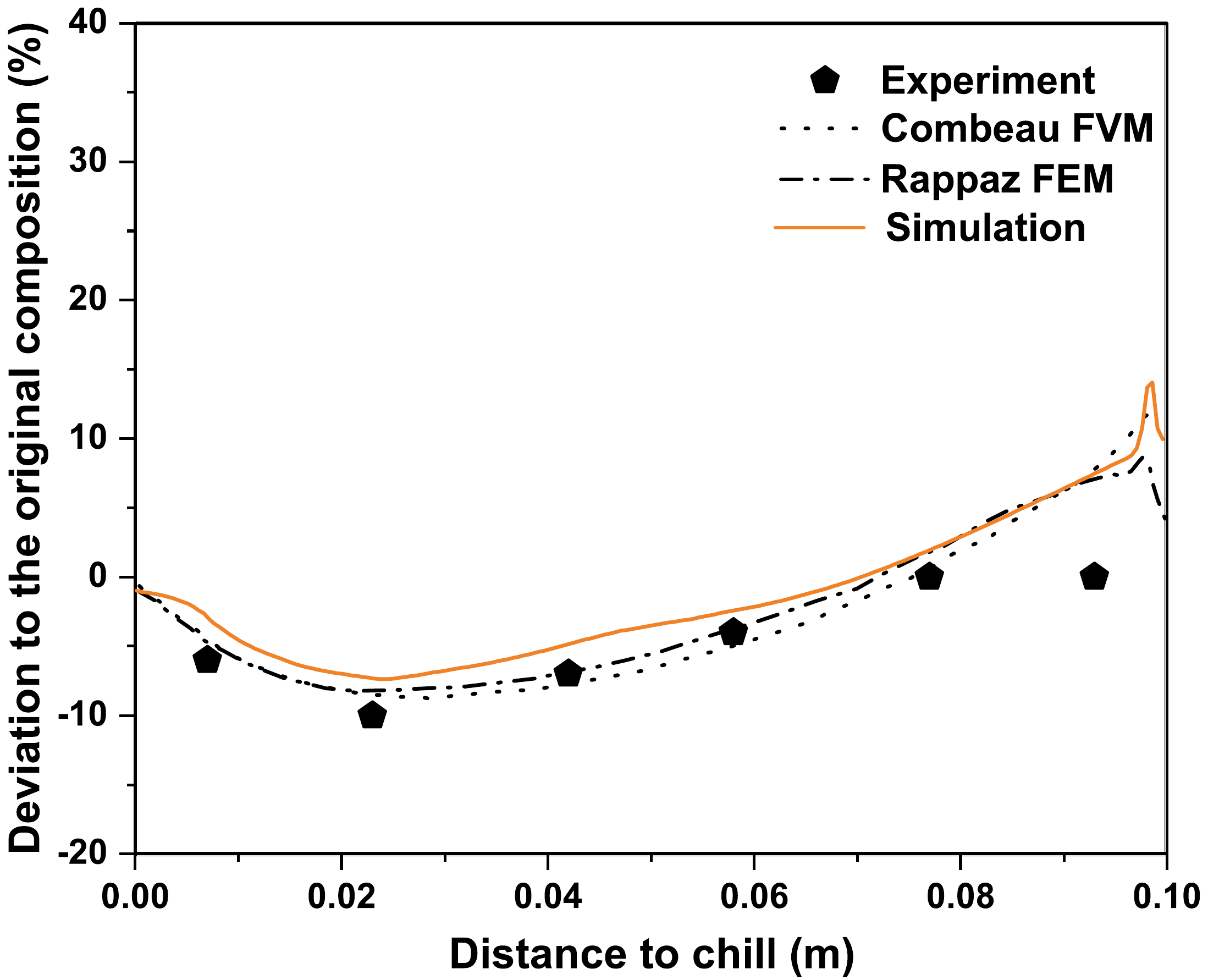}
    \end{minipage}
    }
    \centering \subfigure[55mm]{
    \begin{minipage}[b]{0.35\textwidth}
    \centering
    \includegraphics[width=1.0\textwidth]{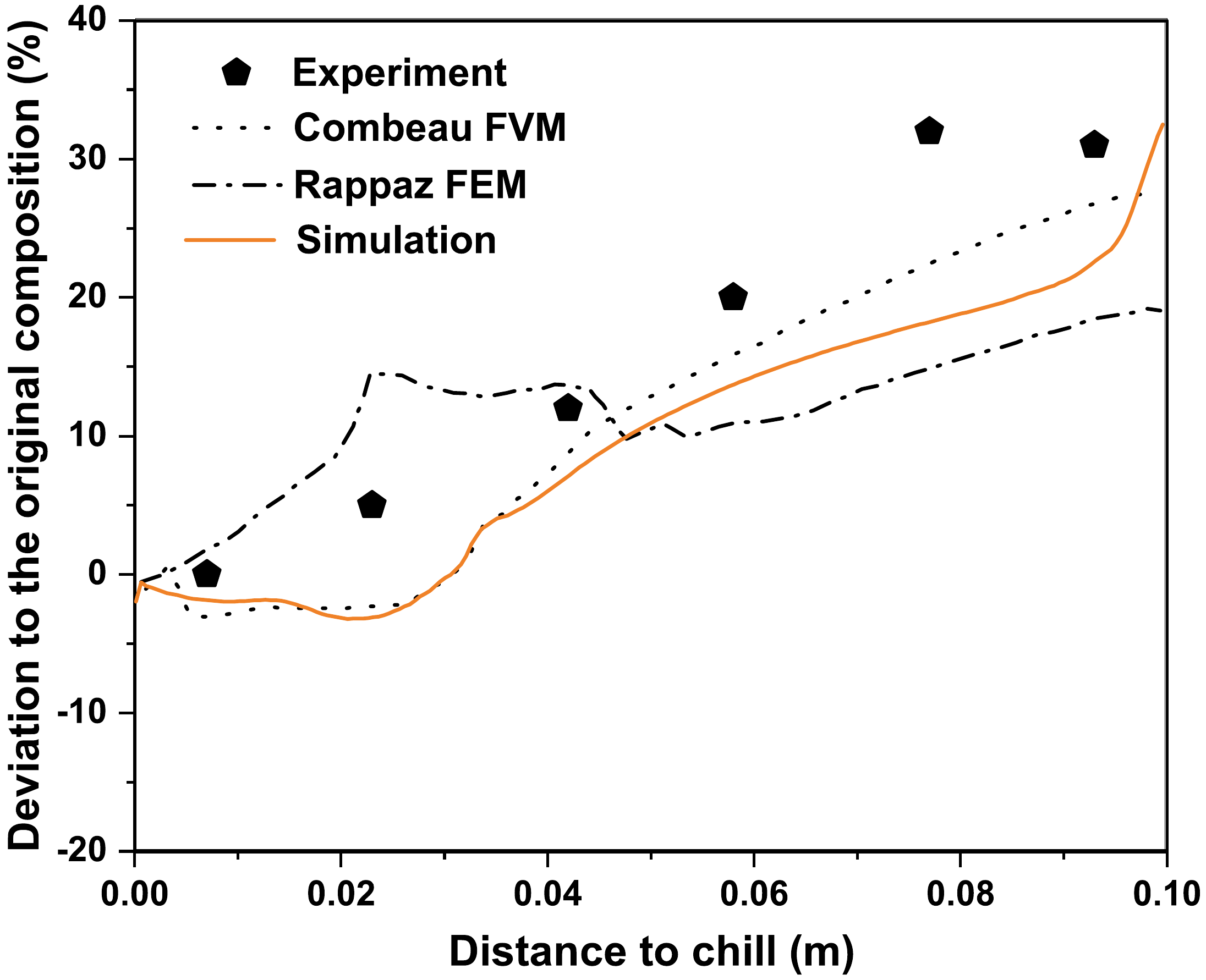}
    \end{minipage}
    }
    \caption{Relative \ch{Sn} concentration variations for the binary alloy at the end of solidification as a function of the distance to the cold wall; the comparison among numerical results from our scheme, from the classic FVM/FEM method, and the measured experiment data, at various heights from the bottom: (a) 5mm, (b) 25mm, (c) 35mm, (d) 55mm.}
    \label{example_2_4lines}
\end{figure*}

Our results are also compared with the data extracted from sample points along these horizontal lines from the real H-H experiment and other simulation trials. In Figure \ref{example_2_4lines}, it can be seen that the results computed by our scheme are in good agreement with the results of the classic FVM or FEM framework for most cases. For the curve describing the horizontal line with height $h = 55mm$, the interval of this curve at a large distance from chill corresponds to the latest region to become fully solid and the most positive segregation. Neither our results nor those of other researchers match up perfectly with the lab data. Previous research attributed the possible cause to the shrinkage and inverse segregation, which were not taken into account in the simulation. Since the data is extracted from the 3D thin block, the 3D effect might be an essential factor resulting in this difference as well. From the results of Example 2, we knew 3D applications would lead to some non-ignorable differences in physical phenomena. It is referred as the ``3D effect''. Then, we use forward equation-based matrix assembly techniques to extend this H–H experiment modeling into 3D as well.

\begin{figure*}[!h]
    \centering \subfigure[velocity field]{
    \begin{minipage}[b]{0.35\textwidth}
    \centering
    \includegraphics[width=1.0\textwidth]{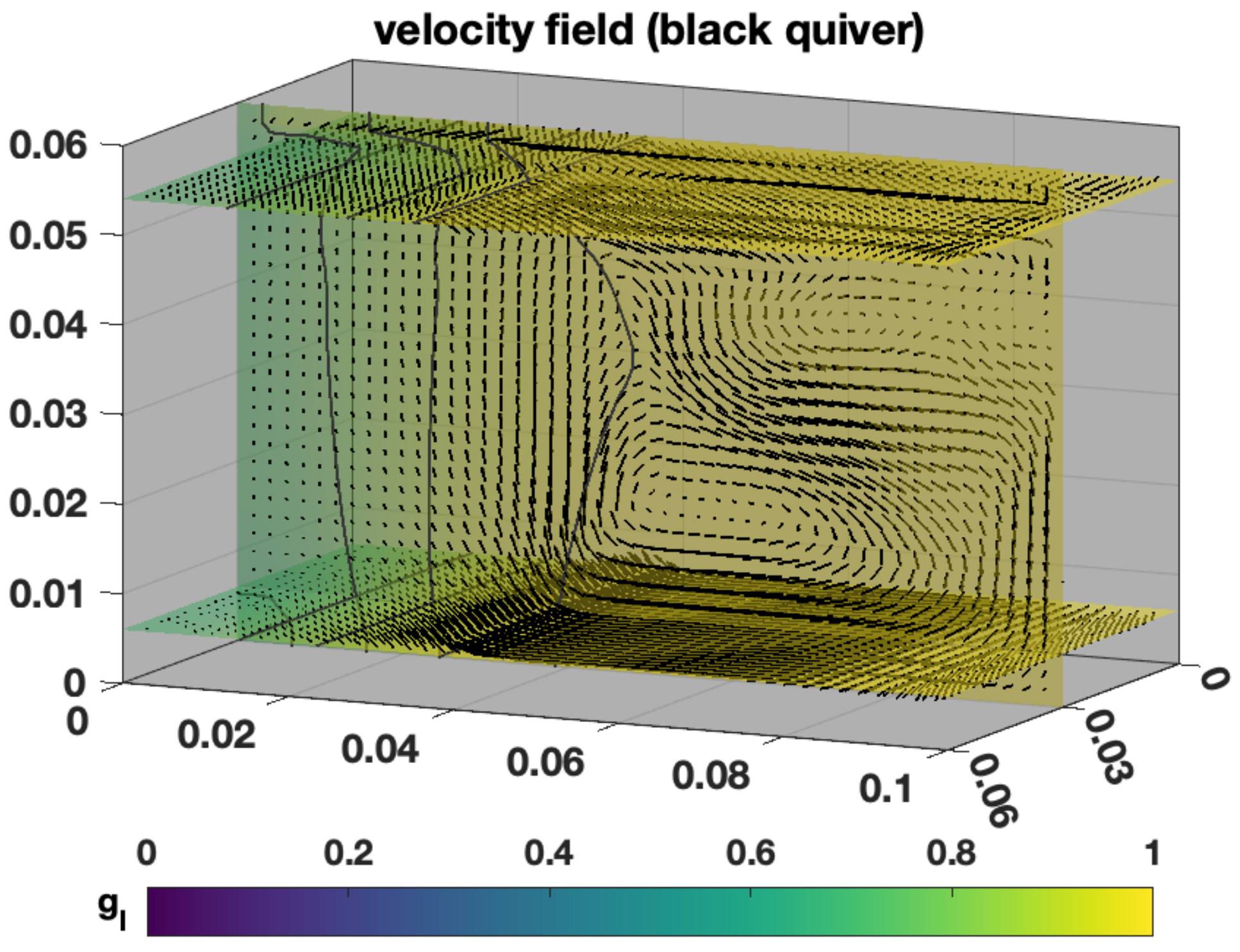} \label{example_3_3dthick50v}
    \end{minipage}
    }
    \centering \subfigure[temperature]{
    \begin{minipage}[b]{0.35\textwidth}
    \centering
    \includegraphics[width=1.0\textwidth]{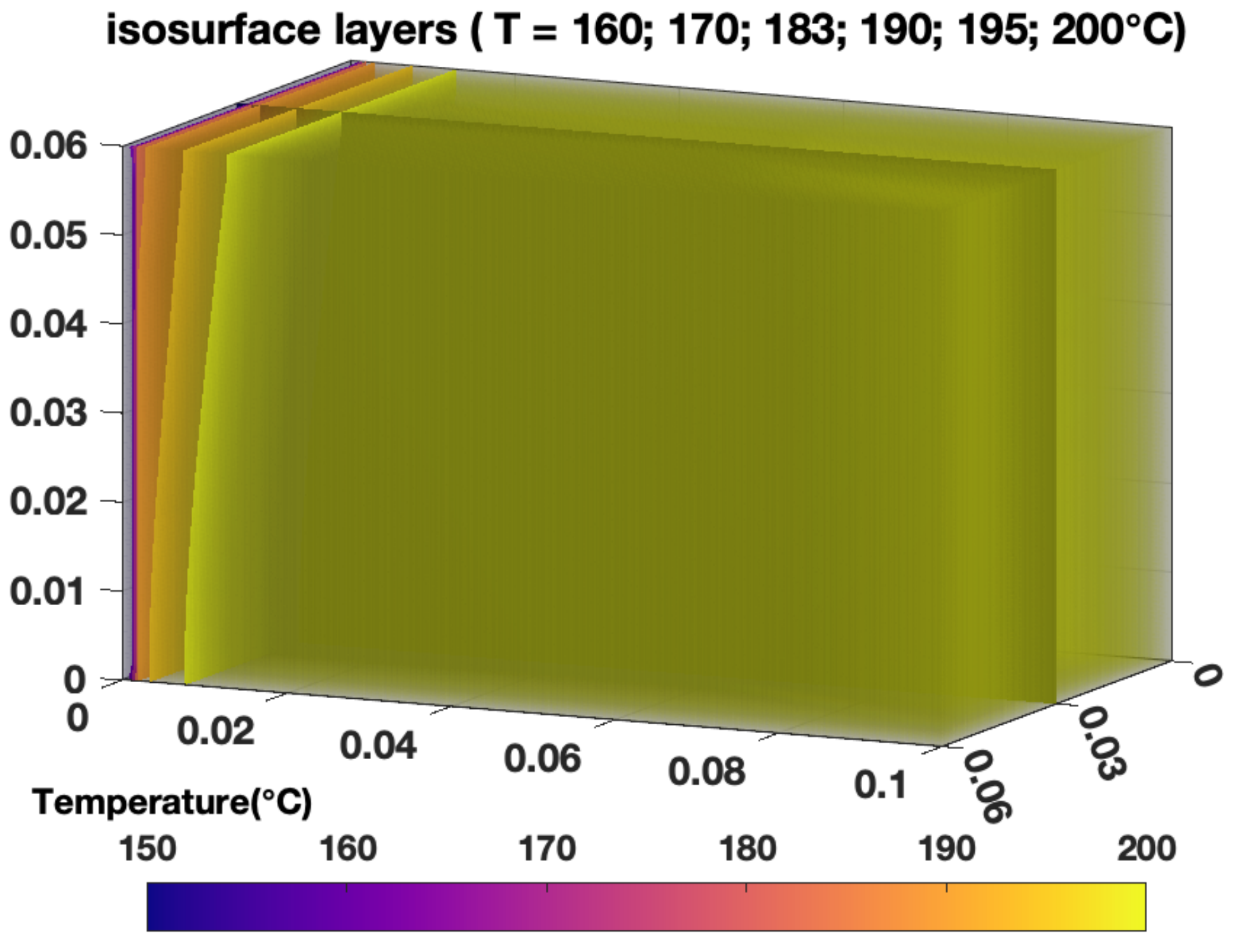}    \label{example_3_3dthick50T}
    \end{minipage}
    }
    \centering \subfigure[liquid fraction]{
    \begin{minipage}[b]{0.35\textwidth}
    \centering
    \includegraphics[width=1.0\textwidth]{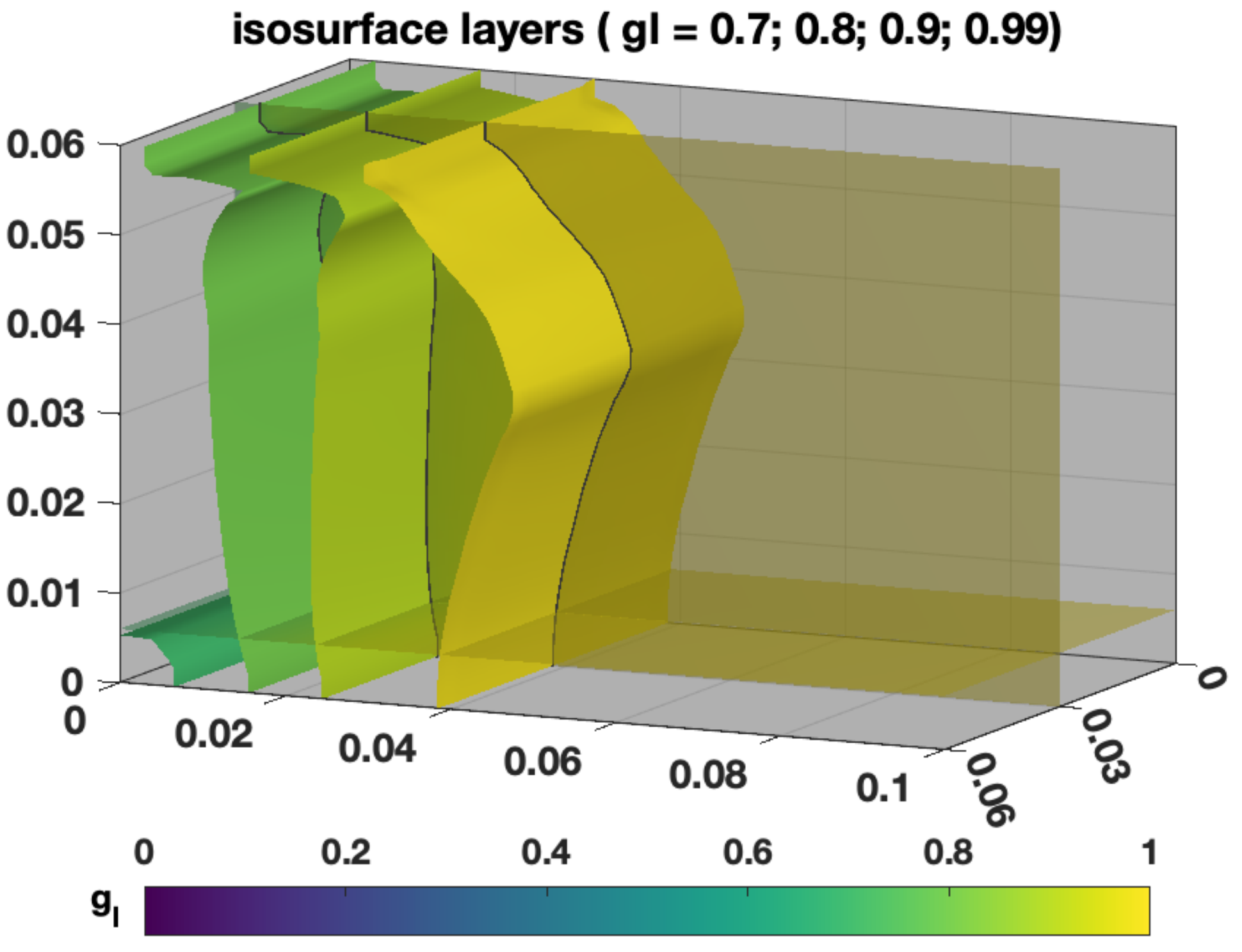}  \label{example_3_3dthick50gl}
    \end{minipage}
    }
    \centering \subfigure[concentration / streamlines]{
    \begin{minipage}[b]{0.35\textwidth}
    \centering
    \includegraphics[width=1.0\textwidth]{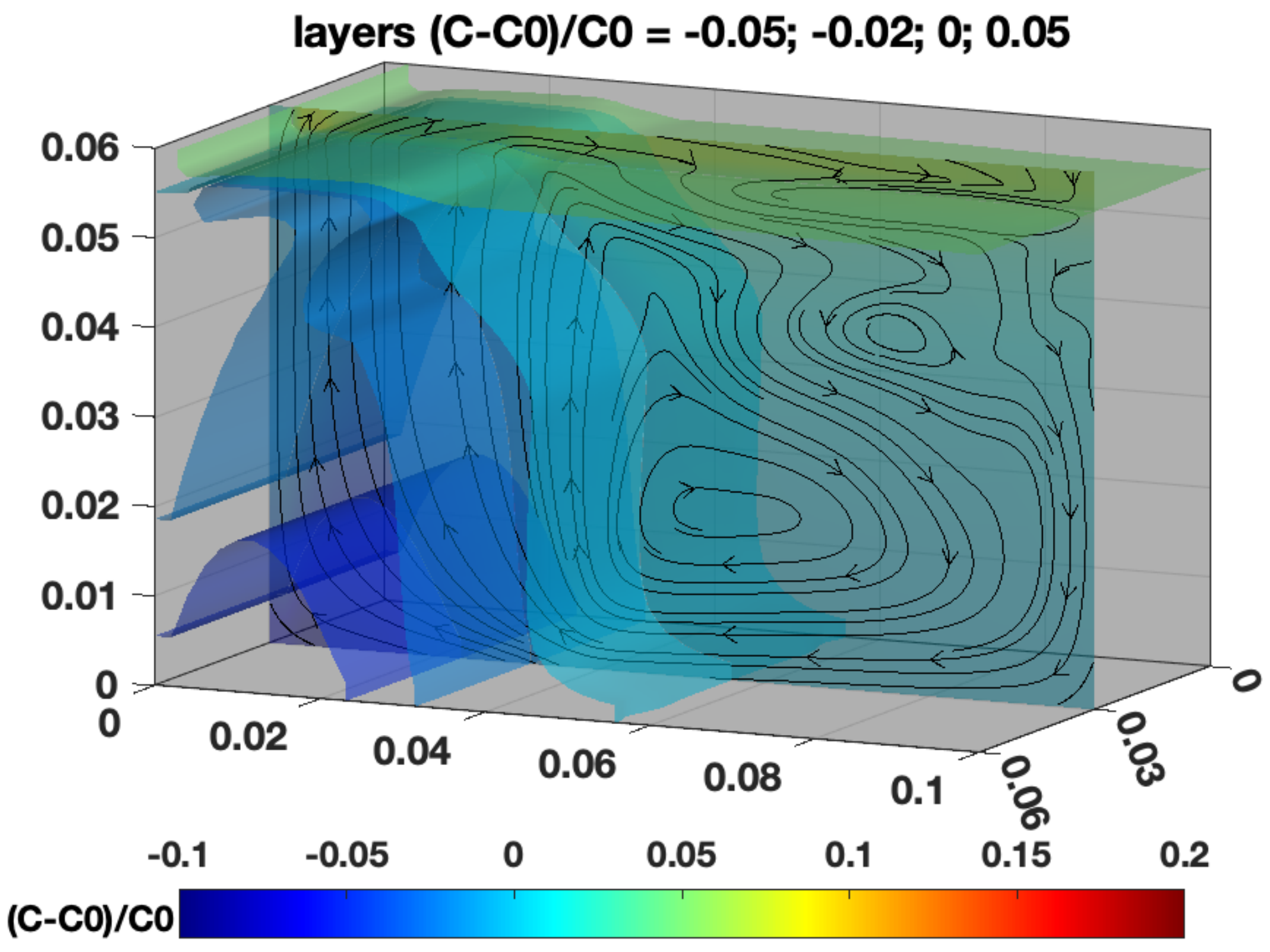}  \label{example_3_3dthick50C}
    \end{minipage}
    }
    \caption{Physical fields of the solidification of the 3D case with 60 mm thickness in Example 3 at time t = 50s}
    \label{example_3_3dthick_50}
 \end{figure*}
   \begin{figure*}[!h]
    \centering \subfigure[velocity field]{
    \begin{minipage}[b]{0.35\textwidth }
    \centering
    \includegraphics[width=1.0\textwidth]{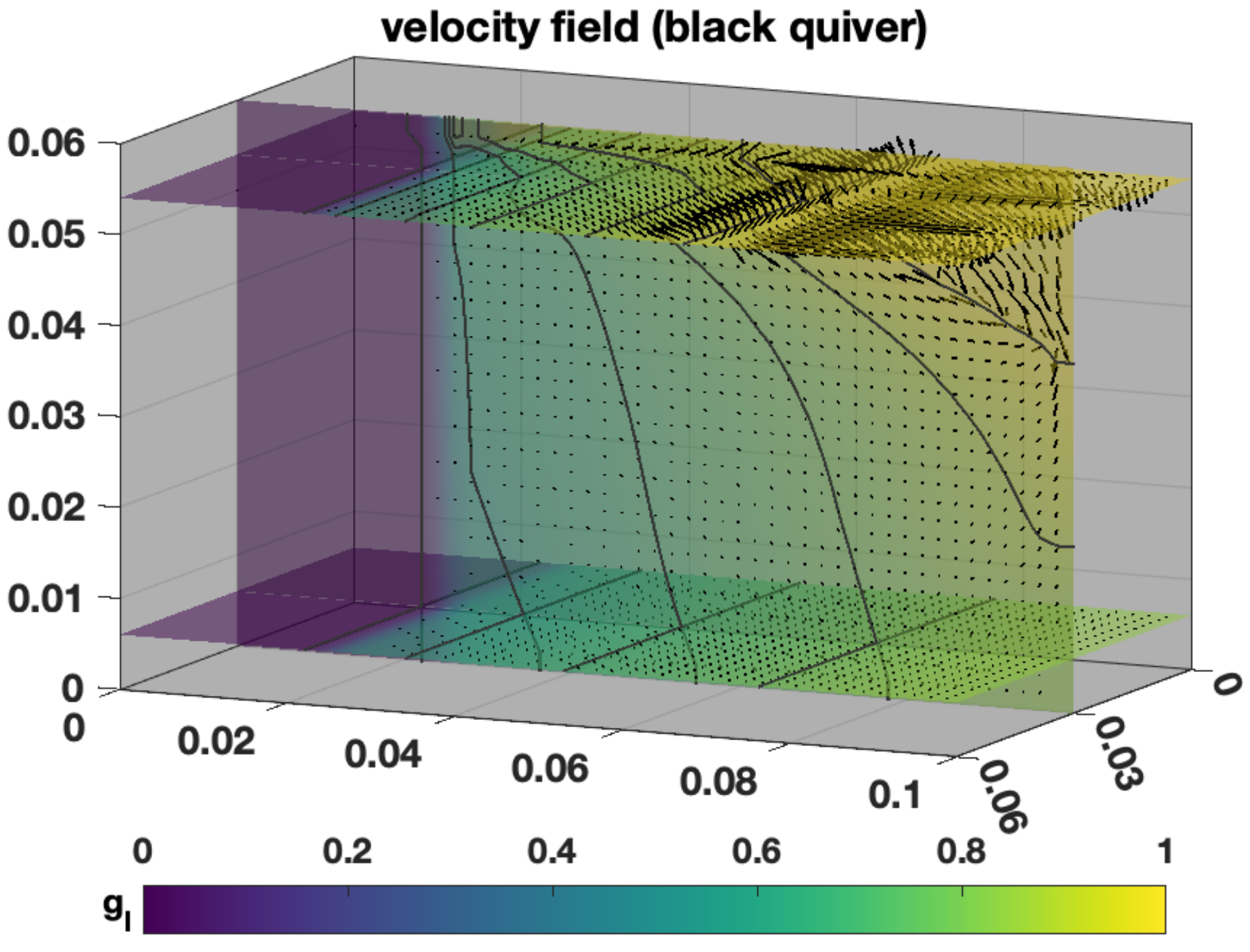}  \label{example_3_3dthick400v}
    \end{minipage}
    }
    \centering \subfigure[temperature]{
    \begin{minipage}[b]{0.35\textwidth}
    \centering
    \includegraphics[width=1.0\textwidth]{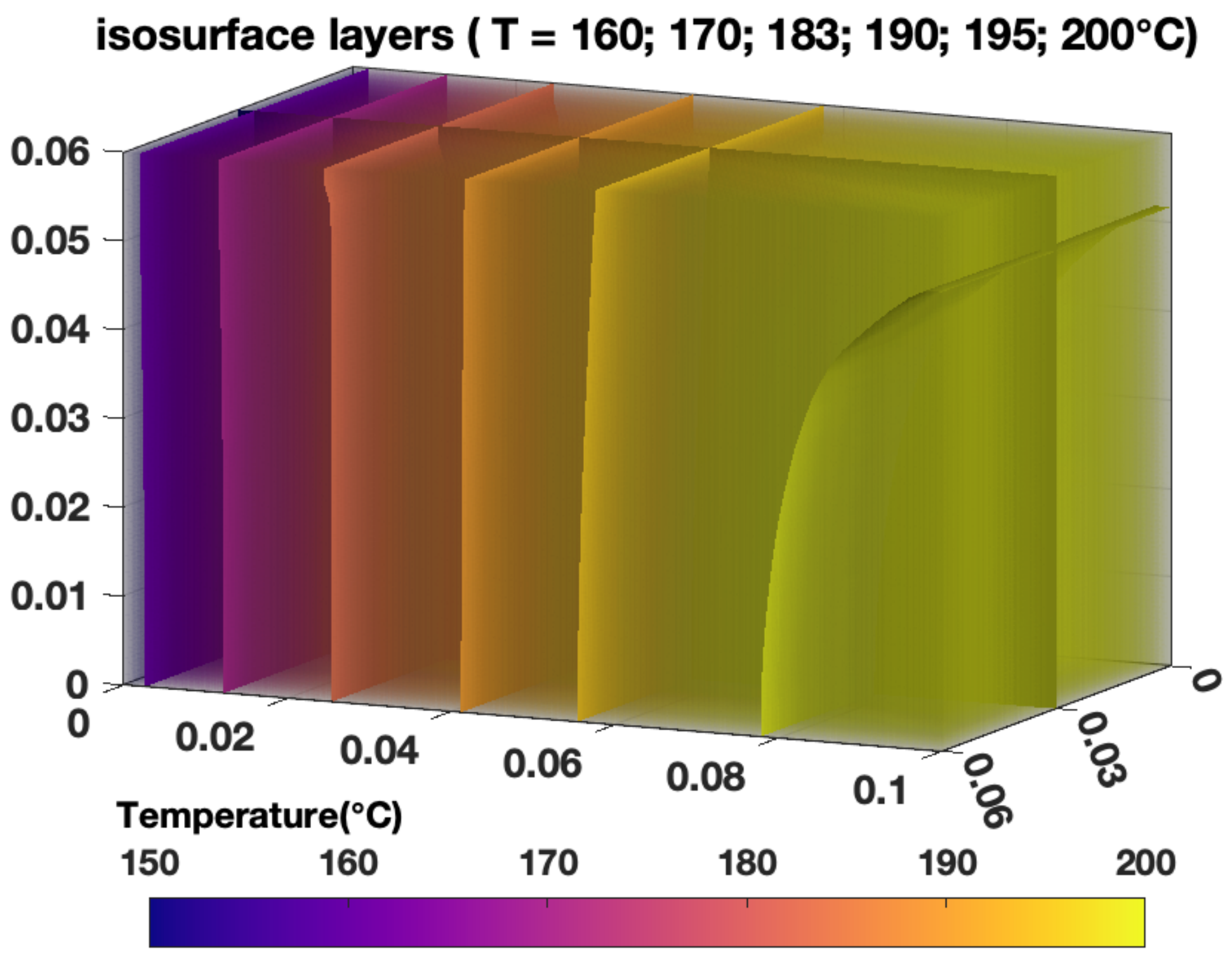} \label{example_3_3dthick400T}
    \end{minipage}
    }
    \centering \subfigure[liquid fraction]{
    \begin{minipage}[b]{0.35\textwidth}
    \centering
    \includegraphics[width=1.0\textwidth]{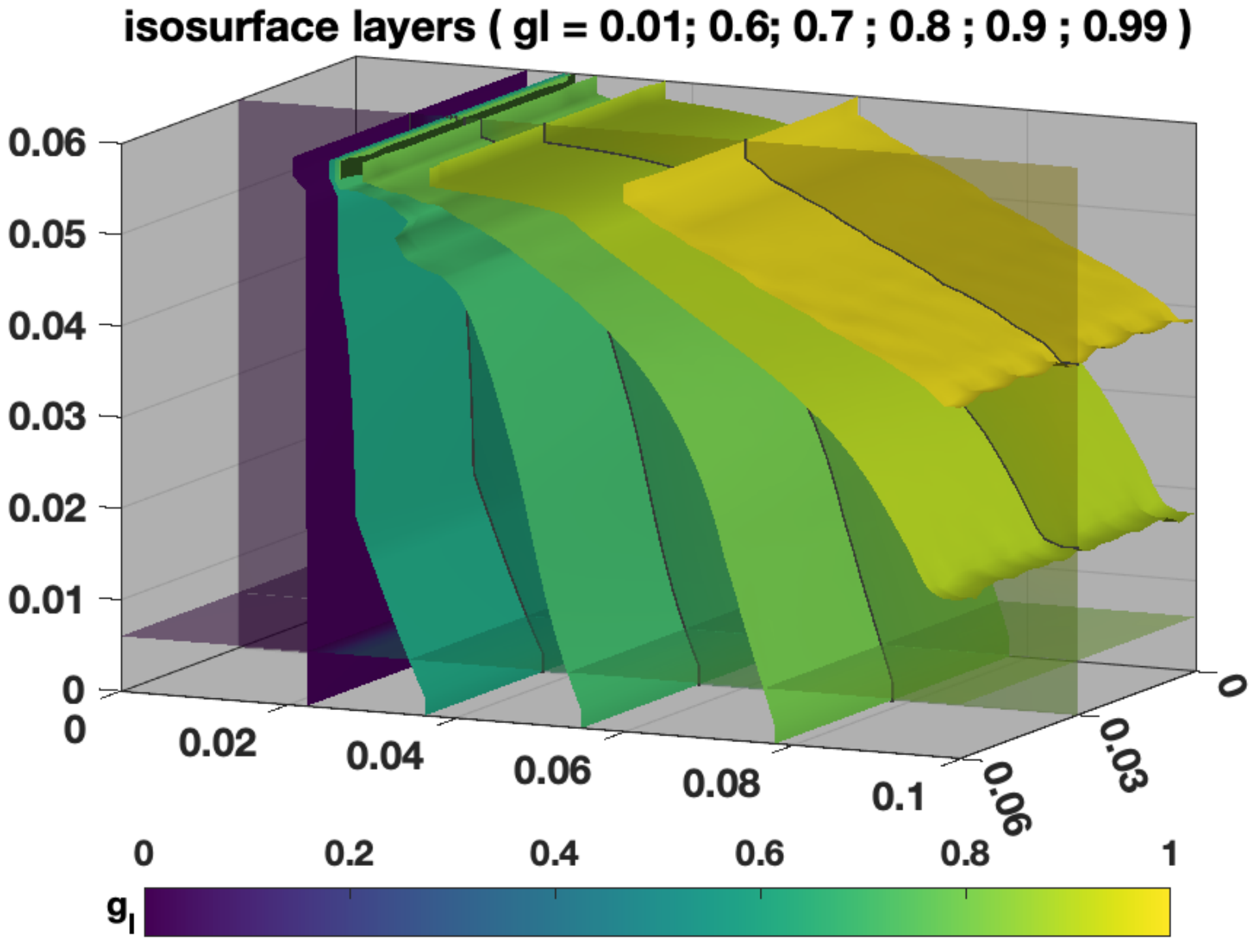}  \label{example_3_3dthick400gl} 
    \end{minipage}
    }
    \centering \subfigure[concentration / streamlines]{
    \begin{minipage}[b]{0.35\textwidth}
    \centering
    \includegraphics[width=1.0\textwidth]{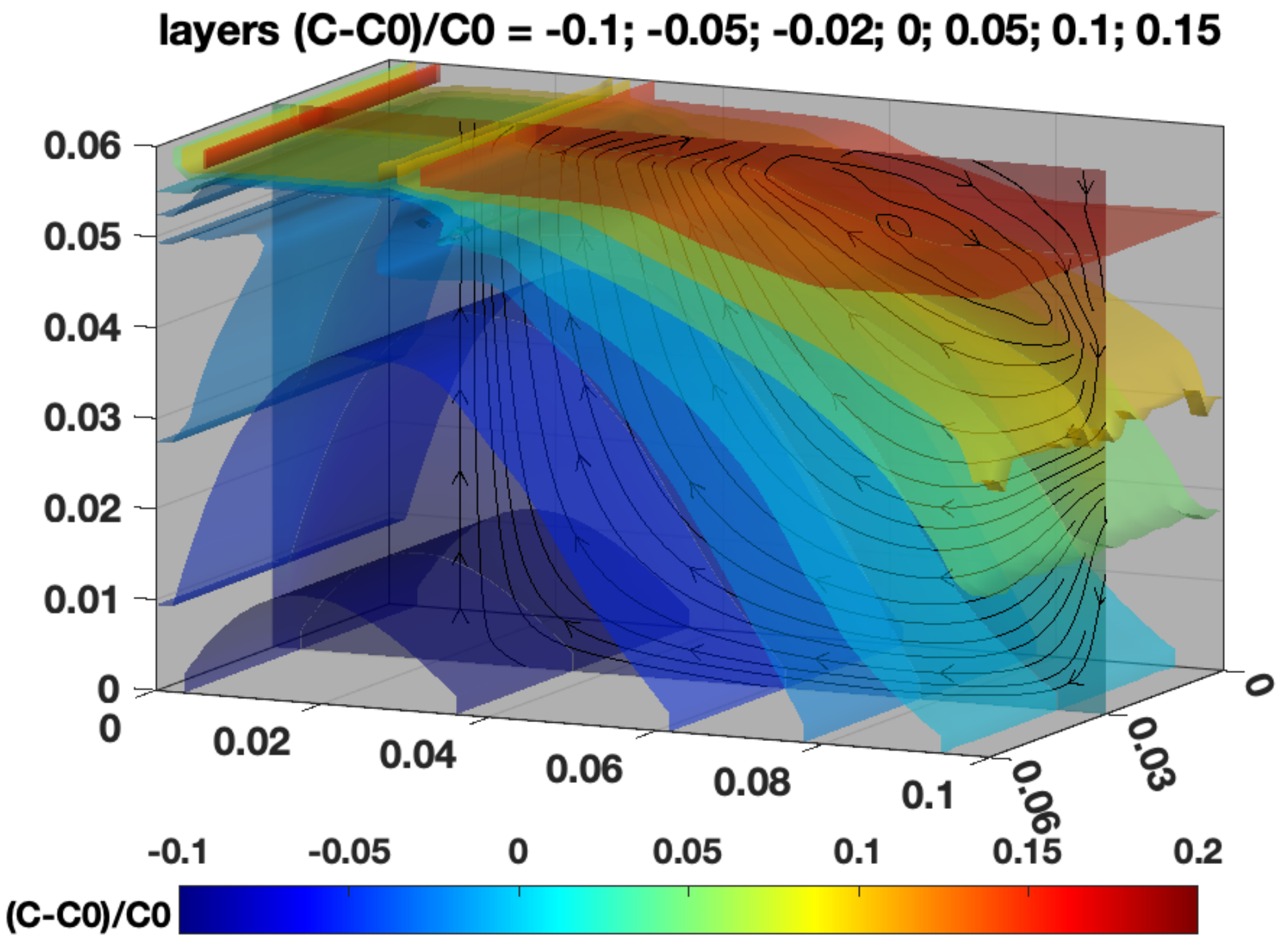}  \label{example_3_3dthick400C} 
    \end{minipage}
    }
    \caption{Physical fields of the solidification of the 3D case with 60 mm thickness in Example 3 at time t = 400s}
    \label{example_3_3dthick_400}
\end{figure*}

The physical fields of the thick 3D case ($y = 60mm$) and the thin 3D case ($y = 13mm$) at $t = 50s$ and $t = 400s$ are also selected to be present for a better comparison with 2D results. For a vivid 3D visualization, some slices are selected to show these physical quantities, such as plane $y = 0.5ly$, plane $z = 6mm$ and plane $z = 54mm$. We show (1) temperature isosurfaces, (2) concentration relative variation, (3) streamline pattern and (4) liquid fraction isosurfaces for all 3D simulation cases. 

\begin{figure*}[!h]
    \centering \subfigure[velocity and liquid fraction]{
    \begin{minipage}[b]{0.35\textwidth }
    \centering
    \includegraphics[width=1.0\textwidth]{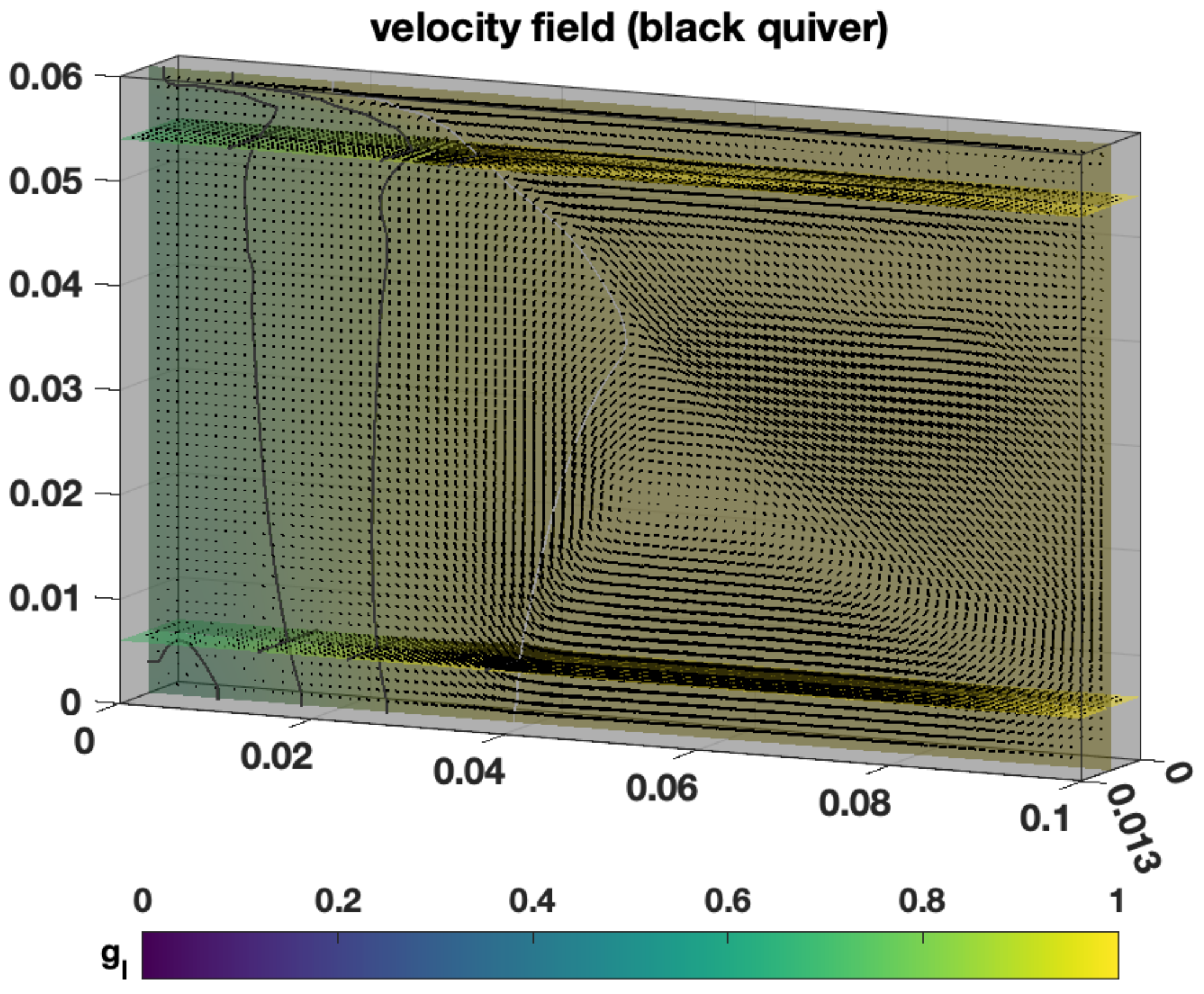}  \label{example_3_3dthin50v}
    \end{minipage}
    }
    \centering \subfigure[temperature]{
    \begin{minipage}[b]{0.35\textwidth}
    \includegraphics[width=1.0\textwidth]{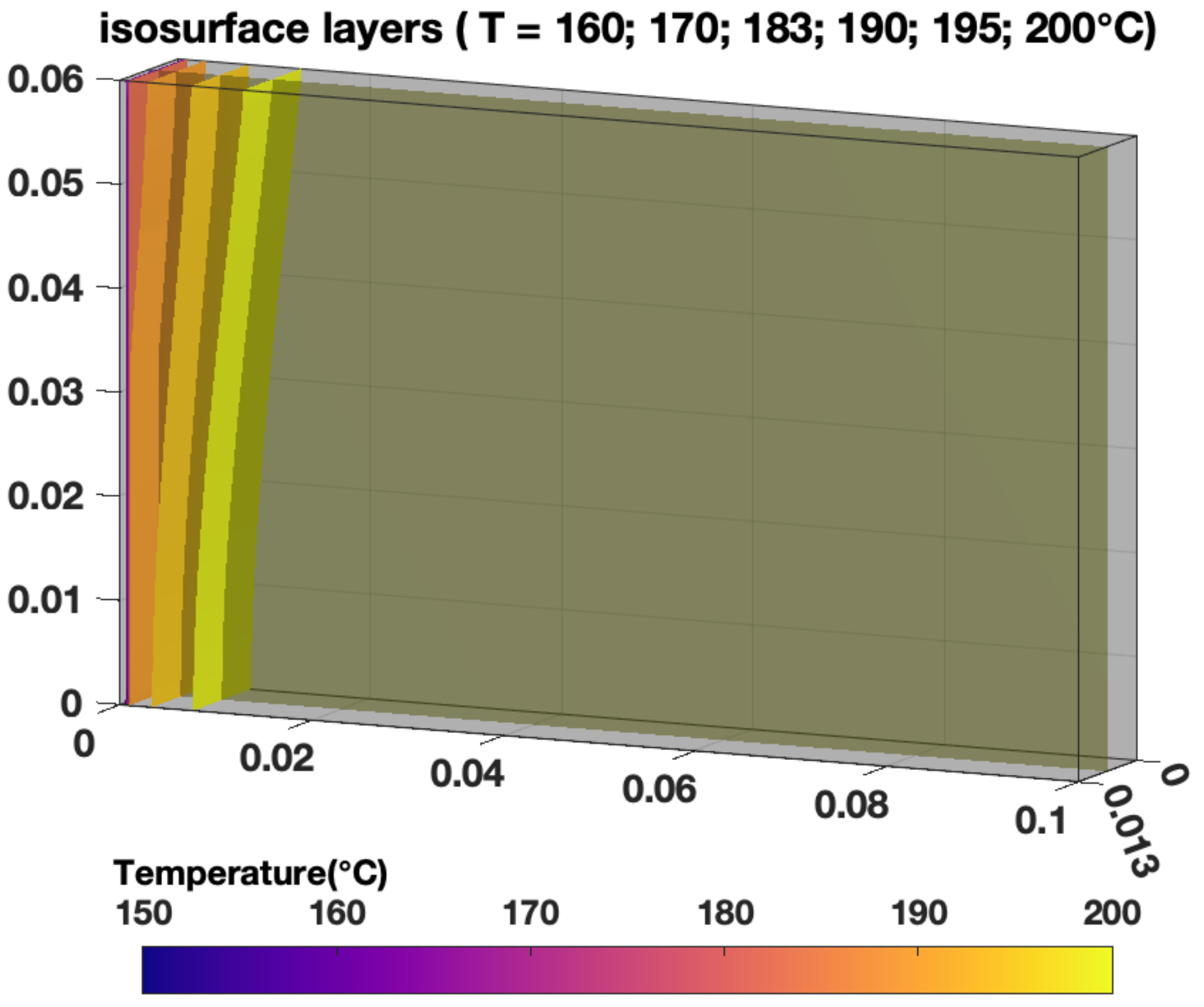} \label{example_3_3dthin50T}
    \end{minipage}
    }
    \centering \subfigure[liquid fraction isosurfaces]{
    \begin{minipage}[b]{0.35\textwidth}
    \centering
    \includegraphics[width=1.0\textwidth]{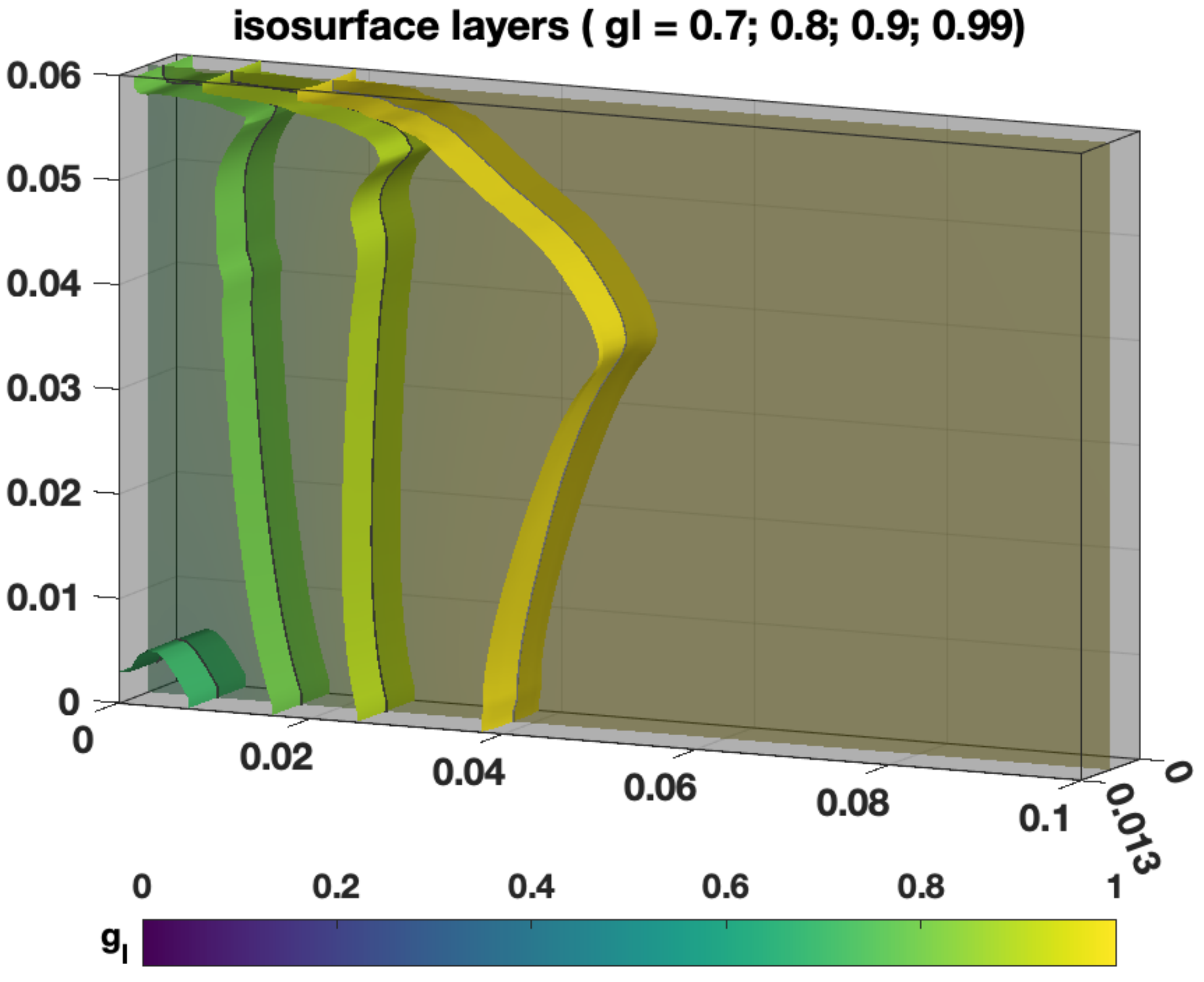} \label{example_3_3dthin50gl}
    \end{minipage}
    }
    \centering \subfigure[concentration / streamlines]{
    \begin{minipage}[b]{0.35\textwidth}
    \centering
    \includegraphics[width=1.0\textwidth]{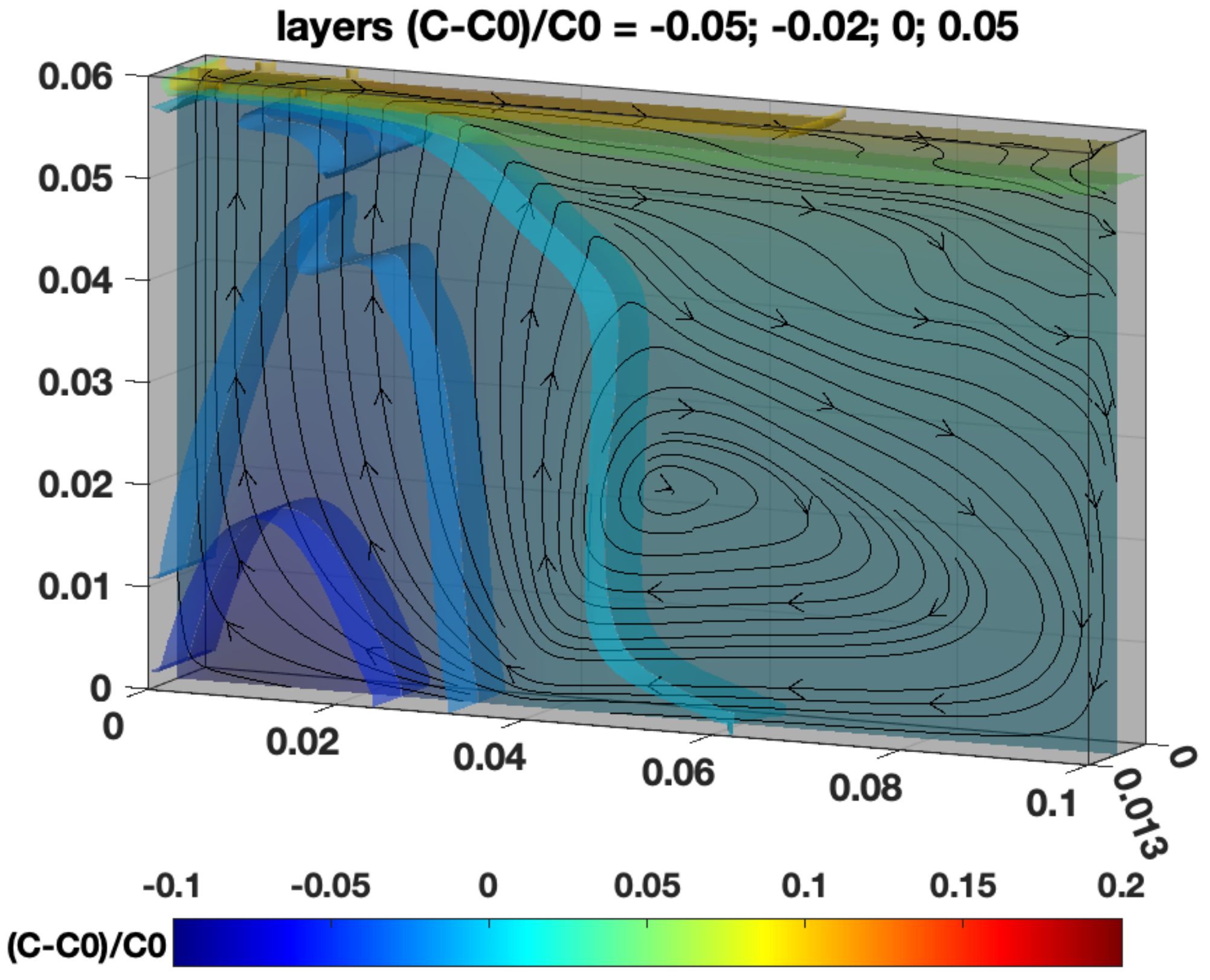} \label{example_3_3dthin50C}
    \end{minipage}
    }
    \caption{Physical fields of the solidification of the 3D case with 13 mm thickness in Example 3 at time t = 50s}
    \label{example_3_3dthin_50}
 \end{figure*}

\begin{figure*}[!t]
    \centering \subfigure[quivers and slices]{
    \begin{minipage}[b]{0.35\textwidth }
    \centering
    \includegraphics[width=1.0\textwidth]{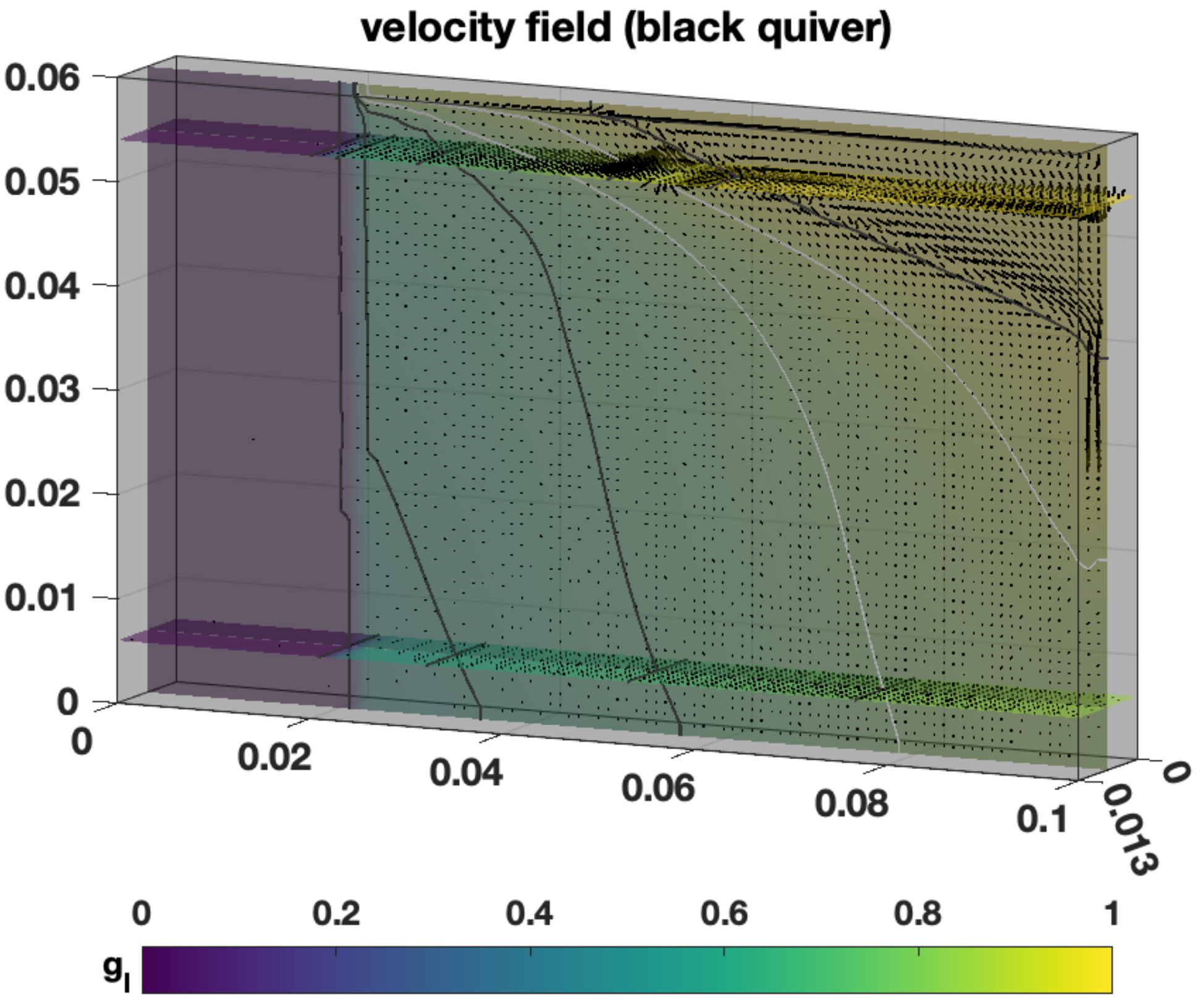}  \label{example_3_3dthin400v}
    \end{minipage}
    }
    \centering \subfigure[temperature isosurfaces ($^{\circ}$C)]{
    \begin{minipage}[b]{0.35\textwidth}
    \includegraphics[width=1.0\textwidth]{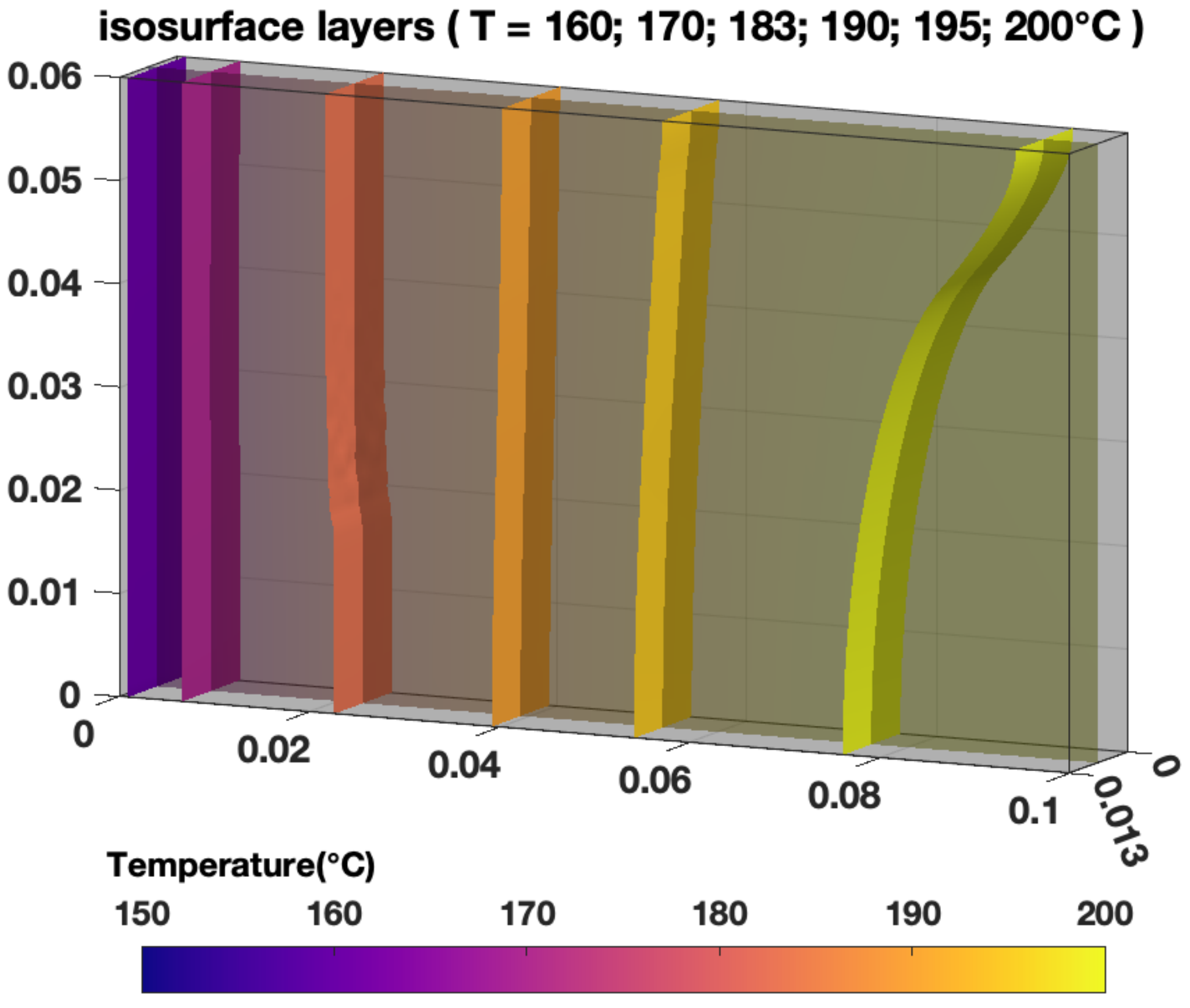}  \label{example_3_3dthin400T}
    \end{minipage}
    }
    \centering \subfigure[liquid fraction isosurfaces]{
    \begin{minipage}[b]{0.35\textwidth}
    \centering
    \includegraphics[width=1.0\textwidth]{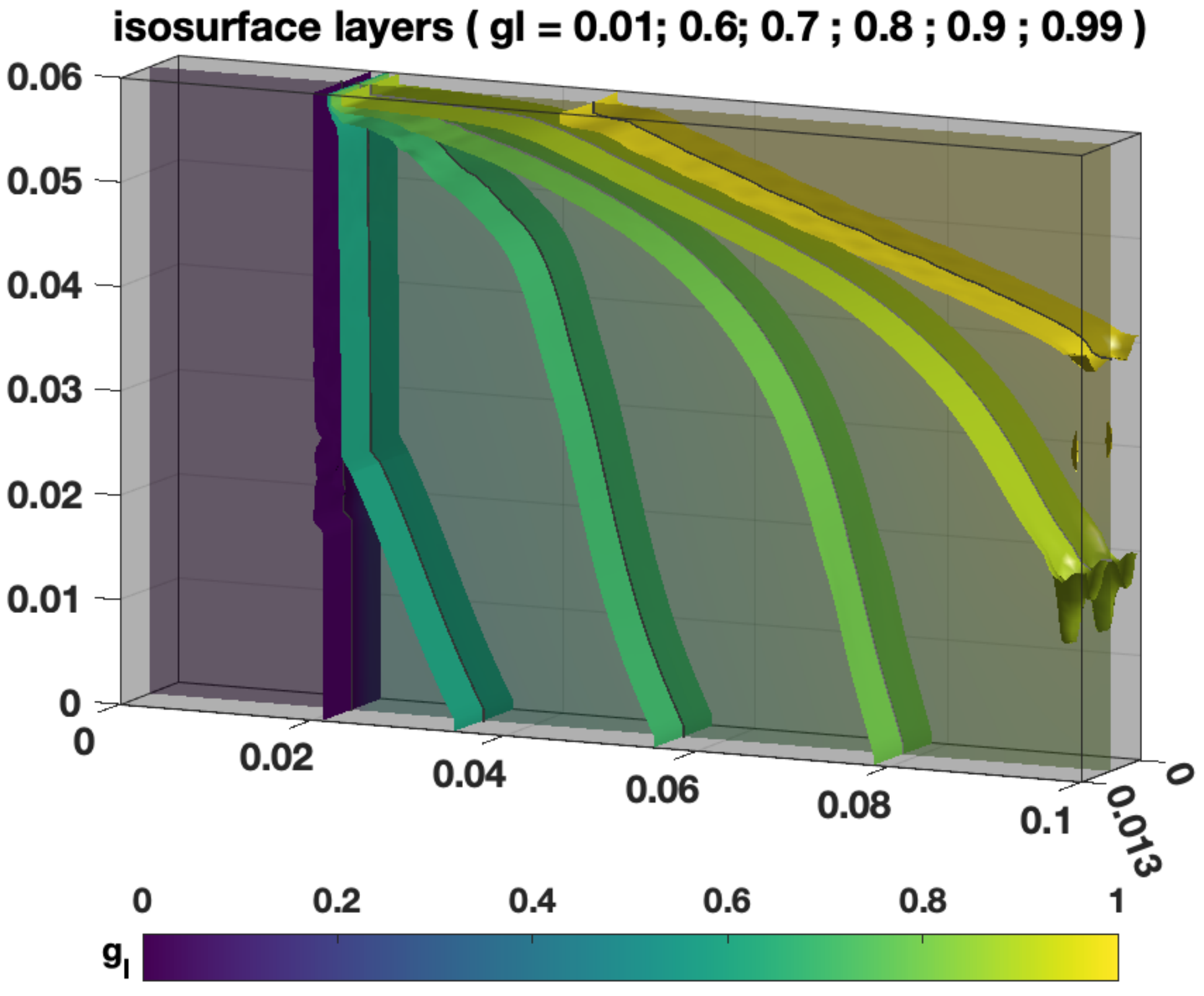}  \label{example_3_3dthin400gl}
    \end{minipage}
    }
    \centering \subfigure[concentration / streamlines]{
    \begin{minipage}[b]{0.35\textwidth}
    \centering
    \includegraphics[width=1.0\textwidth]{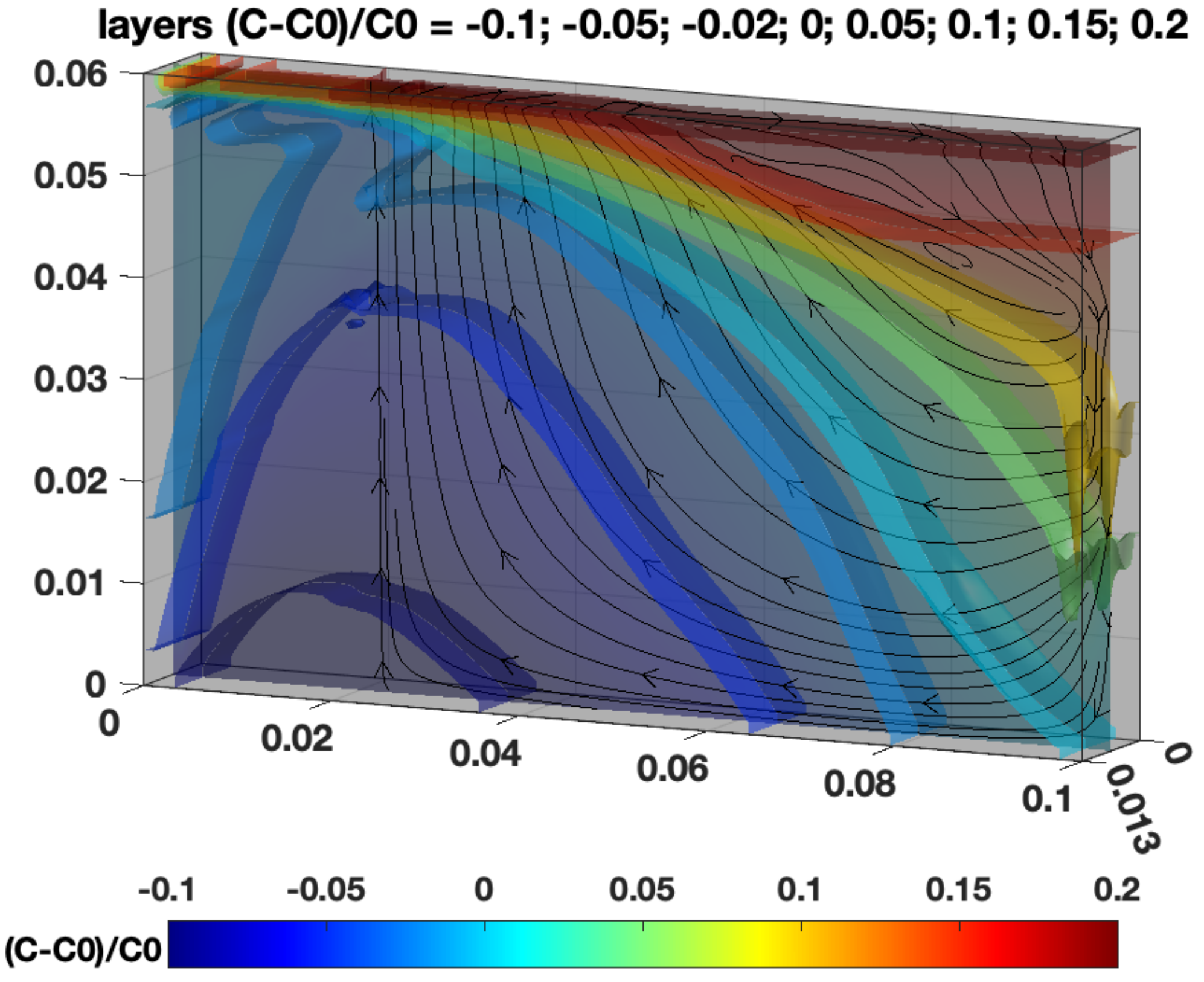}  \label{example_3_3dthin400C}
    \end{minipage}
    }
    \caption{Physical fields of the solidification of the 3D case with 13 mm thickness in Example 3 at time t = 400s}
    \label{example_3_3dthin_400}
 \end{figure*}

Next, the results of the 3D cases with different large thicknesses are compared to those of the 2D case. The physical fields of the thick 3D case simulation (Figure \ref{example_3_3dthick_50}, \ref{example_3_3dthick_400}) are similar to a large extent to those of the 2D case (Figure \ref{example_3_2dCompare}) at time $t = 50s$ and $t = 400s$. While looking at the results (Figure \ref{example_3_3dthin_50}, \ref{example_3_3dthin_400}) obtained from the thin 3D case, some clear differences can be observed. In addition, the most intense flow is also observed at the liquid/mush interface in 3D cases as well (see Figures \ref{example_3_3dthick50v},\ref{example_3_3dthick400v}, \ref{example_3_3dthin50v},\ref{example_3_3dthin400v}).

At $t = 50s$, a large clockwise vortex forms near the liquidus isosurface layer ($g_l = 0.99$) and the concentration isosurface layer with original $C0$ at $t = 50s$ in all three cases (2D, thick 3D, thin 3D). But a small vortex on the northeastern side of the large one only appears in the 2D case and the thick 3D case. The above difference can be checked by the streamline pattern in each case (see Figures \ref{example_3_3dthick50C}, \ref{example_3_3dthin50C} and \ref{example_3_2dstreamC}). At $t = 50s$, the shapes and positions of the temperature isosurfaces, the liquid fraction isosurfaces, and the concentration variation isosurfaces almost coincides with each other for these three cases. 

\begin{figure*}[!h]
\centering
\subfigure[Temperature distribution]{
\begin{minipage}[b]{0.35\textwidth}
\includegraphics[width=1\textwidth]{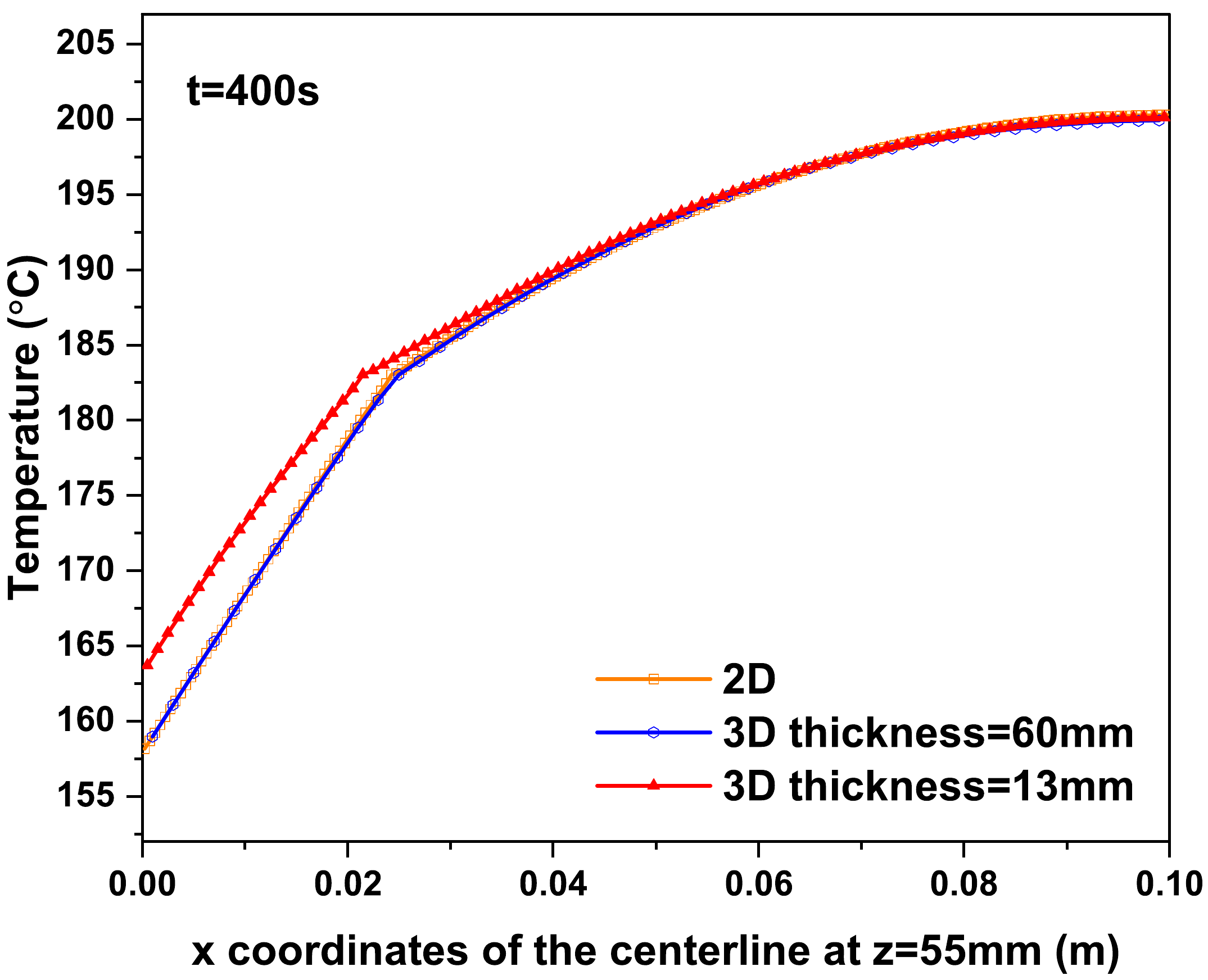}     \label{FIG:COM_T}
\end{minipage}
}
\subfigure[$(C-C0)/C0$]{
\begin{minipage}[b]{0.35\textwidth}
\includegraphics[width=1.\textwidth]{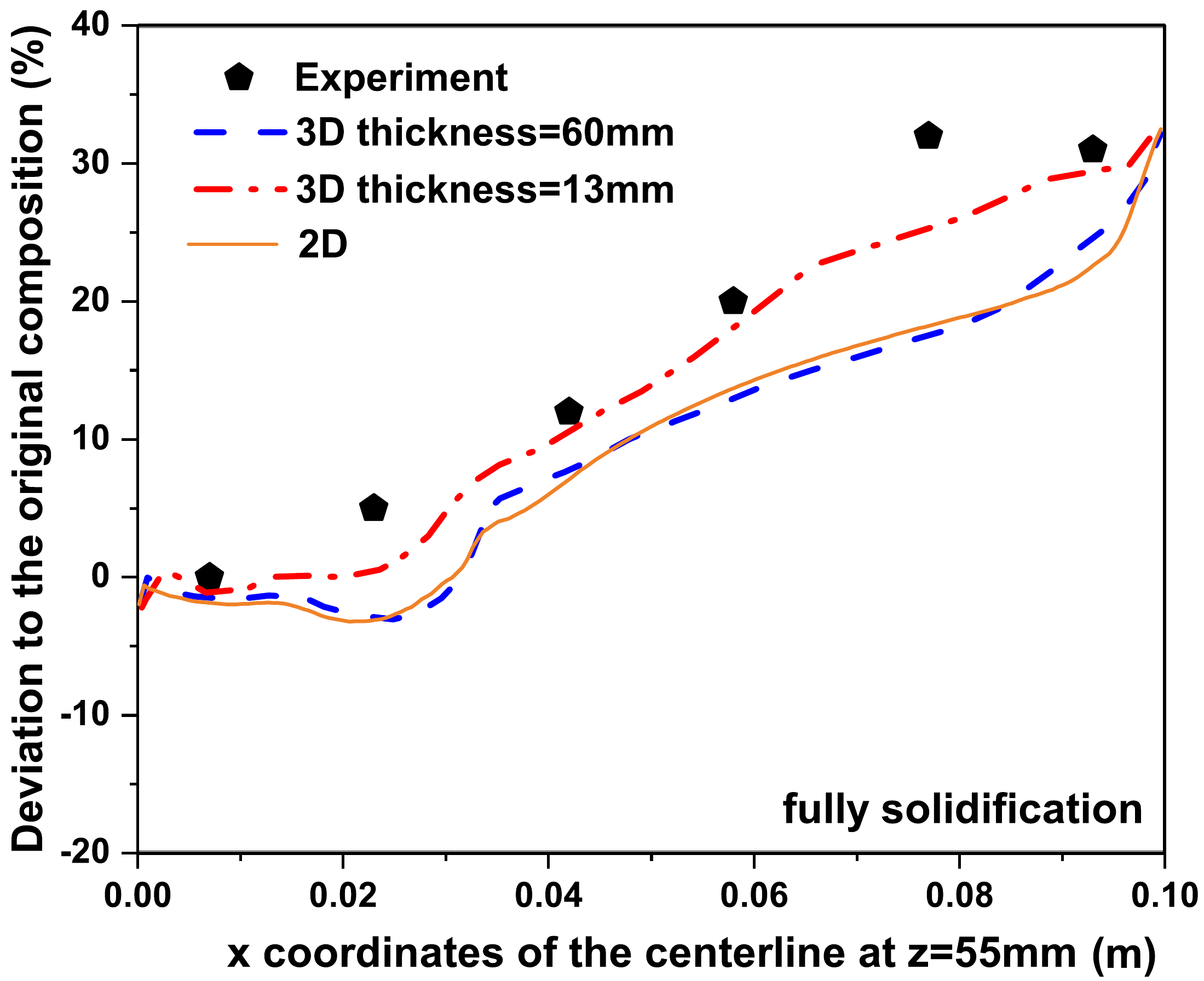}    \label{FIG:COM_C}
\end{minipage}
}
\caption{Comparison of (a) temperature and (b) $(C-C0)/C0$ (\%) on the centerline of the $z = 55mm$ plane among 2D case (orange); thick 3D case (blue); thin 3D case (red)}
\end{figure*}

At t = 400 s, the shape and position of the 200-degree temperature isosurface in the thin 3D case are different from the thick 3D case and the 2D case and the 2D results. This isosurface, slightly different from the other two outcomes, touches the top of the cavity rather than the vertical east border (see Figures \ref{example_3_3dthick400T}, \ref{example_3_3dthin400T} and \ref{example_3_2dTC}). This difference in the temperature field eventually leads to a difference in the distribution of condensation, as shown in Figures \ref{example_3_3dthick400C}, \ref{example_3_3dthin400C} and \ref{example_3_2dTC}. The $(C-C0)/C0$ isosurfaces in three cases (2D, thick 3D, and thin 3D) share a similar tendency. However, in the thin 3D case, the maximum value of $(C-C0)/C0$ is greater than 0.2 at the cavity's northeastern corner, where melts reach the fully solid state at the latest. This most enriched region sticks to the top, making the isosurface layer parallel to the top. For the rest of the two cases (2D and thick 3D), they both get lower concentrations in that region, and the maximum value only surpasses $0.15$ at $t = 400s$. This distinction eventually results in a much larger departure from the original concentration in the northeastern corner region. This implication is supported by the second half of the curve in Figure \ref{FIG:COM_C} for the thin 3D case.
 
For a more direct observation of this kind of difference, we select the simulation data on the horizontal centerline on the plane $h = 55mm$ of the 3D cases to compare with the 2D results at the same height. To compare three cases (2D, thick 3D, thin 3D), the temperature distribution on the height $h = 55mm$ at $t = 400s$ and the distribution of the deviation to the original composition $(C-C0)/C0$ on the height $h = 55mm$ after fully solidification are selected. It can be easily seen that the temperature curve describing the thick 3D case almost overlaps with the 2D case and that the curve for the thin 3D case deviates from the other two (see Figure \ref{FIG:COM_T}). This deviation occurs at the fully solid region. As for concentration distribution, there is overall good agreement between the thick 3D case and the 2D case. Also, an obvious difference can be seen for the thin 3D case in the late interval of the curve (see Figure \ref{FIG:COM_C}). This type of difference indicates the 3D effect once again, and the tendency of the "3D effect" is the same as the conclusion mentioned in the paper \cite{li2012simulation}. A sufficiently long length in one direction can overcome the ``3D effect'' and make the system work more like a 2D domain. The interesting thing is that the deviation of the concentration curve of the thin 3D case is closer to the isolated H-H data points. It should be attributed to this simulation for restoring the original experiment's 3D size information. It also verifies our inference that the inferior agreement on the $h = 55mm$ not only because of the shrinkage but also the ``3D effect'' as well.

%

\section{Conclusions} \label{section:Conclusions}

In this work, the studies of solidification and macro-segregation modeling and their wide application scope are thoroughly reviewed. For modeling the process of solidification, an operator-splitting and matrix-based scheme is proposed. This leads to an alternative, fully decoupled scheme based on the enthalpy-porosity model, which avoids some auxiliary terms. Moreover, a term-wise high-efficiency matrix assembly strategy combining the directly vectorized method with the forward equation-based method is employed to improve the feasibility and implementability of 3D extension applications.

This paper presents a large number of benchmark cases in 2D and 3D to validate good accuracy and reasonable computational efficiency. All these numerical examples show good agreement with the benchmark data from previous numerical simulations and the real experiment. The channel-segregation process, the formation of the freckle, and some interesting physical phenomena such as edge effect, aspect ratio effect, and 3D effect are investigated and analyzed through modeling. Some potential causes of the discrepancy between simulation results and real-world experimental data can be inferred based on our simulation results.

This work was built from scratch using self-written codes, and it focuses on an easy-to-implement, efficient, and accurate direct simulation method. The numerical results demonstrate its ability to precisely describe the details of a complex heat and mass transfer problem involving multi-phase, multi-component, and phase transitions. It provides a practical, decoupled framework for modeling compositional flow or hydrate formation. This work can be further promoted based on a real equation of state model like Peng-Robinson instead of a phase diagram with approximations.

\section*{Acknowledgments}
This work is partially supported by King Abdullah University of Science and Technology (KAUST) through the grants BAS/1/1351-01, URF/1/4074-01, and URF/1/3769-01. The authors also acknowledge support from the National Natural Science Foundation of China (Grant No. 12122115, 11771363, 51936001).

\bibliographystyle{elsarticle-num} 
\bibliography{reference}

\end{document}